\documentclass[11pt]{article}
\usepackage{jheppub}
\pdfoutput=1
\usepackage{epstopdf}
\usepackage{mathrsfs}
\usepackage{color}
\usepackage{hyperref}
\usepackage{amsmath}
\usepackage{amsfonts}
\usepackage{graphicx}
\usepackage{amssymb}
\usepackage{xcolor}
\usepackage{calligra}
\usepackage[T1]{fontenc}
\usepackage{mathtools}

\DeclareMathAlphabet\mathbfcal{OMS}{cmsy}{b}{n}

\newcommand{\f}{\frac}

\newcommand{\eq}{\begin{equation}}

\newcommand{\eqe}{\end{equation}}
\newcommand{\eqa}{\begin{eqnarray}}
\newcommand{\eqae}{\end{eqnarray}}

\title{The EFT-Hedron}
\author{Nima Arkani-Hamed,$^1$}
\author{Tzu-Chen Huang$^{2}$}
\author{Yu-tin Huang,$^{3,4}$}

\affiliation{$^1$ School of Natural Sciences, Institute for Advanced Study, Princeton, NJ 08540, USA}
\affiliation{$^2$ Walter Burke Institute for Theoretical Physics
, California Institute of Technology, Pasadena, CA 91125, USA
}
\affiliation{$^3$ Department of Physics and Astronomy, National Taiwan University, Taipei 10617, Taiwan}
\affiliation{$^4$ Physics Division, National Center for Theoretical Sciences, National Tsing-Hua University, No.101, Section 2, Kuang-Fu Road, Hsinchu, Taiwan}
\abstract{We re-examine the constraints imposed by causality and unitarity on the low-energy effective field theory expansion of four-particle scattering amplitudes, exposing  a hidden ``totally positive" structure strikingly similar to the positive geometries associated with grassmannians and amplituhedra. This forces the infinite tower of higher-dimension operators to lie inside a new geometry we call the ``EFT-hedron". We initiate a systematic investigation of the boundary structure of the EFT-hedron, giving infinitely many linear and non-linear inequalities that must be satisfied by the EFT expansion in any theory.  We illustrate the EFT-hedron geometry and constraints in a wide variety of examples, including new consistency conditions on the scattering amplitudes of photons and gravitons in the real world. 
}

\begin{document}
\begin{flushright}
\vspace{10pt} \hfill{NCTS-TH/2014, CALT-TH 2020-061} \vspace{20mm}
\end{flushright}

\maketitle

%%%%%%%%%%%%%%%%%%%%%%%%%%%%%%%%%%%
\section{Introduction}
%%%%%%%%%%%%%%%%%%%%%%%%%%%%%%%%%%%
%%%%%%%%%%%%%%%%%%%%%%%%%%%%%%%%%%%
There is a long-appreciated, close connection between vacuum stability/ causality/unitarity, and analyticity/positivity properties of scattering amplitudes, going back to the 1960's S-matrix program. In this standard story, there 
are three fundamental origins of positivity: the positivity of energies (vacuum stability), the sharp localization of signals inside the lightcone (causality) and the positivity of probabilities (unitarity).  These basic positivities, together with analyticity properties of scattering amplitudes meant to reflect causality, allow the derivation of more non-trivial positivity constraints on coefficients of higher-dimension operators in low-energy effective field theories (as in \cite{Aharonov:1969vu, Pham:1985cr, Adams:2006sv}). In recent years, a sort of opposite of the S-matrix program has emerged in a number of theories, where notions of positivity take a central role, determining certain ``positive geometries" in the the kinematic space of particle scattering with a fundamentally combinatorial definition, from which the amplitudes are naturally extracted. In this picture, locality and unitarity are not taken as fundamental principles, but instead arise, joined at the hip, from the the study of the boundary structure of the positive geometries. These examples suggest that there is vastly more ``hidden positivity" in scattering amplitudes than meets the eye, with locality and unitarity as {\it derived from}, rather than {\it the origin of}, positivity properties. 

Motivated by these discoveries, in this paper we will revisit the positivity properties of $2 \to 2$ scattering amplitudes, and re-examine the {\it usual} positivity properties dictated by analyticity, causality and unitarity. We will find that there are {\it infinitely many} constraints 
on the coefficients of higher-dimension operators, and that these constraints involve very similar mathematical structures as have already been seen in the story of positivity geometries and amplituhedra. 

To illustrate the nature of the constraints, consider for simplicity the scattering amplitudes for two massless  scalars $a b \to a b$, and suppose we are working in an approximation where we have integrated out massive states but not yet accounted for massless loops in the low-energy theory. Then, the low-energy amplitude has a power-series expansion in the Mandelstam variables $s,t$: 
\begin{equation}
{\cal A}(s,t) = \sum_{\Delta,q} a_{\Delta,q} s^{\Delta-q} t^q
\end{equation}
and all the information in the low-energy effective field theory is captured in the coefficients $a_{\Delta,q}$ which we can organize into a table: 
\begin{equation}
\begin{array}{cccccc}\; & q{=}0 & 1 & 2 & 3& \cdots  \\ \Delta{=}1 & a_{1,0} & a_{1,1} & \; & \;&  \;  \\ \Delta{=}2 & a_{2,0} & a_{2,1} & a_{2,2} & \; &\; \\ \Delta{=}3 & a_{3,0} & a_{3,1} & a_{3,2} & a_{3,3}&\; \\ \vdots & \vdots & \vdots & \vdots & \vdots& \vdots\end{array}\,,
\end{equation}
There are infinitely many constraints on the $a_{\Delta,q}$, forcing this infinite table of coefficients to lie inside ``the EFT-hedron". 

These constraints quantify certain intuitions about ``garden variety" higher dimension operators contributing to $ab \to ab$ scattering, into sharp bounds.  For instance we shouldn't expect  operators of the same mass dimension $\Delta$ to have vastly different coefficients; these correspond to the coefficients in the same row in our table. But we might also think that this is a consequence of ``naturalness", and that by fine-adjustments of the parameters in the high-energy theory, we can engineer any possible relative sizes between these operators we like. The EFT-hedron shows that this is not the case: not everything goes, and indeed the coefficients $a_{\Delta,q}$ for a fixed $\Delta$ must satisfy linear inequalities, that force them to lie inside a certain polytope. We would also expect all operators to be suppressed by a similar scale, i.e. not to have dimension 6 operators suppressed by the TeV scale while dimension 8 operators are suppressed by the Planck scale, though again one might think this can be done with suitable fine-tuning. Again, the EFT-hedron shows this is impossible, and imposes non-linear inequalities between different $a_{\Delta,q}$, which in the simplest case constrain the relative sizes of coefficients at fixed $q$, in a fixed column of the table. 
We will initiate a systematic study of the EFT-hedron in this paper. But before diving in, let us give a high-level overview of the physical and mathematical engines at work. 

The physical starting point is a dispersive representation of $2 \to 2$ scattering amplitudes, as a function of $s$ working at fixed $t$.  To begin with we will assume, as mentioned above, that we integrate out massive states of some typical mass $M$, which generates higher-dimension operators in the low-energy theory, and for the purpose of these introductory comments let us ignore the further running of these higher dimension operators by massless loops in the low-energy theory (we will revisit this point in the body of the paper). Working at fixed $t$ with $|t| \ll M^2$, it can be argued that the amplitudes only have singularities on the real $s$ axis, with discontinuities reflecting particle production in the $s$ and $u$ channels. The discontinuity across these cuts has a partial wave expansion, as a sum over spins with positive coefficients. 
Furthermore, causality is reflected in a bound on the amplitude at large $s$ for fixed $|t|$. In a theory with a mass gap, we have the Froissart bound telling us the amplitude is bounded by ${\cal A} < s$ log$^2 s$. In quantum gravity, we expect that for any UV completion with a weak coupling (like in string theory), the high-energy amplitude in the physical region, with fixed negative $t$, is bounded by ${\cal A} < s^p$ with $p<2$. Thus at fixed $t$, for any theory, we have a dispersive 
representation for the amplitude at fixed $t$, of the form 
\begin{equation}
{\cal A}(s,t) = {\cal A}_0(t) {+}  {\cal A}_1(t) s {+} \int d M^2 \sum_l  p_l(M^2) G_l(1{+} \frac{2t}{M^2}) \left(\frac{1}{s{-} M^2} + \frac{1}{u{-} M^2} \right)
\end{equation}
where $G_l(x)$ are Gegenbauer polynomials.

Now, this dispersive representation has the two basic and crucial long-appreciated positivities we have alluded to: the positivity of energies is reflected in $M^2>0$, and the positivity of probabilities in $p_l(M^2) > 0$. The new surprise we will explore in this paper, are further hidden positive structures associated with the propagator
$1/(s{-}M^2)$, and with the Gegenbauer polynomials $G_l(x)$. It is these new positivities that are responsible for the non-trivial geometry of the EFT-hedron and the associated infinite number of new constraints on the $a_{\Delta,q}$. Here we content ourselves here with summarizing the basic mathematical facts of these hidden positivities, whose consequences we will explore in detail in body of the paper. 

Let's begin with the positivity associated with propagators, which can be illustrated in a simplified setting, where we imagine a dispersive representation for a function $F(s)$ of the form 
\begin{equation} 
F(s) = \int dM^2 \frac{p(M^2)}{M^2 - s} 
\end{equation}
This has a power-series expansion at small $s$, $F(s) = \sum_n f_n s^n$, where
\begin{equation}
f_n = \int dM^2 \frac{p(M^2)}{M^2} (\frac{1}{M^2})^n
\end{equation}
This can be interpreted geometrically as saying that the vector ${\bf f} = (f_0,f_1,f_2, \cdots)$ lies in the convex hull of the continuous moment curve $(1,x,x^2,\cdots)$, 
where here $x=1/M^2$, so we also impose that $x >0$. Thus we have a well-posed mathematical question: what is the region in ${\bf f}$ space that is carved out by the convex hull of the half-moment curve with $x>0$? 
This question has a beautifully simple answer. To begin with, we associate a ``Hankel matrix" ${\bf F}$ with the vector ${\bf f}$ via ${\bf F}_{ij} = {\bf f}_{i+j}$: 
\begin{equation}
{\bf F} = \left(\begin{array}{cccc} f_0 & f_1 & f_2 & \cdots \\ f_1 & f_2 & f_3 & \cdots \\ f_2 & f_3 & f_4 & \cdots \\ f_3 & f_4 & f_5 & \cdots \end{array} \right)
\end{equation}
Then the allowed region in ${\bf f}$ space is completely specified by demanding that {\it all} of the square $k \times k$ minors of the Hankel matrix ${\bf F}$ are positive! This is abbreviated by saying the ${\bf F}$ is a ``totally positive" matrix.  For $k=1$, this just tells us that all the $f_n$ are positive, which is essentially the amplitude positivity found in the early works of \cite{Adams:2006sv}. But there are also infinitely many non-linear positivity conditions. It is striking to see ``all minors of a matrix positive" conditions--earlier seen in the context of the positive grassmannian~\cite{Arkani-Hamed:2016byb} and the amplituhedron~\cite{Amplituhedron} for ${\cal N}=4$ SYM, show up again in a different setting, and in such a basic way, for completely general theories.

Note that all these conditions are homogeneous in the mass dimension of the operators, as they should be, since we have not input any further knowledge of the UV mass scales. But suppose we were also given the gap $M_{gap}$ to the first massive states. In this case, the vector ${\bf f}$ would lie in the convex hull of the moment curve, starting at $x=0$ and cut-off at $x=x_{gap} = 1/M_{gap}^2$. Working in units where $M_{gap} = 1$, the region in ${\bf f}$ space is carved out by looking not only at $f$, but also of its discrete derivatives,
\begin{equation}
\left(\begin{array}{c} f_0 \\ f_1\\ f_2 \\ \vdots \end{array} \right), \, \left(\begin{array}{c} f_1 - f_2 \\ f_2 - f_3 \\ f_3 - f_4 \\ \vdots \end{array} \right), \, \left(\begin{array}{c} (f_2 - f_3)-(f_3 - f_4) \\ (f_3 - f_4)-(f_4 - f_5) \\ (f_4 - f_5) - (f_5 - f_6) \\ \vdots \end{array} \right) \, , \, \cdots
\end{equation}
and demanding that the Hankel matrices associated with all of these vectors are totally positive. A simple illustration of the region in $(f_1/f_0,f_2/f_0)$ space carved out with (patterned region) and without knowledge of the gap is shown in the following plot: 
$$\includegraphics[scale=0.4]{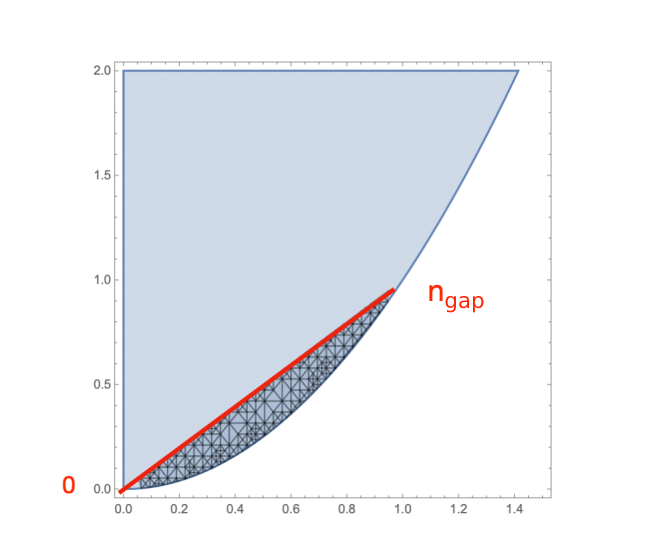}\,.$$

Now to illustrate Gegenbauer positivity, let us again focus on simplest example illustrating the non-trivial point. Consider a dispersive representation for some function $F(s,t)$ only containing $s$-channel (and no $u$-channel) poles: 
\begin{equation}
F(s,t) = \int dM^2 \sum_l \frac{p_l(M^2) G_l(1{+} 2 t/M^2)}{M^2{-}s}
\end{equation}
and consider the low-energy expansion in powers of $s,(2t)$, as $F(s,t) = \sum_{\Delta,q} f_{\Delta,q} s^{\Delta{-}q} (2t)^q$, yielding
\begin{equation}
\left(\begin{array}{c} f_{\Delta,0} \\ f_{\Delta,1} \\ \vdots \\ f_{\Delta,\Delta} \end{array} \right) = \sum_l P_l \left(\begin{array}{c} G_l^{(0)}(x{=}1) \\ G_l^{(1)}(x{=}1) \\ \vdots \\ G_l^{(D)}(x{=}1) \end{array} \right)\, {\rm where} \; P_l = \int dM^2 \frac{p_l(M^2)}{(M^2)^{\Delta+1}}>0\,.
\end{equation}
Here $G_l^{(q)}(x=1)$ are the q'th derivatives of the Gegenbauer polynomials, evaluated at the "forward limit" where $x=1$. The above expression tells us that the projective vector ${\bf f}_\Delta = (f_{\Delta,0},\cdots, f_{\Delta,\Delta})$ lies in the convex hull of all the ``Gegenbauer derivative" vectors.  Finding the space of all consistent ${\bf f}_\Delta$ is then a standard polytope problem: we are given a collection of vectors (an infinite number in this case) whose convex hull specifies some polytope, and we'd like to determine how to characterize the polytope instead by the inequalities that cut out its facets. As we will review in the body of the paper, the facet structure of a $\Delta$-dimensional polytope, in turn, is fully captured by the knowledge of the signs of the all the determinants made from any $(\Delta{+}1)$ vectors of the vertices. In our context, then, we should look at the infinite ``Gegenbauer matrix" $G_{l,q} = G_l^{(q)}(x=1)$, and consider the top $\Delta{+}1$ rows of this matrix and look at all the corresponding $(\Delta{+}1) \times (\Delta{+}1)$ minors. Remarkably, it turn out that all these minors of the Gegenbauer matrix are positive! This is another appearance of the "matrix with all positive minors" phenomenon, and it immediately allows us to fully determine the inequalities cutting out the corresponding polytope in ${\bf f}$ space, which are the famous ``cyclic polytopes". Cyclic polytopes have already made a prominent appearance in the story of ${\cal N}=4$ SYM amplitudes, as the simplest example of ``amplituhedra" for the case of next-to-MHV tree scattering amplitudes. Indeed tree amplituhedra can be thought of as grassmannian generalizations of the notion of cyclic polytopes. It is again interesting to see the same objects show up in the totally different, very general setting of the EFT-hedron. A morally similar geometry was seen in the conformal bootstrap~\cite{CFTHedra}.

We close our introductory remarks with two comments. First, we stress that these constraints on effective field theory are non-trivial statements about any theory, and in particular non-trivial constraints on quantum gravity in the real wold. Of course we don't usually care about relative sizes of very high dimension, ``garden variety" operators, for phenomenological purposes, but we nonetheless find it fascinating that the structure of low-energy dynamics is vastly more constrained than previously appreciated. As a sampling of our results, let's look at some of the constraints for photon and graviton scattering. For the $(-,-,+,+)$ helicity configuration, where it's identical helicity in the $s$-channel, the amplitude for the $D^8F^4$ and $D^8R^4$ operator takes the form:  
\eq
\langle12\rangle^{2h}[34]^{2h}(a_{4,0}s^4{+}a_{4,1}s^3t{+}a_{4,2}s^2t^2\cdots )\,.
\eqe
where $h=1,2$ for photon and graviton respectively. The allowed region for $\frac{a_{4,1}}{a_{4,0}},\,\frac{a_{4,2}}{a_{4,0}}$ is given as:
$${\rm photon} \;\;\vcenter{\hbox{\includegraphics[scale=0.3]{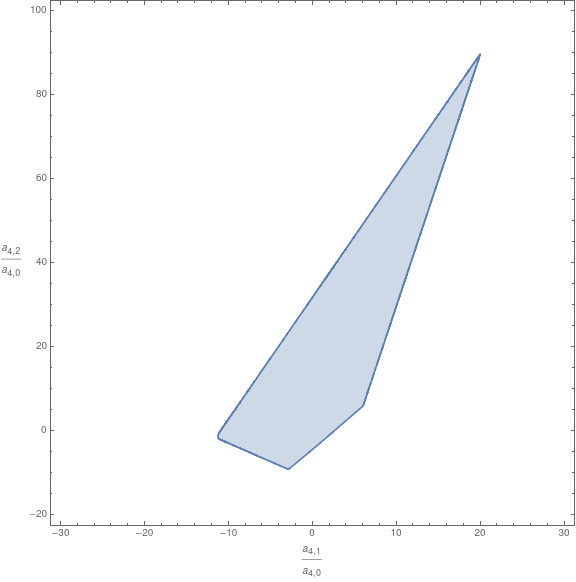}}} {\rm graviton} \;\;\vcenter{\hbox{\includegraphics[scale=0.3]{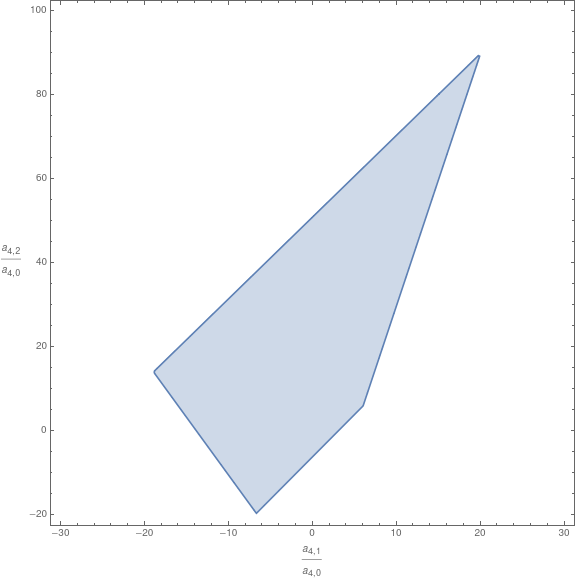}}}$$
Note that the allowed region are bounded.

It is also important to note that, while the EFT-hedron places extremely constraints on the effective field theory expansion, sensible effective field theories do not appear to populate the entire region allowed by the EFT-hedron, but cluster close to its boundaries. The reason is likely that the physical constraints we have imposed, while clearly necessary,  are still not enough to capture consistency with fully healthy UV theories. In particular, our dispersive representation at fixed $t$, does not make it easy to impose the softness of high-energy, fixed-angle amplitudes where both $s,t$ are large with $t/s$ fixed. It would be fascinating to find a way to incorporate this extra information about UV softness into the constraints.  

Having given this high-level overview of the physical and mathematical basis for our results, we proceed to a more systematic discussion. Through sections \ref{Sec0}, \ref{Sec:Dis}, \ref{Appt0}, \ref{sec:TheorySpace} we will present an elementary introduction of EFT amplitudes with explicit examples, the analytic definition of $a_{D,q}$ through dispersion relations and their potential obstructions, and finally the theory space that emerges from the dispersive representation. Next in sec.\ref{OldFashion}, we take a brief sojourn in the positive geometries relevant to our analysis, giving a pedagogical discussion of convex hulls of moment curves and cyclic polytopes. These geometries will be immediately utilized to define the $s$-channel EFT-hedron in sec.\ref{Sec:s-channel}, where we focus on the theory space for scalar EFTs that allow for preferred ordering and hence the absence of $u$-channel thresholds. This will be generalized to include $u$-channel thresholds in sec.\ref{sec:EFT-hedron}, as well as photon and gravitons in sec. \ref{Sec:SpinEFT-hedron}. We will study explicit examples of EFTs and their ``positions" in the EFT-hedron in sec. \ref{sec:Answs}. Finally IR logarithms generated by the massless loops will be incorporated in sec.\ref{RG}.

$$\quad$$

\noindent Many of the results of this paper have been presented in conferences and schools over the past few years \cite{Talks}. As we were preparing our manuscript, a number of independent works appeared on the arxiv overlapping with some of this work. In particular, new positivity constraints involving scale dependent ``arc moments" were introduced in~\cite{Bellazzini:2020cot}, are intimately related to the geometry of the gap discussed in subsection \ref{sec:gap}. These constraints arises from the knowledge of the precise UV cut off, and hence the reach of validity for the EFT description. Bounds involving the combination of positivity away from the forward limit and full permutation invariance was discussed in
~\cite{Tolley:2020gtv} and ~\cite{Caron-Huot:2020cmc}, which have some overlap with the $s$-$u$ polytope discussion in subsection~\ref{sec:supoly}. Other related works can be found in~\cite{RecentWorks}.
\newpage

%%%%%%%%%%%%%%%%%%%%%%%%%%%%%%%%%%%
\section{EFT from the UV}\label{Sec0}
%%%%%%%%%%%%%%%%%%%%%%%%%%%%%%%%%%%
Let's begin by considering a few concrete examples of EFTs emerging from their UV parent amplitudes. We will give a broad stroke description of what types of high energy theories/amplitudes they can arise from, the features that we will be focusing on and their relations to local operators, leaving the detailed analysis to the remainder of the paper. 

%%%%%%%%%%%%%%%%%%%%%%%%%%%%%%%%%%%
\subsection{Explicit EFT amplitudes}\label{Sec1}
%%%%%%%%%%%%%%%%%%%%%%%%%%%%%%%%%%%
\begin{figure}
\begin{center}
\includegraphics[scale=0.6]{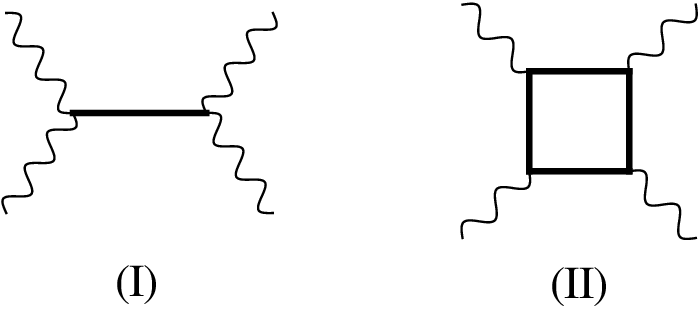}
\caption{Different origins for the EFT: (I) Integrating away massive states in tree exchanges, for example the Higgs for the Sigma model and the infinite tower of higher spin states in string amplitudes, (II) or massive states in the loop, for example the $\varphi X^2$ coupling. }
\label{Dia}
\end{center}
\end{figure}

The amplitude for the low energy degrees of freedom may originate from a UV amplitude where they interact through a tree-level exchanges of massive particles. A simple example is the case of the linear sigma model in the broken phase:
\eq
\mathcal{L}=\frac{1}{2}(\partial_\mu h)^2{-}\frac{m_h^2}{2}h^2+\left(1{+}\frac{h}{v}\right)^2\frac{1}{2}(\partial \pi\cdot\partial \pi)+V(h)
\eqe 
where $v=m_h\sqrt{\frac{2}{\lambda}}$, $\lambda$ is the quartic coupling for the potential in the unbroken phase. As the massless Goldstone boson $\pi$ couples to the massive Higgs via cubic coupling $\pi^2 h$, the following four $\pi$ amplitude in the UV is given by (see fig\ref{Dia}):
\eq\label{LSM4pt}
M(s,t)=-\frac{\lambda}{8m_h^2}\left(\frac{s^2}{s-m_h^2}{+}\frac{t^2}{t-m_h^2}{+}\frac{u^2}{u-m_h^2}\right) \,,
\eqe
where $s=(p_1{+}p_2)^2$, $t=(p_1{+}p_4)^2$ and $u=(p_1{+}p_3)^2$, and as the pions are massless $s{+}t{+}u=0$. In the center of mass frame, we have $s=E_{CM}^2$ the center of mass energy and $t=-\frac{s}{2}(1-\cos\theta)$, where $\theta$ is the scattering angle. At low energies, all Mandelstam variables are small compared to the UV scale $m_h$, and thus the low energy EFT amplitude is obtained by expanding in $\frac{p^2}{m^2_h} \ll 1$,
\eq\label{LSM4ptIR}
M^{\rm IR}(s,t)=\frac{\lambda}{8m_h^2}\left(\frac{s^2{+}t^2{+}u^2}{m_h^2}{+}\frac{s^3{+}t^3{+}u^3}{m_h^4}{+}\cdots\right)=\frac{\lambda}{8}\sum_{n=2}^\infty \frac{\sigma_{n}}{m_h^{2n}}\,,
\eqe 
where $\sigma_n=s^n{+}t^n{+}u^n$. We see that the IR description is given by an infinite series of polynomial terms, reflecting the presence of an infinite number of higher dimensional operators  from integrating out the massive Higgs.

Note that the residues of the poles for the UV amplitude eq.(\ref{LSM4pt}), say in the $s$-channel, are constants. This reflects the fact exchanged particle is spinless. In general a spin-$J$ exchange in the $s$-channel will lead to a residue that is polynomial in $t$ up to degree $J$. For example, consider the four-gluon amplitude of type-I open string theory, given by
\eq\label{TypeIString}
M(1^-2^-3^+4^+)=-g_s\alpha'^2\langle 12\rangle^2[34]^2\frac{\Gamma[{-}\alpha' s]\Gamma[{-}\alpha' t]}{\Gamma[1{-}\alpha' s{-}\alpha' t]}\,,
\eqe   
where we have put the gauge bosons in a four-dimensional subspace and thus the helicity dependence is carried by the spinor brackets. The definition of these brackets as well as their relation to the local operators will be introduced shortly.  Here $g_s$ is the string coupling and in this paper we will set the string scale $\alpha'=1$. The gamma functions in the numerator have poles at $s,t\in \mathbb{N}^+$, reflecting an infinite number of massive states. The residue at $s=n$ is given by  
\eq
g_s\langle 12\rangle^2[34]^2\frac{(-)^n}{n!}\prod_{i=1}^{n{-}1}(t+i)\,,
\eqe
where the non-trivial dependence in $t$ reflects the spinning nature of the exchanged particle. Since $\alpha'=1$ low energy is simply $p^2\ll1$, and the low energy amplitude is given as:
\eq\label{TypeIStringIR}
M^{\rm IR}(1^+2^+3^-4^-)=g_s\langle 12\rangle^2[34]^2\left(-\frac{1}{st}+\zeta_2+\zeta_3(s+t)+\cdots\right)\,,
\eqe
where the leading term contains massless poles corresponding to the field theory Yang-Mills piece. The coefficients for the polynomials are now zeta values $\zeta_n\equiv \sum_{\ell=1}^\infty \frac{1}{\ell^n}$, reflecting the fact that each term in the polynomial expansion receives contribution from the infinite number of UV states at integer values of $m^2$. The same feature can be found for the four-graviton amplitude of type-II closed string theory: 
\eq
M(1^{-2}2^{-2}3^{+2}4^{+2})=g^2_s\langle 12\rangle^4[34]^4\frac{\Gamma[{-}s]\Gamma[{-}t]\Gamma[{-}u]}{\Gamma[1{+}s]\Gamma[1{+}u]\Gamma[1{+}t]}\,,
\eqe
where the low energy expansion gives:
\eq\label{StringAmp}
M^{\rm IR}(1^{-2}2^{-2}3^{+2}4^{+2})=\left.M(s,t)\right|_{\alpha'\rightarrow0}=G_N\langle 12\rangle^4[34]^4\left(\frac{-1}{stu}{+}2\zeta_3{+}\zeta_5\sigma_2{+}2\zeta^2_3stu\cdots\right)\,.
\eqe
The leading piece with the massless poles $\frac{1}{stu}$ correspond to the contribution from the Einstein-Hilbert term and we've identified $G_N=g_s^2$.

Instead of tree-level exchanges, the massive UV states can also contribute via loop process. For example consider a massless scalar $\varphi$ coupled to massive $X$ via $\lambda\varphi X^2$. In the UV four $\varphi$s can interact through a massive $X$ loop, and the amplitude is simply the scalar box-integral (see fig\ref{Dia}):
\eqa
M(s,t)=\lambda^4&&\int \frac{d^4\ell}{(2\pi)^4}\frac{1}{[\ell^2{-}m_X^2][(\ell{-}p_1)^2{-}m_X^2][(\ell{-}p_1{-}p_2)^2{-}m_X^2][(\ell{+}p_4)^2{-}m_X^2]}\nonumber\\
&&{+}perm(2,3,4)\,.
\eqae
The analytic result of the box integral is given as~\cite{MassiveBox}:
\eqa\label{BoxResult}
I_4[s,t]=\frac{1}{(4\pi)^2}\frac{uv}{8\beta_{uv}}\left\{2\log^2\left(\frac{\beta_{uv}+\beta_u}{\beta_{uv}+\beta_v}\right)+\log\left(\frac{\beta_{uv}-\beta_u}{\beta_{uv}+\beta_u}\right)\log\left(\frac{\beta_{uv}-\beta_v}{\beta_{uv}+\beta_v}\right)-\frac{\pi^2}{2}\right.\nonumber\\
\left.+\sum_{i=u,v}\left[2{\rm Li}_2\left(\frac{\beta_i-1}{\beta_{uv}+\beta_i}\right)-2{\rm Li}_2\left(-\frac{\beta_{uv}-\beta_i}{\beta_i+1}\right)-\log^2\left(\frac{\beta_{i}+1}{\beta_{uv}+\beta_i}\right)\right]\right\}\,.
\eqae
where $u=-\frac{4m_X^2}{s}$ and $v=-\frac{4m_X^2}{t}$, and 
\eq
\beta_{u}=\sqrt{1+u},\quad \beta_{v}=\sqrt{1+v},\quad\beta_{uv}=\sqrt{1+u+v}\,.
\eqe 
This gives the following low energy expansion: 
\eqa\label{WLow}
M^{\rm IR}(s,t)&=&\frac{g^4}{2m_X^4}\left(1{+}\frac{1}{5!}\frac{\sigma_2}{m_X^4}{+}\frac{20}{7!3}\frac{\sigma_3}{m_X^6}{+}\frac{2}{7!3}\frac{\sigma^2_2}{m_X^8}{+}\frac{1}{6! 33}\frac{\sigma_3\sigma_2}{m_X^{10}}{+}\cdots\right)\,.
\eqae 
Note that in general for identical scalars, the polynomial part of the four-point amplitude can be expanded on the basis of two permutation invariant polynomials $\sigma_2$ and  $\sigma_3$.
%%%%%%%%%%%%%%%%%%%%%%%%%%%%%%%%%%%
\subsection{From local amplitudes to local operators}\label{Sec2}
%%%%%%%%%%%%%%%%%%%%%%%%%%%%%%%%%%%
\begin{figure}
\begin{center}
\includegraphics[scale=0.7]{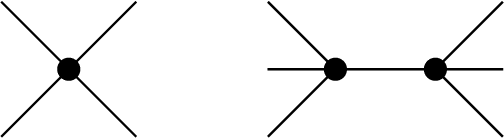}
\caption{An operator of four fields will contribute to the four-point amplitude as a polynomial, and the six-point amplitude as a rational term.}
\label{fig:LocalOp}
\end{center}
\end{figure}
In this paper we are interested in theories whose IR description admits an expansion in terms of local operators, i.e. $\mathcal{L}=\mathcal{L}_{kin}+\mathcal{L}_{I}[\phi,\partial\phi]$, with $\mathcal{L}_{I}[x]$ being polynomial functions.  A local operator that contains $n$ fields, for example $(\partial\phi\cdot \partial\phi)\phi^{n{-}2}$, will contribute to the $n$-point scattering amplitude as a polynomial of Mandelstam invariants $s_{i,j}$. At higher points, it appears in factorization channels, contributing to the residue of rational terms, as illustrated in fig. \ref{fig:LocalOp}.  This translates to the low energy four-point amplitude will taking the form:
\eq\label{eq:PolyRep0}
M^{\rm IR}(s,t)\equiv M(s,t)|_{s,t\rightarrow 0}=\{\rm massless\;\;poles\} {+}\{\rm polynomials\}\,,
\eqe
where $\{\rm massless\;\;poles\}$ reflect the presence of cubic operators, and $\{\rm polynomials\}$ quartic ones. The coefficients of the cubic operators appear  in the residue for the $\{\rm massless\;\;poles\}$, while that of quartic operators are linearly mapped in to the Taylor coefficients in $\{\rm polynomials\}$. Here we have ignored the logarithms arisings massless loops. These effects are of course intimately tied with what we mean by EFT coefficients, as they inevitably run. However, for the sake of simplicity in our presentation, we will focus on tree-level EFT amplitudes for now, and assign section \ref{RG} to discuss how these results extends to the situation where massless loops are present.

Let's begin with operators involving only scalars. First, since the momentum inner products vanish for three-point kinematics,
\eq
p_3^2=(p_1{+}p_2)^2=2p_1\cdot p_2=0\,,
\eqe
the only non-trivial three-point amplitude is a constant. In terms of cubic operators, this is a reflection of the fact that any three-scalar operator with derivatives much vanish via equations of motion:
\eq
(\partial\phi \cdot \partial\phi)\phi \sim \phi^2 \Box\phi=0\,.
\eqe
i.e. it can be removed by a field redefinition. At four-points the amplitude can be expressed as:
\eq\label{eq:PolyRep}
M^{\rm IR}(s,t)=\{{\rm massless\;\;poles}\}\quad {+}\sum_{k,q}\,a_{k,q}\,s^{k{-}q}t^q\,.
\eqe
Here $k$ labels the total degree in Mandelstam variables, $q$ the degree in $t$,.  This labeling will be convenient for considering the expansion near the forward limit, i.e. $t=0$. For fixed $k$ these correspond to dimension $2k{+}4$ operators in four-dimensions. For example, $(\partial \phi \cdot \partial \phi)^2$, $(\partial\phi \cdot \partial \phi)(\partial^2 \phi \cdot \partial^2 \phi)$, translate to 
\eq
(\partial \phi \cdot \partial \phi)^2\rightarrow (2s^2{+}2t^2{+}2st),\quad (\partial \phi \cdot \partial \phi)(\partial^2 \phi \cdot \partial^2 \phi)\rightarrow  -st^2-s^2t\,.
\eqe 
Thus the coefficients of the EFT operators are translated into the coefficients of the polynomials $s^{k{-}q}t^q$.  Note that we do not have an $k=1$ operator $(\partial \phi \cdot \partial \phi)\phi^2$, since on-shell it vanishes by momentum conservation $s{+}t{+}u=0$. Once again, as with the three-point example, this illustrates the important advantage of such ``on-shell basis" eq.(\ref{eq:PolyRep}): it is free from  field redefinition or integration by parts ambiguities.

Generally, it is unnatural for scalars to be massless unless they're Goldstone bosons for some broken symmetry. Thus the degrees of freedom in low energy effective field theories are more naturally associated with photons and gravitons, and the local operators are built out of field strengths and Riemann tensors (Ricci tensor and scalars vanish under Einstein equations). Their imprint on the amplitudes can be more conveniently captured by the spinor-helicity variables, where one express the momenta as:
\eq
p_{i\mu} \rightarrow p_{i\alpha\dot{\alpha}}=p_{i\mu}(\sigma^\mu)_{\alpha\dot{\alpha}}=\lambda_{i\alpha} \tilde{\lambda}_{i\dot{\alpha}}\,.
\eqe
Under the massless U(1) little group, these transforms as $\lambda_{i\alpha}\rightarrow e^{-i\frac{\theta_i}{2}}$ and $\tilde \lambda_{i\dot{\alpha}}\rightarrow e^{i\frac{\theta_i}{2}}\tilde \lambda_{i\dot{\alpha}}$. The polarization vectors are then expressed as 
\eq
\varepsilon^{+}_{i\alpha\dot{\alpha}}=\frac{1}{\sqrt{2}}\frac{\eta_{\alpha}\tilde{\lambda}_{i\dot{\alpha}}}{\langle i\eta\rangle},\quad \varepsilon^{-}_{i\alpha\dot{\alpha}}=\frac{1}{\sqrt{2}}\frac{\tilde{\lambda}_{i\alpha}\eta_{\dot\alpha}}{[i\eta]}
\eqe
where $\langle i j\rangle=\lambda_i^{\alpha}\lambda_{j\alpha}=\epsilon^{\alpha\beta}\lambda_{i\beta}\lambda_{j\alpha}$, and $[i j]=\tilde\lambda_{i\dot\alpha}\tilde\lambda_{j}^{\dot\alpha}=\epsilon_{\dot\alpha\dot\beta}\lambda_{i}^{\dot\beta}\lambda_{j}^{\dot\alpha}$. Here $\eta$ are the reference spinors parameterizing the gauge redundancy associated with the polarization vectors, and drops out for any gauge invariant quantity. Polarization tensors are just the square of these vectors. It is straight forward to see, in terms of these on-shell variables, the field strength and the linear part of Riemann tensor are expressed as: 
\eqa
F_{\mu\nu}&\rightarrow& F_{\alpha \dot{\alpha},\beta \dot{\beta}}=F^{\rm{\tiny +}}_{\dot{\alpha}\dot{\beta}}\,\epsilon_{\alpha\beta}{+}F^{\rm{\tiny -}}_{\alpha\beta}\,\epsilon_{\dot{\alpha}\dot{\beta}},\quad F^{\rm{\tiny +}}_{\dot{\alpha}\dot{\beta}}=\sqrt{2}\tilde\lambda_{\dot{\alpha}}\tilde\lambda_{\dot{\alpha}},\;\;F^{\rm{\tiny -}}_{\dot{\alpha}\dot{\beta}}=\sqrt{2}\lambda_{\alpha}\lambda_{\alpha},\;\;\nonumber\\
R_{\mu\nu\rho\sigma}&\rightarrow& R_{\alpha\dot{\alpha}\beta\dot{\beta}\gamma\dot{\gamma}\delta\dot{\delta}}=\epsilon_{\alpha\beta}\epsilon_{\gamma\delta}\,R^{\rm{\tiny +}}_{\dot{\alpha}\dot{\beta}\dot{\gamma}\dot{\delta}}{+}\epsilon_{\dot\alpha\dot\beta}\epsilon_{\dot\gamma\dot\delta}\,R^{\rm{\tiny -}}_{\alpha\beta\gamma\delta}\nonumber\\
&&R^{\rm{\tiny +}}_{\dot{\alpha}\dot{\beta}\dot{\gamma}\dot{\delta}}=\sqrt{2}\tilde{\lambda}_{\dot{\alpha}}\tilde{\lambda}_{\dot{\beta}}\tilde{\lambda}_{\dot{\gamma}}\tilde{\lambda}_{\dot{\delta}},\quad R^{\rm{\tiny -}}_{\alpha\beta\gamma\delta}=\sqrt{2}\lambda_{\alpha}\lambda_{\beta}\lambda_{\gamma}\lambda_{\delta}\,,
\eqae
where the $\pm$ superscript indicates the $\pm h$ helicity of the polarization (tensors)vector. Indeed up to an overall constant, the above form are uniquely fixed by the little group scaling and dimension analysis.

Thus polynomials of spinor brackets can be straightforwardly translated to local operators of field strengths and Riemann tensors. For example for the three-point amplitude, possible polynomial representation for self interacting spin-1 and 2 particles can be immediately translated into $F^3$ and $R^3$ operators:
\eqa
M_3(1^{-} 2^{-}3^{-})&\rightarrow&2\sqrt{2}\langle12\rangle \langle 23\rangle \langle 31\rangle= (F_1^{\rm{\tiny -}})_\alpha\,^\beta (F_2^{\rm{\tiny -}})_\beta\,^\gamma (F_3^{\rm{\tiny -}})_\gamma\,^\alpha\nonumber\\
M_3(1^{+} 2^{+}3^{+})&\rightarrow&2\sqrt{2} [12][23][31]= (F_1^{\rm{\tiny +}})_{\dot{\alpha}}\,^{\dot{\beta}} (F_2^{\rm{\tiny +}})_{\dot{\beta}}\,^{\dot{\gamma}} (F_3^{\rm{\tiny +}})_{\dot{\gamma}}\,^{\dot{\alpha}}\nonumber\\
M_3(1^{-2} 2^{-2}3^{{-}2}) &\rightarrow&2\sqrt{2} \langle12\rangle^2 \langle 23\rangle^2 \langle 31\rangle^2= (R_1^{\rm{\tiny -}})_{\alpha_1\alpha_2}\,^{\beta_1\beta_2} (R_2^{\rm{\tiny -}})_{\beta_1\beta_2} \,^{\gamma_1\gamma_2}  (R_3^{\rm{\tiny -}})_{\gamma_1\gamma_2}\,^{\alpha_1\alpha_2}\nonumber\\
M_3(1^{+2} 2^{+2}3^{{+}2}) &\rightarrow&2\sqrt{2} [12]^2 [23]^2[ 31]^2= (R_1^{\rm{\tiny +}})_{\dot\alpha_1\dot\alpha_2}\,^{\dot\beta_1\dot\beta_2} (R_2^{\rm{\tiny +}})_{\dot\beta_1\dot\beta_2} \,^{\dot\gamma_1\dot\gamma_2}  (R_3^{\rm{\tiny +}})_{\dot\gamma_1\dot\gamma_2}\,^{\dot\alpha_1\dot\alpha_2}
\eqae
Note that there are no amplitudes associated with $R^2$, reflecting the fact that the Gauss-Bonnet term is a total derivative in four dimensions. Higher dimensional $R^2$ upon dimensional reduction, will reduce to $\phi R^2$ in four-dimensions, and generates the amplitude for a dilaton coupled to two gravitons:
\eqa
 M_3(1^0 2^{+2}3^{+2})&\rightarrow&2[23]^4=(R_1^{\rm{\tiny +}})_{\dot\alpha_1\dot\alpha_2}\,^{\dot\beta_1\dot\beta_2} (R_2^{\rm{\tiny +}})_{\dot\beta_1\dot\beta_2}\,^ {\dot\alpha_1\dot\alpha_2},\;\;\nonumber\\
 M_3(1^0 2^{-2}3^{-2})&\rightarrow&2\langle 23\rangle^4=(R_1^{\rm{\tiny -}})_{\alpha_1\alpha_2}\,^{\beta_1\beta_2} (R_2^{\rm{\tiny -}})_{\beta_1\beta_2}\,^ {\alpha_1\alpha_2}\,,
\eqae
and similar amplitudes for $\phi F^2$.

Extending to four-points we find that there are three possible helicity structures that admit polynomial representations. For spin-1 we have for the lowest mass-dimensions: 
\eqa
M_4(1^{+}2^{+}3^{+}4^+)&\rightarrow&4\left([12]^2[34]^2{+}[13]^2[24]^2{+}[14]^2[23]^2\right)\nonumber\\
&=&(F^{\rm{\tiny +}}_1\cdot F^{\rm{\tiny +}}_2)(F^{\rm{\tiny +}}_3\cdot F^{\rm{\tiny +}}_4)+(F^{\rm{\tiny +}}_1\cdot F^{\rm{\tiny +}}_3)(F^{\rm{\tiny +}}_4\cdot F^{\rm{\tiny +}}_2)+(F^{\rm{\tiny +}}_1\cdot F^{\rm{\tiny +}}_4)(F^{\rm{\tiny +}}_2\cdot F^{\rm{\tiny +}}_3)\;\;\nonumber\\
M_4(1^{+}2^{+}3^{-}4^-)&\rightarrow&4[12]^2\langle 34\rangle^2=(F^{\rm{\tiny +}}_1\cdot F^{\rm{\tiny +}}_2)(F^{\rm{\tiny -}}_3\cdot F^{\rm{\tiny -}}_4)\;\;
\eqae
where $(F^{\rm{\tiny +}}_i\cdot F^{\rm{\tiny +}}_j)\equiv (F^{\rm{\tiny +}}_i)_\alpha\,^\beta (F^{\rm{\tiny +}}_j)_\beta\,^\alpha$ and similar definition for $(F^{\rm{\tiny -}}_i\cdot F^{\rm{\tiny -}}_j)$. We also have $M_4(1^{-}2^{-}3^{-}4^-)$ which is simply changing the square brackets of $M_4(1^{+}2^{+}3^{+}4^+)$ to angles. It is straight forward to translate this back to vector representations, for which the independent $F^4$ contractions are given by: 
\eq
(F^2)^2\equiv (F_{\mu\nu}F^{\mu\nu})^2,\;\;\;(F^2)(F\tilde{F})\equiv (F_{\mu\nu}F^{\mu\nu})(\epsilon^{\mu\nu\rho\sigma}F_{\mu\nu}F_{\rho\sigma}),\;\;\; (F\tilde{F})^2\,.
\eqe
The linear map between to two are given as:
\begin{align}\label{FPolymap}
M_4(1^{+}2^{+}3^{+}4^{+})=&8\left((F^2)^2 - 4 (F\tilde{F})^2+2(F^2)(F\tilde{F})\right)\nonumber\\
M_4(1^{+}2^{-}3^{+}4^{-})=& 8(F^2)^2+32(F\tilde{F})^2\nonumber\\
M_4(1^{-}2^{-}3^{-}4^{-})=&8\left((F^2)^2 - 4 (F\tilde{F})^2-2(F^2)(F\tilde{F})\right)\,.
\end{align}
From the above we immediately see that the combination $(F^2)^2{+}\frac{1}{4}(F\widetilde{F})^2$, which is the square of the Maxwell stress-tensor, only generates the MHV helicity configuration. Similar identification applies to spin-2, where we also have three distinct tensor structure for $R^4$ mapping to the three helicity structures. For higher derivative operators such as $D^{2n}F^4$ or $D^{2n}R^4$, we simply have extra Mandelstam variables multiplying the spinor brackets. For example 
\eq
\sigma_2 \langle 12\rangle^4[34]^4 \rightarrow D^4R^4\,.
\eqe

Thus the EFT amplitude for massless spinning particles, can in general be written in a way such that the spinor brackets are prefactors:
\begin{align}\label{PhotonEFTBasis}
M^{\rm IR}_4(1^{+}2^{+}3^{+}4^{+})=&\frac{[12][34]}{\langle12\rangle\langle34\rangle}\times\left(\sum_{k,q}\,a^{\rm all\,+}_{k,q}\,s^{k{-}q}t^q\right)\nonumber\\
M^{\rm IR}_4(1^{+}2^{+}3^{+}4^{-})=&\frac{[12][23]\langle 24\rangle}{\langle 12\rangle\langle 23\rangle [24]}\times\left(\sum_{k,q}\,a^{\rm single\,{-}}_{k,q}\,s^{k{-}q}t^q\right)\nonumber\\
M^{\rm IR}_4(1^{-}2^{-}3^{+}4^{+})=&\frac{\langle12\rangle^2[34]^2}{stu}\times\left(\sum_{k,q}\,a^{\rm MHV}_{k,q}\,s^{k{-}q}t^q\right)\,
\end{align}
where the spinor prefactors are written in such a way that all possible massless poles are contained and is invariant under the permutation of the same helicity legs. The superscript for the Taylor coefficients $a^{\cdots}_{k,q}$ label the helicity configuration.

Let's consider explicit examples. The low energy expansion for Type-I and II superstring in eq.(\ref{TypeIStringIR}) and eq.(\ref{StringAmp}) gives prime examples of gauge and gravitational EFT amplitudes. However due to being supersymmetric, only MHV configurations are present. For a more general set up, lets consider the open bosonic string amplitude, which contains all three sectors: 
\eqa
f_{{++++}}&=&\frac{[12][34]}{\langle12\rangle\langle34\rangle}stu\left(1{-}\frac{1}{s{+}1}{-}\frac{1}{u{+}1}{-}\frac{1}{t{+}1}\right)\nonumber\\
M^{\rm Bos}(s,t)=\frac{\Gamma[{-}s]\Gamma[{-}t]}{\Gamma[1{+}u]}f_{\{I\}},\quad\quad f_{{+++-}}&=&stu\frac{[12][23]\langle24\rangle}{\langle12\rangle\langle23\rangle[24]}\nonumber\\
f_{{++--}}&=&-[12]^2\langle34\rangle^2\left(1{-}\frac{tu}{s{+}1}\right)
\eqae
The low energy EFT is then given as:
\eqa
M^{\rm IR}(1^+2^+3^+4^+)&=&2u\frac{[12][23][34][41]}{st}{+}2[13]^2[24]^2{-}[12][23][34][41] (\frac{\pi^2}{3}{-}2)+\cdots\nonumber\\
M^{\rm IR}(1^+2^+3^+4^-)&=&[12]^2[23]^2\langle24\rangle^2\left(-\frac{1}{st}+\frac{\pi^2}{6}{-}u\zeta_3{+}\frac{\pi^4}{360}(4s^2{+}st{+}4t^2){+}\cdots\right)\nonumber\\
M^{\rm IR}(1^+2^+3^-4^-)&=&[12]^2\langle34\rangle^2\left(-\frac{1}{st}{+}\frac{u}{s}{+}\frac{\pi^2}{6}{-}u(1{+}\zeta_3){+}\cdots\right)\,,
\eqae
where we've rewritten the spinor brackets in a form that exposes the massless poles. It is instructive to identify local operators in each helicity sector. For the all plus helicity the leading term correspond to the gluon exchange between the Yang-Mills vertex and $F^3$, followed by two types of contractions for $(F^+)^4$. For the single minus sector, we have massless poles associated with the exchange of a vector between $(F^{\rm{\tiny +}})^3$ and a Yang-Mills vertex, while the leading four-point local operator correspond to $D^2(F^{\rm{\tiny +}})^3 F^{\rm{\tiny -}}$. For the MHV sector, we have two sets of massless poles, the leading corresponding to the exchange between the Yang-Mills vertex, and the subleading is between $(F^{\rm{\tiny +}})^3$ and $(F^{\rm{\tiny -}})^3$. The leading four-point local operator is $(F^{\rm{\tiny +}})^2(F^{\rm{\tiny -}})^2$.

%%%%%%%%%%%%%%%%%%%%%%%%%%%%%%%%%%%
\section{Dispersive representation for EFT coefficients}\label{Sec:Dis}
%%%%%%%%%%%%%%%%%%%%%%%%%%%%%%%%%%%
In the previous section, we've seen that given the UV theory, the low energy EFT can be obtained by expanding the UV amplitude in Mandelstam variables, leading to an IR amplitude of the form
\eq\label{IRPolyExp}
M(s,t)|_{s,t \ll m^2}=M^{\rm IR}(s,t)=\{{\rm massless\;\;poles}\}\quad {+}\sum_{k,q}\,a_{k,q}\,s^{k{-}q}t^q\,. 
\eqe
Mapping to on-shell local operators is then a straight forward task. However, it has been long appreciated that general  principles of unitarity and Lorentz invariance imposes non-trivial constraint on the IR description. These constraints arises through the analyticity of the scattering amplitude,  where the poles and branch cuts on the complex Mandelstam variable plane are associated with threshold productions.  For the four-point amplitude, such analytic property allows us to equate the low energy couplings $a_{k,q}$ to the discontinuities of the branch cuts (or residues of poles), giving a dispersive representation for the couplings.

\begin{figure}
\begin{center}
\includegraphics[scale=0.4]{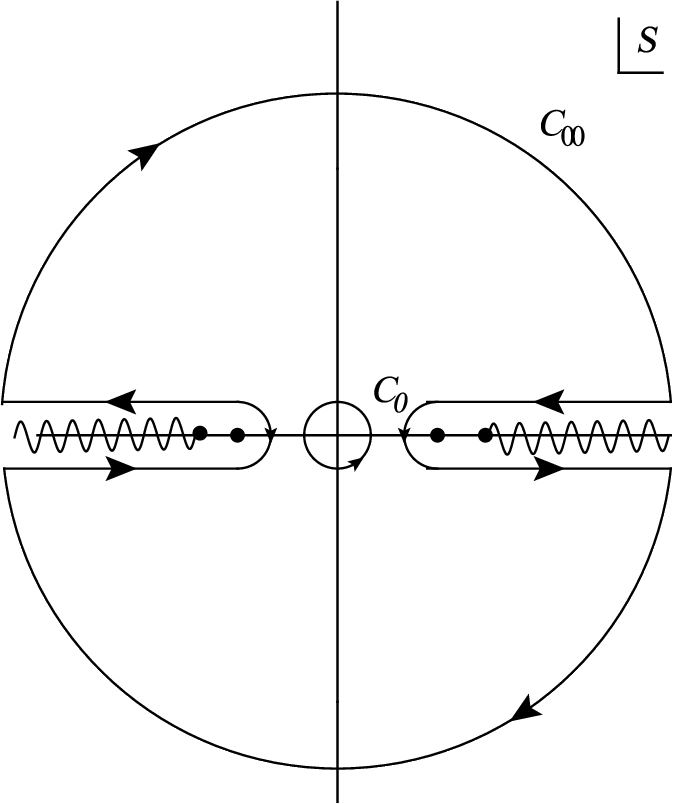}
\caption{We define the low energy couplings through a contour integral on the complex $s$-plane, where the contour $\mathcal{C}_0$ encircles the origin. On the complex plane, if the amplitude only has singularities on the real-$s$ axes, either poles or branch points, then we can deform to contour $\mathcal{C}_\infty$. }
\label{BigContour}
\end{center}
\end{figure}

Let's begin by holding  $t=t^* \ll m^2$ fixed, where $m^2$ is  the characteristic mass associated with the UV completion, and consider four-point amplitude $M(s,t^*)$ as a function of $s$. We will imagine that we are only integrating out the massive states, which generate contact terms in the low-energy effective theory. Of course there will also be calculable massless loops in the low-energy effective theory, which induce logarithmic variation in these coefficients. We will return to discussing this point later in section~\ref{RG}. Note, however, that the very notion of ``higher dimension operators" is only well-defined when there is a weak coupling in the UV theory, so that the contact operators induced by integrating out the massive states dominate over the ones generated by massless loops in the low-energy theory, so that this first-pass analysis captures the most interesting UV physics. In practice, we are assuming that, for small fixed $t \ll m^2$, the amplitude is analytic in the $s$ plane for small s, and develops its first singularity (be it a pole at tree-level, or more generically a branch cut associated with UV particle production) at $s{=}m^2$. 

It is important that when $t$ is $\ll m^2$, the {\it only} singularities of the amplitude are on the real $s$ axis, and correspond to particle production thresholds. This is not true when $t$ is comparable to $m^2$, where new sorts of singularities, simplest amongst them the infamous "anomalous thresholds", with no Lorentzian particle production interpretation, also appear. But for our purposes of controlling EFT coefficients, we only need $t\ll m^2$ and never have to worry about anomalous thresholds. See appendix~\ref{Lehman} for a more detailed discussion of these issues.

As is standard from the study of dispersion relations, we consider the contour integral 
\eq
\frac{i}{2\pi}\int_{\mathcal{C}_0} \frac{ds}{s^{n+1}}M(s,t^*)\,,
\eqe
where $\mathcal{C}_0$ represents the contour that encircles the origin. Since at the origin both $s,t^* \ll m^2$ we know that amplitude takes its low energy form in eq.(\ref{IRPolyExp}), and the residue for the measure $\frac{1}{s^{n+1}}$ will be given by terms in eq.(\ref{IRPolyExp}) proportional to $s^n$. In absence of $t$-channel massless pole, this residue will be a polynomial function of $t$, giving a well defined Taylor expansion around $t=0$. Thus we find that $a_{k,q}$ can be identified as:
\eq\label{gDef}
a_{k,q}=\left.\frac{1}{q!}\left[\frac{\partial^q}{\partial t^q}\frac{i}{2\pi}\int_{\mathcal{C}_0} \frac{ds}{s^{k{-}q{+}1}}M(s,t)\right]\right|_{t=0}\,.
\eqe 
In other words, the low energy couplings can be analytically defined through the on-shell amplitude. Note that taking the residue is equivalent to taking derivatives, and the result of this action is often referred to as the subtracted amplitude.
Now instead of $\mathcal{C}_0$ we deform to the contour encircling infinity $\mathcal{C}_\infty$. If the non-analyticities are associated with particle production, they occur on the real axes where depending their origin as $s$ or $u$-channel threshold, they will lie on the positive or negative real $s$-axes respectively. Thus the contour $\mathcal{C}_\infty$ takes the form shown in fig.\ref{BigContour}, where one picks up the discontinuity on the real axes as well as boundary contributions.  At large $s$, if the amplitude falls of faster than $s^{k{-}q}$ then the latter simply yields zero, and we would have an identity between  $a_{k,q}$ and the residues or discontinuities.

Let us consider the linear sigma model as an explicit example. Once again the UV tree-amplitude is given as:
\eq\label{LSM4pt1}
M(s,t)=-\frac{\lambda}{8m_h^2}\left(\frac{s^2}{s-m_h^2}{+}\frac{t^2}{t-m_h^2}{+}\frac{u^2}{u-m_h^2}\right) \,.
\eqe
As $s\rightarrow \infty$ the amplitude grows linearly in $s$,  the contour deformation of eq.(\ref{gDef}) will have no boundary contributions when $k{-}q\geq2$. Focusing on the couplings with $q=0$, i.e. those that survive in the forward scattering limit $t=0$, we find eq.(\ref{gDef}) implies:
\eq
a_{k,0}= - \frac{1}{(m^2_h)^{k{+}1}}\left(Res_{s} M(s,0){+}({-})^{k}Res_{u} M(s,0)\right)\,.
\eqe
That is, the coupling $a_{k,0}$ is given by the residue of the Higgs pole in the $s$ and $u$ channel. Plugging in $Res_{s=m^2_h} M(s,0)=-\frac{\lambda m^2_h}{8}$ and $Res_{s={-}m^2_h} M(s,0)=-\frac{\lambda m^2_h}{8}$, we have
\eq
a_{k,0}=\frac{\lambda }{4(m^2_h)^{k}},\quad k\in even, 
\eqe
and $0$ for $k\in odd$. Indeed this reproduces the low energy couplings in eq.(\ref{LSM4ptIR}), for $k\geq2$.

In general for theories whose four-point amplitude admits a convergent partial wave expansion, causality and unitarity dictates that the four-particle amplitude at $t=0$ is bounded by $s\log^{D{-}2}s$, i.e. the Froissart bound~\cite{Froissart:1961ux, Martin:1962rt}. When massless particles are present, such as in gravity, the $t$-channel singularity obstructs a convergent polynomial expansion in $t$ and the Froissart analysis no longer holds. However, assuming a weakly coupled UV completion for gravity, causality consideration requires the presence of an infinite tower of massive higher spin states, leading to the forward amplitude behaving as $s^{p}$ for $p<2$ at large $s$ for fixed negative $t$~\cite{Camanho:2014apa}. From now on we will assume that for $|t|\ll m^2$ the amplitude is bounded by $s^2$ at large $s$. For a more detailed discussion, see Appendix~\ref{Froissart}.

For general tree-level UV completions it is obvious that all poles lies on the real $s$-axes. More generally, the amplitude admits a dispersive representation
\eq
M(s,t)|_{t\ll m^2}=M^{\rm Sub}+\int^\infty_{M^2_s} \;dM^2\;\frac{\rho_s(M^2)}{s-M^2}+\int^\infty_{M^2_u} \;dM^2\;\frac{\rho_u(M^2)}{u-M^2}
\eqe 
where $M^{\rm Sub}$ represents the appropriate subtraction terms, representing the contributions from infinity in the dispersion relation. Note again the importance of keeping $t \ll m^2$ here. In general, we don't have good control on the analytic structure even of 4pt amplitudes in general theories. But we do have good control on the analytic structure of 2-pt functions as restricted by causality and unitarity. Intuitively, by keeping $t\ll m^2$, our 4-pt amplitude is close to forward scattering and hence a 2-pt function. A standard justification that the only singularities for $t\ll m^2$ are associated with usual particle production is given by studying Landau equations. In appendix~\ref{Lehman} we give a different, more direct derivation following directly from Feynman/Schwinger parametrization of loop integrals. Putting everything together,  we conclude that for $k{-}q\geq2$:
\eq\label{Relate0}
a_{k,q}=\left.-\frac{1}{q!}\frac{\partial^q}{\partial t^q}\left(\sum_a \quad \frac{Res_{s=m^2_a} M(s,t)}{(m^2_a)^{k{-}q{+}1}} +\int_{4m_a^2} \frac{ds'}{s'^{k{-}q{+}1}}Dis M(s,t)\right)\right|_{t=0}{+}\{u\}\,,
\eqe
here $a$ labels all the massive states and $\{u\}$ represents the $u$-channel contributions.

Let us study the above identity with two explicit examples, the infinite resonance of a string theory tree level exchange and the one-loop massive bubble in three-dimensions. 

\noindent \textbf{Tree-level dispersive representation}:  Let's begin with the type-I string amplitude introduced in eq.(\ref{TypeIString}), where the $s$-channel residue is given as:
\eq\label{StringResI}
Res_{s=n}\left[-\frac{\Gamma[-s]\Gamma[-t]}{\Gamma[1+u]}\right]=-\frac{(t+1)(t+2)\cdots (t{+}n{-}1)}{n!}
\eqe
Now using eq.(\ref{Relate0}) we have, 
\eq\label{Demo0}
a_{k,q}=\frac{1}{q!}\frac{\partial^q}{\partial t^q}\left(\sum_{n=1}^\infty \frac{1}{n!}\; \frac{(t+1)(t+2)\cdots (t{+}n{-}1)}{n^{k{-}q{+}1}}\right)\,.
\eqe
First consider the coefficients relevant to the strict forward limit, $a_{k,0}$, which corresponds to setting $t=0$ in the above, and we find:
\eq
a_{k,0}=\sum_{n=1}^{\infty}\frac{1}{n^{k+2}}=\zeta_{k+2}\,.
\eqe 
Indeed this is the reproduces the $\zeta_2$ and $\zeta_3$ for the constant and the coefficient  for $s$ in eq.(\ref{TypeIStringIR}) respectively. Now let's move away from the strict forward limit and consider coefficients of $t$ to the first power. From eq.(\ref{Demo0}) we have,
\eq\label{SummationForm}
a_{k,1}=\sum_{n=2}^\infty \frac{1}{n^{k{+}1}}\left(1{+}\frac{1}{2}{+}\frac{1}{3}{+}\cdots{+}\frac{1}{n{-}1}\right),\quad \,.
\eqe 
Explicitly expanding eq.(\ref{TypeIString}) to the fifth power in Mandelstam variables one find, 
\eq
a_{5,1}=-\frac{1}{90}  (\pi^4 \zeta_3 {+}15 \pi^2 \zeta_5{-} 270 \zeta_7)\,,
\eqe
which once again agrees with eq.(\ref{SummationForm}).

\noindent \textbf{Loop-level dispersive representation}:  Consider a three-dimensional theory with a massless scalar $\phi$ and a massive one $X$,  interacting via the quartic coupling $\lambda \phi^2X^2$. At low energies we have an effective action for $\phi$, generated by integrating away the massive $X$ loops.  For example at leading order in $\lambda$, operators of the form $\partial^{2n}\phi^4$ are obtained by integrating out $X$ from the one-loop bubble diagrams: 
\eqa
\vcenter{\hbox{\includegraphics[scale=0.5]{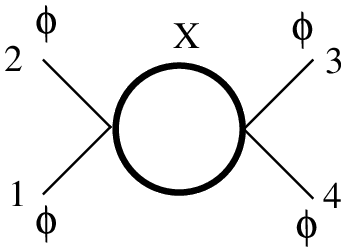}}}\,.
\eqae
This yields the following UV amplitude, 
\eqa\label{exp0}
M(s,t)&=&\lambda^2\left[\mathcal{I}^3_{bubble}(s)+\mathcal{I}^3_{bubble}(t)+\mathcal{I}^3_{bubble}(u)\right]\,,\nonumber\\
\mathcal{I}^3_{bubble}(s)&=&\int \frac{d\ell^3}{(2\pi)^3}\frac{1}{(\ell^2-m^2)((\ell+p_{12})^2-m^2)}=\frac{1}{8 \pi\sqrt{s}}\log\left(\frac{2m{+}\sqrt{s}}{2m{-}\sqrt{s}}\right)\,.\nonumber\\
\eqae
The low energy expansion yields,
\eqa\label{exp2}
M^{\rm IR}(s,t)&=&\frac{\lambda^2}{8\pi m}\left(3{+}\frac{\sigma_2}{80 m^4}+\frac{\sigma_3}{448 m^6}{+}\frac{\sigma_4}{2304 m^8}\right)+\mathcal{O}\left(\frac{1}{m^{11}}\right)\,.
\eqae
Now since the UV amplitude eq.(\ref{exp0}) behaves as $\sim s^{0}$ as $s\rightarrow \infty$, we expect that through eq.(\ref{Relate0}) we can recover all low energy coefficients in eq.(\ref{exp2}) with degree 1 and higher in $s$ from the discontinuity of the bubble integrals. For fixed $t$, only the $s$- and $u$-channel bubble integrals contain branch cuts. The $\mathcal{I}^3_{bubble}(s)$ has a branch cut starting from $4m^2$ to $\infty$, with the discontinuity given by $\frac{i}{4\sqrt{s}}$, while the branch cut for $\mathcal{I}^3_{bubble}(u)$ is on the negative real $s$-axes from  $-4m^2-t$ to $-\infty$, with discontinuity $\frac{i}{4 \sqrt{-t-4m^2}}$. Thus from eq.(\ref{Relate0}), we find
\eq
a_{n+q,q}=\left.\frac{1}{q!}\frac{\partial^q}{\partial t^q}\left[\frac{1}{2\pi i}\left(\int_{4m^2}^{\infty}\frac{1}{s^{n+1}} \frac{i}{4 \sqrt{s}}+\int_{-t-4m^2}^{-\infty}\frac{1}{s^{n+1}} \frac{i}{4 \sqrt{-t-4m^2}}\right)\right]\right|_{t=0}\,.
\eqe
For example to reproduce the coefficients of $s^2t^q$, we take $n=2$ in the square bracket above, yielding:
\eqa
&&\frac{1}{640 m^5 \pi}+\frac{1}{64 \pi t^{5/2}}\left( 3 \pi  - 6  \tan^{-1} \left(\frac{2 m}{\sqrt{t}}\right)-\frac{  
    4 m\sqrt{t} (12 m^2 + 5 t)}{(4 m^2 + t)^2}\right)\nonumber\\
&=&\frac{1}{320 m^5 \pi}-\frac{3t}{3584 m^7 \pi}+\frac{t^2}{3072 m^9 \pi}+\mathcal{O}(t^3)\,.
\eqae  
Indeed the first three terms in the $t$ expansion matches with the coefficients of $s^2, s^2 t$ and $s^2t^2$ in eq.(\ref{exp2}) respectively.

Before closing this section, we comment on two potential obstructions in utilizing the dispersive representation:
\begin{itemize}
  \item  The the residue at $s=0$ contains $t$-channel singularity.
  \item  The presence of massless cuts, which leads to branch point singularity at the origin.  
\end{itemize}
A $1/t$ pole in the residue at $s=0$ renders the Taylor expansion in eq.(\ref{gDef}) ill defined. More precisely since  by Cauchy theorem the $t$-channel pole must be reproduced by the sum over residues and branch cuts, the singularity in the $t\rightarrow 0$ limit indicates that the sum is not convergent. The graviton pole mentioned previously is a famous example of such obstruction. We will discuss this in great detail in the following section.

At loop-level there are two forms of non-analyticity at the origin for massless theories, IR singularities and massless cuts. For those with massless three-point interactions, such as gravity, loop-corrections are accompanied by collinear divergences. However, if we assume that the UV completion occurs while the self-coupling of the massless states are still perturbative, these divergences can be suppressed or computed order by order.  The presence of massless cuts imply that one can no longer define the EFT couplings via the contour at $\mathcal{C}_0$. As previously mentioned this is reflecting the subtlety in what we mean by EFT couplings when log runnings are present. As we will see in sec. \ref{RG}, the choice of ``scale" against which the couplings run, are naturally introduced by moving the contour off the origin. After introducing such ``generalized coupling" the remaining analysis are almost identical of the tree amplitude.

%%%%%%%%%%%%%%%%%%%%%%%%%%%%%%%
\section{Obstructions from the massless poles}\label{Appt0}
%%%%%%%%%%%%%%%%%%%%%%%%%%%%%%%%
The presence of massless poles in the four-point amplitude, can potentially forbid a near forward limit dispersion representation. Take for an example an IR amplitude that behaves as
 \eq
M^{\rm IR}(s,t)|_{s,t\rightarrow 0}\sim \frac{s^n}{t}+a_{n,0} s^n{+}\mathcal{O}(t)\,.
\eqe
Applying the dispersive representation for $a_{n,0}$ in eq.(\ref{Relate0}), we find:
\eq
\left. \frac{1}{t}+a_{n,0}{+}\sum_a\frac{Res_{s=m^2_a}\,M(s,t)}{(m^2_a)^{n{+}1}} \right|_{t=0}=0\,.
\eqe 
 Now since the above equality holds in the limit where $t\rightarrow 0$, the divergent behaviour of the $\frac{1}{t}$ pole tells us that the remaining summation cannot be convergent. For a concrete example, let's consider the four gluon amplitude in type-I super string. Stripping off the spinor factors, the following contour integral yields,
\eq
\frac{i}{2\pi}\int_{\mathcal{C}_0} ds \frac{M^{\rm TypeI}(1^+2^+3^-4^-)}{[12]^2\langle34\rangle^2}=\frac{1}{t}\,.
\eqe  
This isolates the field theory contribution $\frac{1}{st}$ in the low energy amplitude. Now at large $s$ and small $t$, the amplitude scales as 
\eq\label{eq:Exp}
\left.\frac{M^{\rm TypeI}(1^+2^+3^-4^-)}{[12]^2\langle34\rangle^2}\right|_{s\rightarrow \infty}<s^{-1}\,.
\eqe
Thus if we deform the contour to $\mathcal{C}_\infty$, there are no boundary contributions and one only picks up poles on the real axes, whose residue is given by eq.(\ref{StringResI}). Thus we have
\eq\label{tpolecancel}
\frac{1}{t}-\sum_{n=1}^\infty \frac{(t{+}1)(t{+}2)\cdots (t{+}n{-}1)}{n!}=0\,,
\eqe 
and setting $t=0$ we indeed find that the summation is non convergent, $\sum_{n=1}^\infty \frac{1}{n}=\infty$! 

In this paper, we will focus on $a,b\rightarrow a,b$ scattering where $a,b$ may or may not be of the same type. When embedded in a gravitational theory one inevitably encounters the $t$-channel graviton exchange. For example consider the four-dilaton amplitude of type-II string theory
\eq\label{DilatonString0}
M^{\rm Type{-}II}(1^02^03^04^0)=g_s^2(s t+ tu +s u)^2\frac{\Gamma[{-}s]\Gamma[{-}u]\Gamma[{-}t]}{\Gamma[1{+}s]\Gamma[1{+}u]\Gamma[1{+}t]}\,.
\eqe
At low energies, beyond the tree-level graviton exchange the leading local amplitude is associated with $D^8 \phi^4$,
\eq\label{DilatonString}
M^{\rm IR}(1^02^03^04^0)=G_N\left({-}\frac{s t}{u}{-}\frac{tu}{s}{-}\frac{su}{t}{+}2\zeta_3(s t+ tu +s u)^2+\cdots\right)\,.
\eqe
Note that there are no four derivative couplings $D^4 \phi^4$, which appears to violate the positivity bound $a_{2,0}>0$ introduced long ago~\cite{Adams:2006sv}. The resolution precisely lies in the presence of the $t$ graviton pole! Let us see how this play out in detail.  First, as the amplitude enjoy $s\leftrightarrow u$ symmetry, we manifest this symmetry by switching to 
\eq\label{suz}
z=s{+}\frac{t}{2}\,,
\eqe
then $s\leftrightarrow u$ translates to $z\leftrightarrow {-}z$. We take the contour integral in $z$-plane, and defining the low energy coupling via its degree in $z,t$.  Now let's compare the dispersive representation for the coupling of the four- and eight-derivative couplings, $a_{4,0}$ $a_{2,0}$. The integrals of interest are then:
\eqa\label{eq:tpole}
\frac{i}{2\pi}\int_{\mathcal{C}_0}\frac{dz}{z^3}\,M^{\rm Type{-}II}(1^02^03^04^0)&=&-\frac{1}{t}{+}\sum_{q=0}^{\infty}a_{q{+}2,q}t^{q},\quad \nonumber\\
\frac{i}{2\pi}\int_{\mathcal{C}_0}\frac{dz}{z^5}\,M^{\rm Type{-}II}(1^02^03^04^0)&=&\sum_{q=0}^{\infty}a_{q{+}4,q}t^{q}
\eqae
Note that the contour $\mathcal{C}_0$ picked up residues  at $z=0,\pm t/2$, since $t\rightarrow0$.  Comparing the two integrals we see that the dispersive representation should be convergent for $a_{q{+}4,q}$ (including $a_{4,0}$), but not for $a_{q{+}2,q}$ (including $a_{2,0}$). As the representation is not convergent for $a_{2,0}$, positivity based on such dispersive arguments are no longer applicable.

However, the presence of massless $t$-poles in the field theory amplitude \textit{does not} necessarily imply an obstruction. Consider a gravitational EFT whose low energy limit is given by the Einstein-Hilbert action and no modification to the graviton cubic couplings (i.e. no $R^3$). The low energy amplitude for $M(1^{+2}2^{+2}3^{-2}4^{-2})$ is given by 
\eq
M^{\rm IR}(1^{+2}2^{+2}3^{-2}4^{-2})=[12]^4\langle34\rangle^4\left(\frac{1}{stu}{+}\sum_{k,q}\,a_{k,q}s^{k{-}q}t^q\right)\,.
\eqe
Even though the low energy amplitude contains massless $t$ poles, the $\mathcal{C}_0$ contour actually picks up multiple $1/t$ that cancels
\eq
\int_{\mathcal{C}_0} \frac{ds}{s^{n}}\frac{M^{\rm IR}(1^{+2}2^{+2}3^{-2}4^{-2})}{[12]^4\langle34\rangle^4}={-}\frac{1}{t^{n{+}2}}{+}\frac{1}{t^{n{+}2}}{+}\sum_{q}\,a_{q{+}n{-}1,q}t^q\,.
\eqe
This can be tied to the massless poles coming in the combination $\frac{1}{stu}$. This result is deeply tied to the fact that the amplitude for minimally coupled self-interacting massless particles are ``$3$-particle constructible", i.e. consistent factorization in one channel automatically enforces consistency in all other channels.

Thus in summary, while graviton exchanges can introduce $t$-channel singularity, \textit{if the four-point amplitude is  $3$-particle constructible, then the combined contributions cancel each other and we are free of $t$-channel obstruction}. Examples include four-graviton amplitude of pure Einstein-Hilbert gravity, as well as the gravitational Compton amplitude for minimally coupled particles. If we have extra symmetry which relates the amplitude to a $3$-particle constructible partner, or that it suppresses the $t$-channel exchange, one can similarly avoid the  $t$-channel obstruction. Let us go through explicit examples for spin-0, 1 and 2 amplitudes with graviton exchange.

\noindent \textbf{Scalars} We have discussed identical scalars in eq.(\ref{eq:tpole}). For distinct scalars, we can arrange the scalars such that there are no $t$-channel exchanges. For example a pair of complex scalars with  U(1) symmetry, the graviton exchange is given by:
\eq
M^{IR}(\phi_1\overline{\phi}_2\overline{\phi}_3\phi_4)=\frac{tu}{s}+\frac{s t}{u}\,,
\eqe
where there would be no $t$-channel poles and free from obstructions.

%%%%%%%%%%%%%%%%%%%%%%%%%%%%%%%%%%%%%%%%%%%%%
\noindent \textbf{Photons}
%%%%%%%%%%%%%%%%%%%%%%%%%%%%%%%%%%%%%%%%%%%%%
The graviton poles and its residues are dictated by its minimal coupling, $F^2\phi$ and $RF^2$ operators. Let's start by choosing the same helicity to be in the $t$-channel, one has:
\eq
M^{\rm IR}(1^-2^+3^+4^-)=[23]^2\langle14\rangle^2\left(\frac{1}{s}+\frac{1}{u}+\alpha_1\frac{1}{t}+\alpha_2\frac{su}{t}{+}\cdots\right)\,,
\eqe
where $\alpha_1$ and $\alpha_2$ represents contribution from $\phi F^2$ and $RF^2$ respectively. Note that due to the helicity arrangements, the contribution from the latter only appears in $t$-channel. Factoring out the universal helicity factor and taking the contour integral near the origin we find,
\eq
\int \frac{dz}{z^{n{+}1}}\left(\frac{4 t}{4 z^2-t^2}+\frac{\alpha_1}{t} + \alpha_2\left(\frac{t}{4} - \frac{z^2}{t}\right) \right)=\left\{\begin{array}{cc}& \frac{\alpha_1}{t} + \frac{\alpha_2 t}{4} \quad {\rm for}\;n=0 \\& -\frac{\alpha_2}{t} \quad {\rm for}\;n=2\end{array}\right.
\eqe
while the integral vanishes for other $n$. Thus we see that minimal coupling does not introduce $t$-channel poles, while the presence of $\phi F^2$ and $RF^2$ leads to $t$-channel obstruction for the four and eight derivative terms respectively. Following our scalar example, let's arrange the helicity such that contributions from these higher-derivative operators only appear in the $s$-channel, as:
\eq\label{RefEq}
M^{\rm IR}(1^-2^-3^+4^+)=[34]^2\langle12\rangle^2\left(\frac{1}{t}+\frac{1}{u}+\alpha_1\frac{1}{s}+\alpha_2\frac{tu}{s}{+}\cdots\right)\,,
\eqe  
This time we find,
\eq\label{t0pole}
\int \frac{du}{u^{n{+}1}}\left(\frac{1}{t}+\frac{1}{u}-\alpha_1\frac{1}{u+t}-\alpha_2\frac{tu}{u+t} \right)=\frac{1}{t}- \alpha_2 t\quad {\rm for}\;n=0
\eqe
and zero otherwise. Since we've factored out the spinor brackets, we see that $t$-channel singularities from minimal coupling obstructs the convergence of four derivative operators.

Let's consider the case where we wish to apply dispersive representation to the coefficient of $F^4$ operators, relevant for the analysis of weak gravity conjecture. After factoring out the spinor brackets, the coefficient of $F^4$ is mapped to $a_{0,0}$. For helicity $(1^-2^-3^+4^+)$ the spinor brackets are $s^2$ and thus we can bound $a_{0,0}$. However due to eq.(\ref{t0pole}) we see that  $a_{0,0}$ suffers the $t$-pole obstruction. One might attempt to use the configuration $(1^-2^+3^+4^-)$, where there are no  $t$-pole obstruction for the four-derivative term. However in this case the spinor prefactor  is simply $t^2$ up to a phase, thus the coefficient for $F^4$ is mapped to the coefficient of $s^0$ for which the dispersive representation is not applicable due to boundary contributions.

%%%%%%%%%%%%%%%%%%%%%%%%%%%%%%%%%%%%%%%%%%%%%
\noindent \textbf{Gravitons}
%%%%%%%%%%%%%%%%%%%%%%%%%%%%%%%%%%%%%%%%%%%%%

For external gravitons, the analysis is parallel to the photon case except that the relevant couplings are now the Einstein-Hilbert term, $\phi R^2$ and $R^3$.  For the MHV amplitude, with equal helicity in the $s$-channel we have 
\eq
M^{IR}(1^{-2}2^{-2}3^{+2}4^{+2})=[12]^4\langle34\rangle^4\left(-\frac{1}{stu}+\alpha_1\frac{1}{s}+\alpha_2\frac{tu}{s}\right)\,,
\eqe
where now $\alpha_1$ and $\alpha_2$ represents $\phi R^2$ and $R^3$ respectively. Since as previously discussed summing over the massless residues cancels for the Einstein-Hilbert term, there are no potential $t$-channel singularities. If we were to choose the other two channels, then from $t$-channel exchanges between $\phi R^2$ or $R^3$, we would have encounter the similar obstruction as the photon case for the eight and twelve derivative terms respectively.

%%%%%%%%%%%%%%%%%%%%%%%%%%%%%%%%%%%%%%%%%%%%%
\noindent \textbf{The $t$-channel pole and Reggie behaviour }
%%%%%%%%%%%%%%%%%%%%%%%%%%%%%%%%%%%%%%%%%%%%%
In cases where the $t$-channel singularity implies non-convergence of the dispersive representation, it is instructive to see how the singularity is analytically reproduced. Let's reexamine the summation eq.(\ref{tpolecancel}) in the $t\rightarrow 0$ limit. In such case it can be approximated as
\eq
\sum_{n=1}^\infty \frac{(t{+}1)(t{+}2)\cdots (t{+}n{-}1)}{n!}\sim \sum_{n=1}^\infty \frac{1+t+\frac{t}{2}+\cdots \frac{t}{n-1}}{n}\sim\sum_{n=1}^\infty \frac{1+t\log n}{n}\sim \sum_{n=1}^\infty \frac{e^{t\log n}}{n}
\eqe 
Finally, the last line simply becomes $ \sum_{n=1}^\infty n^{t-1}$ which after approximating the sum as an integral, yields $\frac{1}{t}$. Recall that the summation is over the residues of the amplitude at $s=n$, which is the dominant contribution for the amplitude as $s$ nears threshold. The fact that at small $t$ the residue is approximated by $n^t$, implies that the amplitude behaves as $s^t$ in the near forward limit. This is nothing but the linear Regge behaviour of string theory, except that it holds true for large but finite values of $s$. Of course this is not surprising given that in order for equation eq.(\ref{tpolecancel}) to hold, the amplitude is required to die off at $s\rightarrow \infty$, which is true precisely due to such Regge behaviour.

%%%%%%%%%%%%%%%%%%%%%%%%%%%%%%%%%%%%%%%%%%%%%
\section{Theory space as a convex hull}\label{sec:TheorySpace}
%%%%%%%%%%%%%%%%%%%%%%%%%%%%%%%%%%%%%%%%%%%%%
As we have reviewed, there is a simple expression for the coefficients of low-energy effective field theory coefficients in terms of the spectrum and discontinuities of the high-energy amplitude: 
\eq
a_{k,q}=\left.-\frac{1}{q!}\frac{\partial^q}{\partial t^q}\left(\sum_a \quad \frac{Res_{s=m^2_a} M(s,t)}{(m^2_a)^{k{-}q{+}1}} +\int_{4m_a^2} \frac{ds'}{s'^{k{-}q{+}1}}Dis M(s,t)\right)\right|_{t=0}{+}\{u\}\,.
\eqe
Since optical theorem tells us that the sum of residue and discontinuity of the forward amplitude is proportional to the total cross-section $\sigma(s)$, Im $M(s,0)=-s\sigma(s)$, one immediately concludes that $a_{k,0}>0$.

However, this is not the whole story since the optical theorem is really a ``coarse grained" description of the residues and discontinuity. Lorentz invariance and factorization tells us vastly more than just the positivity in the forward limit. In particular when combined with unitarity, Lorentz invariance tells us that the discontinuities are positively expandable on a preferred polynomial basis! To see this, consider the $2\rightarrow 2$ scattering of scalar particles $M(1^a,2^b,3^b,4^a)$, where $a,b$ labels the distinct species. Let's consider the general form of the residue from a tree-level spin-$\ell$ exchange:
\eq
\includegraphics[scale=0.5]{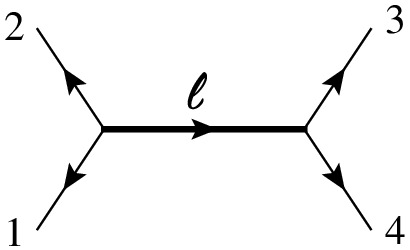}
\eqe
The residue is given by the product of three-point amplitudes for two scalars $a,b$ coupled to the spin-$\ell$ state. The amplitude is fixed by Lorentz invariance to be: 
\eq
M_3(1^a,2^b,\epsilon_I)=ic_{\ell}(p_1-p_2)^{\mu_1}\cdots(p_1-p_2)^{\mu_\ell}\epsilon_{I\mu_1\cdots\mu_\ell}\,,
\eqe
where $c_\ell$ is the coupling constant, $\epsilon_{I\mu_1\cdots\mu_\ell}$ is the polarization tensor and $I$ labels the components of the spin-$\ell$ representation of the SO($D$-1) massive Little group. The residue is then:
\eq\label{Product}
 \sum_{I} M_3(1^a,2^b,\epsilon_I)M_3(3^b,4^a,\epsilon_I)\,.
\eqe
Denoting $(p_1{-}p_2)$ and $(p_3{-}p_4)$ as $(X,Y)$, in the center of mass (c.o.m) frame these are ($D{-}1$){-}dimensional vectors. The sum over the $I$ converts the product of polarization tensors into a polynomial of $\eta_{\mu\nu}$s, which is symmetric and traceless in the Lorentz indices on both sides of the factorization pole. This suggests that eq.(\ref{Product}) is simply a polynomial function of $(X^2,Y^2,X\cdot Y)$ that is of degree $\ell$ in $X$ and $Y$ respectively, and vanishes under the Laplacian $\nabla^2_X$ and $\nabla^2_Y$. The last constraint is a reflection of the traceless condition. In other words, one can read off the polynomial from the $D{-}1$ dimension solution to the Laplace equation:
\eq
\frac{1}{(X^2-2 X\cdot Y+Y^2)^{\frac{D{-}3}{2}}}\,.
\eqe
Without loss of generality, we can scale $|X|=1$, $|Y|=r$, and $X\cdot Y=r\cos\theta$, where $\theta$ is the scattering angle. Then the polynomial can be identified through 
\eq\label{GegDef}
\frac{1}{(1-2r\cos\theta+r^2)^{\frac{D{-}3}{2}}}=\sum_{\ell=0}^\infty r^\ell G^{\textrm{\tiny (D)}}_\ell(\cos\theta)\,.
\eqe
which is the generating function for the Gegenbauer polynomials. For $D=4$ this reduces to Legendre polynomial, while the three-dimensional counter part is the Chebyshev polynomials. From now on we will suppress the superscript $(D)$ unless needed.

We've seen that the residue is simply a sum of Gegenbauer polynomials. Now due to our specific choice of external states, $M(1^a,2^b,3^b,4^a)$, the three-point couplings on both sides of the ($u$) $s$-channel exchange are identical, i.e. the coupling constants squared $c_{\ell}^2$. Thus we see that the residue is a function that is \textit{positively expandable} on the Gegenbauer basis:
\eqa\label{Res1}
\framebox[8cm][c]{$\displaystyle Res_{s{=}m^2} M(s,t)= -\sum_\ell \textsf{p}_\ell\, G_{\ell}(\cos\theta),\quad \textsf{p}_\ell\geq 0\,,$}
\eqae
where $\cos\theta=1{+}\frac{2t}{m^2}$.  Functions that have such property are referred to as positive functions, and they enjoy the feature that such positivity is preserved under multiplication and differentiation. Note that since Gegenbauer polynomials are positive when $\theta=0$, the optical theorem is simply a corollary of eq.(\ref{Res1}). Gegenbauer polynomials is a particular example of orthogonal polynomials that are orthogonal to each other under prescribed integration measure.  Gegenbauer polynomials are orthogonal with respect to SO($D{-}1$) invariant measure $(\sin\theta)^{D{-}4}d\cos\theta$. Since SO($D{-}1$) symmetry is simply a reflection of our kinematic setup, it is applicable for discontinuities as well. Indeed as we will demonstrate in appendix \ref{ApennA}, when combined with unitarity, the discontinuity in the near forward limit is again given by a positive sum of Gegenbauer polynomials:
\eqa\label{Dis1}
\framebox[10cm][c]{$\displaystyle Dis_{s\geq 4m^2} M(s,t)=- \sum_{\ell}\textsf{p}_{\ell}(s)\, G_{\ell}(\cos\theta)\,,\quad \textsf{p}_{\ell}(s)\geq 0\,.$}
\eqae
Here, $\textsf{p}_{\ell}(s)$ is the positive ``spinning" spectral function. Note that at weak couplings, $\textsf{p}_{\ell}>0$ is all we can say. The full non-linear constraint implied by unitarity, $Im[\mathbf{a}_{\ell}(s)]\geq |\mathbf{a}_{\ell}(s)|^2$ where $\mathbf{a}_\ell$s are the partial wave coefficients, is only relevant for theories where the amplitudes becomes genuinely large/the theory is genuinely strongly coupled in the UV. 

While the discussion so far is applicable the scattering amplitude of scalars, and hence scalar EFT, one can easily generalize when ever the three-point couplings of two massless one massive state are kinematically unique. This is the case in four-dimensions with external helicity states~\cite{MassiveTree}, where the corresponding orthogonal polynomials are Jacobi polynomials. We will review and discuss its property in great detail in sec.\ref{Spinning}.

Now that we see the residue/discontinuity of the four-point amplitude is given by a special class of functions, positive functions, we would like to extract the image of this property on the space of low energy couplings. Naturally this can be done through eq.(\ref{Relate0}).  In other words, we would like to explore the full implication of: 
\eq\label{Master10}
\,a_{k,q}=\left.\frac{1}{q!}\frac{d^q}{d t^q}\left(\sum_{a}\frac{\textsf{p}_{a} G_{\ell_a}(1+2\frac{t}{m_a^2})}{(m^2_a)^{k{-}q{+}1}}+\sum_{b}\int ds'\textsf{p}_{b,\ell}(s')\frac{G_{\ell}(1+2\frac{t}{s'})}{(s')^{k{-}q{+}1}}+\{u\}\right)\right|_{t=0}\,,
\eqe
where the equality is understood to hold as a Taylor series in $t$. i.e. $|t| \ll m^2$. More precisely, \textit{coefficients of the higher dimensional operators as an expansion away from the forward limit, must be given as a positive sum of  the Taylor expansion of Gegenbauer polynomials.} Note that since the difference between contributions from residues and discontinuities is simply whether the spectrum of mass is discrete or continuous, by not assuming discreteness we will cover both. In this context,  the previous forward limit positivity constraint at is really the $q=0$ ``tip" of the iceberg. It is coarse grained because it did not fully exploit the fact that the residue and discontinuity is a positive function.

Collecting the low energy couplings, eq.(\ref{Master10}) is equivalent to:
\eq\label{Master1}
\sum_{k,q}\,a_{k,q}s^{k{-}q}t^q=-\sum_{a}\textsf{p}_{a}G_{\ell_a}\left(1{+}\frac{2t}{m_a^2}\right)\left(\frac{1}{s{-}m^2_a}{-}\frac{1}{s{+}t{+}m^2_a}\right)\,,
\eqe
where again the equality is understood in the sense of Taylor expansion in $t,s$. In other words, the near forward limit low energy expansion is captured by the $s$ and $u$-channel factorizations alone. Now eq.(\ref{Master1}) is gives us a relation between $a_{k,q}$ and the Taylor coefficients of the Gegenbauer polynomials expanded around 1, 
\eq
G_{\ell}(1+2\delta)=\sum_{q=0}v_{\ell,q}\;\delta^q\,,%,\quad v^{\alpha}_{\ell,q}=\frac{1}{q!}\left.\frac{d^q}{d x^q}G^{\alpha}_{\ell}(x)\right|_{x=1}%
\eqe
If we only have $s$-channel contribution, eq.(\ref{Master1}) implies:
\eq\label{Master0a}
\framebox[8cm][c]{$\displaystyle s\;{\rm channel}:\;\;a_{k,q}=\sum_{a}\textsf{p}_{a}\frac{v_{\ell_a,q}}{(m^2_a)^{k+1}} \quad \textsf{p}_{a}\geq0$}
\eqe
If $u$-channel contributions are present, we redefine the coupling in terms of expanding in $(t,z)$, i.e. $a_{k,q}z^{k{-}q}t^q$, we find eq.(\ref{Master1}) can instead be rewritten as:
\eq\label{Master1a}
\framebox[8cm][c]{$\displaystyle  s{-}u\;{\rm channel}:\;\; a_{k,q}=\sum_{a}\textsf{p}_{a}\frac{u_{\ell_a,k,q}}{(m^2_a)^{k+1}} \quad \textsf{p}_{a}\geq0$}
\eqe
where $u_{\ell,k,q}$ is a linear combination of $v_{\ell,q}$ with its explicit form  given in eq.(\ref{Sum}). For $q=0$, $u_{\ell,k,0}>0$  and we are back to the old forward limit positivity constraint. For $q\neq0$, $u_{\ell,k,q}$ can have either sign and we no longer have strict positive bounds for individual $a_{k,q}$, and naively there is no constraint. However, while there may no longer be constraint for individual $a_{k,q}$ with $q\neq0$, there are non-trivial constraints as a collective. For example collecting the coefficients with fixed $k$ but distinct $q$ into a vector $\mathbf{a}_k$, we find
\eq
\mathbf{a}_k\equiv\left(\begin{array}{c}a_{k,0} \\ a_{k,1} \\ a_{k,2} \\ \vdots\end{array}\right), \quad \vec{u}_{\ell,k}\equiv\left(\begin{array}{c}u_{\ell,k,0} \\ u_{\ell,k,1} \\ u_{\ell,k,2} \\ \vdots\end{array}\right)\quad \Rightarrow \;\; \mathbf{a}_k=\sum_{a}\textsf{p}_{a}\vec{u}_{\ell_a,k}\quad  \textsf{p}_{a}\geq0\,,
\eqe
where we absorbed the positive factors $(m^2_a)^{k+1}$ into $\textsf{p}_{a}$. In other words, $\mathbf{a}_k$ must be in the \textit{convex hull} of the vectors $\vec{u}_{\ell,k}$! That is the boundary of ``theory space", the space of allowed $\mathbf{a}_k$, is given by the boundaries of the hull.

Let us ``see" explicitly examples of what this space looks like. For simplicity consider color ordered EFT amplitude whose  UV completion does not include $u$-channel contributions. Taking $k=1$ we find that eq.\ref{Master1a} tells us: 
\eq
\mathbf{a}_2=\left(\begin{array}{c}a_{1,0} \\ a_{1,1}  \end{array}\right)=\sum_{a}\textsf{p}_{a}\left(\begin{array}{c}v_{\ell_a,0} \\ v_{\ell_a,1} \end{array}\right)\,.
\eqe
Since $\textsf{p}_{a}$ is positive, the equality is projective in nature and we can rescale the top component of each vector to be $1$. This then implies the following inequality, 
\eq
\frac{a_{2,1}}{a_{2,0}}\geq Min\left[\frac{v_{\ell,1}}{v_{\ell,0}}\right]
\eqe 
Taking $D=4$, we have $v_{\ell,0}=1$ and $v_{\ell,1}=\ell (\ell{+}1)$, and we conclude that $\frac{a_{2,1}}{a_{2,0}}\geq0$. For $k=2$, the vector $\mathbf{a}_3$ lives in $\mathbb{P}^2$
\eq
\mathbf{a}_3=\left(\begin{array}{c}a_{2,0} \\ a_{2,1}\\  a_{2,2}\end{array}\right)=\sum_{a}\textsf{p}_{a}\left(\begin{array}{c}v_{\ell_a,0} \\ v_{\ell_a,1}\\ v_{\ell_a,2} \end{array}\right)\quad\rightarrow\quad \left(\begin{array}{c} a_{2,1}/a_{2,0}\\  a_{2,2}/a_{2,0}\end{array}\right)=\sum_{a}\textsf{p}_{a}\left(\begin{array}{c} v_{\ell_a,1}\\ v_{\ell_a,2} \end{array}\right)\,.
\eqe
where after the rescaling, besides  $\textsf{p}_{a}\geq0$, we further have $\sum_{a}\textsf{p}_{a}=1$. Using $v_{\ell,2}=\frac{(1)_{\ell{+}2}}{4(\ell{-}2)!}$, the allowed region is now given as: 
$$\left.\begin{array}{c} \;\\ \;\\ \;\\ \frac{a_{2,2}}{a_{2,0}} \\ \ \;\\ \;\\ \;\\ \; \end{array}\right.\vcenter{\hbox{\includegraphics[scale=0.35]{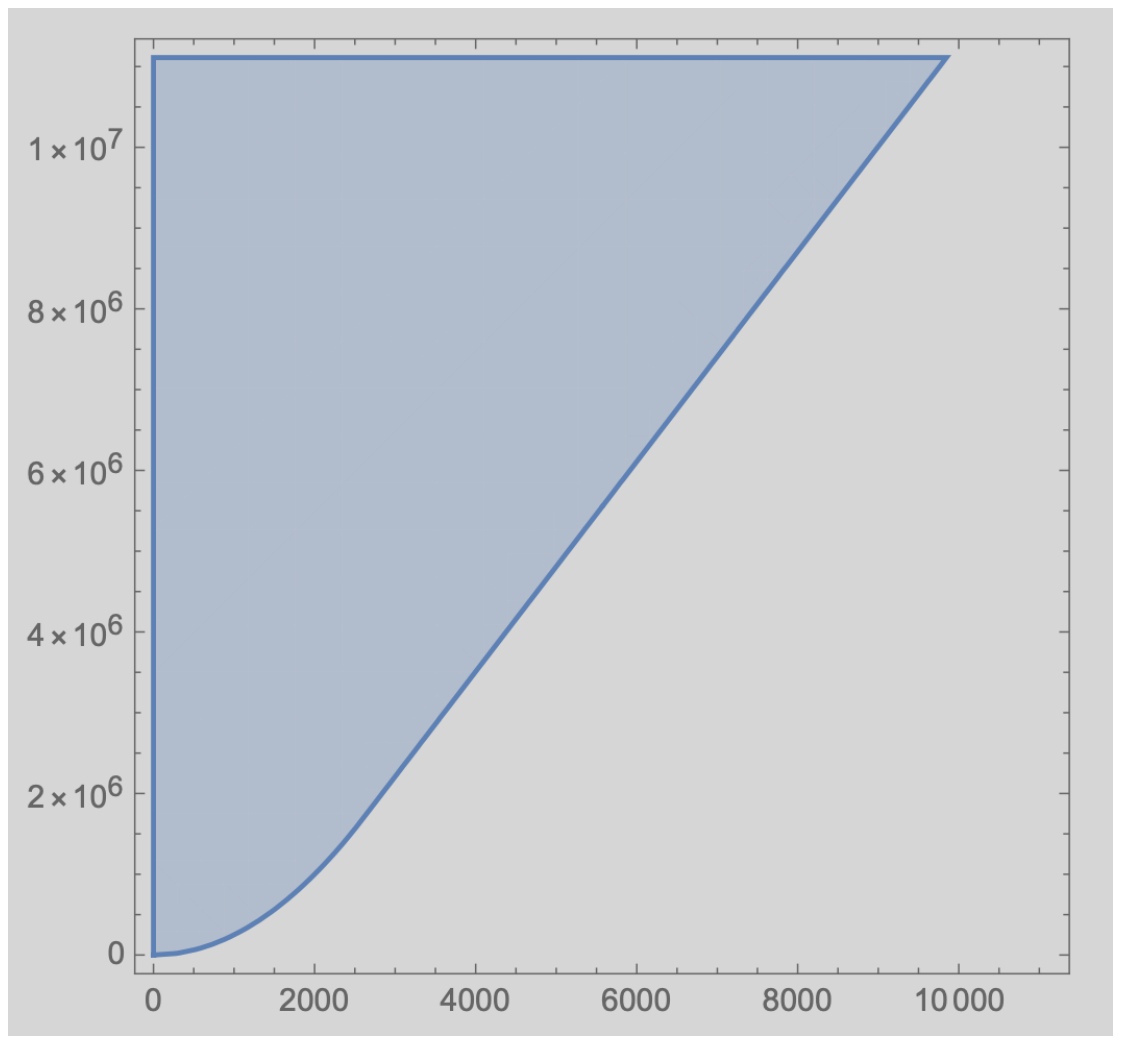} }}\left.\begin{array}{c} \;\\ \;\\ \;\\ \;\\ \;\\ \;\\ \;\\ \;\\ \;\\  \frac{a_{2,1}}{a_{2,0}} \end{array}\right.\,$$
Once again, the positivity bound of ~\cite{Adams:2006sv} simply tells us that $a_{2,0}>0$ and thus has no constraint for the above plot.  As we extend to higher degree in $k$, eq.(\ref{Master0a}) and eq.(\ref{Master1a}) becomes the statement that $a_{k,q}$ lives in the convex of vectors $\vec{v}_\ell$ and $\vec{u}_{\ell,k}$ for fixed $k$, and the relevant question is what are the boundaries of this hull.

In general the spin is unbounded especially when the UV completion involves massive loops, thus the number of vectors that constitute the hull is infinite. Naively determining the boundaries of such space is computationally prohibitive. Note that these polytopal constraints, being for fixed $k$, bound operators of the same dimension. At the same time, we should expect non-trivial constraints that are cross dimensional since operators of different dimension are constrained by the same UV completion.  As we will see these fascinating questions have a beautiful geometric answer to be explored in the remaining sections.

%%%%%%%%%%%%%%%%%%%%%%%%%%%%%%%
\section{Hidden total positivity from unitarity and locality}\label{OldFashion}
%%%%%%%%%%%%%%%%%%%%%%%%%%%%%%%
In this section we briefly review the positive geometries relevant for our analysis. The spaces that we will be interested in are invariantly constructed as a \textit{positive} sum of a fix set of vectors $\{\mathbf{V}_a\}$:
\eq\label{Example}
\mathbf{a}\in\sum_{a} \textsf{p}_{a} \mathbf{V}_a,\quad \textsf{p}_{a}>0\,.
\eqe  
Such construction are referred to as convex hulls and the resulting geometry convex polytopes. Given a convex polytope, we will seek the complete set of inequalities that defines its interior. In other words we would like to ``carve out"  the subspace satisfying eq.(\ref{Example}) through equations of the form:
\eq
f_i(\mathbf{a})>0\,.
\eqe
In the above $i$ labels the distinct constraints. Depending on the nature of vectors,  we will find that $f_i$ can be either linear or non-linear functions of $\mathbf{a}$. In the context of constraints for EFT, $\mathbf{a}$ is identified with the space of EFT couplings $\{a_{k,q}\}$ and the vectors $\mathbf{V}_a$ are determined by Lorentz invariance and locality, properties that we assume for the UV completion.

%%%%%%%%%%%%%%%%%%%%%%%%%%%%%%%
\subsection{Convex hulls and Cyclic polytopes}
\begin{figure}
\begin{center}
\includegraphics[scale=0.5]{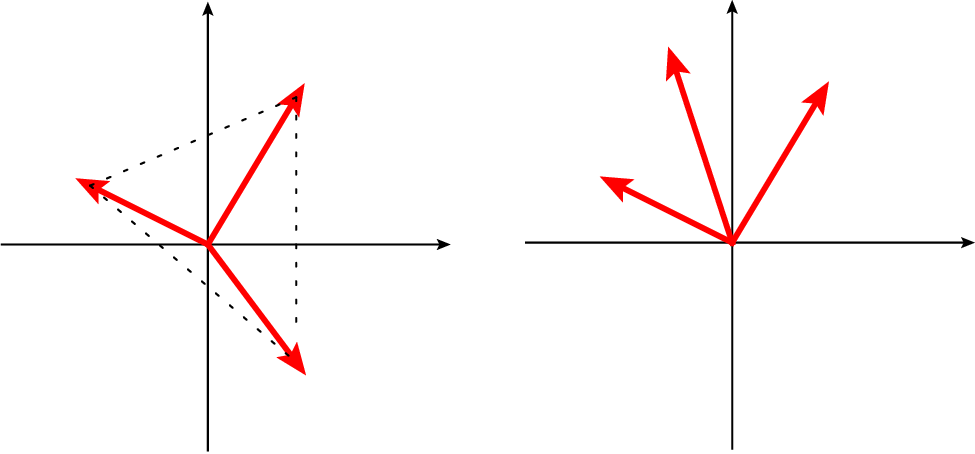}
\caption{The convex hull of these three vectors encloses the origin, and hence trivially covers the entire two-dimensional plane.}
\label{trivial}
\end{center}
\end{figure}

%%%%%%%%%%%%%%%%%%%%%%%%%%%%%%%
Let us begin with the definition of convex hull. Given a set of $d{+}1$-dimensional vectors $\mathbf{V}_a$, consider the subspace spanned by its positive weighted sum:
\eq\label{Def31}
\mathbf{a} \in \sum_{a} \textsf{p}_{a} \mathbf{V}_a, \quad \textsf{p}_{a}>0\,.
\eqe
The number of vectors will in general be greater than the dimension, and one must first determine whether this span the whole space. For example consider three vectors in two dimensions as in fig.(\ref{trivial}). In the first case the three vectors span the whole space, as any point on the two-dimensional plane can be written as some positive sum of the three vectors. This is not the case for the second configuration since all vectors are on one side of the horizontal axes. Thus in order for the hull to be non-trivial, all the vectors must be on the same side of some hyper plane, or equivalently there are no non-trivial solutions to 
\eq\label{ceq}
\sum_a \textsf{p}_{a} \mathbf{V}_a,=0 \quad \textsf{p}_{a}>0\,,
\eqe
i.e. the vectors do not enclose the origin.

Clearly for any $\mathbf{a}$ that satisfies eq.(\ref{Def31}), so will $\rho\mathbf{a}$ with $\rho>0$. Thus the solution space is naturally projective, and we identify  $\mathbf{a}\sim \rho\mathbf{a}$ and $\mathbf{V}_a\sim \rho_a \mathbf{V}_a$. Since all the vectors lie on the same side of some hyperplane, we can choose our coordinates such that the top component is always positive, which we choose to normalize to $1$:
\eq
\mathbf{V}_a=\left(\begin{array}{c}1 \\ \vec{v}_a\end{array}\right), \quad \mathbf{X}=\left(\begin{array}{c}1 \\ \vec{x}\end{array}\right)\,.
\eqe
In terms of $(\vec{v}_a, \vec{x})$ the canonical definition of convex hull is written as:
\eq
Conv[\vec{v}_a]=\left\{\left.\sum_a \textsf{p}_a\; \vec{v}_a,\quad \right|(\forall\,a: \textsf{p}_a>0)\wedge \sum_a \textsf{p}_a=1\right\}\,.
\eqe
As we will see it will be useful to retain the use of homogeneous coordinates, i.e. considering the vectors in its full $(d{+}1)$- component, and consider the hull as a projective polytope in $\mathbb{P}^d$:
\eq\label{CanonicalDef}
Conv[\mathbf{V}_a]=\left\{\left.\sum_a \textsf{p}_a\; \mathbf{V}_a,\quad \right|(\forall\,a: \textsf{p}_a>0)\right\}\,.
\eqe
The advantage of this is that it allows us to define various co-plane or incidence conditions projectively with the help of of the $d{+}1$-dimensional Levi-Cevita tensor, $\epsilon_{I_1I_2\cdots I_{d{+}}1}$. For example, for the 3 vectors to be on a line in $\mathbb{P}^2$ we have 
\eq
\langle a, b, c\rangle\equiv\epsilon_{I_1I_2I_3}V^{I_1}_aV^{I_2}_bV^{I_3}_c=0\,.
\eqe 
, where $I_i=1,2,3$. Similarly for $d{+}1$ vectors to lie on a $d{-}1$-dimensional plane in $\mathbb{P}^d$, tells us that the bracket $\langle a_1, a_2,\cdots,a_{d{+}1}\rangle=0 $. In this paper, the dimension of the angle brackets $\langle \cdots \rangle$ will be implicit from the number of entires or the surrounding discussions.

While eq.(\ref{CanonicalDef}) gives us a $d$-dimensional polytope, not all vectors in $\mathbf{V}_a$ are vertices of the polytope,  some might be \textit{inside}. Thus given a convex hull, one needs to identify the vectors that constitute the vertices which ultimately defines the polytope. The polytope can equivalently be defined through its boundaries, which are a set of co-dimension one hyper-planes or facets. The advantage of such facet point of view is that the polytope can be carved out successively one facet at a time. Not surprisingly, these facets can also be defined through the vertices of the polytope. More precisely, a co-dimension one plane is defined by a set of $d$ distinct vectors, say $(\mathbf{V}_{a_1},\mathbf{V}_{a_2},\cdots,\mathbf{V}_{a_d})$. We can represent this plane as a $d{+}1$ component dual vector $\mathbf{W}_{i}$, where $i$ labels the set of $\{a_i\}$ that defined the plane, and its components given by:
\eq\label{Wdef}
(W_i)_I\equiv \epsilon_{II_1I_2\cdots I_d}V^{I_1}_{a_1},V^{I_2}_{a_2},\cdots,V^{I_d}_{a_d}=\langle *,a_1,a_2,\cdots, a_d\rangle\,.
\eqe 
Then the inside of polytope is then given by the condition that $\mathbf{a}$ lies on one side of the facet $\mathbf{W}_{i}$. This constraint can be phrased in terms of a positivity condition:  
\eq\label{BoundariesCond}
\mathbf{W}_{i}\cdot \mathbf{a}=(W_i)_I a^I=\langle \mathbf{a},{a_1}, {a_2},\cdots ,{a_d}\rangle>0,\quad \forall \mathbf{a}\in Conv[\mathbf{V}_a]\,.
\eqe
It is useful to see how such constraint arrises in simple setup. Consider a polygon in $\mathbb{P}^2$:
$$\includegraphics[scale=0.5]{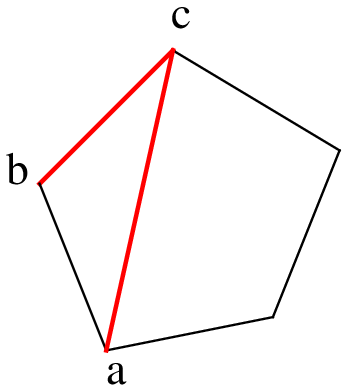}\,.$$
The line $\overline{bc}$ is a boundary since the interior of the polygon is on one side of the line. This is not the case for $\overline{ac}$. Not only does points of the interior lie on both sides, it can be \textit{on} the line, i.e.  collinear with $(a,c)$. Since collinear means $\langle \mathbf{a},a ,c\rangle=0$, this implies that $\langle \mathbf{a},a ,c\rangle$ is positive on one side of $\overline{ac}$, and negative one the other. Thus if $\mathbf{W}_{i}$ is a boundary, $\mathbf{W}_{i}\cdot\mathbf{a}$ must have the same sign for all $\mathbf{a}$, which we can always chose to be positive by appropriately arranging the sequence of vectors in $\{a_i\}$ eq.(\ref{Wdef}).

Given the complete set of $\{\mathbf{W}_i\}$, we now have a set of inequalities $f_{i}(\mathbf{a})>0$ that carves out the space. The function $f_i$ in this case is linear in $\mathbf{a}$:
\eq\label{Polytopal}
f_i(\mathbf{a})=\mathbf{W}_i\cdot \mathbf{a}\geq0\,.
\eqe 
The equal sign refers to points that are \textit{on} the boundary. Now one can see that given a set of vectors $\mathbf{V}_a$, to determine the full set of $\{\mathbf{W}_i\}$, one would need the to compute the sign of $\langle a_1,a_2,\cdots, a_{d{+}1}\rangle$ for all $d{+}1$-tuples. The sign patterns will tell us which vectors are vertices that form facets, and which ones are inside. For $n$ vectors, this involves the computation of $\left(\begin{array}{c} n \\ d{+}1 \end{array}\right)$ number of $d{+}1\times d{+}1$ determinants, which becomes intractable for large $n$. In the context of our EFT setup, $n$ is associated with the number of Gegenbauer polynomials which is infinite. Thus the problem appears intractable, unless some reasonable truncation can be established. As we will now see, if the vectors satisfy special positivity conditions, the boundary and the vertices can be straight forwardly determined before hand. Remarkably, for us these properties are readily satisfied as a consequence of Lorentz invariance and locality of the UV completion!

\noindent \textbf{Cyclic polytopes} Let's start with a set of vectors $\mathbf{V}_{a}$ that are endowed with some preferred ordering. If all ``ordered" $d{+}1\times d{+}1$ determinants are positive:  
\eq\label{PositiveDet}
\langle a_1,a_2,\cdots,a_{d{+}1}\rangle>0,\quad \forall a_1>a_2>\cdots>a_{d{+}1}\,,
\eqe
then the convex hull $Conv[\mathbf{V}_{a}]$ yields a  \textit{cyclic polytope}.  The canonical example for a cyclic polytope is the convex hull of points on a moment curve. A moment curve is the embedding of the real line in $d$-dimensional space, such that each point on the line maps to a $d$-component vector with successive ``moments", i.e.  $(z,z^2,\cdots,z^d)$, with $z\in\mathbb{R}$. The convex hull of points on a moment curve is then a positive weighted sum of vectors taking the form: 
\eq
\mathbf{V}_a=\left(\begin{array}{c}1\\z_a \\z^2_a \\ \vdots \\z^d_a \end{array}\right)\,.
\eqe
Naturally, $\mathbf{V}_a$ can be ordered by the value of $z_a$. In such case $\langle a_1,a_2,\cdots,a_{d{+}1}\rangle$ is simply the the determinant of the Vandermonde matrix:
\eq
{\rm Det}\left[\begin{array}{cccc}1 & 1 & \cdots & 1 \\ z_1 & z_2 & \cdots & z_{d{+}1} \\ (z_1)^2 & (z_2)^2 & \cdots & (z_{d{+}1})^2 \\ \vdots & \vdots & \vdots & \vdots \\ (z_1)^d & (z_2)^d & \cdots & (z_{d{+}1})^d\end{array}\right]=\prod_{i<j}(z_j-z_i)\,.
\eqe
Indeed this determinant is positive for \textit{ordered} points, $z_1<z_2<\cdots<z_{d{+}1}$.

 Given eq.(\ref{PositiveDet}) one can straight forwardly see that the boundaries for a cyclic polytope in $\mathbb{P}^d$ are simply given as:
\eqa\label{CPolyBoundaries}
d\;\in\;{\rm even}\;&&\rightarrow \mathbb{P}^2:\;\; \langle * ,i ,i{+}1\rangle,\quad \mathbb{P}^4:\;\; \langle * ,i ,i{+}1,j,j{+}1\rangle,\nonumber\\
d\;\in\;{\rm odd}\;&&\rightarrow \mathbb{P}^3:\;\; \langle 0, * ,i ,i{+}1\rangle,\;\;\langle * ,i ,i{+}1,\infty\rangle,\nonumber\\
&&\quad\; \mathbb{P}^5:\;\; \langle 0, * ,i ,i{+}1,j ,j{+}1\rangle,\;\;\langle * ,i ,i{+},j ,j{+}1,\infty\rangle,\;\;
\eqae
where $i, i{+}1$ represents vectors that are adjacent in the ordering, and $0$, $\infty$ is the first and final vector. To see that these are true boundaries, we must show for each of the walls in eq.(\ref{CPolyBoundaries}), any point inside the hull $\mathbf{a}\in Conv[\mathbf{V}_a]$ will satisfy $\langle \mathbf{a},\cdots\rangle\geq0$ or $\langle 0,\mathbf{a},\cdots\rangle\geq0$. Let's take $\langle 0, * ,i ,i{+}1,j ,j{+}1\rangle$ as an example: 
\eq
\langle 0, \mathbf{a} ,i ,i{+}1,j ,j{+}1\rangle=\sum_a\textsf{p}_{a}\langle 0, a ,i ,i{+}1,j ,j{+}1\rangle\,,
\eqe
since each bracket in the sum is even permutation away from canonical ordering, they are positive due to eq.(\ref{PositiveDet}). As $\textsf{p}_{a}>0$ the RHS is a sum of positive terms and thus establishes $\langle 0, * ,i ,i{+}1,j ,j{+}1\rangle$ being a boundary of $ Conv[\mathbf{V}_a]$.  Note that similar argument also tells us that there are no other boundaries.

Thus in summary, if the vectors $\mathbf{V}_a$ satisfy eq.(\ref{PositiveDet}), then the boundaries for $Conv[\mathbf{V}_a]$ is completely determined and constructed from consecutive pairs as illustrated in eq.(\ref{CPolyBoundaries}). Furthermore since eq.(\ref{CPolyBoundaries}) are boundaries for any $i,j,\cdots$,  all vectors are vertices.

 \subsection{Hankel matrix total positivity}\label{Sec:Hankel}
 
 Let us consider a simple example where the positive geometry of cyclic polytopes arises in our EFT discussion. Take the following four point amplitude:
 \eq\label{SampleM}
 M(s)=\sum_a -\frac{\textsf{p}_{a}}{s-m_a}\,.
 \eqe 
 This arrises naturally as the dispersive representation of the four-point amplitude in the forward limit. Note that the positivity of  $\textsf{p}_{a}$ is a reflection of unitarity and the simple pole in $s$ is a reflection of locality. Thus the geometry that arrises from eq.(\ref{SampleM}) will have its origin in the union of unitarity and locality. 
 
 Expanding eq.(\ref{SampleM}) in small $s$ we find 
\eq\label{SchematicRep3a}
\sum_{k}a_{k}\;s^{k}\;=\sum_a \frac{\textsf{p}_a}{m^{2}_a}\left(1+\frac{s}{m_a^2}+\frac{s^2}{m_a^4}+\cdots\right)\,.
\eqe   
Matching both sides of the above equation we immediately see that the $a_{k}$s are positive. But there is more!  If we collect the couplings into a vector $\vec{a}$, eq.(\ref{SchematicRep3a}) becomes: 
\eq\label{HalfMomCurve}
\mathbf{a}=\left(\begin{array}{c}1\\ a_{1}/a_0 \\ a_{2}/a_0 \\ \vdots  \\ a_{k}/a_0\end{array}\right)=\sum_{a}\textsf{p}'_a\left(\begin{array}{c}1 \\ x_a \\ \vdots  \\ x^{k}_a\end{array}\right)\,,\quad x_a\equiv \frac{1}{m^2_a}\,,
\eqe
where we've used the projective nature of the problem to rescale the top component to be $1$. We find that eq.(\ref{SchematicRep3a}) tells us that $\vec{a}$ lies in the convex hull of moment curves! Note that since $m^2_a>0$, we are really considering the ``half" moment curve where $x_a\in \mathbb{R}^+$. Using what we've learned in the previous subsection, we have 
\eq
\mathbf{W}_i\cdot \mathbf{a}\geq0
\eqe
where $\mathbf{W}_i$ are the boundaries listed in  eq.(\ref{CPolyBoundaries}) with $V_a$ determined by $x_a$ and we have an infinite number of constraint on the couplings! However these constraints are not ideal as they rely on the explicit vectors $V_a$ and for a low energy theorist we are not privy to the information of the UV spectrum, i.e. we do not know what the $x_a$s are. It would be desirable to find constraints $f_i(\mathbf{a})\geq0$ , such that the functions $f_i$ do not depend on the explicit values $x_a$, while reflecting the fact that $x_a\in\mathbb{R}^+$.

Let's start by assuming the knowledge of the spectrum and see if we can rewrite $\mathbf{W}_i\cdot \mathbf{a}>0$ in such a way that the information of the spectrum decouples. We can assume the spectrum to be continuous without lost of generality, since any of the $\textsf{p}_{a}$s can be set to be arbitrarily to match with any specific spectrum. Beginning with $d=1$, we have $\mathbf{a}=(1,\frac{a_1}{a_0})$ and there is only one boundary $\mathbf{W}=(1,0)$. Thus we have: 
\eq
\mathbf{W}\cdot \mathbf{a}=\langle 0, \mathbf{a}\rangle =\frac{a_1}{a_0}>0,
\eqe  
which is trivial since we know that $a_0,a_1>0$. For $d=2$, $\mathbf{a}=(1,\frac{a_1}{a_0}, \frac{a_2}{a_0})$ and the constraint is
\eq\label{Posg1}
\langle \mathbf{a}, a,a+1\rangle>0\,.
\eqe
Since the spectrum is continuous, given a point $x_a$ on the moment curve we can take $a{+}1$ to be arbitrarily close to $a$, such that $\langle *, a,a+1\rangle\rightarrow \langle *, a,\dot{a}\rangle$, where $\dot{a}$ represents the derivative. The determinant then becomes 
\eq
\langle \mathbf{X}, a,a+1\rangle={\rm Det}\left[\begin{array}{ccc} 1 & 1 & 0 \\ \frac{a_1}{a_0} & x_a & 1 \\ \frac{a_2}{a_0} & x_a^2 & 2x_a \end{array}\right]=\frac{a_2 - 2 a_1 x_a + a_0 x_a^2}{a_0}\,.
\eqe
We see that the minimum occurs at $x_a = \frac{a_1}{a_0}$, and thus for eq.(\ref{Posg1}) to hold we must have:
\eq\label{exp1a}
a_0a_2-a_1^2={\rm Det}\left[\begin{array}{cc}a_0 & a_1 \\ a_1 & a_2\end{array}\right]>0\,.
\eqe 
Note that is non-linear in $\mathbf{a}$ and no longer depends on the point $x_a$! Moving on to $d=3$, the analysis for $\langle 0, \mathbf{a}, a, a+1\rangle$ is identical to that for the $d=2$ case, leading to 
\eq\label{exp2a}
a_1a_3-a_2^2={\rm Det}\left[\begin{array}{cc}a_1 & a_2 \\ a_2 & a_3\end{array}\right]>0\,.
\eqe 
Two comments are in order. First note that we have not considered constraints involving the  infinity vertex. This is because projectively, the infinity vector is simply $(0,\cdots,0,1)$ and  when plugged into $\langle \cdots, a, a{+}1,\infty\rangle$, it reduces to the constraint one dimension lower. Second, as we move from even to odd dimensions, we obtain the same constraint as before only with $a_{i}\rightarrow a_{i{+}1}$, for example eq.(\ref{exp1a}) and eq.(\ref{exp2a}). This can be understood as follows: the facets in both cases are comprised of the same set of vertices, just with the inclusion of the origin $0$ for the odd case. In taking the determinant, $0$ removes the first component of each vector, and the remaining part is proportional to the vector one dimension lower. Thus the condition in the odd dimension is simply and overall factor multiplying that of one dimension lower. Importantly since we are on a half moment curve, the overall prefactor will be positive. For example:
\eq
\langle 0, \mathbf{a}, a,a{+}1\rangle={\rm Det}\left(\begin{array}{cccc}x_0 & 1 & 1 & 1 \\ a_1 & 0 & x_a & x_{a{+}1} \\ a_2 & 0 &  x^2_a & x^2_{a{+}1} \\a_3 & 0 & x^3_a & x^3_{a{+}1}\end{array}\right)=x_{a{+}1}x_{a}{\rm Det}\left(\begin{array}{ccc}  a_1  & 1 & 1 \\ a_2  &  x_a & x_{a{+}1} \\ a_3  & x^2_a & x^2_{a{+}1}\end{array}\right)\,.
\eqe
Since $x_a, x_{a{+}1}>0$, the fact that the very LHS is positive translate to the positivity on the very RHS, i.e. in eq.(\ref{exp2a}). Let's consider one more example before moving on to the general constraint. For $d=4$ we have 
\eq
\langle \mathbf{a}, a,a{+}1, b, b{+}1\rangle={\rm Det}\left[\begin{array}{ccccc}1 & 1 & 0 & 1 & 0 \\ \frac{a_1}{a_0} & x_a & 1 & x_b & 1 \\ \frac{a_2}{a_0} & x_a^2 & 2x_a & x_b^2 & 2x_b \\ \frac{a_3}{a_0} & x_a^3 & 3x_a^2 & x_b^3 & 3x_b^2 \\ \frac{a_4}{a_0} & x_a^4 & 4x_a^3 & x_b^4 & 4x_b^3\end{array}\right]=(x_a{-} x_b)^4 (a_4 {-} 2 \alpha a_3  {+} a_2 (\alpha^2 {+} 2 \beta) {+} 
   \beta (a_0 \beta {-} 2 a_1 \alpha))\,,
\eqe 
where $\alpha=(x_a{+}x_b)$ and $\beta=x_ax_b$. The minima in terms of $\alpha$ occurs at $\alpha=\frac{\beta a_1+a_3}{a_2}$. Plugging into the RHS of the above and requiring it to be positive leads to:
\eq
\quad {\rm Det}\left[\begin{array}{ccc}a_0 & a_1 & a_2 \\ a_1 & a_2 & a_3 \\ a_2 & a_3 &a_4\end{array}\right]>0\,.
\eqe

We are now ready to give the result for general $d$. Collecting the coefficients of $\vec{a}$ into the symmetric Hankel matrix:
\eq\label{Hankel}
K(\vec{a})=\left(\begin{array}{cccc}a_0 & a_1 & \cdots & a_{p{-}1} \\ a_1 & a_2 & \cdots & a_{p} \\ \vdots & \vdots & \vdots & \vdots \\  a_{p{-}1} & a_{p} & \cdots & a_{2p{-}2}\end{array}\right)\,,
\eqe
\textit{then the coefficients are in the convex hall of the half-moment curve if and only if the Hankel matrix is a totally positive matrix!} A totally positive matrix has the property that all of its minors are non-negative. This is the well known solution to the Stieltjes moment problem. Note that due to $K$ being a symmetric matrix, not all minors are independent. The independent constraints are the positivity of the principle minors of $K(\vec{a})$ and $K(\vec{a})_{i\rightarrow i{+}1}$. That is   
\eq\label{Constraint}
\framebox[16cm][c]{$ i\in even:\quad {\rm Det}\left[\begin{array}{cccc}a_0 & a_1 & \cdots & a_{\frac{i}{2}} \\ a_1 & a_2 & \cdots & a_{\frac{i}{2}+1} \\ \vdots & \vdots & \vdots & \vdots \\  a_{\frac{i}{2}} & a_{\frac{i}{2}+1} & \cdots & a_{i}\end{array}\right]\geq0,\quad \quad i\in odd:\quad {\rm Det}\left[\begin{array}{cccc}a_1 & a_2 & \cdots & a_{\frac{i+1}{2}} \\ a_2 & a_3 & \cdots & a_{\frac{i+3}{2}} \\ \vdots & \vdots & \vdots & \vdots \\  a_{\frac{i+1}{2}} & a_{\frac{i+3}{2}} & \cdots & a_{i}\end{array}\right]\geq0$}
\eqe
Its validity can be seen by the analytic representation of eq.(\ref{Constraint}): 
\eqa\label{Det}
i\in even:&&\sum_{\{b_1,b_2,\cdots,b_{\frac{i}{2}{+}1}\}}\left(\prod_{k=1}^{\frac{i}{2}{+}1}\;\textsf{p}_{b_k}\right)\prod_{1\leq k<l\leq \frac{i}{2}{+}1 }\left(x_{b_k}-x_{b_l}\right)^2,\nonumber\\
i\in odd:&&\sum_{\{b_1,b_2,\cdots,b_{\frac{i+1}{2}}\}}\left(\prod_{k=1}^\frac{i+1}{2}\;\textsf{p}_{b_k}x_{b_k}\right)\prod_{1\leq k<l\leq \frac{i+1}{2} }\left(x_{b_k}-x_{b_l}\right)^2
\eqae
For $i\in even$ it is manifestly positive, thus must hold for the convex hull of general moment curves. Indeed this was already noted in~\cite{Orbit}. For $i\in odd$, its positivity then relies on $x_a>0$, and thus only hold for the convex hull of \textit{half} moment curves.

%%%%%%%%%%%%%%%%%%%%%%%%%%%%%%%%%%%%%%%%%%%%%%%%%%%%%%%%%%%%%%%%%%%%%%%%
\subsection{The Gegenbauer cyclic polytopes}
%%%%%%%%%%%%%%%%%%%%%%%%%%%%%%%%%%%%%%%%%%%%%%%%%%%%%%%%%%%%%%%%%%%%%%%%
We now turn to the positivity associated with the Gegenbauer polynomials. From the its definition from the generating function in eq.(\ref{GegDef}), it is straight forward to see that $G^{(n)}_\ell(1)\equiv\frac{1}{n!}\partial^n_zG_\ell(z)|_{z=1}\geq0$. However, just as the case with moments and Vandermonde determinants, further positive properties can be found when the components are organized into matrices. Let us consider the following Gegenbauer matrix
\eq\label{PositionPos}
{\rm Det}\left[\begin{array}{cccc}G_{\ell_1}(z_1) & G_{\ell_2}(z_1) & \cdots & G_{\ell_n}(z_1) \\ G_{\ell_1}(z_2) & G_{\ell_2}(z_2) & \cdots & G_{\ell_n}(z_2) \\ \vdots & \vdots & \vdots & \vdots \\ G_{\ell_1}(z_n) & G_{\ell_2}(z_n) & \cdots & G_{\ell_n}(z_n)\end{array}\right]\,.
\eqe
It turns out, the above matrix is totally positive if $1\leq z_1<z_2<\cdots z_n$ and $\ell_1<\ell_2<\cdots<\ell_n$. For Chebychev polynomials, which are the Gegenbauer polynomials in $D=3$, this can be straightforwardly proven, and we present the result in  appendix \ref{PositiveProof}. For general $D$, the proof follows from that presented by Karlin and McGregor for general orthogonal polynomials~\cite{GeneralProof}.  In appendix \ref{PositiveProof}, we also give a direct computation of the relevant determinants for the Gegenbauer case of interest to us, allowing us to see the positivity explicitly

Such ``position space" positivity, where the $z_i$s are evaluated at separate points, is not convenient for our EFT analysis. In anticipating the Taylor expansion in eq.(\ref{Master10}), we would like to instead extract conditions on the derivatives of the polynomials. This can be done by taking the positions to be close to some common point, say $1$. Then the determinant of the Gegenbauer matrix becomes that for derivatives of Gegenbauer polynomial evaluated at $z_i=1$. For example, defining 
\eq
\mathbf{G}_\ell\equiv\left(\begin{array}{c}G^{(0)}_\ell(1) \\ G^{(1)}_\ell(1) \\ G^{(2)}_\ell(1)  \\ \vdots \\ G^{(n)}_\ell(1)  \end{array}\right)\,,
\eqe
 the determinant of the Gegenbauer matrix with $1\leq z_1<z_2<\cdots z_n<1+\epsilon$ becomes the determinant of the  ``Taylor" scheme matrix
\eqa\label{GegnebauerPosFun}
(\mathbf{G}_{\ell_1}(1),\,\mathbf{G}_{\ell_2}(1),\,\cdots,\,\mathbf{G}_{\ell_{n{+}1}}(1))\,.
\eqae
Thus the positivity of the Gegenbauer matrix in position space will imply the determinant of the above matrix is positive. Let's write out the explicit Taylor coefficients:
\eq\label{TaylorDef}
G_\ell(1+2\delta)=\sum_{q=0}^\ell v_{\ell,q}\delta^q\,,\quad v_{\ell,q}=\left\{\begin{array}{cc} \frac{2^q}{q!(\ell-q)!}\frac{(\alpha)_{\ell+q}}{\prod_{a=1}^q(\alpha+2a-1)} &\;{\rm for}\,q\leq\ell\\ 0&\;{\rm for}\,q>\ell\end{array}\right.\,,
\eqe
where $\alpha=D{-}3$. Note that the coefficients are all positive, which reflects the fact that the derivative of $G_\ell(x)$ is again a positive function.\footnote{This can be deduced by taking the derivative on the generating function. Such extended positivity away from the forward limit was suggested long ago in~\cite{Martin:1965jj}, and utilized as consistency conditions for EFT in~\cite{Nicolis:2009qm}, deriving bounds in ~\cite{Bellazzini:2017fep}. } Using this one can show that the determinant of eq.(\ref{GegnebauerPosFun}) is (see appendix \ref{PositiveProof}):
\eq\label{Proof1}
{\rm Det}\left[\mathbf{G}_{\ell_1},\,\cdots,\,\mathbf{G}_{\ell_{n{+}1}}\right]=2^{\frac{n(1{+}n)}{2}}\left(\prod^{n{+}1}_{i=1}\frac{(\alpha)_{ \ell_i}}{\ell_i!}\frac{1}{\prod_{a=1}^{i-1}(\alpha{+}2a{-}1)a!}\right)\prod_{i<j}(\ell_j{-}\ell_i)(\alpha{+}\ell_j{+}\ell_i)\,,
\eqe
which is manifestly positive for ordered spins, $\ell_1<\ell_2<\cdots<\ell_{d{+}1}$. This immediately tells us that
\eq
 \framebox[10cm][c]{$ \textit{the convex hull of the $\mathbf{G}_{\ell}$ is a cyclic polytope}! $}
 \eqe
Thus just as for the convex hull of points on the moment curve, the boundaries for $Conv[\mathbf{G}_{\ell}]$ are simply given by:
\eqa\label{GegPolyBoundaries}
&&d\;\in\;{\rm even}\;\rightarrow \mathbb{P}^2:\;\; \langle * ,\ell_i ,\ell_i{+}1\rangle,\quad \mathbb{P}^4:\;\; \langle * ,\ell_i ,\ell_i{+}1,\ell_j,\ell_j{+}1\rangle,\;\;\nonumber\\
&&d\;\in\;{\rm odd}\;\rightarrow \mathbb{P}^3:\;\; \langle 0, * ,\ell_i ,\ell_i{+}1\rangle,\;\;\langle * ,\ell_i ,\ell_i{+}1,\infty\rangle,\nonumber\\
&&\quad\quad\quad\quad\quad\;\;\; \mathbb{P}^5:\;\; \langle 0, * ,\ell_i ,\ell_i{+}1,\ell_j ,\ell_j{+}1\rangle,\;\;\langle * ,\ell_i ,\ell_i{+},\ell_j ,\ell_j{+}1,\infty\rangle\,.\nonumber\\
\eqae

Going back to the position space Gegenbauer matrix, instead of setting all of the positions close to $1$, lets have $z^*\leq z_1<z_2<\cdots<z_{n}<z^*+\delta$, with $1<z^*$, the eq.(\ref{PositionPos}) becomes
\eq
{\rm Det}\left[\begin{array}{cccc}G_{\ell_1}(z_1) & G_{\ell_2}(z_1) & \cdots & G_{\ell_n}(z_1) \\ G_{\ell_1}(z_2) & G_{\ell_2}(z_2) & \cdots & G_{\ell_n}(z_2) \\ \vdots & \vdots & \vdots & \vdots \\ G_{\ell_1}(z_n) & G_{\ell_2}(z_n) & \cdots & G_{\ell_n}(z_n)\end{array}\right]={\rm Det}\left[\mathbf{G}_{\ell_1}(z^*),\,\cdots,\,\mathbf{G}_{\ell_{n}}(z^*)\right]>0\,.
\eqe
Thus the convex hull of $\mathbf{G}_{\ell}(z^*)$ is in fact a cyclic polytope for all $z^*\geq1$! Now consider a series of cyclic polytope,
\eq\label{PolyZiDef}
Poly_i=Conv[\mathbf{G}_{\ell}(z_i)]\,.
\eqe
defined with with $1\leq z_1<z_2<\cdots$. Since the derivative of $G_\ell(z)$ is a positive function, i.e. 
\eq
\frac{dG_\ell(z)}{dz}=\sum_{\ell'}c_{\ell\ell'} G_{\ell'}(z)\quad  c_{\ell\ell'} \geq0
\eqe
we can deduce
\eq
\mathbf{G}_{\ell}(z+\delta)=\left(\begin{array}{c} \mathbf{G}_{\ell}(z)+\delta \mathbf{G}'_{\ell}(z) \\ \mathbf{G}'_{\ell}(z)+\delta \mathbf{G}''_{\ell}(z) \\ \vdots\end{array}\right){+}\mathcal{O}(\delta^2)=\mathbf{G}_{\ell}(z){+}\sum_{\ell'}c_{\ell\ell'} \mathbf{G}_{\ell'}(z){+}\mathcal{O}(\delta^2)\,.
\eqe
That is, a positively shifted $\mathbf{G}_{\ell}(z)$ can be positively re-expanded on $\mathbf{G}_{\ell}(z)$. Now starting with  $z_1<z_2$, since we've concluded $\mathbf{G}_{\ell}(z_2)$ is positively expanded on $\mathbf{G}_{\ell}(z_1)$, its convex hull is \textit{inside} the polytope $Pol_1$. Thus given a series of ordered points, $z_1<z_2<z_3$, the corresponding $Poly_i$ defined in eq.(\ref{PolyZiDef}) satisfies:
\eq
\framebox[10cm][c]{$ Poly_3 \subset Poly_2 \subset Poly_1\,\quad {\rm for}\; z_1<z_2<z_3$}
\eqe
In other words, as we push $z$ away from $1$, not only is the convex hull of $\mathbf{G}_{\ell}(z)$ a cyclic polytope, it goes deeper and deeper \textit{inside} the original polytope!

%%%%%%%%%%%%%%%%%%%%%%%%%%%%%%%%%%%%%%%
\noindent\textbf{Spinning Gegenbauer cyclic polytope}\label{Spinning}
%%%%%%%%%%%%%%%%%%%%%%%%%%%%%%%%%%%%%%%
Recall that the Gegenbauer polynomial being the unique polynomial for scalar amplitude with a spin-$\ell$ exchange is rooted in the three-point amplitude of two scalars and a spin-$\ell$ particle is unique. For general three-point amplitudes with spins this is no longer true. However as discussed in~\cite{MassiveTree}, in four-dimensions given the helicities of the two massless particles and the spin of the massive particle, the amplitude is fixed. This allows one to define a set of ``spinning" Gegenbauer polynomial basis.

To see this, lets consider the three-point amplitude involving a massive spin-$\ell$ particle and massless particles with helicity $h_1,h_2$. We again have a polarization tensor $\epsilon_{\mu_1\mu_2\cdots\mu_\ell}$ needing $\ell$ vectors to contract. Due to $h_1,h_2\neq0$, besides from $p_{12}$ we now have two new vectors $q=\lambda_1\tilde{\lambda}_2$ and $\tilde{q}=q^*=\lambda_2\tilde{\lambda}_1$, that can be used to contract with the polarization tensor. Up to an overall constant,  the amplitude is fixed by $\{h_1,h_2,\ell\}$ to be:  
\eqa
q^{\mu_1}q^{\mu_2}\cdots q^{\mu_{h_2{-}h_1}}(p_{12})^{\mu_{h_2{-}h_1{+}1}}\cdots (p_{12})^{\mu_\ell}\epsilon_{\mu_1\cdots\mu_\ell},\quad {\rm for}\;h_2{-}h_1>0\nonumber \\
\tilde{q}^{\mu_1}\tilde{q}^{\mu_2}\cdots \tilde{q}^{\mu_{h_1{-}h_2}}(p_{12})^{\mu_{h_1{-}h_2{+}1}}\cdots (p_{12})^{\mu_\ell}\epsilon_{\mu_1\cdots\mu_S},\quad {\rm for}\;h_1{-}h_2>0\,.
\eqae
We can now glue the two three-point amplitudes together to construct the residue for a spin-$\ell$ exchange. As discussed in~\cite{MassiveTree}, since the polarization tensors form irreps of the little group, the gluing of the three-point amplitude is simplified by first rewriting it in SL(2,C) irreps as:
\eqa\label{3pt}
[12]^{\ell+h_1+h_2}\left(\lambda_1^{\ell+h_2-h_1}\lambda_2^{\ell+h_1-h_2}\right)_{\{\alpha_1\cdots\alpha_{2\ell}\}}\,,
\eqae
then contract the SL(2,C) indices between both sides of the factorization channel. In the center of mass frame, we can parameterize the spinors as: 
\eq
 \lambda_1=m^{\frac{1}{2}}\begin{pmatrix}1\\0 \end{pmatrix}, \lambda_2=m^{\frac{1}{2}}\begin{pmatrix}0\\1 \end{pmatrix},\lambda_3=im^{\frac{1}{2}}\begin{pmatrix}\sin\f{\theta}{2}\\ -\cos\f{\theta}{2} \end{pmatrix},\lambda_4=im^{\frac{1}{2}}\begin{pmatrix}\cos\f{\theta}{2}\\ \sin\f{\theta}{2} \end{pmatrix}.
 \eqe 
We can identify the the three-point coupling in eq.(\ref{3pt}) involving legs $1,2$ as a spin-$\ell$ state with ``$J_z$" quantum number $m=h_1-h_2$. Replacing $1,2$ with $3,4$ we then have a spin-$\ell$ state with quantum number $m=h_3-h_4$, acted upon a rotation matrix in the ``$y$"-axes by $\theta$. The gluing of the three-point amplitude on both sides then simply corresponds to computing the overlap of the two states, which is nothing by the Wigner $d$-matrix! Thus we see that for general spinning particles the polynomial is simply:
\eq
d^{\ell}_{h_1-h_2,h_3-h_4}(\theta)\,.
\eqe 
where $d_{m',m}^j(\theta)$ is the Wigner $d$-matrix defined by $d_{m',m}^j(\theta)=\langle j,m'|e^{-i\theta \mathcal{J}_y}|j,m\rangle$.

Let us consider as an example the residue for a spin-$\ell$ exchange in the helicity configuration $({+}h,{-}h,{+}h,{-}h)$. Writing it as a product of three point amplitudes, we find:
\eqa\label{reference}
n^{\{{+}h,{-}h,{+}h,{-}h\}}_{\ell}&=& A(1^{+h}2^{-h} \mathbf{P}^\ell)A(3^{+h}4^{-h} \mathbf{P}^\ell)\nonumber\\
&=& \frac{|c_\ell|^2([12][34])^{\ell}}{m^{4\ell{-}2}}\left(\lambda_1^{\ell{-}2h}\lambda_2^{\ell{+}2h}\right)^{\{\alpha_1\cdots\alpha_{2\ell}\}}\left(\lambda_3^{\ell{-}2h}\lambda_4^{\ell{+}2h}\right)_{\{\alpha_1\cdots\alpha_{2\ell}\}}\nonumber\\
&=&|c_\ell|^2[13]^{2h}\langle 24\rangle^{2h}\frac{([12][34])^{\tau}}{m^{4\ell{-}2{-}4h}}\left[\left(\lambda_1^{\tau}\lambda_2^{\tau}\right)\cdot\left(\lambda_3^{\tau}\lambda_4^{\tau}\right)\right]\,,\nonumber\\
&=& |c_\ell|^2\,m^2 d^{\ell}_{2h,2h}(\theta)\,,
\eqae
where $\mathbf{P}^\ell$ indicates a spin-$\ell$ state with $P^2=(p_1{+}p_2)^2=m^2$, $\tau=\ell{-}2h$ and we've normalized the amplitudes such that the coupling constant $c_\ell$ is dimensionless. Note that the $\ell$-independent prefactor $[13]^{2h}\langle 24\rangle^{2h}$ is required from helicity constraints, indicating that $d^{\ell}_{2h,2h}(\theta)\propto \cos^{4h} \frac{\theta}{2}$. Exchanging $3,4$ one obtains the residue for other helicity configurations:
\eq
n^{\{{+}h,{-}h,{-}h,{+}h\}}_{\ell}=\sum_\ell |c_\ell|^2\,m^2 (-1)^\ell d^{\ell}_{2h,-2h}(\theta)\,.
\eqe
Note that $n^{\{{+}h,{-}h,{-}h,{+}h\}}_{\ell}=(-1)^\ell n^{\{{+}h,{-}h,{+}h,{-}h\}}_{\ell}|_{\theta\rightarrow \theta+\pi}$.\footnote{We thank Z. Bern, A. Zhiboedov,
and D. Kosmopoulos for pointing out this relation.} For example, the polynomials for the first few spins in $n^{\{{+}1,{-}1,{+}1,{-}1\}}_{\ell}$ are:
\eqa
d^{2}_{2,2}(\theta)&=&\cos^4 \frac{\theta}{2}\nonumber\\
d^{3}_{2,2}(\theta)&=&\cos^4 \frac{\theta}{2}(3 \cos\theta{-}2)\nonumber\\
d^{4}_{2,2}(\theta)&=&\cos^4 \frac{\theta}{2}(1{-}7 \cos\theta{+}7 \cos^2\theta)\nonumber\\
d^{5}_{2,2}(\theta)&=&\cos^4 \frac{\theta}{2}(1{+} 3 \cos\theta {-}18 \cos^2\theta {+}15 \cos^3\theta)\,.
\eqae
Note that one starts from $\ell=2$ a reflection of Landau-Yang's theorem.

Now following the previous discussion, since the Wigner d-matrices are also orthogonal polynomials, we expect that their Taylor vectors yield a positive definite matrix when the spins are ordered. Indeed consider the Taylor vectors for $d^{\ell}_{2,2}(\theta)$ expanded around $\theta=0$. The Taylor vectors for spins $2,3,\cdots,9$ are given as: 
\eqa\label{SpinVecs}
h{=}1:\quad \left(\begin{array}{cccccccc}\frac{1}{4}& \frac{1}{4} & \frac{1}{4} & \frac{1}{4} & \frac{1}{4} & \frac{1}{4} & \frac{1}{4} & \frac{1}{4} \\0 &\frac{3}{4}& \frac{7}{4} & 3 & \frac{9}{2} & \frac{25}{4} & \frac{33}{4} & \frac{21}{2} \\0 & 0 & \frac{7}{4} & \frac{27}{4} & \frac{135}{8} & \frac{275}{8} & \frac{495}{8} & \frac{819}{8} \\0 & 0 & 0 & \frac{15}{4} & \frac{165}{8} & \frac{275}{4} & \frac{715}{4} & \frac{3185}{8} \\0 & 0 & 0 & 0 & \frac{495}{64} & \frac{3575}{64} & \frac{15015}{64} & \frac{47775}{64} \\0 & 0 & 0 & 0 & 0 & \frac{1001}{64} & \frac{9009}{64} & \frac{5733}{8} \\0 & 0 & 0 & 0 & 0 & 0 & \frac{1001}{32} & \frac{10829}{32} \\0 & 0 & 0 & 0 & 0 & 0 & 0 & \frac{1989}{32}\end{array}\right)\quad\nonumber\\
\eqae
It is straight forward to verify that, just as the vectors from Gegenbauer polynomials, the above is a totally positive matrix. Thus we see that the convex hull of the Taylor vectors from the spinning polynomial yields a cyclic polytope.

%%%%%%%%%%%%%%%%%%%%%%%%%%%%%%%
\section{The $s$-channel EFT-hedron}\label{Sec:s-channel}
%%%%%%%%%%%%%%%%%%%%%%%%%%%%%%%
In the previous section we've seen that for $\mathbf{a}$ to reside inside a convex hull, the geometry set up in eq.(\ref{Master0a}, \ref{Master1a}), it can be cast into a (infinite) set of positivity conditions:
\eq
f_i(\mathbf{a})\geq0\,.
\eqe
The explicit function $f_i$ depends on the vectors that constitute the hull, and can be linear or non-linear functions of $\mathbf{a}$. Let us now explore the geometry for the simplest class of EFTs where the massless degrees of freedom are colored state. We can then focus on color ordered four-point amplitude and assume the absence of UV states in the $u$-channel.  In such case we have eq.(\ref{Master0a})
\eq\label{sMaster}
a_{k,q}=\sum_{a}\textsf{p}_{a}\frac{v_{\ell_a,q}}{(m^2_a)^{k+1}}\,\quad \textsf{p}_{a}\geq 0\,. 
\eqe
where once again $v_{\ell,q}$ is the $q$-th Taylor coefficient in expanding $G_{\ell}(1+2\delta)$.  The couplings $a_{k,q}$ are naturally dimensionful, but since our bounds will be projective in nature, only dimensionless ratios will be constrained. Note that since we are considering color ordered amplitudes, cyclic symmetry implies that the amplitude is symmetric under $s\leftrightarrow t$. Translated to the EFT couplings we have that they must lie on the ``cyclic plane" $\mathbf{X}_{\rm Cyc}$ defined by  
\eq
a_{k,q}=a_{k,k-q}\,.
\eqe
Thus the geometry of interest will be the intersection of the convex hull in eq.(\ref{sMaster}), with the cyclic plane $\mathbf{X}_{\rm Cyc}$.

Recall that the origin of eq.(\ref{sMaster}) is the fact that the low energy amplitudes can be reproduced from the $s$-channel singularities. This can be recast into the following equivalence:     
\eq\label{SchematicRep}
\sum_{k,q}a_{k,q}\;\;s^{k{-}q}t^q\;=\;\sum_a-\frac{\textsf{p}_a\; G_{\ell_a}\left(1+2\frac{t}{m_a^2}\right)}{s-m_a^2}\;\;\; {\rm for}\; s, t \ll m^2 \,,
\eqe 
where the equality is understood as the matching of Taylor series in $s,t$ on both sides, with $n\geq2$. Thus the sum on the RHS is only expected to reproduced $a_{k,q}$ with $q\leq k{-}2$.  Writing out the Taylor series for the RHS,
\eq\label{SchematicRep2}
\sum_{k,q}a_{k,q}\;\;s^{k{-}q}t^q\;=\sum_a \frac{\textsf{p}_a}{m^{2}_a}\left(1+\frac{s}{m_a^2}+\frac{s^2}{m_a^4}+\cdots\right)\left(v_{\ell_a,0}+v_{\ell_a,1}\frac{t}{m_a^2}+v_{\ell_a,2}\left(\frac{t}{m_a^2}\right)^2+\cdots\right)\,,
\eqe   
we immediately see the emergence of two types of geometries, one is the coefficients associated with the expansion in $t$ and the other is the expansion in $s$. The geometry encoded in the former is a reflection of UV Lorentz invariance, since the convex hull depends on the details of the Gegenbauer polynomials, while the geometric series of the later reflects locality, i.e. that the only singularities of the four-point amplitude are in the Mandelstam variables.  We will begin our analysis by disentangling the two geometry, taking the point of view of either fixed $k$ or fixed $q$, and end in the geometry that is defined by its union.

%%%%%%%%%%%%%%%%%%%%%%%%%%%%%%%
\subsection{Fixed $k$: the Gegenbauer cyclic polytope}
%%%%%%%%%%%%%%%%%%%%%%%%%%%%%%%
Identifying the coefficient for $s^{k{-}q}t^q$ on both sides of eq.(\ref{SchematicRep2}), we have
\eq\label{Prop}
a_{k,q}=\sum_{a}\textsf{p}_a\left[\,x^{k{+}1}_a\,  v_{\ell_a,q}  \right]\quad x_a\equiv\frac{1}{m^2_a}.
\eqe
Now consider terms with the same mass-dimension, corresponding to fixed $k$. We write
\eq
\mathbf{a}_k=\left(\begin{array}{c}a_{k,0} \\ a_{k,1}\\ \vdots \end{array}\right)=\sum_{a}\;\textsf{p}_a x^{k{+}1}_a \left(\begin{array}{c}v_{\ell_a,0} \\ v_{\ell_a,1} \\ \vdots \end{array}\right)\,.
\eqe
Since $\textsf{p}_a, x_a>0$, this says that  
\eq
\mathbf{a}_k\in Conv[\mathbf{G}_\ell]\,,
\eqe
that is, the coefficients for the distinct polynomials associated with the mass-dimension $2k{+}4$ operator must live inside the Gegenbauer cyclic polytope!  We will refer to $Conv[\mathbf{G}_\ell]$ as the unitary polytope $\textbf{U}_k$, where the subscript $k$ indicates that the polytope is in $\mathbb{P}^{k{-}2}$. The dimension is projectively $k{-}2$, since there are $k{+}1$ distinct polynomials at given $k$, with $a_{k,k}$ and $a_{k,k{-}1}$ not subject to the constraints implied by eq.(\ref{sMaster}).  

\begin{figure}
\begin{center}
\includegraphics[scale=0.5]{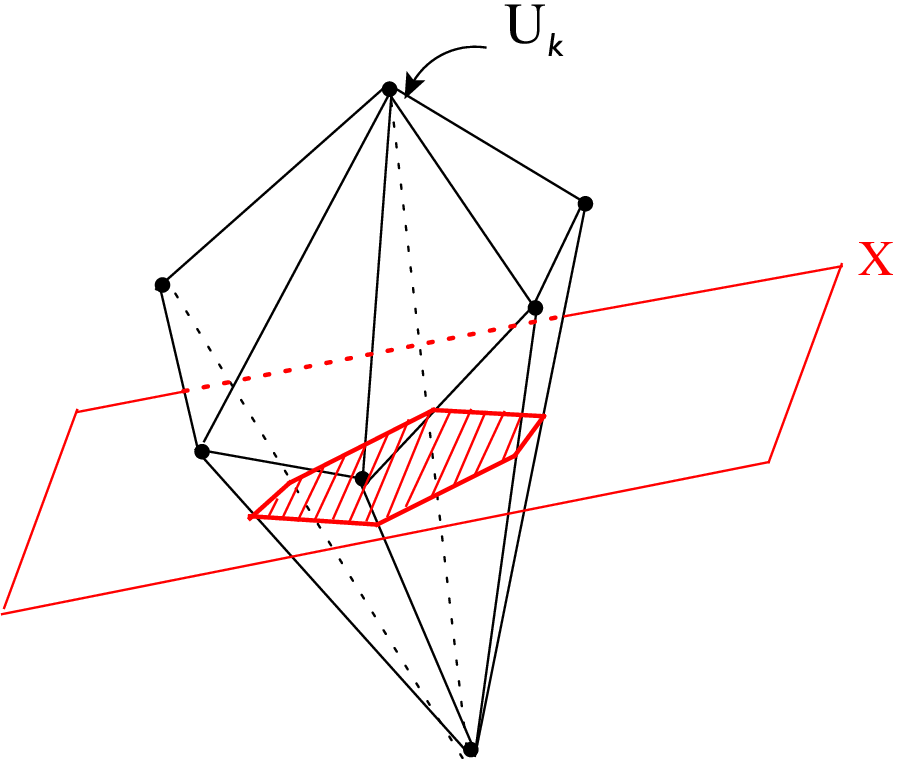}
\caption{The $s$-channel geometry at fixed $k$. The vector $\mathbf{a}_k$ must live on the intersection between the cyclic plane $\mathbf{X}_{\rm cyc}$ with the unitary polytope $\textbf{U}_k$. }
\label{Intersection}
\end{center}
\end{figure}

Furthermore, cyclic symmetry requires that the couplings lie on the cyclic plane $\mathbf{X}_{\rm cyc}$. For $k<5$ cyclic symmetry simply relates the coefficients $a_{k,k}$ and $a_{k,k{-}1}$ to those that are constrained by $\textbf{U}_k$. For $k\geq 5$ the cyclic plane $\textbf{X}_{\rm cyc}$ defines a $\lceil \frac{k{+}1}{2}\rceil{-}1$-dimensional subspace inside $\textbf{U}_k$, i.e.  the space of allowed couplings are now given by the intersection of the  cyclic plane $\textbf{X}_{\rm cyc}$ with the unitary polytope $\textbf{U}_k$, i.e. $\textbf{U}_k\cap\textbf{X}_{\rm cyc}$, as illustrated in fig.(\ref{Intersection}). In the following, we will consider explicit examples up to $k=5$.

\begin{itemize}
\item $k{=}2:D^4\phi^4$:

\eq
M_{D^4\phi^4}(s,t)=(a_{2,0}s^2{+}a_{2,1}st{+}a_{2,2}t^2)
\eqe
we will only be able to bound $a_{2,0}$ and the geometry is $\mathbb{P}^0$. From the fact that $v_{\ell,0}$ is a positive number, we simply have  $a_{2,0}>0$, the forward limit positivity bound discussed in~\cite{Adams:2006sv}.

\item $k{=}3:D^6\phi^4$
\eq
M_{D^6\phi^4}(s,t)=(a_{3,0}s^3{+}a_{3,1}s^2t{+}\cdots)
\eqe
where from now on we'll suppress listing the couplings that cannot be bounded. The geometry is now $\mathbb{P}^1$, and $\mathbf{a}_3=(1,\frac{a_{3,1}}{a_{3,0}})$ is bounded by the minimum and maximum value of $\frac{v_{\ell,1}}{v_{\ell,0}}$, which is $0$ and $\infty$ respectively. Thus we simply have $a_{3,0},a_{3,1}>0$. 
  
\item $k{=}4:D^8\phi^4$ 
\eq
M_{D^8\phi^4}(s,t)=(a_{4,0}s^4{+}a_{4,1}s^3t{+}a_{4,2}s^2t^2{+}\cdots)
\eqe
we have $\mathbf{a}_4=(1,\frac{a_{4,1}}{a_{4,0}},\frac{a_{4,2}}{a_{4,0}})\equiv(1,x,y)$. The boundaries of the two-dimensional polygon are given by $(i,i{+}1)$, and the constraint on $\mathbf{a}_4$ is given by $\langle\mathbf{a}_4,i,i{+}1\rangle>0$ and $\langle \mathbf{a}_4,\infty,0\rangle>0$, where 
\eq
\langle\mathbf{a}_4,i,i{+}1\rangle={\rm Det}\left(\begin{array}{ccc}1 & v_{i,0} & v_{i{+}1,0} \\ x & v_{i,1} &  v_{i{+}1,1} \\y & v_{i,2} &   v_{i{+}1,2}\end{array}\right)
\eqe 
Listing the first sets of constraint:
\eq
\langle\mathbf{a}_4,0,1\rangle>0\Rightarrow y>0,\; \langle\mathbf{a}_4,1,2\rangle>0\Rightarrow 6 {-} 3 x {+}2 y > 0,\;\langle\mathbf{a}_4,2,3\rangle>0\Rightarrow 18 {-} 4 x {+}  y > 0\,.
\eqe
The combined constraint is plotted in fig.\ref{splotk=4}.

\begin{figure}
\begin{center}
\includegraphics[scale=0.4]{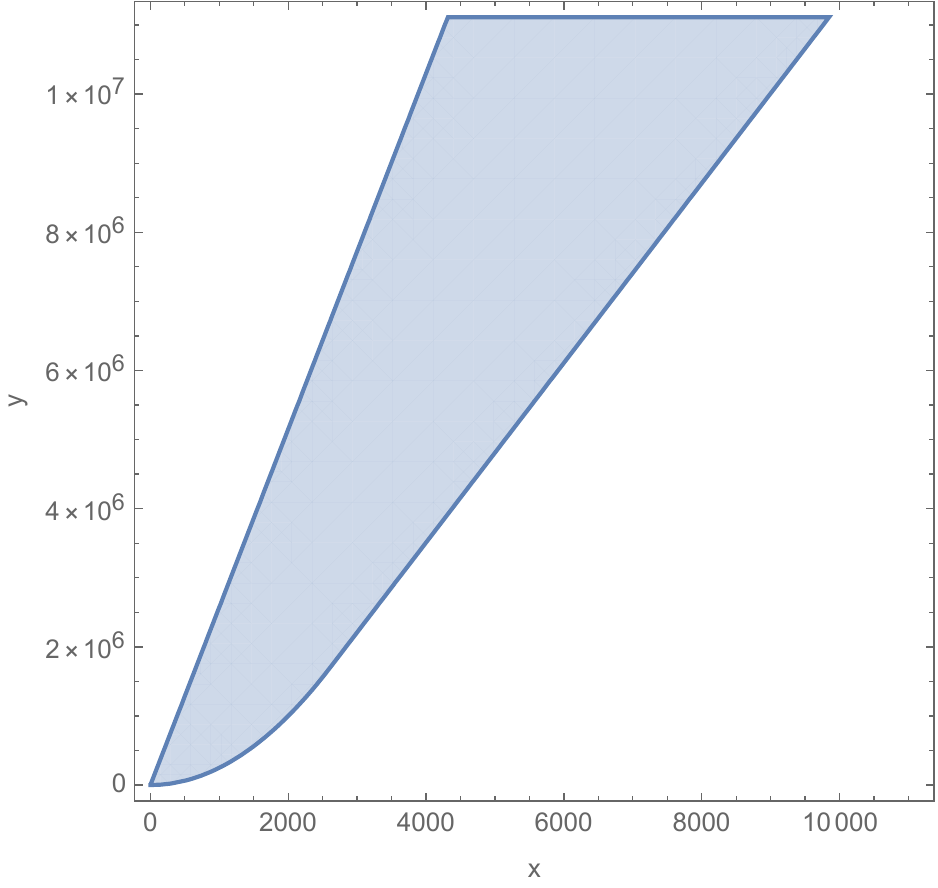}
\caption{The allowed region satisfying $\langle\mathbf{a}_4,i,i{+}1\rangle>0$. We have plotted the combined constraint for $i\leq 40$. For larger $i$s, the constraint does not appear for the range of $(x,y)$ displayed in the plot.   }
\label{splotk=4}
\end{center}
\end{figure}

\item $k{=}5:D^{10}\phi^4$ 
\eq
M_{D^{10}\phi^4}(s,t)=(a_{5,0}s^5{+}a_{5,1}s^4t{+}a_{5,2}s^3t^2{+}a_{5,3}s^2t^3{+}\cdots)
\eqe
In this case, the cyclic plane $\mathbf{a}_5\in\textbf{Y}=(1,x,y,y)$ is two dimensional and thus represent a subspace of the three-dimensional unitary polytope $\textbf{U}_5$. There are two sets of constraint coming from $\langle 0,\mathbf{a}_5, i,i{+}1\rangle>0$ and $\langle \mathbf{a}_5, i,i{+}1,\infty\rangle>0$, given as:
\eqa
\langle 0,\mathbf{a}_5 ,i,i{+}1\rangle=\left(\begin{array}{cccc}1 & 1 &  v_{i,0} & v_{i{+}1,0} \\ 0 &x & v_{i,1} & v_{i{+}1,1}  \\ 0 &y &  v_{i,2} & v_{i{+}1,2}   \\ 0& y &   v_{i,3} & v_{i{+}1,3}  \end{array}\right),\quad  \langle \mathbf{a}_5 ,i,i{+}1,\infty\rangle=\left(\begin{array}{cccc} 1 &  v_{i,0} & v_{i{+}1,0} & 0 \\ x & v_{i,1} & v_{i{+}1,1}& 0  \\ y &  v_{i,2} & v_{i{+}1,2} &0  \\  y &   v_{i,3} & v_{i{+}1,3} &1  \end{array}\right)
\eqae 
The first set of constraints simply leads to $y \geq 0, x \geq \frac{y}{3}$, while the second set is shown in fig.~\ref{Constraint12}. The combined constraint leads to finite region comprised of boundaries $(i,i{+}1,\infty)$ with $i=0,1,\cdots,4$ and $(0,4,5)$ as shown in fig.\ref{ComboConstraint}.

\begin{figure}
\begin{center}
\includegraphics[scale=0.42]{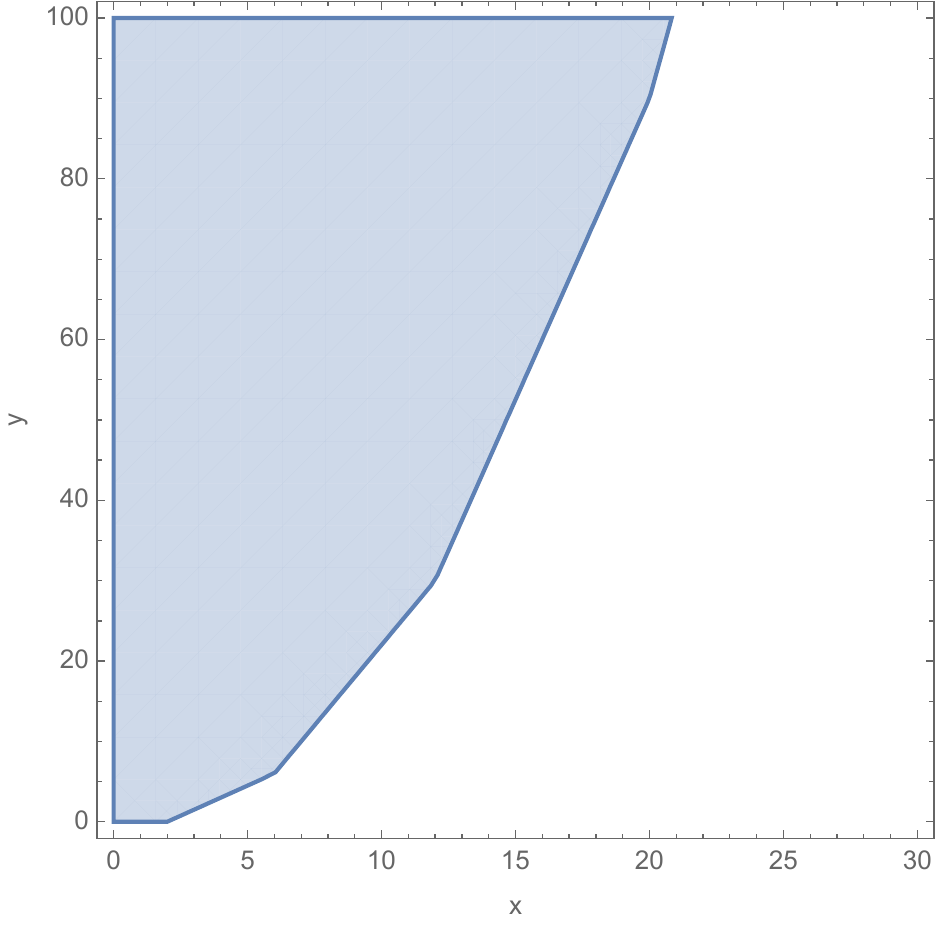}
\caption{The constraints curved out from $\langle \mathbf{a}_5, i,i{+}1,\infty\rangle>0$.}
\label{Constraint12}
\end{center}
\end{figure}

\begin{figure}
\begin{center}
\includegraphics[scale=0.35]{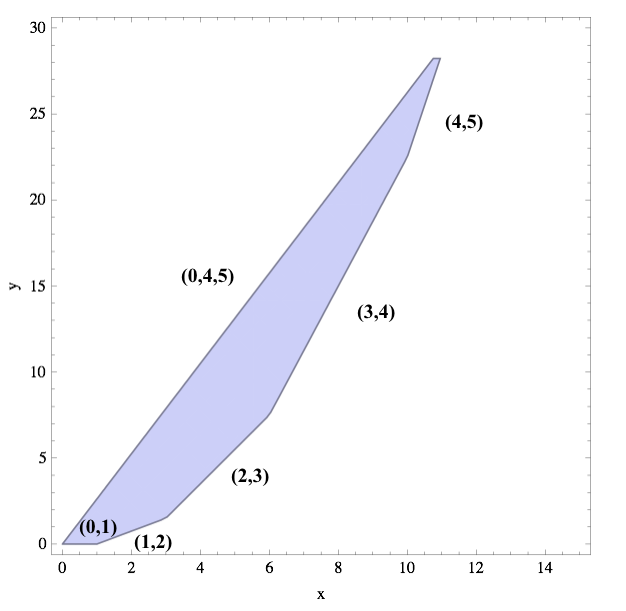}
\caption{The projection of the unitary polytope onto the cyclic plane at $k=5$. The boundary is given by $(0,4,5)$ as well as $(i,i{+}1,\infty)$ for $i=0,\cdots,4$, displayed as $(i,i{+}1)$.}
\label{ComboConstraint}
\end{center}
\end{figure}
\end{itemize}

The fact that the ratio of coefficients $\frac{a_{k,q}}{a_{k,0}}$ are bounded within finite regions tells us that, in the on-shell basis, it is not only unnatural to have two distinct operators with the same dimension yet large differences in their coupling constants, \textit{unitarity in the UV tells us that it is impossible to do so !}

\begin{figure}
\begin{center}
$\text{\footnotesize k{=}4}\quad\vcenter{\hbox{\includegraphics[scale=0.35]{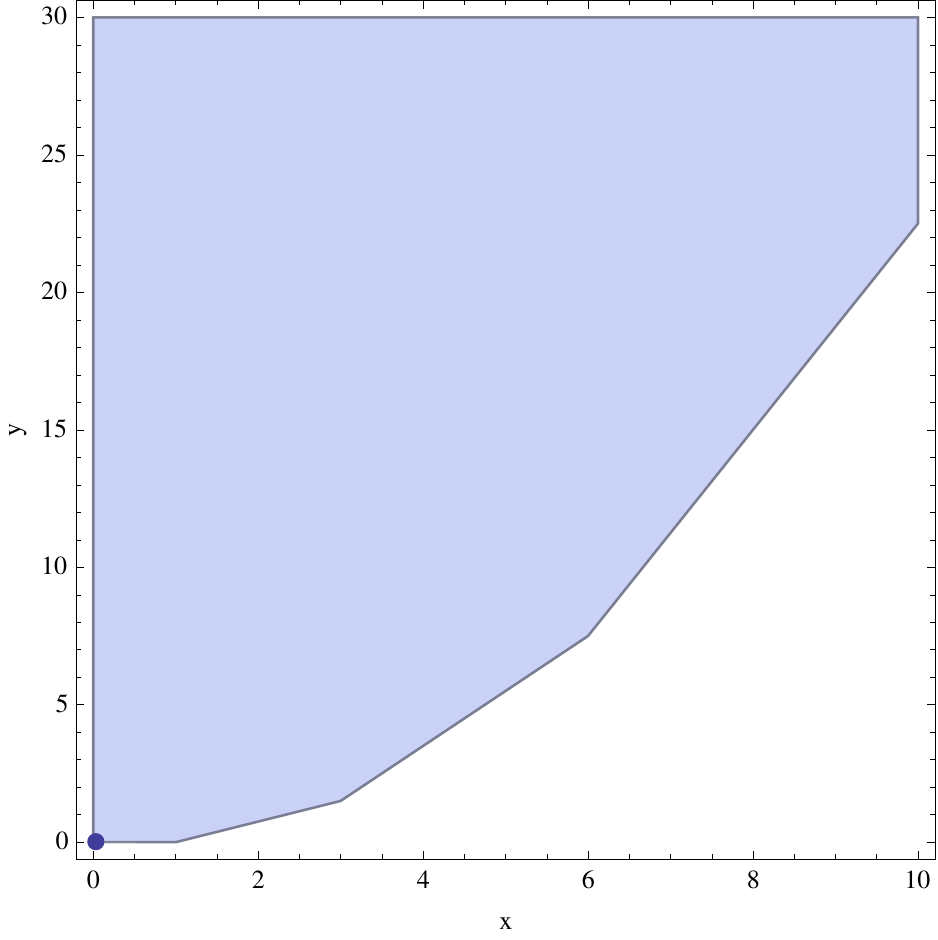}}}\quad \text{\footnotesize k{=}5}\quad\vcenter{\hbox{\includegraphics[scale=0.35]{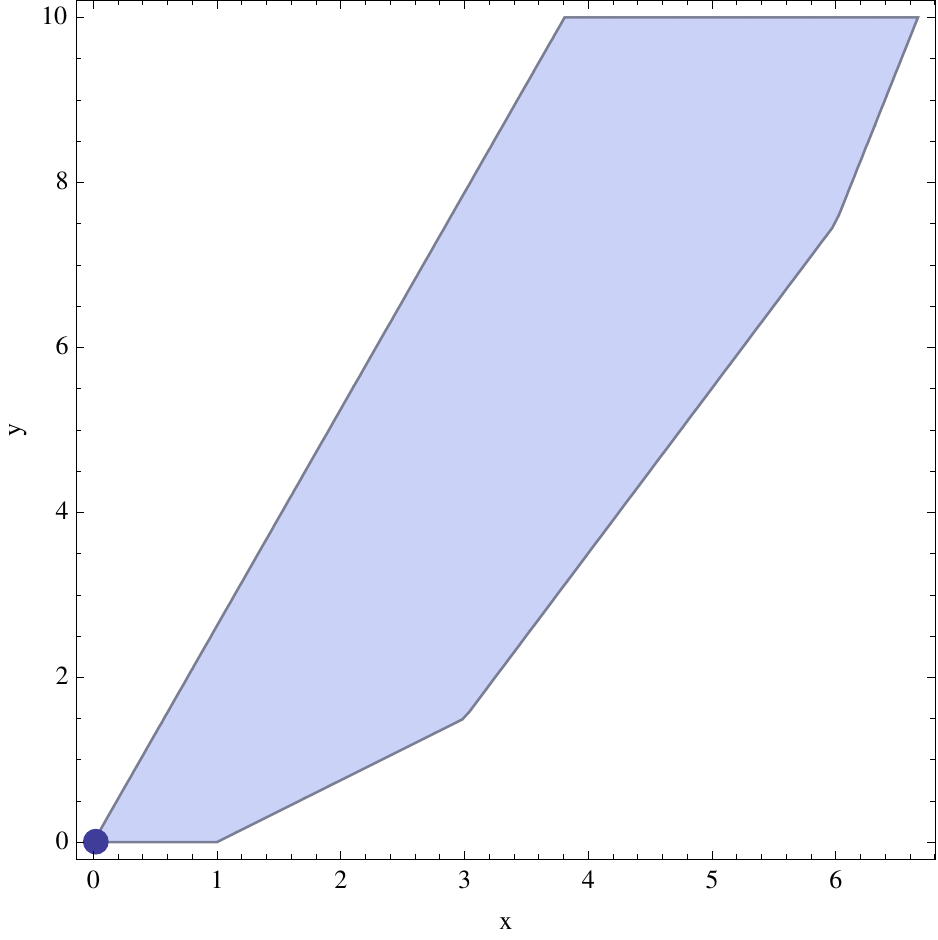}}}$
\caption{The position of the string theory coefficients given in eq.(\ref{StringTest}) inside the region $\textbf{U}_k\cap\textbf{Y}$, for $k=4,5$ respectively.}
\label{Sacks}
\end{center}
\end{figure}

Let's see where explicit EFTs sit inside $\textbf{U}_k\cap\textbf{X}_{\rm cyc}$. Consider the open superstring four-gluon amplitude in eq.(\ref{TypeIString}), where its low-energy expansion is given in eq.(\ref{TypeIStringIR}). Stripping off the spinor brackets and consider the expansion up to $k=5$ we find,
\eqa\label{StringTest}
k=2:&&\;\; a_{2,0}=\frac{2\zeta_2^2}{5},\quad k=3:\;\; a_{3,0}=\zeta(5),\quad a_{3,1}=2\zeta(5){-}\zeta(3)\zeta(2)\nonumber\\
k=4:&&\;\;(x,y)=\left(\frac{a_{4,1}}{a_{4,0}},\frac{a_{4,2}}{a_{4,0}}\right)=\left(\frac{3}{4} - \frac{945 \zeta_3^2}{2 \pi^6},\frac{23}{20160}-\frac{3\zeta^2_3}{4\pi^6}\right)\nonumber\\
k=5:&&\;\;(x,y)=\left(\frac{a_{5,1}}{a_{5,0}},\frac{a_{5,2}}{a_{5,0}}\right)=\left(3 - \frac{\pi^4 \zeta_3+15 \pi^2\zeta_5}{90 \zeta_7},5-\frac{\pi^4\zeta_3+24\pi^2\zeta_5}{72\zeta_7}\right)\,.
\eqae  
For $k=2,3$ the coefficients are not only inside $\textbf{U}_k$, it close to the ``boundary". This behaviour is more prominent for $k=4,5$ where the EFT couplings are close to the boundary comprised of low spins, as we display in fig.(\ref{Sacks}). This indicates that the $\textsf{p}_{a}$s in eq.(\ref{sMaster}) is dominated by contributions from low spin sector. In fact, in section~\ref{sec:Answs} we will see that such behaviour is common amongst all known EFTs.

%%%%%%%%%%%%%%%%%%%%%%%%%%%%%%%
\subsection{Fixed $q$: Hanekl matrix constraints}
%%%%%%%%%%%%%%%%%%%%%%%%%%%%%%%
Instead of fixed $k$ and considering the constraint on $\mathbf{a}_k$, let's now examine the geometry associated with fixed $q$, i.e. that associated with the first parenthesis on the RHS of eq.(\ref{SchematicRep2}). First taking $q=0$, we have 
\eq\label{SchematicRep3}
a_{k,0}\;=\sum_a \textsf{p}'_a (x_a)^k\,,
\eqe   
where $\textsf{p}'_a=x_a \textsf{p}_a\,v_{\ell,0}$, and the equality holds for the $k\geq2$. Since $v_{\ell,0}=G_{\ell}(1)$ is positive, $\textsf{p}'_a>0$. We immediately see that eq.(\ref{SchematicRep3}) implies $a_{k,0}>0$, which is the forward limit positivity bound discussed in  \cite{Adams:2006sv} extended to higher derivatives. We've seen this before in section \ref{Sec:Hankel}, the vector 
\eq\label{HalfMomCurve}
\tilde{\mathbf{a}}_0=\left(\begin{array}{c} 1\\  \frac{a_{3,0}}{a_{2,0}} \\ \frac{a_{4,0}}{a_{2,0}} \\ \vdots \end{array}\right)=\sum_{a}\textsf{p}_a\left(\begin{array}{c}1 \\ x_a \\ \vdots  \\ x^{k}_a\end{array}\right)\,,\quad x_a\equiv \frac{1}{m^2_a}\,,
\eqe
lies in the convex hull of points on a half moment curve, and thus the Hankel matrix of its entries $K[\tilde{a}_0]$ is a totally positive matrix. Note that since $v_{\ell,q}>0$ for all $q$, the same holds true for any fixed $q$. Thus in general we have: 
\eq
\framebox[8cm][c]{$K\left[\tilde{\mathbf{a}}_q\right]\in {\rm Total\; positive\; matrices\;}\;\forall q\,.$}
\eqe

Once again, lets us demonstrate this for the Type-I string amplitude. Collecting the coefficients as
\eq
\vec{a}_0=\left(\begin{array}{c}\frac{2 }{5} \zeta_2^2\\ \zeta_5 \\ \frac{8}{35}\zeta_2^3 \\\zeta_7 \\ \frac{24 }{175}\zeta_2^4\end{array}\right),\quad \vec{a}_1=\left(\begin{array}{c} 2\zeta_5{-}\zeta_2\zeta_3\\ \frac{6}{35}\zeta_2^3{-}\frac{1}{2}\zeta_3^2 \\ 3\zeta_7{-}\zeta_2\zeta_5{-}\frac{2}{5}\zeta_2^2\zeta_3 \\\frac{6}{35}\zeta_2^4{-}\zeta_3\zeta_5 \\ 4\zeta_9{-}\zeta_2\zeta_7{-}\frac{2}{5}\zeta_2^2\zeta_5{-}\frac{8}{35}\zeta_2^3\zeta_3\end{array}\right)\,,
\eqe
The corresponding Hankel matrix are,
\eqa
K\left[\vec{a}_0\right]&{=}&\left(\begin{array}{ccc} \frac{2 }{5} \zeta_2^2 &\zeta_5  &\frac{8}{35}\zeta_2^3  \\ \zeta_5 &\frac{8}{35}\zeta_2^3& \zeta_7  \\ \frac{8}{35}\zeta_2^3& \zeta_7 & \frac{24 }{175}\zeta_2^4\end{array}\right)\nonumber\\
K\left[\vec{a}_1\right]&{=}&\left(\begin{array}{ccc} 2\zeta_5{-}\zeta_2\zeta_3  & \frac{6}{35}\zeta_2^3{-}\frac{1}{2}\zeta_3^2 & 3\zeta_7{-}\zeta_2\zeta_5{-}\frac{2}{5}\zeta_2^2\zeta_3 \\ \frac{6}{35}\zeta_2^3{-}\frac{1}{2}\zeta_3^2 & 3\zeta_7{-}\zeta_2\zeta_5{-}\frac{2}{5}\zeta_2^2\zeta_3 & \frac{6}{35}\zeta_2^4{-}\zeta_3\zeta_5 \\ 3\zeta_7{-}\zeta_2\zeta_5{-}\frac{2}{5}\zeta_2^2\zeta_3 & \frac{6}{35}\zeta_2^4{-}\zeta_3\zeta_5 & 4\zeta_9{-}\zeta_2\zeta_7{-}\frac{2}{5}\zeta_2^2\zeta_5{-}\frac{8}{35}\zeta_2^3\zeta_3\end{array}\right)
\eqae
It is straight forward to check that all minors of the above Hankel matrix are indeed positive. A more detailed study of the Hankel matrix constraint for superstring amplitude was recently done in~\cite{Green:2019tpt}.

It is interesting to ask which theories lie on boundaries of the Hankel constraints, i.e., for which theories do all the minors of the Hankel matrix greater than some size all vanish? The answer is extremely simple and satisfying. Only UV amplitudes with a finite number of poles satisfy this property; that is, only UV theories with $N$ massive states exchanged at tree-level lie on the boundary of the Hankel constraints.  This can be seen from the analytic expression of the determinants in eq.(\ref{Det}), where it is proportional to the Vandermonde determinant of the masses of the UV state $x_a$. 
 This gives us a way to “detect” the number of massive states: if there are  $a$ massive states, then the $(a+1) \times (a+1)$ determinant vanishes.

%%%%%%%%%%%%%%%%%%%%%%%%%%%%%%%
\subsection{The $s$-channel EFT-hedron}
\begin{figure}
\begin{center}
\includegraphics[scale=0.4]{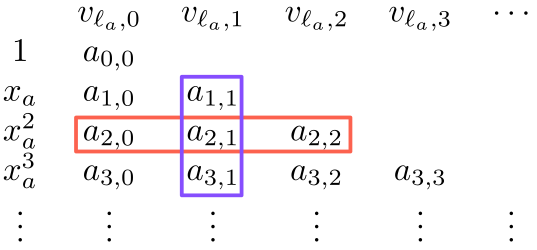}
\caption{We organize the information that each state contributes to the determination of $a_{k,q}$. For each fixed row (fixed $k$), for example the red box, each state's contribution is proportional to a Gegenbauer vector multiplied by a universal factor $x_a^k$.  For a fixed column (fixed $q$), the purple box, each state contributes to a point on a half moment curve multiplied by  universal factor $v_{\ell_a,q}$.  }
\label{Organized}
\end{center}
\end{figure}

%%%%%%%%%%%%%%%%%%%%%%%%%%%%%%%
 Up to now, we've been considering the constraints from the two parenthesis in eq.(\ref{SchematicRep2}) separately. These, however, are not the full set of  constraints. To see this it is useful to organize the information each state contributes to $a_{k,q}$ as in fig.\ref{Organized}. For a given row, each state contributes a fixed positive factor $x_a^k$ multiplying the Gegenbauer vector, which led to the constraint that the row vectors must lie in the convex hull of a cyclic polytope. For a fixed column, each state contributes a point on the half moment-curve weighted by a positive factor $v_{\ell,q}$, and thus implying the constraint that the Hankel matrix of the column vector is a totally positive matrix.

\begin{figure}
\begin{center}
\includegraphics[scale=0.4]{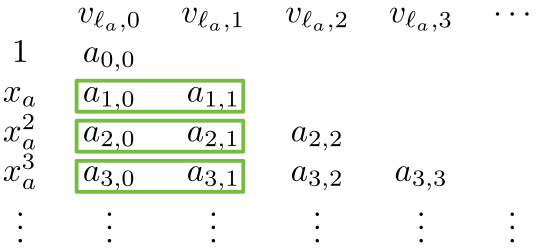}
\caption{For a given state, its contribution to each row is the same vector $(v_{\ell_a,0}, v_{\ell_a,1})$ after scaling away the moment factor $x_a^k$. }
\label{Organized2}
\end{center}
\end{figure}

As one can see from the above description, these are not the complete constraints. For example, the cyclic polytope constraint does not tell us that the positive proportionality factor takes the form $x_a^k$, which is only visible if we consider different $k$s at the same time. Put in another way, if we truncate our expansion of $t$ to a fix order, say the first order, we should see that for different moments $(x^k_a)$, each state contribute the \textbf{same vector} $(v_{\ell_a,0}, v_{\ell_a,1})$, as illustrated in fig.\ref{Organized2}. In other words, not only does each row must lie in the cyclic polytope, but it must be the same point after scaling away the moment factors !

To recap, the space of higher dimensional operator is given by the tensor product of two positive geometries, the Gegenbauer cyclic polytope and convex hull of half moment curve, and we would like to find the \textit{full} set of inequalities that carve out this space. This is reminiscent to the (tree) Amplituhedron which gives the scattering amplitude of $\mathcal{N}=4$ SYM~\cite{Amplituhedron}. There we have a subspace of $k$-planes in $k{+}4$ dimensions, $Y_{\alpha}^I$, given by the product of two positive geometries   
\eq\label{AmpHed}
Y_{\alpha}^I=\sum_{i=1,n}C_{\alpha,i} Z_i^{I},\quad\quad C_{\alpha,i}\in Gr_{>0}(k,n),\quad Z_i^{I}\in M_{+}(n,k{+}4)
\eqe
where the $C_{\alpha,i}$ is in the positive Grassmannian $Gr_{>0}(k,n)$, a $k\times n$ matrix with all ordered minors positive mod GL(k), and $Z_i^{I}$ is a $n\times k{+}4$ positive matrix with positive ordered minors. The $Z$s are the ``external data" that is given and already in the positive region. Note that for $k=1$, this is simply a polytope in $\mathbb{P}^4$. To carve out this space via inequalities, we require that $Y$ satisfies:
\eq
\langle Y_{1}Y_{2}\cdots Y_{k} Z_{i}Z_{i{+}1}Z_{j}Z_{j{+}1}\rangle>0\,.
\eqe
To see this note that we can interpret eq.(\ref{AmpHed}) as expanding $Y_{\alpha}^I$ on the the ``basis" $Z_i^{I}$, with coefficients $C_{\alpha,i}$. Then the above condition implies 
\eq
\langle Y_{1}Y_{2}\cdots Y_{k} Z_{i}Z_{i{+}1}Z_{j}Z_{j{+}1}\rangle=\sum_{i_1<i_2<\cdots <i_k}\langle C_{i_1}C_{i_2}\cdots C_{i_k}\rangle\langle Z_{i_1} Z_{i_2}\cdots Z_{i_k}Z_{i}Z_{i{+}1}Z_{j}Z_{j{+}1}\rangle>0\quad.
\eqe
For this to hold for any choice of $Z_i^{I}\in M_{+}(n,k{+}4)$, forces $C_{\alpha,i}\in Gr_{>0}(k,n)$.

For our case, the fixed external data is the Gegenbauer vectors, which automatically yield positive matrices. This motivates us to first organize all the states with the same spin together and rewrite eq.(\ref{Prop}) as:
\eq\label{Def1}
a_{k,q}=\sum_{a}\textsf{p}_{a} \left[x^{k{+}1}_a\,v_{\ell_a,q}\right]\equiv \sum_{\ell}     C_{k,\ell}V_{\ell,q}\,.
\eqe
Here $V_{\ell,q}=v_{\ell,q}$, and $C_{k,\ell}=\sum_{\{a:\ell_a=\ell\}}p_{a}\,x^{k{+}1}_a$, where one sums over all the states with the same fixed spin $\ell$. Collecting the $C$s into a column vector $\mathbf{C}_{\ell}=\{C_{1,\ell},C_{2,\ell},\cdots,C_{k,\ell}\}$, we see that $\mathbf{C}_{\ell}$ is inside the convex hull of half moment curve.  We are now ready to define the EFT-hedron: the space of consistent coefficients of higher dimension operators are given by the product (with $k\geq q$)
\eq
a_{k,q}=\sum_{\ell} C_{k,\ell}V_{\ell,q}\quad
\eqe 
where $C_{k,\ell}$ is positive in the sense that $K[\mathbf{C}_{\ell}]$ is a totally positive matrix for each $\ell$, and $V_{\ell,q}$ is positive in that any ordered minor of the vectors are positive. Let us make a comparison with the amplituhedron~\cite{Amplituhedron}. For the EFT-hedron the positivity property in $C$ is defined for each column (spin) independently, while for the amplituhedron the $C$ being in $Gr_{>0}(k,n)$, the positivity condition mixes the columns. For the amplituhedron $I$ is locked in with $k$ being $4{+}k$ dimensional, while for the EFT-hedron $q$ can be any dimension independent of $k$.    

Now let us carve out the space via inequalities. Consider a set of ``walls", which are dual vectors $\mathcal{W}^q_I$, labelled by $I$, satisfying
\eq\label{Pos}
\sum_{q} \mathcal{W}^q_IV_{\ell,q}\geq0,\quad \forall \ell.
\eqe
Unit vectors $\{0,0,1,\cdots,0\}$ trivially satisfies this criteria due to the positivity of the Gegenbauer Taylor coefficients. We denote these as $\mathcal{W}^q_{I_{\mathbb{I}}}$. There are also walls comprised of the facets of $Conv[V_{\ell}]$, taking the form 
$(i, i{+}1)$, $(1,i, i{+}1)$, e.t.c, which in dual vector form is given by $\langle * ,i, i+1\rangle$, $\langle * ,1,i,i+1\rangle$. We denote these as $\mathcal{W}^q_{I_b}$. Given these walls we take the inner product with the higher dimension operators. Defining:
\eq
A_{k,I}\equiv\sum_q a_{k,q}\mathcal{W}^q_I,\quad \forall\;\; \mathcal{W}^q_I\in\{\mathcal{W}^q_{I_{\mathbb{I}}},\mathcal{W}^q_{I_b}\}
\eqe
then the EFT-hedron is carved out by the inequality:
\eq\label{EFT-hedron}
\framebox[8cm][c]{$\displaystyle K[\vec{A}_I] \quad {\rm is\; a\; totally\;positive\;matrix}\,$}~\,.
\eqe
where $\vec{A}_I=(A_{0,I},A_{1,I},\cdots)$. In other words, for any of one of the walls $\mathcal{W}^q_I$, the $A_{k,I}$s satisfy the following infinite set of constraints
\eqa
A_{0,I}\geq0,\quad A_{1,I}\geq0,\quad {\rm Det}\left(\begin{array}{cc} A_{0,I} & A_{1,I}  \\   A_{1,I} & A_{2,I}\end{array}\right)\geq0,\quad {\rm Det}\left(\begin{array}{cc} A_{1,I} & A_{2,I}  \\   A_{2,I} & A_{3,I}\end{array}\right)\geq0\nonumber\\
{\rm Det}\left(\begin{array}{ccc} A_{0,I} & A_{1,I}& A_{2,I} \\   A_{1,I} & A_{2,I} & A_{3,I} \\   A_{2,I} & A_{3,I} & A_{4,I} \end{array}\right)\geq0,\quad {\rm Det}\left(\begin{array}{ccc} A_{1,I} & A_{2,I}& A_{3,I} \\   A_{2,I} & A_{3,I} & A_{4,I} \\   A_{3,I} & A_{4,I} & A_{5,I} \end{array}\right)\geq0, \cdots e.t.c.
\eqae

Before closing, let us confirm that the inequalities in eq.(\ref{EFT-hedron}), combined with the information of the walls, indeed carves out the space in eq.(\ref{Def1}). First take the walls to be the unit vectors, then $K[\vec{A}_{I_{\mathbb{I}}}]$ being a totally positive matrix simply implies 
\eq
a_{k,q}=\sum_{a}\;p_{a,q} (x_{a})^k,\quad \;p_{a,q}>0\quad x_a>0\,,
\eqe
i.e. for each fixed $q$, the vector $\vec{a}_q=(a_{1,q},a_{2,q},\cdots)$ lies in the convex hull of half moment curves. Next, we consider the walls that are the boundaries of the $Conv[V_{\ell}]$. The positivity of individual $A_{k,I_b}$ tells us that each row $a_{k,q}$ is inside $Conv[V_{\ell}]$. This combined with the previous result tells us that 
 \eq
 a_{k,q}=\sum_{a,\ell} \;p_{a}  (x_{a})^k\;\mathcal{O}_{a,k,\ell}\; V_{\ell,q}, \quad  p_{a}>0\,,\quad x_a>0\,,\quad \mathcal{O}_{a,k,\ell}>0\,.
 \eqe
Finally, the total positivity of $K[\vec{A}_{I_b}]$ then tell us that $\mathcal{O}_{a,k,\ell}$ must be such that 
$(x_{a})^k\;\mathcal{O}_{a,k,\ell}=(x'_{a,\ell})^k$ in other words: 
 \eq
 a_{k,q}=\sum_{a,\ell} \; p_{a}  (x'_{a,\ell})^k\; V_{\ell,q}, \quad  \;p_{a}>0\,\quad x'_{a,\ell}>0\,
 \eqe
we see that indeed eq.(\ref{Def1}) is recovered.

%%%%%%%%%%%%%%%%%%%%%%%%%%%%%%%
\subsection{The geometry of the gap }\label{sec:gap}
%%%%%%%%%%%%%%%%%%%%%%%%%%%%%%
Let's suppose we have the extra information of the scale of the UV completion, i.e. the UV spectrum starts at $M_{\rm Gap}$ above the massless modes. This allows us to write 
\eq
a_{k,0}=\sum_{a}\frac{\textsf{p}_a}{M_{\rm Gap}^{2(k{+}1)}} \left(\frac{M_{\rm Gap}}{m_a}\right)^{2(k{+}1)}=\frac{1}{M_{\rm Gap}^{2(k{+}1)}}\sum_{a}\textsf{p}_a\, x^{k{+}1}_{a}, \quad x_{a}\leq1\,.
\eqe
Now since $x_{a}\leq1$, we see that the gap implies
\eq
a_{2,0}\;\geq M_{\rm Gap}^{2} a_{3,0}\;\geq \cdots \geq M_{\rm Gap}^{2(k{-}2)}a_{k,0}\geq 0\,.
\eqe
The fact that $x_{a}\leq1$ also tells us that the convex hull of $\mathbf{a}_k$ is now over a restricted region of the half-moment curve:
\eqa
&&\mathbf{a}_k=\sum_{a}\textsf{p}_a\left(\begin{array}{cccc}1 & 0 & 0 & 0 \\0 & \frac{1}{M^2_{\rm Gap}} & 0 & 0 \\0 & 0 & \vdots & 0 \\0 & 0 & 0 &   \frac{1}{M^{2(k{-}2)}_{\rm Gap}} \end{array}\right)\left(\begin{array}{c}1 \\ \left(\frac{M_{\rm Gap}}{m_a}\right)^2 \\ \vdots  \\ \left(\frac{M_{\rm Gap}}{m_a}\right)^{2(k{-}2)}  \end{array}\right)\nonumber\\
&&\;\rightarrow \; \left(\begin{array}{cccc}1 & 0 & 0 & 0 \\0 & M^2_{\rm Gap} & 0 & 0 \\0 & 0 & \vdots & 0 \\0 & 0 & 0 &   M^{2(k{-}2)}_{\rm Gap}\end{array}\right)\mathbf{a}_k=\sum_{a}\textsf{p}_a\left(\begin{array}{c}1 \\ x_a \\ \vdots  \\ x_a^{k{-}2}  \end{array}\right)\,,\quad \textsf{p}_a>0, \; x_a\leq 1\,,
\eqae
that is, instead of  $x\in R^+$ we now have  $x\in[0,1]$.  For simplicity we set $M^2_{\rm Gap}=1$ from now on, and we write:
\eq
\mathbf{a}_k=\sum_{a}\textsf{p}_a\left(\begin{array}{c}1 \\ x_a \\ \vdots  \\ x^{k{-}2}_a\end{array}\right)\,,\quad \textsf{p}_a>0, \; x_a\leq 1\,.
\eqe
where the components of $\mathbf{a}_k$ have been rescaled by appropriate factors of $M^2_{\rm Gap}$ to be dimensionless. Now since the curve is bounded by $x_a=1$, we now have a new boundary vertex  
\eq
n_{\rm Gap}=\left(\begin{array}{c}1 \\1 \\ \cdots \\1\end{array}\right)\,.
\eqe
The change in geometry is fully illustrated in the following $\mathbb{P}^2$ example 
$$\includegraphics[scale=0.4]{P2Gap}$$
where the convex hull now has a new boundary consists of $(0,n)$, with $0$ denoting the spin-$0$ vector. Extending to higher dimensions we now have a new set of boundary consists of $(0,i,i{+}1,\cdots,n)$, thus besides the usual Hankel matrix constraints, $\mathbf{a}$ now must also respect 
\eq\label{GapConst}
\langle  0, \mathbf{a},i,i{+}1,\cdots, j,j{+}1,n\rangle>0\,.
\eqe
where we recall $(i,i{+}1)\rightarrow (i,\dot{i})$.

Now the new constraint eq.(\ref{GapConst}) can be translated to the geometry projected through the line $(0,n)$. To see this geometry cleanly, we take a GL transformation $G$ that keeps $0$ fixed and rotate $n$ to: 
\eq
G\; 0=\left(\begin{array}{c}1 \\0 \\ 0\\ \vdots \\0\end{array}\right),\;\;G\; n=\left(\begin{array}{c}0 \\1 \\ 0\\\vdots \\0\end{array}\right),\rightarrow\quad G=\left(\begin{array}{ccccc}1 & {-}1 & 0 & 0 & 0 \\0 & 1 & 0 & 0 & 0 \\0 & 1 & {-}1 & 0 & 0 \\ \vdots & \vdots & \vdots & \vdots & \vdots \\0 & 0 & 0 & 1 & {-}1\end{array}\right),\quad , 
\eqe
The action of $G$ on the moment curve yields 
\eq
G\; \left(\begin{array}{c}1 \\x \\ x^2\\x^3\\\vdots \\ x^d\end{array}\right)=\left(\begin{array}{c}1{-}x \\x \\ x(1{-}x)\\x^2(1{-}x)\\\vdots \\ x^{d-1}(1{-}x)\end{array}\right)\,.
\eqe
Thus after the the GL transformation, the presence of $(0, n)$ in the determinant $\langle  0,\mathbf{a},i,i{+}1,\cdots, j,j{+}1,n\rangle$ simply knocks out the first two component of the other vectors, and eq.(\ref{GapConst}) becomes 
\eq
\langle  \tilde{\mathbf{a}},\tilde{i},\tilde{i}{+}1,\cdots, \tilde{j},\tilde{j}{+}1\rangle>0
\eqe
where the ``$\tilde{\;\;\;}$'' represents the GL transformed vector with the first two components removed. For example 
\eq
G \,\mathbf{a}= \left(\begin{array}{c}a_2{-}a_3\\a_3 \\  a_3{-}a_4 \\ a_4{-}a_5 \\ a_5{-}a_6\\ \vdots\end{array}\right)\rightarrow \; \tilde{\mathbf{a}}= \left(\begin{array}{c}  a_3{-}a_4 \\ a_4{-}a_5 \\ a_5{-}a_6\\ \vdots\end{array}\right)\,.
\eqe

Now  $\tilde{i}$ takes the form :  
\eq
\left(\begin{array}{c} x_i(1{-}x_i)\\x_i^2(1{-}x_i)\\\vdots \\ x_i^{d-1}(1{-}x_i)\end{array}\right)=x_i(1{-}x_i) \left(\begin{array}{c} 1\\x_i\\ \vdots \\ x_i^{d-2}\end{array}\right)
\eqe
which, since $0<x\leq1$, up to a positive factor is once again a moment curve! In other words, the constraint $\langle  \tilde{\mathbf{a}},\tilde{i},\tilde{i}{+}1,\cdots, \tilde{j},\tilde{j}{+}1\rangle>0$ implies that $ \tilde{\mathbf{a}}$, which are twisted sum of $a_i$s,  also satisfies the non-linear Hankel matrix constraint! For example, starting with $\mathbf{a}\in\mathbb{P}^4$, we have  $\tilde{\mathbf{a}}=( a_4{-}a_3 ,a_5{-}a_4,  a_6{-}a_5)$, and the Hanel matrix constraint implies $a_{i}>a_j$ for $i>j$ and  
\eq
(a_3-a_4 )(a_5-a_6)-(a_4-a_5)^2>0\,.
\eqe

The above argument is not all!  We have just noted that  $\tilde{i}$ is positively proportional to a moment curve, but once again since $x\leq1$, it is a capped moment curve and we can reiterate our analysis! The above argument gives an intuitive explanation for the additional gapped Hankel constraints, but with hindsight it is also easy to derive them even more directly. We simply note that if $(a_2,a_3,a_4,a_5,\cdots)$ is in the convex hull of $(1,x,x^2,\cdots)$, then $(a_2{-}a_3,a_3{-}a_4,a_4{-}a_5,\cdots)$ is the the convex hull of $x(1{-}x) \times (1, x, x^2,\cdots)$. Since $x(1{-}x) \geq 0 $ for $0 \leq x \leq 1$, this is the same as the hull of $(1,x,^2,\cdots)$. Thus the discrete derivative $(a_2{-}a_3,a_3{-}a_4,a_4{-}a_5,\cdots)$ must have a totally positive Hankel matrix!

In summary, with a known gap, we can find that the following sequence of ``twisted" couplings satisfies the positive Hankel matrix constraint:
\eq
\left(\begin{array}{c}a_2\\a_3 \\  a_4 \\ a_5 \\ a_6\\a_7\\ \vdots\end{array}\right),\quad\left(\begin{array}{c}  a_3{-}a_4 \\ a_4{-}a_5 \\ a_5{-}a_6\\a_6{-}a_7\\ \vdots\end{array}\right),\quad \left(\begin{array}{c} (a_4{-}a_5)-(a_5{-}a_6)\\(a_5{-}a_6)-(a_6{-}a_7) \\  \vdots\end{array}\right)\,.
\eqe
This is known as the Hausdorff moment problem~\cite{Hausdorff}. The extra constraints from the knowledge of the gap are interesting, however, they are obviously  of 
 only academic interest to the low-energy observer that has no knowledge of the gap. Any higher-dimension operator measured by a low-energy observer could be produced by arbitrarily weakly coupled, arbitrarily low-mass states, and in the limit where the masses and couplings go to zero we recover the pure Hankel constraints. Note that the pure Hankel constraints are homogeneous in mass dimensions, comparing sums of products of couplings with the same total mass dimension, which are the only sorts of constraints we can talk about without knowledge of an absolute mass scale (such as the gap).  For this reason, in the rest of this paper, we will focus on these types of universal constraints on that can be sensibly formulated in the low-energy theory, assuming no knowledge of the gap.  

\newpage

%%%%%%%%%%%%%%%%%%%%%%%%%%%%%%%
\section{Scalar EFT-hedron}\label{sec:EFT-hedron}
%%%%%%%%%%%%%%%%%%%%%%%%%%%%%%
So far we have restricted ourselves to the geometry arising from singularities on the positive real $s$-axis. For a general $2\rightarrow 2$ process, $M(1^a,2^b,3^b,4^a)$, the amplitude will have poles and discontinuities on \textit{both} positive and negative real $s$-axes, reflecting $s$ and $u$-channel exchanges:
$$\includegraphics[scale=0.7]{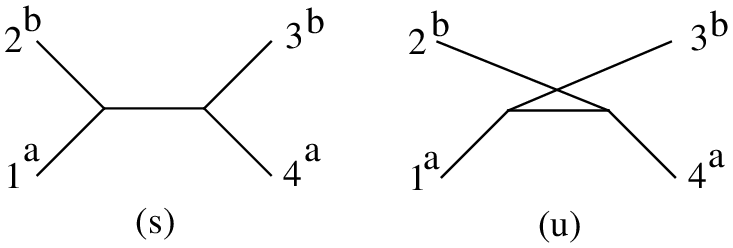}$$
The residue or discontinuity on the $s$-channel as a function of $t$ will be identical to that in the $u$-channel since the two diagrams are related via $2\leftrightarrow 3$ exchange. However, while the residues are the same, the $u$-channel singularities lie on the negative $s$-axes with a $t$-dependent shift: $u-m^2=-s-(t{+}m^2)$. In other words, the low energy couplings are now governed by the Taylor expansion of:
\eq
{-}\sum_{a}\textsf{p}_{\ell_a}\left[\frac{1}{s-m_a^2}{+}\frac{1}{-s-t-m_a^2}\right]G_{\ell_a}\left(1{+}\frac{2t}{m_a^2}\right)
\eqe
Recall that in the previous section, the $s$-channel EFT-hedron is the direct product of the positive geometry of the Gegenbauer vectors and that of the moment curve.  Compared to the above one can see that we now have a new feature: upon Taylor expansion, the $t$ in the $u$-channel will mix with that from $G_{\ell}(1+\frac{2t}{m^2})$, and the two geometry is no longer a direct product, but  ``entangled".

Due to the $s,u$ symmetry, it will be more convenient to parameterize our kinematics as 
\eq
s=-\frac{t}{2}+z,\quad u=-\frac{t}{2}-z\,,
\eqe
and the four-point amplitude is a function of $z,t$, $M(z,t)$. The low energy couplings are now extracted from the Taylor expansion of:
\eqa
&&-\sum_a\textsf{p}_a\left(\frac{1}{-\frac{t}{2}-z-m^2_a}+\frac{1}{-\frac{t}{2}+z-m^2_a}\right)G_{\ell_a}\left(1+\frac{2t}{m^2_a}\right)\nonumber\\
\eqae
The resulting Taylor expansion only has even powers of $z$, which is a reflection of the underlying $s\leftrightarrow u$ symmetry. If we consider the geometry associated with fixed $k$ or fixed $q$, then the geometry here is the Minkowski sum of the $s$- and $u$-channel convex hull.  Thus we have\footnote{Here we define the couplings $a_{k,q}$ with respect to powers of $z,t$. To avoid proliferation of new couplings, we will continue to use the notation $a_{k,q}$ where the context is obvious.}
\eq\label{suConvex}
a_{k,q}z^{k{-}q} t^q=\sum_{a}\textsf{p}_a\left[x^{k+1}_a\; u_{\ell_a,k,q}\right]z^{k{-}q} t^q\,\quad k{-}q\in {\rm even}\,.
\eqe 
where the coefficients $u_{\ell,k,q}$ are linear combinations of Gegenbauer Taylor coefficients $v_{\ell,q}$s:
\eq\label{Sum}
u_{\ell,k,q}=\sum_{a+b=q}(-)^a\frac{(k{-}q+1)_a}{a!}2^{b-a}v_{\ell,b}\,.
\eqe
Thus for fixed $k$, the couplings must live inside $Conv[\vec{u}_{\ell,k}]$, where
\eqa
k\in even: &&\vec{u}_{\ell,k}=(u_{\ell,k,0},\,u_{\ell,k,2},\cdots,\,u_{\ell,k,k} )\nonumber\\
k\in odd:  &&\vec{u}_{\ell,k}=(u_{\ell,k,1},\,u_{\ell,k,3},\cdots,\,u_{\ell,k,k})\,.
\eqae
Importantly, the vectors $\vec{u}_{\ell,k}$ are labeled by both the spin and $k$. This $k$-dependence was absent in the $s$-channel analysis, where  $Conv[\vec{v}_{\ell}]$ only depends on spin. This new feature leads to an important distinction between $s$-channel and full EFT-hedron. 

Due to the absence of $z^{odd}$ terms, at fixed $k$ the dimensionality of $\vec{u}_{\ell,k}$ is smaller than  $\vec{v}_{\ell}$ (half for $k\in$ odd). More precisely, $\vec{u}_{\ell,k}$ is obtained by a GL rotation of $\vec{v}_{\ell}$ that projects away the odd components. For example for $k\in even$ we have:
\eq\label{uvec}
\vec{u}_{\ell,k}=\left(\begin{array}{c}u_{\ell,k,0} \\0 \\u_{\ell,k,2}\\ \vdots \\0\\ u_{\ell,k,k}\end{array}\right)=\left(\begin{array}{cccccc}1 & 0 & 0 & 0 & 0 & 0 \\0 & 0 & 0 & 0 & 0 & 0 \\ \frac{(k{-}1)_{2}}{2}\frac{1}{2^2}& (k{-}1)_{1} & 1 & 0 & 0 & 0 \\ \vdots & \vdots & \vdots & \vdots & \vdots & \vdots \\0 & 0 & 0 & 0 & 0 & 0 \\\frac{(1)_k}{k!}2^{-k} & -\frac{(1)_{k{-}1}}{k{-}1!}2^{2-k} & \frac{(1)_{k{-}2}}{k{-}2!}2^{4-k} & \cdots & -2^{k{-}2} & 1\end{array}\right)\left(\begin{array}{c}v_{\ell,0} \\ v_{\ell,1} \\v_{\ell,2}\\ \vdots \\v_{\ell,k{-}1}\\ v_{\ell,k}\end{array}\right)\,.
\eqe 
Due to this projection,  $Conv[\vec{u}_{\ell,k}]$  does not inherit the positivity of $Conv[\vec{v}_{\ell}]$, and thus we cannot conclude that $Conv[\vec{u}_{\ell,k}]$ is a cyclic polytope. Similarly for fixed $q$, comparing the coefficient of $x_a^{k{+}1}$ in eq.(\ref{suConvex}) with the $s$-channel eq.(\ref{Prop}), we see that the $k$-dependence of $u_{\ell_a,k,q}$ results in each moment $x_a^{k{+}1}$ being weighted differently, and we no longer have a momentum curve. Thus naively, the positivity geometry that defined the $s$-channel EFT-hedron is lost, and we no longer have control over the geometry.  As we will now see, there is in fact a hidden positivity that retains most of the structure of the $s$-channel cyclic polytope, and thus allowing us to carve out \textit{the} EFT-hedron.

%%%%%%%%%%%%%%%%%%%%%%%%%%%%%%%
\subsection{The $s{-}u$ polytope}\label{sec:supoly}
%%%%%%%%%%%%%%%%%%%%%%%%%%%%%%
Let us consider the boundaries of the  $(s{-}u)$ polytope, i.e. $Conv[\vec{u}_{\ell,k}]$. We will be interested in the sign for the determinant of ordered  $\vec{u}_{\ell,k}$s. Setting $k=4$ as an example, we find:
\eqa\label{udet}
&&{\rm Det}\left(\begin{array}{ccc}\vec{u}_{\ell_1,4} &\vec{u}_{\ell_2,4}& \vec{u}_{\ell_3,4}  \end{array}\right) ={\rm Det}\left(\begin{array}{ccc}v_{\ell_1,0} &\; & \;   \\ v_{\ell_1,2}-\frac{3}{4}v_{\ell_1,1} & \{\ell_2\}  &  \{ \ell_3\} \\ v_{\ell_1,4}-\frac{1}{4}v_{\ell_1,3}+\frac{1}{16}v_{\ell_1,2}-\frac{1}{64}v_{\ell_1,1} &\; &\; \end{array}\right)\nonumber\\
&=&{\rm Det}\left(\begin{array}{ccc}v_{\ell_1,0} &\; &\;   \\ v_{\ell_1,2}  &\{\ell_2\} & \{ \ell_3\} \\ v_{\ell_1,4}  &\; & \;\end{array}\right)-\frac{3}{4}{\rm Det}\left(\begin{array}{ccc}v_{\ell_1,0} &\;  & \;   \\ v_{\ell_1,1}  &\{ \ell_2\} & \{ \ell_3\} \\ v_{\ell_1,4}  &\; & \; \end{array}\right)-\frac{1}{32}{\rm Det}\left(\begin{array}{ccc}v_{\ell_1,0} &\; & \;   \\ v_{\ell_1,1}  &\{\ell_2\} & \{ \ell_3\} \\ v_{\ell_1,2}  &\; & \; \end{array}\right)\nonumber\\
&-&\frac{1}{4}{\rm Det}\left(\begin{array}{ccc}v_{\ell_1,0} &\; & \;   \\ v_{\ell_1,2}  &\{\ell_2\} & \{ \ell_3\} \\ v_{\ell_1,3}  &\; & \; \end{array}\right)
+\frac{3}{16}{\rm Det}\left(\begin{array}{ccc}v_{\ell_1,0} &\; & \;   \\ v_{\ell_1,1}  &\{\ell_2\} & \{ \ell_3\} \\ v_{\ell_1,3}  &\; & \; \end{array}\right)+\cdots\,,
\eqae
where $\{\ell_i\}$ represent the same as the first column just with $\ell_1\rightarrow\ell_i$, and $\ell_1<\ell_2<\ell_3$. We see that the determinant for ordered $\vec{u}_{\ell,k}$ is given by a sum of determinant for ordered $\vec{v}_{\ell,k}$ with mixed signs, and thus the positivity of the later do not imply that for the former.

\begin{figure}
\begin{center}
\includegraphics[scale=0.3]{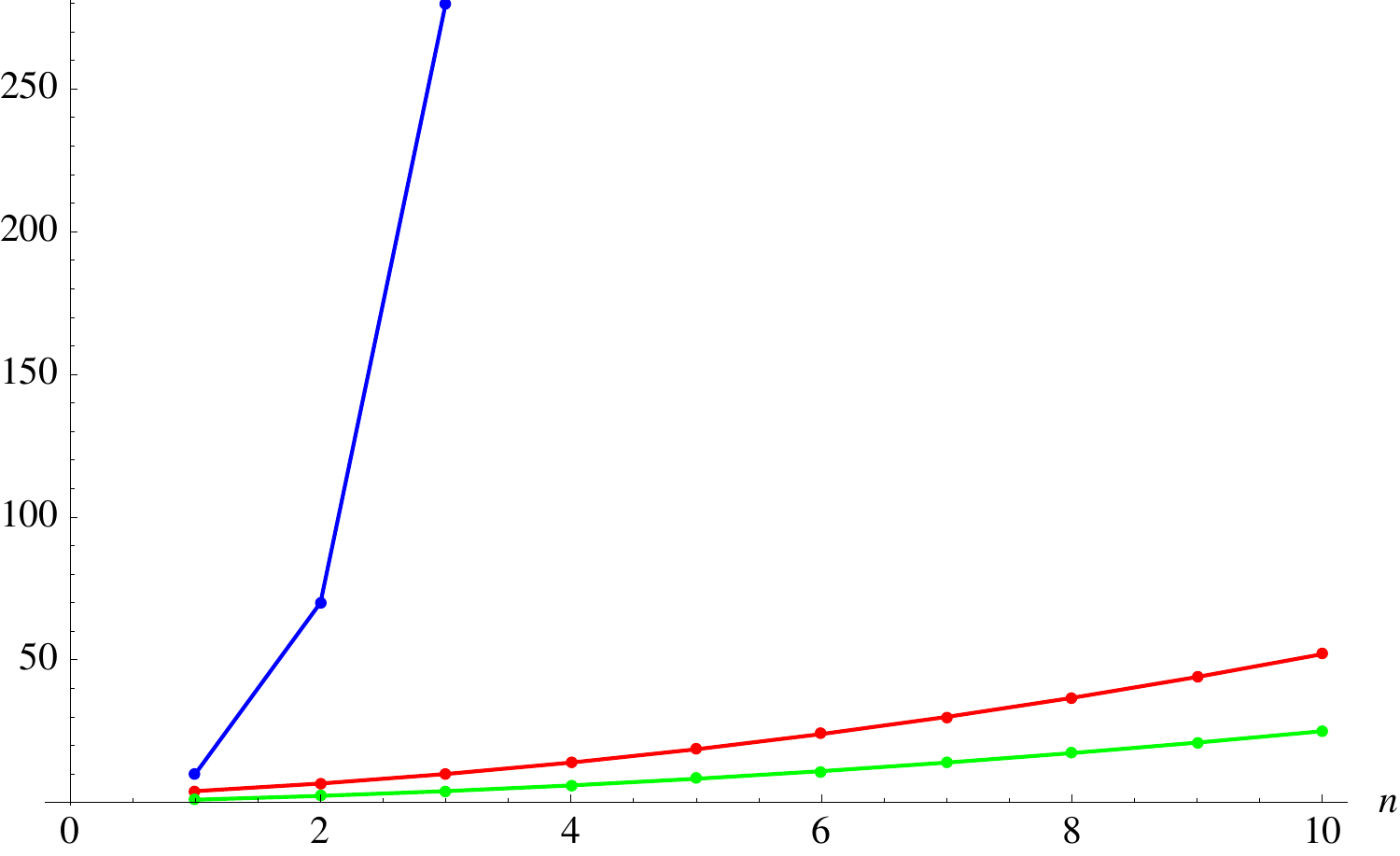}
\caption{We take the ratio of the four determinants in the second and third line in eq.(\ref{udet}), denoted as $m_i(\ell_1,\ell_2,\ell_3)$, for $i=1,\cdots,4$. We plot $\frac{m_1(1{+}n,2{+}n,3{+}n)}{m_2(1{+}n,2{+}n,3{+}n)}$ (red), $\frac{m_1(1{+}n,2{+}n,3{+}n)}{m_3(1{+}n,2{+}n,3{+}n)}$ (blue), and $\frac{m_1(1{+}n,2{+}n,3{+}n)}{m_4(1{+}n,2{+}n,3{+}n)}$ (green), with $n=1,\cdots$. As we can see, $m_1(\ell_1,\ell_2,\ell_3)$ is the largest and the ratio is an increasing function with spins.}
\label{RatioPlots}
\end{center}
\end{figure}

Amazingly, explicit evaluations of eq.(\ref{udet}) reveals that the determinant is positive so long as $\{\ell_i\}$s are larger than some critical spin ! That is, above some critical spin, $\ell_c$,
\eq
{\rm Det}[\{\vec{u}_{\ell_1,k},\vec{u}_{\ell_2,k},\cdots\}]>0,\quad \forall \;\ell_c\leq\ell_1<\ell_2<\cdots\,.
\eqe   
In other words the convex hull of Gegenbauer vectors above the critical spin yields a cyclic polytope.\footnote{A fun ``historic" note, the authors actually first observed the positivity of the ordered determinants for $\vec{u}_{\ell,k}$, not $\vec{v}_{\ell}$.}  For example, focusing on four-dimensions, we find the critical spin at different $k$  given as:
\eq\label{CriticalTable}
\begin{tabular}{|c|c|c|c|c|c|c|c|c|c|}
\hline
  $k$ & 2 & 3 & 4 & 5 & 6 &7 &8 &9&10\\
  \hline
   $\ell_c$  & 1 & 2  & 2 & 3 & 3 &4 &4 &5 &5 \\
\hline
\end{tabular}\,.
\eqe
It is intriguing to understand how such positivity emerged. In the RHS of eq.(\ref{udet}), each term can be identified as a minor of the Gegenbauer matrix with half of the rows removed. Consider the ratio of the first term on the RHS of eq.(\ref{udet}), against the next three. The first term has the property that it retains only even Taylor expansion terms. We plot these ratios for spins $(\ell_1,\,\ell_2,\,\ell_3)=(1+n,2+n,3+n)$ in fig.(\ref{RatioPlots}). As we can see, the leading term is dominant to the others as we increase in spin. Thus the even though the other determinants in eq.(\ref{udet}) may have negative coefficients, their contributions are overwhelmed by the leading term which leads to the observed positivity. In other words, the minors with all even (or odd depending on the dimensions) Taylor coefficients take the maximal value!

The fact that $\vec{u}_{\ell,k}$ form a cyclic polytope above the critical spin, indicates that for our $s{-}u$ polytope, most of the boundaries are known except for those involving spins below the critical spin, which can be computed straightforwardly. For coefficients that we can reliably bound, i.e. those proportional to $z^n$ with $n\geq 2$. For $k{=}2$ ($D^4\phi^4$), we have 
\eq
M_{D^4\phi^4}=(a_{2,0}z^2+a_{2,2}t^2)
\eqe
which simply gives us $a_{2,0}>0$. For higher $k$, we have:
\begin{itemize}
  \item $k{=}3:D^6\phi^4$ 
 \eq
M_{D^6\phi^4}=(a_{3,1}z^2t{+}a_{3,3}t^3)
\eqe
  Here we again have a single coefficient $a_{3,1}$ to bound. Since
  \eq
  u_{\ell,3,1}=\{-3,1,9,21,...\}\,,
  \eqe
  due to the first entry being negative, the positive span of these numbers will cover the whole real line, meaning we have \textit{no bound} for the coefficient $a_{3,1}$.
  
   \item $k{=}4:D^8\phi^4$
 \eq
M_{D^8\phi^4}=( a_{4,0}z^4+a_{4,2}z^2t^2+a_{4,4}t^4)
\eqe
 We can hope to bound $(a_{4,0},a_{4,2})$. The $\vec{u}_{l,k}$ for each spin is
   \eq
   \left(\begin{array}{c}u_{\ell,4,0} \\ u_{\ell,4,2}\end{array}\right)=\left(\begin{array}{c}2 \\ 3\end{array}\right), \left(\begin{array}{c}2 \\ {-}3\end{array}\right),\left(\begin{array}{c}2 \\ {-}3\end{array}\right),\left(\begin{array}{c} 2  \\ 27\end{array}\right),\;\cdots 
   \eqe
  Projectively these are points in $\mathbb{P}^1$, and the boundaries are given by the minimum and maximum value for the ratio $ \frac{u_{\ast,4,2}}{u_{\ast,4,0}}$, which is given by $-\frac{3}{2}$ and $\infty$ respectively. Thus we simply have the bound: 
   \eq
   \frac{a_{4,2}}{a_{4,0}}\geq -\frac{3}{2}\,.
   \eqe

   \item $k{=}5:D^{10}\phi_4$
   \eq
   M_{D^{10}\phi_4}(s,t)=(a_{5,1}z^4t{+}a_{5,3}z^2t^3{+}\cdots)
   \eqe 
 where we've suppressed the couplings that we cannot bound. We would like to bound $(a_{5,1},a_{5,3})$ and the space is $\mathbb{P}^1$. However, listing the relevant contributions from each spin  
   \eq
    \frac{u_{\ell,5,3}}{u_{\ell,5,1}}=\left\{\frac{1}{2},{-}\frac{7}{2},-\frac{5}{14},-\frac{33}{38},...\right\},
   \eqe
we see that just as in the $k=3$ case, the positive span will cover the entire $\mathbb{P}^1$, and thus the bound is trivial. 
 
   \item $k{=}6:D^{12}\phi_4$
   \eq
    M_{D^{12}\phi_4}=(a_{6,0}z^6{+}a_{6,2}z^4t^2{+}a_{6,4}z^2t^4{+}\cdots)
   \eqe
  we can bound $\mathbf{a}_6{=}(a_{6,0},a_{6,2},a_{6,4})$ and the geometry is  $\mathbb{P}^2$. The boundaries are given by:
  \eq
  \langle \mathbf{a}_6, 2, 1\rangle,\quad 
  \langle \mathbf{a}_6, 1, 4\rangle,\quad
  \langle \mathbf{a}_6, i,i{+}1\rangle_{i\geq 4},\quad
  \langle \mathbf{a}_6, \infty, 2\rangle.
  \eqe
  
We see that $Conv[\vec{u}_{\ell,6}]$ retains most of the boundaries of a cyclic polytope. Note that since the spin-$0$ and $3$ vector are not involved with any boundary, they are inside the hull.\footnote{Here, the critical spin is $4$ instead of $3$ as listed in table \ref{CriticalTable}. This is because here we are only keeping the first three components of $\vec{u_{\ell,6}}$, i.e. $u_{\ell,6,0},u_{\ell,6,2},u_{\ell,6,4}$.}

\end{itemize}

Moving to higher-$k$s, in general there are no bounds for $k\in odd$, while for $k \in even$ we have the familiar cyclic polytope boundaries above a critical spin and a few additional boundaries involving spins below the critical spin.

\noindent \textbf{Identical scalars: intersecting with the permutation symmetry plane}

\begin{figure}
\begin{center}
\includegraphics[scale=0.5]{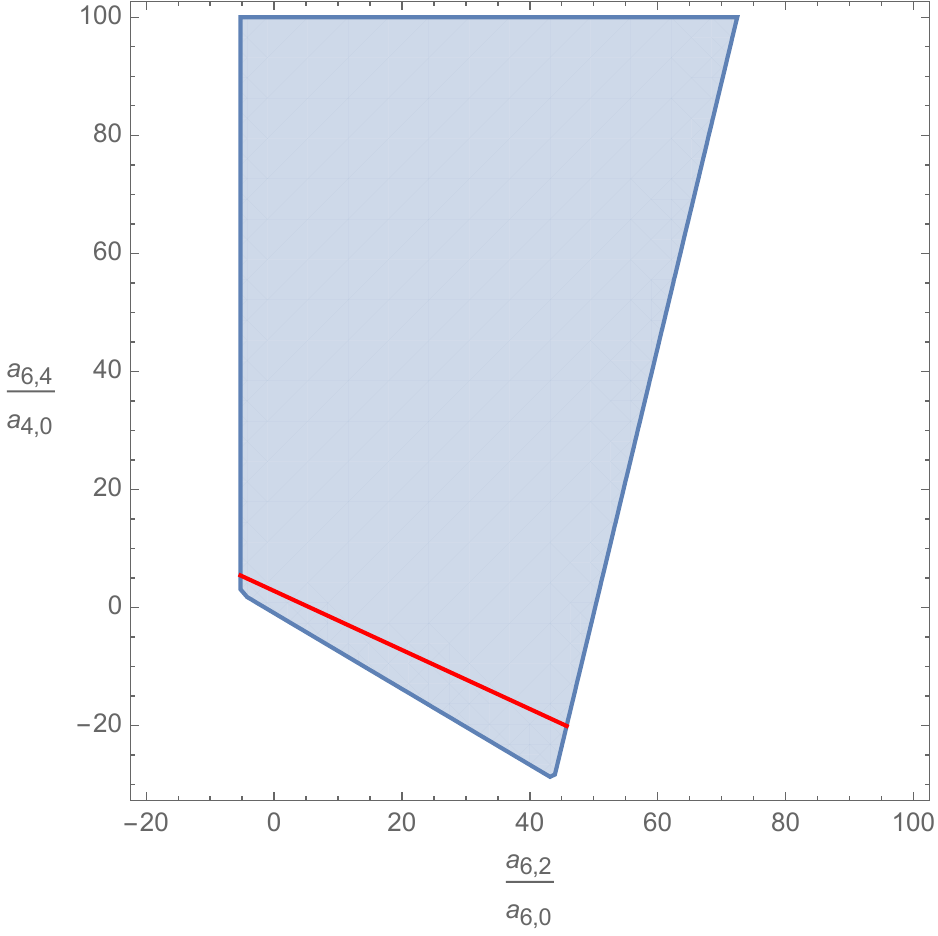}
\caption{The space of allowed $(\frac{a_{6,2}}{a_{6,0}}, \frac{a_{6,4}}{a_{6,0}})$. The shaded region is carved out  by the unitary polygon, whose boundary is comprised of $(\ell,\ell+2)$ with $\ell\geq2$, and $(2,\infty)$. Note that spin-0 is inside the hull and thus not part of the boundary. Finally the red-line represents in the intersection of the permutation ``line" $\mathbf{X}_{\rm perm}$, and the unitary polygon. }
\label{PermFig}
\end{center}
\end{figure}

When the scalars are identical, the amplitude further respects permutation invariance, and at low energies will be given as a polynomial in $\sigma_2=(s^2+t^2{+}u^2)$ and $\sigma_3=(s^3{+}t^3{+}u^3)$. This translate to the couplings $a_{k,q}$ living on the permutation plane  $\mathbf{X}_{\rm perm}$, defined through,
\eq\label{PermXDef}
\mathbf{X}_{\rm perm}: \quad M\left(z,t\right)=M\left( \frac{z}{2} {+} \frac{3 t}{4}, {-}\frac{t}{2} {+} z\right)\,.
\eqe 
Thus the geometry of interest is the intersection between $\mathbf{X}_{\rm perm}$ and the unitary polytope, where the later is now constructed from \textit{even spins} only. The dimensionality of $\mathbf{X}_{\rm perm}$ is the number of independent polynomials built from $\sigma_3$ and $\sigma_2$. For $k=2,4$ the polynomial is unique, and the first place where there are two possibilities is $k=6$:  $\sigma_3^2$ and $\sigma_2^3$. On $\mathbf{X}_{\rm perm}$ the couplings are parameterize as: 
\eq\label{perminvplane}
\left(\begin{array}{cccccc}a_{2,0} & a_{2,2} & \; & \;& \; & \; \\ a_{4,0} & a_{4,2} & a_{4,4} & \; & \; & \;\\ a_{6,0} & a_{6,2} & a_{6,4} & a_{6,6}& \; & \; \\ a_{8,0} & a_{8,2} & a_{8,4} & a_{8,6}&  a_{8,8} & \;  \end{array}\right)\rightarrow\left(\begin{array}{cccccc}e_2 & \frac{3}{4}e_2 & \; & \;& \; & \; \\ e_4 & \frac{ 3}{ 2}e_4 & \frac{9}{16}e_4 & \; & \; & \;\\ e_6 & f_6 & \frac{45}{16}e_6 - \frac{1}{2}f_6 & \frac{9}{32}e_6 + \frac{1}{16}f_6& \; & \; \\ e_8 & f_8 & \frac{21}{8}e_8  + \frac{1}{4} f_8 & \frac{21}{8} e_8- \frac{5}{16}f_8 & \frac{45}{256}e_8  + \frac{3}{64} f_8 & \;  \end{array}\right)\,.
\eqe
For $k=2,4$ we simply have the bound $e_2,e_4>0$. At $k=6,8$, the boundaries bound the ratio $\frac{f}{e}$ to be:
\eq\label{UpLowBound}
k=6:\quad-\frac{21}{4} < \frac{f_6}{e_6} < \frac{183}{4},\quad k=8:\quad-8 < \frac{f_8}{e_8}< \frac{223}{4}\,.
\eqe
In fig.\ref{PermFig} we display the intersection geometry in $\mathbb{P}^2$ for $k=6$.

These can be explicitly checked against the spinor-bracket stripped type-II closed string amplitude:
\eq
\frac{\Gamma[{-}s]\Gamma[{-}u]\Gamma[{-}t]}{\Gamma[1{+}s]\Gamma[1{+}u]\Gamma[1{+}t]}\,.
\eqe
We can then identify:
\eq
k=6:\;\; \frac{f_6}{e_6}=\frac{(8 \zeta_3^3 + 31 \zeta_9)}{12 \zeta_9}=3.73895,\quad k=8:\;\;\frac{f_8}{e_8}=\frac{2(2 \zeta_{11} +  \zeta_5\zeta^2_3)}{ \zeta_{11}}=6.99512\,. 
\eqe 
We see that it indeed resides in the bounds given by eq.(\ref{UpLowBound}).

%%%%%%%%%%%%%%%%%%%%%%%%%%%%%%%
\subsection{Deformed moment curves and the EFT-hedron}
%%%%%%%%%%%%%%%%%%%%%%%%%%%%%%%
We've seen the $k$-dependence of $\vec{u}_{\ell,k}$ leads to a deformation of the cyclic polytope discussed in the $s$-channel geometry. Now we would like to see how such mixing modifies the Hankel constraints, and the EFT-hedron.

\noindent \textbf{Deformed moment curves}

Let's again collect the coefficient with different $k$s and fixed $q$ in to a column vector:
 \eq\label{DeformedMoment}
 \left(\begin{array}{c} a_{2,q} \\ a_{4,q}\\ a_{6,q} \\ \cdots  \\ a_{k,q}\end{array}\right)=\sum_{a}\textsf{p}_{a} \left(\begin{array}{c} u_{\ell_a,2,q} \\ u_{\ell_a,4,q}\,x_a \\ u_{\ell_a,6,q}\,x^2_a\\ \cdots  \\ u_{\ell_a,k,q}\,x^{\frac{k{-}2}{2}}_a\end{array}\right)\,.
 \eqe 
For $q=0$ as $u_{\ell,k,0}=v_{\ell,0}\geq0$, the vectors on the RHS are just points on a moment curve multiplied by an overall positive factor and the usual  Hankel matrix constraint applies.  For $q\neq 0$, the $k$ dependence of $u_{\ell,k,q}$ spoils this overall proportionality. This leads us to consider a generalization of moment curves:  given a set of distinct  positive factors $\alpha_i$, we define a \textit{deformed} moment curve $(1,x,\alpha_1 x^2,\cdots, \alpha_{n{-}1} x^n)$.  Note that the convex hull of such deformed moment curve can be straight forwardly carved out by the total positivity of the \textit{rescaled} Hankel matrix:
 \eq
 \left(\begin{array}{cc}a_{4,q} & a_{6,q}  \\ a_{6,q} & \frac{a_{8,q}}{\alpha_1}  \end{array}\right),\quad\left(\begin{array}{cc}a_{6,q} & \frac{a_{8,q}}{\alpha_1}  \\ \frac{a_{8,q}}{\alpha_1} & \frac{a_{10,q}}{\alpha_2}  \end{array}\right),\quad   \left(\begin{array}{ccc}a_{4,q} & a_{6,q}  & \frac{a_{8,q}}{\alpha_1} \\ a_{6,q}  & \frac{a_{8,q}}{\alpha_1}  &  \frac{a_{10,q}}{\alpha_2}  \\  \frac{a_{8,q}}{\alpha_1}  & \frac{a_{10,q}}{\alpha_2}  & \frac{a_{12,q}}{\alpha_3} \end{array}\right)\cdots\,.
  \eqe
However, this is not sufficient to describe eq.(\ref{DeformedMoment}) for two reason: 1. while each vector on the RHS of eq.(\ref{DeformedMoment}) is a point on a rescaled moment curve, the scaling factors are \textit{distinct} for different spins, and 2. the rescaled factor $u_{\ell,k,q}$ is not necessarily positive. 

Let's instead collect the different $q$s into row vectors $\vec{u}_{\ell,k}$ and $\vec{a}_{k}$, and rewrite eq.(\ref{DeformedMoment}) as:  
\eq\label{DeformC}
 \left(\begin{array}{c} \vec{a}_{2} \\ \vec{a}_{4} \\ \vec{a}_{6} \\ \cdots  \\ \vec{a}_{k}\end{array}\right)=\sum_{a}\textsf{p}_{a} \left(\begin{array}{c} \vec{u}_{\ell_a,2} \\ \vec{u}_{\ell_a,4}\,x_a\\ \vec{u}_{\ell_a,6}\,x^2_a\\ \cdots  \\ \vec{u}_{\ell_a,k}\,x^{\frac{k{-}2}{2}}_a\end{array}\right)\,.
 \eqe 
Here each vector $\vec{u}_{\ell,k}$ will be of the same dimension. Now denote the boundaries of $Conv[\vec{u}_{\ell,k}]$ as $\vec{\mathcal{W}}^{k}_I$. The inner product $(\vec{u}_{\ell,k}\cdot\vec{\mathcal{W}}^{k'}_I)$ by construction will give a positive factor when $k=k'$, but no longer guaranteed for $k'\neq k$. If we find some wall such that $(\vec{u}_{\ell,k}\cdot\vec{\mathcal{W}}_I)$ is always positive, then we are in business. Thus the task at hand is to find the boundary for $Conv[\vec{u}_{\ell,2}, \vec{u}_{\ell,4}, \cdots]$, i.e. we will be interested in  the boundary of the Minkowski sum.  Remarkably, numerical analysis so far has shown that the boundaries of $Conv[\vec{u}_{\ell,2}, \vec{u}_{\ell,4}, \cdots]$ are simply that of the highest $k$. 
\eq\label{MaxBound}
Conv[\vec{u}_{\ell,k_1}]\subset Conv[\vec{u}_{\ell,k_2}],\quad \forall k_1<k_2\,,
\eqe 
in other words the inner product of $\vec{u}_{\ell,k}$ with $\vec{\mathcal{W}}^{k'}_I$ is guaranteed do be positive for $k\geq k'$.

 Let us take eq.(\ref{DeformC}) and dotted into the boundaries of the highest $k$:
\eq\label{DeformCurve}
\left(\begin{array}{c} \vec{a}_{2}\cdot \vec{\mathcal W}^k_I \\ \vec{a}_{4}\cdot \vec{\mathcal W}^k_I  \\ \vec{a}_{6}\cdot \vec{\mathcal W}^k_I \\ \cdots  \\ \vec{a}_{k}\cdot \vec{\mathcal W}^k_I\end{array}\right)=\sum_{a}\textsf{p}_{a}\left(\begin{array}{c}  (\vec{u}_{\ell_a,2}\cdot \vec{\mathcal W}^k_I)\, \\ (\vec{u}_{\ell_a,4}\cdot \vec{\mathcal W}^k_I)\,x_a\\ (\vec{u}_{\ell_a,6}\cdot \vec{\mathcal W}^k_I)\,x^2_a \\ \cdots  \\ (\vec{u}_{\ell_a,k}\cdot \vec{\mathcal W}^k_I)\,x_a^{\frac{k{-}2}{2}}\end{array}\right)
\eqe
Since by construction $\vec{u}_{\ell,k}\cdot \vec{\mathcal W}^k_I\geq 0$, the RHS gives a sum over points on a set of deformed moment curves, with the  deformation parameters given as $\{\vec{\alpha}_\ell\}=\{\vec{u}_{\ell,2}\cdot \vec{\mathcal W}^k_I, \vec{u}_{\ell,4}\cdot \vec{\mathcal W}^k_I, \cdots\}$. Note that the $\{\vec{\alpha}_\ell\}$s are distinct for each spin.

Now we have arrived at a well posed positive geometry: the convex hull of an infinite number of deformed half moment curves. To proceed we will construct a ``\textit{principle deformed curve}" such that the deformed curves defined by $\{\vec{\alpha}_\ell\}$ resides in the hull of the former, i.e. we will like to find a set of parameters $\{\tilde{\alpha}_i\}$ that defines a deformed moment curve whose convex hull encapsulates the RHS of eq.(\ref{DeformCurve}) for all $\ell$. Note that since $\{\vec{\alpha}_\ell\}$ depends on the boundary $\vec{\mathcal W}^k_I$, so will $\{\tilde{\alpha}_i\}$. Let us see how this work in practice. 

\begin{itemize}
  \item \textbf{k=6}: 
  Beginning with eq.(\ref{DeformCurve}) and setting $k=6$, we would like to find a deformed moment curve 
\eq
(1,x,\alpha_1 x^2)
\eqe
such that the RHS of eq.(\ref{DeformCurve}) lies inside its convex hull. Since the being inside its hall translates to total positivity of the deformed Hankel matrix, we conclude that we need to find $\alpha_1$ such that
\eq\label{conditions1}
 \left(\begin{array}{cc}\vec{u}_{\ell,2}\cdot \vec{\mathcal W}^6_I & \vec{u}_{\ell,4}\cdot \vec{\mathcal W}^6_I  \\ \vec{u}_{\ell,4}\cdot \vec{\mathcal W}^6_I & \frac{\vec{u}_{\ell,6}\cdot \vec{\mathcal W}^6_I}{\tilde\alpha_1}  \end{array}\right),
\eqe
is totally positive for all $\ell$, or
\eq
\frac{(\vec{u}_{\ell,6}\cdot \vec{\mathcal W}^6_I)(\vec{u}_{\ell,2}\cdot \vec{\mathcal W}^6_I)}{(\vec{u}_{\ell,4}\cdot \vec{\mathcal W}^6_I)^2}\geq \tilde\alpha_1,\quad\forall \ell\,, 
\eqe
Thus there is a maximal value for $\tilde{\alpha}_1$ corresponding to the minimal value of the RHS of the above. Importantly, since some of the vectors $\vec{u}_{\ell,6}$ will inevitably be on the boundary $\vec{\mathcal W}^6_I$, the upper bound for $\tilde{\alpha}_1$ is actually zero! To this end, it will be natural to consider boundaries that are \textit{outside} of $Conv[\vec{u}_{\ell,6}]$, which we will denote as $\vec{\mathcal W}^{6'}_I\equiv \vec{\mathcal W}^6_I+\Delta w$. The value for $\tilde{\alpha}_i$ now becomes $\Delta w$ dependent. 

  \item \textbf{k=8}:  taking $k=8$ on the RHS of eq.(\ref{DeformCurve}) for fixed $\vec{\mathcal W}^8_I$, the independent positivity constraint will be the total positivity of  
 \eq\label{conditions2}
 \left(\begin{array}{cc}\vec{u}_{\ell,2}\cdot \vec{\mathcal W}^{8'}_I & \vec{u}_{\ell,4}\cdot \vec{\mathcal W}^{8'}_I  \\ \vec{u}_{\ell,4}\cdot \vec{\mathcal W}^{8'}_I & \frac{\vec{u}_{\ell,6}\cdot \vec{\mathcal W}^{8'}_I}{\tilde\alpha_1}  \end{array}\right), \quad \left(\begin{array}{cc}\vec{u}_{\ell,4}\cdot \vec{\mathcal W}^{8'}_I & \frac{\vec{u}_{\ell,6}\cdot \vec{\mathcal W}^{8'}_I}{\tilde\alpha_1}  \\ \frac{\vec{u}_{\ell,6}\cdot \vec{\mathcal W}^{8'}_I}{\tilde\alpha_1} & \frac{\vec{u}_{\ell,8}\cdot \vec{\mathcal W}^{8'}_I}{\tilde\alpha_2}  \end{array}\right)\,,
\eqe
where once again $\vec{\mathcal W}^{8'}_I=\vec{\mathcal W}^{8}_I+\Delta w$. To find a set of suitable $(\tilde\alpha_1,\tilde\alpha_2)$, we first solve total positivity for the first matrix to determine $\tilde\alpha_1$, and use the result to solve the second matrix to determine $\tilde\alpha_2$. 

\end{itemize}

For general $k$ one iteratively solves the $\tilde\alpha_i$ in sequence. As a final example, for $k=10$ we simply iteratively solve total positivity of the following three matrices 
\eq\label{conditions}
 \left(\begin{array}{cc}\vec{u}_{\ell,2}\cdot \vec{\mathcal W}^{10'}_I & \vec{u}_{\ell,4}\cdot \vec{\mathcal W}^{10}_I  \\ \vec{u}_{\ell,4}\cdot \vec{\mathcal W}^{10'}_I & \frac{\vec{u}_{\ell,6}\cdot \vec{\mathcal W}^{10'}_I}{\tilde\alpha_1}  \end{array}\right),\quad  \left(\begin{array}{cc}\vec{u}_{\ell,4}\cdot \vec{\mathcal W}^{10'}_I & \frac{\vec{u}_{\ell,6}\cdot \vec{\mathcal W}^{10'}_I}{\tilde\alpha_1}  \\ \frac{\vec{u}_{\ell,6}\cdot \vec{\mathcal W}^{10'}_I}{\tilde\alpha_1} & \frac{\vec{u}_{\ell,8}\cdot \vec{\mathcal W}^{10'}_I}{\tilde\alpha_2}  \end{array}\right),\quad  \left(\begin{array}{ccc}\vec{u}_{\ell,2}\cdot \vec{\mathcal W}^{10'}_I & \vec{u}_{\ell,4}\cdot \vec{\mathcal W}^{10'}_I & \frac{\vec{u}_{\ell,6}\cdot \vec{\mathcal W}^{10'}_I}{\tilde\alpha_1} \\ \vec{u}_{\ell,4}\cdot \vec{\mathcal W}^{10'}_I  & \frac{\vec{u}_{\ell,6}\cdot \vec{\mathcal W}^{10'}_I}{\tilde\alpha_1}  &  \frac{\vec{u}_{\ell,8}\cdot \vec{\mathcal W}^{10'}_I}{\tilde\alpha_2}  \\  \frac{\vec{u}_{\ell,6}\cdot \vec{\mathcal W}^{10'}_I}{\tilde\alpha_1}  & \frac{\vec{u}_{\ell,8}\cdot \vec{\mathcal W}^{10'}_I}{\tilde\alpha_2}  & \frac{\vec{u}_{\ell,10}\cdot \vec{\mathcal W}^{10'}_I}{\tilde\alpha_3} \end{array}\right)\,.
  \eqe
In all cases, we need to choose a deformed boundary $\vec{\mathcal W}^{k'}_I=\vec{\mathcal W}^{k}_I+\Delta w$.

\noindent \textbf{The EFT-hedron}
 
We now turn to the full EFT-hedron. Again begin with 
\eq\label{AIDef}
\vec{A}_{I}=\left(\begin{array}{c} A_{2,I} \\ A_{4,I} \\ \cdots  \\ A_{k,I} \end{array}\right)=\left(\begin{array}{c} \vec{a}_{2}\cdot \vec{\mathcal W}_I \\ \vec{a}_{4}\cdot \vec{\mathcal W}_I \\ \cdots  \\ \vec{a}_{k}\cdot \vec{\mathcal W}_I\end{array}\right)\,,
\eqe
where we've taken $k$ to be even. Firstly $A_{k,I}$ is positive, whenever  $\vec{\mathcal{W}}_{I}$ is one of the facets of $Conv[\vec{u}_{\ell,k}]$. Furthermore we require total positivity of the deformed Hankel matrix of $\vec{A}_{I}$, given as 
\eq\label{DefHankel}
 \left(\begin{array}{cc}A_{2,I} & A_{4,I}  \\ A_{4,I} & \frac{A_{6,I}}{\tilde\alpha_1}  \end{array}\right),\quad\left(\begin{array}{cc}A_{4,I} & \frac{A_{6,I}}{\tilde\alpha_1}  \\ \frac{A_{6,I}}{\tilde\alpha_1} & \frac{A_{8,I}}{\tilde\alpha_2}  \end{array}\right),\quad   \left(\begin{array}{ccc}A_{2,I} & A_{4,I}  & \frac{A_{6,I}}{\tilde\alpha_1} \\ A_{4,I}  & \frac{A_{6,I}}{\tilde\alpha_1}  &  \frac{A_{8,I}}{\tilde\alpha_2}  \\  \frac{A_{6,I}}{\tilde\alpha_1}  & \frac{A_{8,I}}{\tilde\alpha_2}  & \frac{A_{10,I}}{\tilde\alpha_3} \end{array}\right),\,e.t.c.
\eqe
where $\vec{\mathcal{W}}_{I}$ is now the deformed boundary of maximal $k$, $\vec{\mathcal{W}}^{k'}_{I}$, and the deformation parameters $\{\tilde\alpha_i\}$s defined through the total positivity of eq.(\ref{conditions}). These two constraints are encapsulated as:
\eq\label{AlphaEFT}
\framebox[8cm][c]{$\displaystyle K[\vec{A}_I]_{\{\tilde\alpha_i\}} \quad {\rm is\; a\; totally\;positive\;matrix}\,$}~\,.
\eqe

Let us compare side by side the $s$-channel EFT-hedron and the general EFT-hedron: starting with $\vec{A}_I$ given in eq.(\ref{AIDef}), they are defined by:

$$\begin{tabular}{c|c|c}

% after \\ : \hline or \cline{col1-col2} \cline{col3-col4} ...
   & $s$-ch EFT-hedron & EFT-hedron \\
    \hline
Hankel matrix   &  Canonical $K[X]$  &  Deformed $K[X]_{\{\tilde\alpha_i\}}$   \\
   \hline
 $\mathcal{W}_{I}$  & boundaries of $Conv[\vec{v}_{\ell}]$  &  boundaries of $Conv[\vec{u}_{\ell,k}]$\\
 \end{tabular}$$

 In the following we will consider the  $\mathbb{P}^1$ geometry.  
\noindent \textbf{Example:}

Let's consider the explicit example for $k=4,6,8$, where 
\eq\label{1Dsu}
\left(\begin{array}{cc} a_{4,0} & a_{4,2} \\ a_{6,0} & a_{6,2} \\ a_{8,0} & a_{8,2}\end{array}\right)=\sum_a\textsf{p}_{a}\left(\begin{array}{cc} x_a^4 \vec{u}_{\ell_a,4} \\ x_a^6\vec{u}_{\ell_a,6} \\ x_a^8\vec{u}_{\ell_a,8}\end{array}\right)\;\quad \vec{u}_{\ell,k}= (u_{\ell,k,0}, u_{\ell,k,2})\,,
\eqe
Since $u_{\ell,k,0}$ is positive for all $\ell,k$, we can use it to positively rescale the first entry to 1 and define $u^{(k)}_{\ell}=\frac{u_{\ell,k,2}}{u_{\ell,k,0}}$. Then $Conv[\vec{u}_{\ell,k}]$ is simply a line segment in $\mathbb{P}^1$ with its boundary determined by the minimum value of $u^{(k)}_{\ell}$. From eq.(\ref{Sum}) one can check that the minimum value of $u^{(k)}_{\ell}$ for fixed $k$ and arbitrary spin is given as:
\eq\label{Umin}
 Min\left[u^{(4)}_{\ell}\right]=-\frac{3}{2}\;\;(\ell=1,2),\quad Min\left[u^{(6)}_{\ell}\right]=-\frac{21}{4}\;\; (\ell=2),\quad Min\left[u^{(8)}_{\ell}\right]=-8\;\;(\ell=2)\;.
\eqe 
Note the above agrees with eq.(\ref{MaxBound}), which states that the boundary of the Minkowski sum is given by that of the largest $k$, here $8$. Rescaling $\left(a_{k,0} ,a_{k,2}\right)=a_k\left(1, \beta_k\right)$, the above tells us that the boundaries of $Conv[\vec{u}_{\ell,k}]$ for each $k$ translates to 
\eq\label{Bounds}
a_4\geq0,\quad a_6\geq0,\quad a_8\geq0,\quad \beta_4\geq -\frac{3}{2},\quad\beta_6\geq -\frac{21}{4}\quad,\beta_8\geq -8\,.
\eqe
Furthermore, we also have that $a_{k,0}$ is inside the convex hull of half-moment curve:
\eq
a_6^2-a_4 a_8\geq 0\,.
\eqe
These inequalities corresponds to $A_{4,I}, A_{6,I}, A_{8,I}$ being positive with $\mathcal{W}_I$ is chosen to be the boundary of $Conv[\vec{u}_{\ell,k}]$, and 
\eq\label{Red}
\left(\begin{array}{cc}A_{4,I} & A_{6,I}  \\ A_{6,I} & \frac{A_{8,I}}{\tilde\alpha_1}  \end{array}\right)
\eqe
being totally positive, where $\mathcal{W}_I=(1,0)$ and $\tilde{\alpha}_1=1$. 

Next, we consider the positivity of Det$[eq.(\ref{Red})]$ where $\mathcal{W}_I$ is the boundary of the Minkowski sum. Since the boundary of $Conv[\vec{u}_{\ell,4},\vec{u}_{\ell,6},\vec{u}_{\ell,8}]$ is given by $(1,-8 )$, the upper bound for $\tilde\alpha_1$ is such that 
\eq\label{ineq468}
\frac{(u^{(4)}_{\ell}+8+\Delta w)(u^{(8)}_{\ell}+8+\Delta w)}{\tilde\alpha_1}-\left(u^{(6)}_{\ell}+8+\Delta w\right)^2\geq0,\quad\forall \ell \,.
\eqe
Note that we have add a small deformation $\Delta w$. This is needed since here $\mathcal{W}_I$ is identified with $u^{(6)}_{2}$, which would cause the first term in the above (with $\Delta w=0$) to be zero for $\ell=2$ and invalidate the inequality. Picking $\Delta w=\frac{1}{100}$ we find $\tilde\alpha_1\leq 0.0085$. Equipped with this the positivity of the determinant eq.(\ref{Red}) translate to 
\eq\label{ineq468beta}
\frac{(\beta_4+8+\frac{1}{100})(\beta_8+8+\frac{1}{100})}{0.0085}-\left(\beta_6+8+\frac{1}{100}\right)^2\geq0 \,.
\eqe

Note that in the above it is necessary to consider walls that are deformed away from the boundary of $Conv[\vec{u}_{\ell,4},\vec{u}_{\ell,6},\vec{u}_{\ell,8}]$, and $\tilde\alpha_1$ as well as the non-linear constraint that follows depends on the choice of deformation parameter $\Delta w$. As we will see in appendix~\ref{EFTP1}, the most stringent non-linear constraint does no necessarily correspond to $\Delta w$ being small ! In other words, the true boundary of the EFT-hedron is actually defined by a new wall that can be far from the boundaries of the cyclic polytope. A  more complete understanding of the true boundaries will be left to future studies.

When the external particles are identical, we should consider even spins only. However, since the minimum in \eqref{Umin} is given by spin-$2$, the optimal value for $\tilde{\alpha}_1$ remains the same. Thus the problem simply reduces to the intersection of the permutation plane defined in \eqref{PermXDef} with our $\mathbb{P}^1$ geometry. From \eqref{perminvplane}, we see that $\beta_4=\frac{a_{4,2}}{a_{4,0}}$ is fixed to $\frac{3}{2}$. This turns \eqref{ineq468beta} into a quadratic bound for $\beta_6$ and $\beta_8$. Thus for identical scalars, the EFT-hedron bounds are given by eq.(\ref{Bounds}) and
\eq
(\frac{19}{2}+\frac{1}{100})(\beta_8+8+\frac{1}{100})-0.0085\left(\beta_6+8+\frac{1}{100}\right)^2\geq0\,.
\eqe

%%%%%%%%%%%%%%%%%%%%%%%%%%%%%%%
\subsection{Multiple Species}
%%%%%%%%%%%%%%%%%%%%%%%%%%%%%%%
Let us now return to the scattering of $a,b$, but now consider the amplitude $M(a,b,b,a)$ in combination with all $a$ and all $b$ scattering. For simplicity we will assume each of $a,b$ have a $\mathbb{Z}_2$ symmetry, so the only non-vanishing amplitude involves even number of $a$'s and $b$'s. Now we can get constraints mixing the $a^4$, $a^2b^2$ and $b^4$ amplitudes, if we consider $ABBA$ scattering of general states $A=\alpha a{+}\beta b$,and  $B=\gamma a{+}\rho b$. These must satisfy the EFT-constraints for all $(\alpha,\beta,\gamma,\rho)$; in the special case of $A=B$ ($\alpha=\gamma$, $\beta=\rho$) we intersect with the crossing symmetry plane as well. A systematic exploration of the geometry associated with this envelope of constraints is left for future work, but it is easy and illuminating to look at the simplest example. 

Consider the leading $4$-derivative amplitudes 
\eq
M(a^4)=c_a(s^2{+}t^2{+}u^2),\;M(b^4)=c_b(s^2{+}t^2{+}u^2),\;M(abba)=c(s^2{+}u^2){+}\frac{d}{2}t^2\,.
\eqe
Note our analysis of $M(abba)$ just tells us that $c>0$; $d$ can have any sign. But we will now see that magnitude of $d$ is bounded by $c_{a,b}$ as 
\eq
c_a c_b-d^2>0\,.
\eqe
To whit, the amplitude for $M(ABBA)$ is given by 
\eqa
M(ABBA)&=&(\alpha\gamma)^2M(a^4){+}(\beta\rho)^2M(b^4){+}(\gamma\beta)^2M(baab){+}(\alpha\rho)^2M(abba)\nonumber\\
&{+}&(\alpha\beta\gamma\rho)\left[M(aabb){+}M(baba){+}M(abab){+}M(bbaa)\right]\,.
\eqae
Note that while the term proportional to $d$ in $M(abba)$ drops out in the forward limit as $t\rightarrow 0$, this is not the case e.g. for $M(aabb)=c(u^2{+}t^2){+}\frac{d}{2}s^2$ which becomes $s^2(c+d/2)$ in the forward limit. 

Taking the $t\rightarrow 0$ limit, the coefficient of $s^2$ in the $M(ABBA)$ amplitude, which must be positive, is given by 
\eqa
(\alpha\gamma)^2c_a{+}(\beta\rho)^2c_b{+}(\alpha\beta\gamma\rho)(2d{+}4c){+}2c((\gamma\beta)^2{+}(\alpha\rho)^2)\nonumber\\
=(\alpha\gamma)^2c_a{+}(\beta\rho)^2c_b{+}2d(\alpha\beta\gamma\rho){+}2c(\gamma\beta{+}\alpha\rho)^2\,.
\eqae
Now of course if we put $\alpha=1,\beta=0,\gamma=0,\rho=1$, we go back to $A=a$, $B=b$, and we learn that $c>0$. But now let's put $\gamma\beta{+}\alpha\rho=0$. We then have $x^2c_{a} + y^2c_b{+}2xyd>0$, where $x=-\alpha^2\rho/\beta$, $y=\beta\rho$; note that varying over $\alpha,\beta,\rho$, $(x,y)$ can be any real numbers. Thus we learn that $c_{a,b}>0$ and $c_ac_b-d^2>0$, or the positivity of the matrix in
\eq
\left(\begin{array}{cc}x & y\end{array}\right)\left(\begin{array}{cc}c_a & d \\d & c_b\end{array}\right)\left(\begin{array}{c}x \\y\end{array}\right)\,.
\eqe
Note it was important in this analysis to allow general $AB$ states; had we taken only $A=B\rightarrow\alpha=\gamma, \beta=\rho$, we would find no constraints on $d>0$.

This can be straightforwardly generalized to any number of species labelled by the index $i$. Again assuming $\mathbb{Z}_2$ symmetry for each species, writing
\eq
M(i^4)=c_i(s^2+t^2+u^2), \;M(ijji)=c_{ij}(s^2{+}u^2){+}d_{ij}t^2\,,
\eqe
we find that $c_{ij}\geq 0$, and that the matrix 
\eq
\left(\begin{array}{cccc}  c_{11}& d_{12} & d_{13}&\cdots \\ d_{12} & c_{22} & d_{23} &\cdots\\  d_{13} & d_{23} & c_{33} &\cdots\\ \vdots & \vdots& \vdots &\vdots \end{array}\right)\,,
\eqe
is positive. The positivity of a symmetric matrix $S$ is equivalent to the positivity of all the leading principle minors (determinant of all upper left square matrices) of the matrix (the Sylvester’s criterion ). As an example we have
\eqa
det\left(\begin{array}{ccc}  c_{11}& d_{12} & d_{13} \\ d_{12} & c_{22} & d_{23} \\  d_{13} & d_{23} & c_{33}  \end{array}\right)\geq0,\quad det\left(\begin{array}{ccc}  c_{11}& d_{12}  \\ d_{12} & c_{22}  \end{array}\right)\geq0, \quad c_{11}\geq0\,.
\eqae
%%%%%%%%%%%%%%%%%%%%%%%%%%%%%%%
\section{The spinning EFT-hedron}\label{Sec:SpinEFT-hedron}
%%%%%%%%%%%%%%%%%%%%%%%%%%%%%%%
So far we have examined constraints on amplitudes with external scalars. The analysis can be readily extended to external spinning states such as gluons, photons and gravitons, where the higher dimensional operators of the EFT will be given in terms of field strengths, Riemann tensors and derivatives there of.  In subsection \ref{Spinning} we've seen that the Taylor vectors of spinning polynomials also generate cyclic polytopes, and thus we can simply retrace all of the previous discussion, with $v_{\ell,q}$ replaced by the Taylor coefficient of the spinning polynomials.

An important question is which helicity configuration should one select for the dispersive representation. The choice should be such that one is expanding around a forward process, i.e. the $t\rightarrow 0$ limit corresponds to $a, b \rightarrow b, a$ scattering. Take for example $M(1^{+},2^{-},3^{+},4^{-})$. In the $s$-channel threshold where $1,2$ are incoming and $3,4$ outgoing, the process corresponds to $1^+2^- \rightarrow 3^- 4^+$. Note that the helicity of legs $3$ and $4$ are flipped since we've defined the helicity for $M$ with all momenta incoming. For it to be forward, we should identify the state on leg $1$ with $4$, so we set $p_4=p_1$ and $p_2=p_3$ which indeed corresponds to $t=0$. For the $u$-channel threshold one instead has $1^+3^+ \rightarrow 2^+4^+$, which once again correspond to a forward process with $p_4=p_1$ and $p_2=p_3$. Similarly $M(1^{+},2^{+},3^{-},4^{-})$ also admits a positive expansion. This is in contrast with $M(1^{+},2^{-},3^{-},4^{+})$, where in the $s$-channel we have $1^{+}2^{-} \rightarrow 3^{+} 4^{-}$. In order for this to be forward, we need to take $p_1=p_3$ and $p_2=p_4$ which corresponds to $u=0$ instead of $t=0$. So in this case the small $t$ expansion of the residue is not an expansion around a forward process, and does not enjoy the positivity properties we wish to exploit. 

\begin{figure}
\begin{center}
\includegraphics[scale=0.5]{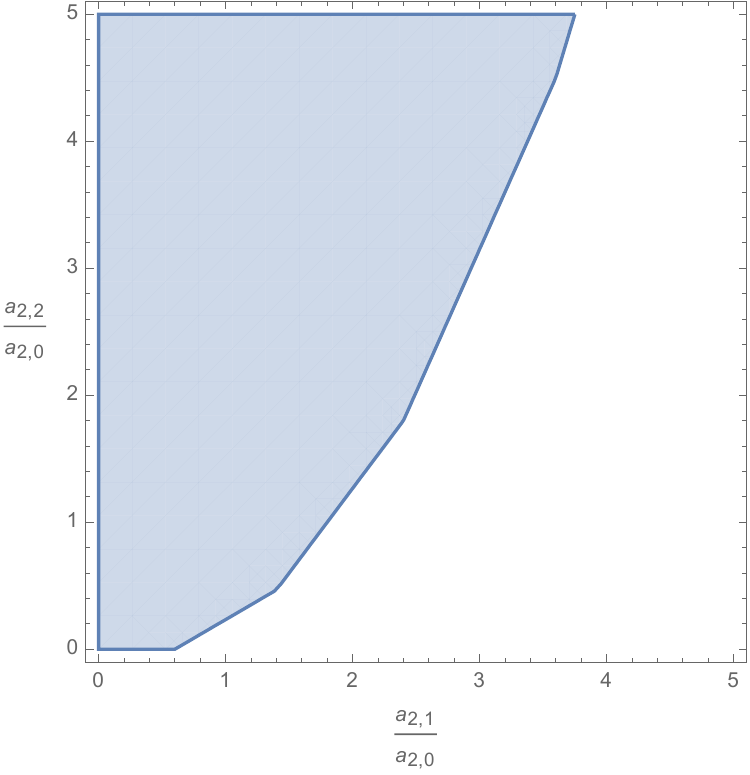}
\caption{The $k=2$ polygon for $(+-+-)$ gluon scattering.}
\label{SpinFig}
\end{center}
\end{figure}

As a simple example, the $s$-channel EFT hedron can be generalized to color ordered states. From the previous discussion, we've seen that expanding in $t$ for $M(1^{+},2^{-},3^{+},4^{-})$  corresponds to an expansion around the forward limit. Thus the $s$-channel residue can be positively expanded on  $d^\ell_{2,2}(\theta)$ (see eq.(\ref{reference}))
\eq
Res_s[M(1^{+},2^{-},3^{+},4^{-})]=\sum_{\ell} \textsf{p}_\ell d^\ell_{2,2}(\theta)\quad \textsf{p}_\ell \geq0.
\eqe
Removing the overall spinor bracket mandated by the helicity weights, we have:
\eq
\left.\langle 24\rangle^2[13]^2\left(\sum_{k,q}\; a_{k,q} s^{k{-}q}t^q \right)=-\langle 24\rangle^2[13]^2\left(\sum_{a} \textsf{p}_{\ell_a} \frac{ \tilde{d}^{\ell_a}_{2,2}(\theta)}{s-m_a^2}\right)\right|_{\theta=\arccos(1{+}2t/m_a^2) }
\eqe
where once again the equality is understood in terms of Taylor expansion, and $\tilde{d}^{\ell_a}_{2,2}(\theta)=\frac{d^{\ell_a}_{2,2}(\theta)}{\cos^4\frac{\theta}{2}}$. We can then bound operators using the boundaries of the cyclic polytopes, as an example, for $k=2$, which corresponds to $D^{4}F^4$, we have 
\eq
\langle \mathbf{a}_{2}, \ell, \ell+1\rangle\geq0, \quad \mathbf{a}_{2}=(a_{2,0},a_{2,1}, a_{2,2})\,.
\eqe
The  two-dimensional region is then given in fig.\ref{SpinFig}. Imposing cyclic symmetry sets $a_{2,2}/{a_{2,0}}=1$ and the region becomes a one dimensional line, and the bound becomes 
\eq
 0\leq a_{2,1}/{a_{2,0}}\leq\frac{9}{5}\,.
\eqe
For open super-string, we have $\frac{a_{2,1}}{a_{2,0}}=\frac{1}{4}$ and are thus inside the bound.

For photons and gravitons, we need to consider the contributions from both $s$ and $u$-channel. Here we choose the amplitude $M(1^{{+}h},2^{{+}h},3^{{-}h},4^{{-}h})$, and the $s$-channel residue for a spin-$\ell$ exchange is written as:
\eqa
Res_s\left[M(1^{{+}h},2^{{+}h},3^{{-}h},4^{{-}h})\right]&=&g^{{++}}_{\ell}g^{{--}}_{\ell}[12]^{2h}\langle34\rangle^{2h}\;d^\ell_{0, 0}(\theta)\,,
\eqae
where $g_\ell^{{++}/{--}}$ is the coupling constant of a real spin-$\ell$ state to a pair of plus/minus helicity photon. CPT requires $g_\ell^{++}=(g_\ell^{{--}})^*$, and the above yields a positive expansion as expected. Furthermore under $3,4$ exchange $d^\ell_{0, 0}(\theta)\rightarrow d^\ell_{0, 0}({-}\theta)=({-})^\ell d^\ell_{0, 0}(\theta)$, thus bose symmetry requires $\ell\in even$. The $u$-channel residue is given as:
\eqa
Res_u\left[M(1^{{+}h},2^{{+}h},3^{{-}h},4^{{-}h})\right]&=&(g^{{+-}}_{\ell})^2[12]^{2h}\langle 34\rangle^{2h}\tilde{d}^{\ell_a}_{2,2}
\eqae
where now CPT simply requires $g^{{+-}}_{\ell}$ to be real. Thus we arrive at the following dispersive representation\footnote{The first version of this paper had an error in the residue polynomials in the spinning dispersion relation, which we correct here, modifying the obtained bounds. We thank Zvi Bern, Alexander Zhiboedov, and Dimitrios Kosmopoulos for pointing out this mistake to us. } 
\eq\label{SpinDisper}
\framebox[15cm][c]{$\displaystyle [12]^{2h}\langle 34\rangle^{2h}\left(\sum_{k,q}\; a_{k,q} s^{k{-}q}t^q \right)=-[12]^{2h}\langle 34\rangle^{2h}\left(\sum_{a} \textsf{p}_{\ell_a} \frac{ d^{\ell_a}_{0,0}(\theta)}{s-m_a^2}{+}\sum_{b} \tilde{\textsf{p}}_{\ell_b} \frac{\tilde{d}^{\ell_b}_{2,2}}{u-m_b^2}\right)$}\,,
\eqe
where $\textsf{p}_{\ell_a}$ and $ \tilde{\textsf{p}}_{\ell_b}$ are distinct positive coefficients and $\ell_a\in even$.

In the following, we will analyze external photons and gravitons separately. For $k=even$ the bounds are listed as:
\begin{center}
({-}h,{-}h,{+}h,{+}h):\quad\quad 
\begin{tabular}{l|c|c}
&photon&graviton\\ \hline
$k=2$& $D^4F^4$\;\eqref{photonNonSym2}&$D^4R^4$\;\eqref{gravitonNonSym2}\\ \hline
$k=4$&$D^8F^4$\;\eqref{photonNonSym4}&$D^8R^4$\;\eqref{gravitonNonSym4}\\
\end{tabular}
\end{center}

%%%%%%%%%%%%%%%%%%%%%%%%%%%%%%%%%%%%%%%%%
\subsection{Photon EFT}\label{sec:photoneft}
%%%%%%%%%%%%%%%%%%%%%%%%%%%%%%%%%%%%%%%%%
For photons, our analysis can be separated into whether or not gravity decouples. For EFTs whose gravitational dynamics are irrelevant, such as the Euler-Heisenberg  theory,  one can bound  operators of degree 2 or higher in $s$. If gravity does not decouple, as discussed in sec.\ref{Appt0} the forward limit graviton pole will obstruct any bound on $s^2$. In practice, starting with the geometry for gravitationally decoupled EFTs, one can incorporate gravity simply by projecting the geometry onto the directions perpendicular to $a_{k,k-2}$.\footnote{We will assume that $RF^2$ is not relevant for the analysis, although it is straightforward to incorporate.}

\begin{figure}
\begin{center}
\includegraphics[scale=0.5]{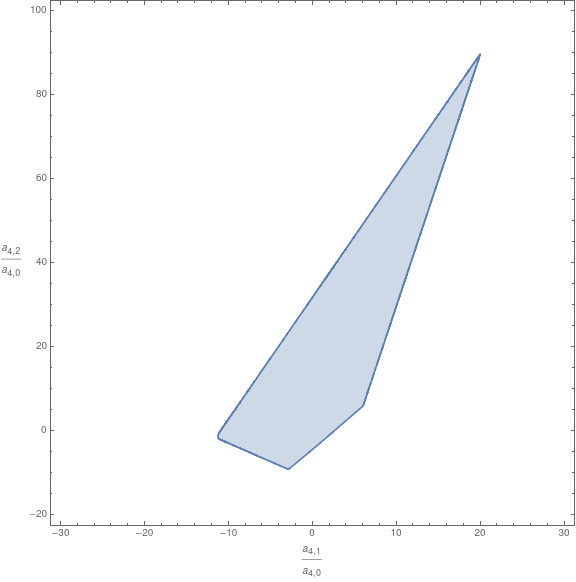}
\caption{The intersection of the $\mathbb{P}^4$ polytope defined by the boundaries in eq.(\ref{photonNonSym4}) with $\mathbf{X}_{sym}$. }
\label{FigPhoton2DPlot}
\end{center}
\end{figure}

Note that now the $s$- and $u$-channel have distinct polynomials, we will label the vectors from the $s$ and $u$ channel in eq.(\ref{SpinDisper}) as $\ell_s$ and $\ell_u$ respectively, and the unitary polytope is the Minkowski sum of the two polytopes. Furthermore, this helicity configuration is invariant under $t\leftrightarrow u$ exchange, and thus the amplitude must lie on the ``symmetry plane" $\mathbf{X}_{\rm sym}$ parameterized as:
\eq\label{SymPlane}
\left(\begin{array}{ccccc}a_{1,0} & a_{1,1} & \; & \; & \;  \\ a_{2,0} & a_{2,1} & a_{2,2} & \; & \;  \\ a_{3,0} & a_{3,1} & a_{3,2} & a_{3,3} & \; \\ a_{4,0} & a_{4,1} & a_{4,2} & a_{4,3} & a_{4,4}  \end{array}\right) \quad \rightarrow \quad \left(\begin{array}{ccccc}x& 0 & \; & \; & \;  \\ x  & y & y & \; & \;  \\ x & y & y & 0 & \; \\ x & y & z & 2(z{-}y) & (z{-}y)  \end{array}\right)\,.
\eqe
We now give the intersection of $\mathbf{X}_{\rm sym}$ with the unitary polytope:

\begin{itemize}
\item $k{=}2:D^4F^4$
\eq
M_{D^4F^4}=\langle12\rangle^{2}[34]^{2}(a_{2,0}s^2+a_{2,1}st+a_{2,2}t^2).
\eqe
Now we would like to bound $\mathbf{a}_{2}=(a_{2,0},a_{2,1}, a_{2,2})$ which live in $\mathbb{P}^2$. The edge of the polygon is given by
\eq\label{photonNonSym2}
\langle *, i_u{+}1, i_u,\rangle_{ i_u\geq 2}, \quad \langle *, i_s, i_s{+}2\rangle_{ i_s\geq 2}, \quad \langle *, 2_u, 2_s\rangle\,,
\eqe
where $i_s, i_{u}$ represents the Taylor vectors from $d^{i_s}_{0,0}$ and $d^{i_u}_{-2,-2}$ respectively. Note that the majority of the edges for the $s$- and $u$-channel cyclic polytope remains a facet for the Minkowski sum. The polygon is presented in projective coordinates $\left(\frac{a_{2,1}}{a_{2,0}},\frac{a_{2,2}}{a_{2,0}}\right)$ in fig.\ref{MinkFig}, where we've labeled the vertices from the (purple)$s$ and (red)$u$ channels explicitly. 

On $\mathbf{X}_{\rm sym}$ we have $\frac{a_{2,1}}{a_{2,0}}=\frac{a_{2,2}}{a_{2,0}}$ and the geometry reduces to $\mathbb{P}^1$. The region of intersection is given as:
\eq
-\frac{30}{7}\leq\frac{a_{2,1}}{a_{2,0}}=\frac{a_{2,2}}{a_{2,0}}\leq 6\,.
\eqe
Note that similar to the intersection of the scalar $s{-}u$ polytope with the permutation plane, here the intersection yields leads to EFT coefficients being bounded from both sides.
\begin{figure}
\begin{center}
\includegraphics[scale=0.5]{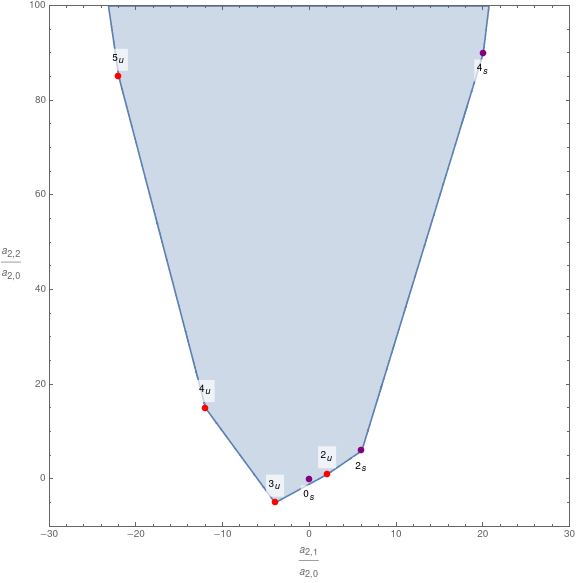}
\caption{The $k=2$ polygon for $(--++)$ photon scattering. It is bounded by the Minkowski sum of the vectors originated from the $s$-channel (purple dots) and $u$-channel (red dots).}
\label{MinkFig}
\end{center}
\end{figure}

\item $k=4:D^8F^4$
\eq
M_{D^8F^4}=\langle12\rangle^{2}[34]^{2}(a_{4,0}s^4+a_{4,1}s^3 t+a_{4,2}s^2t^2+a_{4,3}s t^3+a_{4,4} t^4).
\eqe
The coupling $\mathbf{a}_4=(a_{4,0}, a_{4,1}, a_{4,2},a_{4,3},a_{4,4})$ lives in $\mathbb{P}^4$, and is bounded by
\eqa\label{photonNonSym4}
\langle \mathbf{a}_4, 2_u, 3_u,4_u, 5_u\rangle,\quad \langle \mathbf{a}_4, i_u, i_u{+}1,j_u, j_u{+}1\rangle_{ i_u,j_u\geq 3},\nonumber\\
  \langle \mathbf{a}_4, i_s, i_s{+}2,j_s, j_s{+}2\rangle_{ i_s,j_s\geq 2},
\quad    \langle \mathbf{a}_4, i_s{+}2,  i_s,j_u, j_u{+}1\rangle_{ i_s,\geq 4, j_u\geq 3},\nonumber\\
\langle \mathbf{a}_4, 4_s,2_s, 3_u, 2_u\rangle,\quad\langle \mathbf{a}_4, 4_s, 2_s,2_u, 5_u\rangle,\quad
\langle \mathbf{a}_4, 4_s, 2_s,i_u, i_u+1\rangle_{i_u\geq 5}\nonumber\\
\langle \mathbf{a}_4, i_s+2, i_s,\infty_u, \infty_s\rangle_{i_s\geq 2},\quad\langle \mathbf{a}_4, i_u, i_u+1,\infty_u, \infty_s\rangle_{i_u\geq 3}\nonumber\\
\langle \mathbf{a}_4, 2_s,\infty_s, 3_u, \infty_u\rangle,\quad
\langle \mathbf{a}_4, 4_s,2_u,3_u,4_u\rangle,\quad
\langle \mathbf{a}_4, 4_s,2_u,4_u,5_u\rangle,\quad\nonumber\\
\langle \mathbf{a}_4, i_s+2,i_s,2_s,3_u\rangle_{i_s\geq 4},\quad
\langle \mathbf{a}_4, 2_s,3_u,2_u,5_u\rangle_{i_s\geq 4},\quad
\langle \mathbf{a}_4, 2_s,3_u,i_u,i_u+i\rangle_{i_u\geq 5},\quad
.
\eqae
being non-negative. Note that the boundary of the Minkowski sum consists of almost all the boundaries of the individual cyclic polytope, label by a pair of consecutive spins, as well as the tensor products of consecutive pair from both sides. At lower spin region we have some irregular boundaries as well. The intersection of the above with $\mathbf{X}_{sym}$ is illustrated in fig.\ref{FigPhoton2DPlot}. 
\end{itemize}

%%%%%%%%%%%%%%%%%%%%%%%%%%%%%%%%%%%%%%%%%
\subsection{Graviton EFT}
%%%%%%%%%%%%%%%%%%%%%%%%%%%%%%%%%%%%%%%%%
For gravity the analysis is a straight forward extension of the photon EFT: simply set $h=2$ in the polynomial basis. From the discussion in sec.(\ref{Appt0}), we've seen that the tree-level four-graviton amplitude does not introduce any $t$-channel massless obstructions, thus here we will be able to bound operators proportional to $s^n$ with $n\geq2$. Once again, we will consider the intersection of the unitary polytope with the symmetry plane $\mathbf{X}_{\rm sym}$ defined in eq.(\ref{SymPlane}): 

\begin{itemize}
\item $k{=}0: R^4$
\eq
M_{R^4}=\langle 12\rangle^4[34]^4a_{0,0}.
\eqe
and we simply have $a_{0,0}>0$.

\item $k{=}2: D^4R^4$
\eq 
M_{D^4R^4}=\langle 12\rangle^4[34]^4(a_{2,0}s^2+a_{2,1}st+a_{2,2}t^2).
\eqe
The facets are again given by that of the individual cyclic polytope in the $s$- and $u$-channel. The bounds are then given by:
\eq\label{gravitonNonSym2}
\langle \mathbf{a}_2, i_u{+}1, i_u,\rangle_{ i_u\geq 4}, \quad \langle \mathbf{a}_2, i_s, i_s{+}2\rangle_{ i_s\geq 2}, \quad \langle \mathbf{a}_2, 4_u, 1_s\rangle\,.
\eqe
being non-negative, with $\mathbf{a}_2=(a_{2,0},a_{2,1},a_{2,2})$. On $\mathbf{X}_{\rm sym}$ we have
\eq
-\frac{90}{11}\leq\frac{a_{2,1}}{a_{2,0}}=\frac{a_{2,2}}{a_{2,0}}\leq 6\,.
\eqe

\item $k{=}4: D^8R^4$
\eq
M_{D^8R^4}=\langle 12\rangle^4[34]^4(a_{4,0}s^4+a_{4,1}s^3 t+a_{4,2}s^2t^2+a_{4,3}s t^3+a_{4,4} t^4).
\eqe

The facets are:
\eqa\label{gravitonNonSym4}
\langle \mathbf{a}_4, 4_u, 5_u,6_u, 7_u\rangle, \quad
\langle \mathbf{a}_4, i_u, i_u{+}1,j_u, j_u{+}1\rangle_{ i_u,j_u\geq 5},
\nonumber\\
\langle \mathbf{a}_4, i_s, i_s{+}2,j_s, j_s{+}2\rangle_{ i_s,j_s\geq 2},\quad
\langle \mathbf{a}_4, i_s{+}2,i_s, i_u, i_u{+}1\rangle_{ i_s\geq 4,i_u\geq 5},
\nonumber\\
\langle \mathbf{a}_4, i_s{+}2,i_s, \infty_u, \infty_s\rangle_{ i_s\geq 4},\quad
\nonumber\\
\langle \mathbf{a}_4, 4_s, 2_s, 6_s,5_u\rangle,\quad
\langle \mathbf{a}_4, 4_s, 2_s, 5_u,4_u\rangle,\quad
\langle \mathbf{a}_4, 4_s, 2_s, 4_u,7_u\rangle,
\nonumber\\
\langle \mathbf{a}_4, 4_s, 2_s, i_u,i_u{+}1\rangle_{i_u\geq 7},\quad
\langle \mathbf{a}_4, 4_s, 2_s, \infty_u,\infty_s\rangle,
\nonumber\\
\langle \mathbf{a}_4, 2_s, 5_u, 4_u,7_u\rangle,\quad
\langle \mathbf{a}_4, 2_s, 5_u, i_u,i_u{+}1\rangle_{i_u\geq 7},
\nonumber\\
\langle \mathbf{a}_4, 2_s, 5_u, \infty_u,\infty_s\rangle,\quad
\langle \mathbf{a}_4, 2_s, 5_u, i_s{+}2,i_s\rangle_{i_s\geq 4},\quad
\nonumber\\
\langle \mathbf{a}_4, 4_s, 4_u, 5_u,6_u\rangle,\quad
\langle \mathbf{a}_4, 4_s, 4_u, 6_u,7_u\rangle,\quad
\langle \mathbf{a}_4, i_u, i_u{+}1, \infty_u,\infty_s\rangle_{i_u\geq 5}
\eqae
Once again, the facets maintain a cyclic structure at higher spins, while some irregularities occur at lower spin region. Its intersection with the symmetry plane $\mathbf{X}_{\rm sym}$ is displayed in fig.\ref{SymGravi2DPlot}.

\begin{figure}
\begin{center}
\includegraphics[scale=0.5]{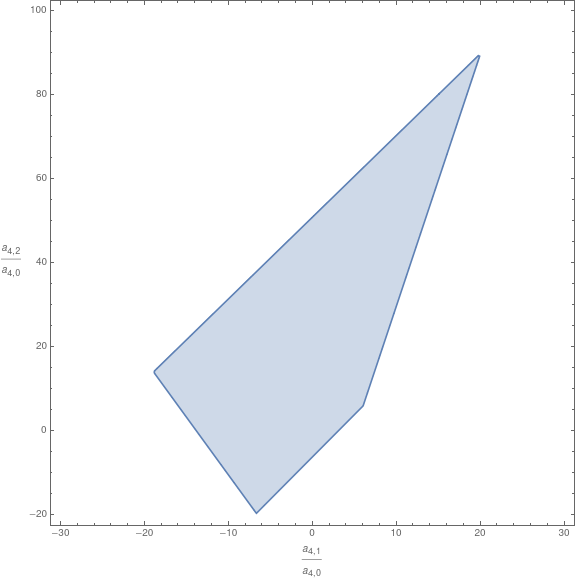}
\caption{The intersection of the $\mathbb{P}^4$ polytope defined by the boundaries in eq.(\ref{gravitonNonSym4}) with $\mathbf{X}_{sym}$. }
\label{SymGravi2DPlot}
\end{center}
\end{figure}

\end{itemize}

In this section we have focused for simplicity on the scattering of a single species--
photons or gravitons--but it is easy to constrain photon-graviton couplings as well. The amplitude $M(1^{-1}2^{+2}3^{-2} 4^{+1})$ is forward as $t\rightarrow 0$ in both the $s$- and $u$- channels, and so has a positive expansion. Thus considering the Gegenbauer constraints, the coefficients must lie inside the unitarity polytopes; but we don't have the extra crossing symmetry constraints enjoyed by  pure photon/graviton scattering. While this is all we can say considering only photon-graviton scattering, as with our multi-species discussion for the scalar case, there are clearly constraints relating the pure photon and pure graviton scattering coefficients to those of photon-graviton scattering, considering the scattering of general linear combinations of different species, which would be interesting to further explore.

%%%%%%%%%%%%%%%%%%%%%%%%%%%%%%%
\section{Explicit EFTs in the EFT-hedron}\label{sec:Answs}
%%%%%%%%%%%%%%%%%%%%%%%%%%%%%%%
So far we have been mostly discussing bounds on general EFTs, derived from the analyticity and unitarity in the UV. In this section we will discuss in more detail how realistic EFTs with explicit UV completions, satisfy these bounds. 
 
%%%%%%%%%%%%%%%%%%%%%%%%%%%%%%%
\subsection{$s$-channel EFT-hedron}
%%%%%%%%%%%%%%%%%%%%%%%%%%%%%%%
Let's begin with the $s$-channel constraints. We will use the tree-level massless open superstring amplitude as an example eq.(\ref{TypeIString}), which we display again here:
\eq
M(1^+2^-3^+4^-)=-\langle 24\rangle^2[13]^2\frac{\Gamma[{-}s]\Gamma[{-}t]}{\Gamma[1{-}s{-}t]}=\langle 24\rangle^2[13]^2\left[-\frac{1}{st}+\sum_{k,q}a_{k,q}s^{k{-}q}t^q\right]\,
\eqe 
and the coupling constants, up to $k=4$, are given as:
\eq
\left(\begin{array}{ccccc}a_{0,0} & \; & \; & \; & \; \\ a_{1,0} & a_{1,1} & \; & \;& \; \\ a_{2,0} & a_{2,1} & a_{2,2} & \;& \; \\ a_{3,0} & a_{3,1} & a_{3,2} & a_{3,3}& \; \\ a_{4,0} & a_{4,1} & a_{4,2} & a_{4,3}& a_{4,4}\end{array}\right)=\left(\begin{array}{ccccc}\zeta(2) & \; & \; & \; & \;  \\ \zeta(3) & \zeta(3) & \; & \; & \;  \\ \frac{\pi^4}{90} & \frac{\pi^4}{360} & \frac{\pi^4}{90} & \; & \;  \\ \zeta(5) & \;\;2\zeta(5){-}\zeta(3)\zeta(2) & \;\;2\zeta(5){-}\zeta(3)\zeta(2)  & \;\;\zeta(5) & \; \\  \frac{\pi^6}{945} &\frac{\pi^6{-}630\zeta^2(3)}{1260} & \frac{23\pi^6}{15120}-\zeta^2(3)  & \frac{\pi^6{-}630\zeta^2(3)}{1260} & \frac{\pi^6}{945} \end{array}\right)\,.
\eqe
The $s$-channel EFT-hedron defined in eq.(\ref{EFT-hedron}) says that the Hankel matrix for $A_{k,I}=\vec{a}_k\cdot \mathcal{W}_I$ must be a totally positive matrix, where $\mathcal{W}_I$ is the facets.

Let us first consider the facets $\mathcal{W}_{I_{\mathbb{I}}}$, the unit vectors.The Hankel matrix for these facets are 
\eqa
\mathcal{W}^q_{I_{\mathbb{I}}}=\delta^{q0}:\;\;\left(\begin{array}{ccc} \zeta_2 & \zeta_3 & \frac{\pi^4}{90} \\ \zeta_3 &  \frac{\pi^4}{90} & \zeta_5 \\  \frac{\pi^4}{90} & \zeta_5  & \frac{\pi^6}{945}\end{array}\right),\quad \mathcal{W}^q_{I_{\mathbb{I}}}&=&\delta^{q1}:\;\;\left(\begin{array}{cc} \zeta_3 & \frac{\pi^4}{360} \\ \frac{\pi^4}{360} & 2\zeta_5{-}\zeta_3\zeta_2\end{array}\right),\;\; \left(\begin{array}{cc} \frac{\pi^4}{360} & 2\zeta_5{-}\zeta_3\zeta_2 \\ 2\zeta_5{-}\zeta_3\zeta_2 & \frac{\pi^6{-}630\zeta_3^2}{1260}\end{array}\right)\nonumber\\
\mathcal{W}^q_{I_{\mathbb{I}}}&=&\delta^{q2}:\;\;\left(\begin{array}{cc} \frac{\pi^4}{90} & 2\zeta_5{-}\zeta_3\zeta_2  \\ 2\zeta_5{-}\zeta_3\zeta_2 & \frac{23\pi^6}{15120}{-}\zeta_3^2\end{array}\right)\,.
\eqae  
It is straight forward to check that these matrices are positive semi-definite. 

Next we consider facets of the cyclic polytope $\mathcal{W}_{I_b}$. For this we utilize the Taylor vectors for spinning polynomials of $h=1$ listed in eq.(\ref{SpinVecs}), and denote each column as $\vec{\nu}_\ell$. Recall that due to Yang's theorem, $\ell$ starts at $2$. Since the Taylor vectors forms a cyclic polytope, the boundaries for the $\mathbb{P}^1$, $\mathbb{P}^2$, and $\mathbb{P}^3$ geometry are given by:
\eq
\mathbb{P}^1: (2),\quad \mathbb{P}^2: (i,i{+}1),\quad \mathbb{P}^3: (2,i,i{+}1)\;.
\eqe 
When written in terms of dual vectors, they are given by contracting the $d$ vectors with the $d{+}1$ component Levi-Cevita tensor. Explicitly they are given as:
\eq\label{Boundaries}
\langle *,2\rangle=det\left(\begin{array}{cc}1 & * \\ 0 & *\end{array}\right),\quad \langle*,i,i{+}1 \rangle=det\left(\begin{array}{ccc}* & 1 & 1 \\ * & \frac{\nu_{\ell,1}}{\nu_{\ell,0}} & \frac{\nu_{\ell{+}1,1}}{\nu_{\ell{+}1,0}} \\ * & \frac{\nu_{\ell,2}}{\nu_{\ell,0}} & \frac{\nu_{\ell{+}1,2}}{\nu_{\ell{+}1,0}}\end{array}\right),\quad  \langle2,*,i,i{+}1 \rangle=det\left(\begin{array}{cccc}1&* & 1 & 1 \\ 0&* & \frac{\nu_{\ell,1}}{\nu_{\ell,0}} & \frac{\nu_{\ell{+}1,1}}{\nu_{\ell{+}1,0}} \\0& * & \frac{\nu_{\ell,2}}{\nu_{\ell,0}} & \frac{\nu_{\ell{+}1,2}}{\nu_{\ell{+}1,0}}\\0& * & \frac{\nu_{\ell,3}}{\nu_{\ell,0}} & \frac{\nu_{\ell{+}1,3}}{\nu_{\ell{+}1,0}}\end{array}\right)\,.
\eqe 
When taking the inner product with some vector $X$, then the $*$s denote the position where components of $X$ should be placed. For example for $\mathbb{P}^1$, the coupling constants are organized as
\eq
\vec{a}_{k}=\left(\begin{array}{c}1 \\ \frac{a_{k,1}}{a_{k,0}}\end{array}\right)
\eqe
and identify $\mathcal{W}_I$ as the boundary for $\mathbb{P}^1$ in eq.(\ref{Boundaries}), we find (again with $A_{k}\equiv \vec{a}_{k}\cdot \mathcal{W}_I$)
\eq
\left(\begin{array}{c} A_1 \\ A_2 \\ A_3\end{array}\right)= \left(\begin{array}{c} \frac{a_{1,1}}{a_{1,0}} \\ \frac{a_{2,1}}{a_{2,0}} \\ \frac{a_{3,1}}{a_{3,0}}\end{array}\right)
\eqe
Then from eq.(\ref{EFT-hedron}), we see that being inside the s-channel EFT-hedron requires
\eq
K[\vec{A}]=\left(\begin{array}{cc}A_1 & A_2 \\A_2 & A_3\end{array}\right)=\left(\begin{array}{cc}1 & \frac{1}{4} \\ \frac{1}{4} & 2-\frac{\zeta(2)\zeta(3)}{\zeta(5)}\end{array}\right)
\eqe
to be a totally positive matrix. Indeed one can straightforwardly verify that each component and the determinant of the above matrix is positive. Next let's consider the constraint in $\mathbb{P}^2$. Choosing $\mathcal{W}_I$ from eq.(\ref{Boundaries}) to be $\langle *, 6, 7\rangle$, we find,
\eq
\left(\begin{array}{c} A_2 \\ A_3 \\ A_4 \end{array}\right)= \left(\begin{array}{c} \frac{7(45a_{2,0}-20a_{2,1}+6a_{2,2})}{30a_{2,0}} \\ \frac{7(45a_{3,0}-20a_{3,1}+6a_{3,2})}{30a_{3,0}} \\  \frac{7(45a_{4,0}-20a_{4,1}+6a_{4,2})}{30a_{4,0}}\end{array}\right) = \left(\begin{array}{c}  \frac{161}{15} \\  \frac{7}{90} \left(51 + \frac{7 \pi^2 \zeta(3)}{\zeta(5)} \right)\\  \frac{721}{80} + \frac{882 \zeta^2(3)}{\pi^6}\end{array}\right)\,.
\eqe
One again finds that the matrix $\left(\begin{array}{cc}A_2 & A_3 \\A_3 & A_4\end{array}\right)$, is totally positive.

%%%%%%%%%%%%%%%%%%%%%%%%%%%%%%
 \subsection{Full EFT-hedron}
%%%%%%%%%%%%%%%%%%%%%%%%%%%%%%
Now let's consider the tree-level closed superstring amplitude in four-dimensions, with $M(1^{+2}2^{-2}3^{+2}4^{-2})$:
\eq
-\langle 24\rangle^4[13]^4\frac{\Gamma[{-}s]\Gamma[{-}t]\Gamma[{-}u]}{\Gamma[1{+}s]\Gamma[1{+}t]\Gamma[1{+}u]}=\langle 24\rangle^2[13]^2\left[-\frac{1}{stu}+\sum_{k,q}a_{k,q}z^{k{-}q}t^q\right]\,,
\eqe 
whose low energy effective coupling constants are:
\eq\label{closedstringcouplings}
\left(\begin{array}{cccc}a_{0,0} & \; & \; & \; \\ a_{2,0} & a_{2,2} & \; & \; \\ a_{4,0} & a_{4,2} & a_{4,4}& \; \\ a_{6,0} & a_{6,2} & a_{6,4} & a_{6,6}\end{array}\right)=\left(\begin{array}{cccc}2\zeta(3) & \; & \; & \; \\ 2\zeta(5) & \frac{3}{2}\zeta(5) & \; & \; \\ 2\zeta(7) & 3\zeta(7) & \frac{9}{8}\zeta(7) & \; \\ 2\zeta(9) & \;\;\frac{1}{6} (8 \zeta^3(3) {+} 31 \zeta(9)) & \;\;\frac{1}{24} ({-}16 \zeta^3(3) {+} 73 \zeta(9))  & \;\; \frac{1}{96} (8 \zeta^3(3) {+} 85 \zeta(9)) \end{array}\right)\,.
\eqe
Since the UV states now appear in both $s{-}u$ channels, the couplings should satisfy the constraints of the full EFT-hedron.

Now let's consider the simplest EFT-hedron constraint in $\mathbb{P}^1$, which was discussed in detail in Appendix.\ref{EFTP1}. The difference is that we will use spinning polynomials for our facets. Furthermore, due to the helicity configuration, the $s$-channel and $u$-channel will contribute independently and a Minkowski sum over polytopes will be taken. To simplify the discussion, we will assume permutation invariance for the space of amplitudes that we want to constrain here. The absence of $a_{2,1}, a_{4,1}, \cdots$ terms in the above is then just a direct consequence of this, and other amplitudes in this space can be compared with the closed superstring amplitude on equal footing. For each $k$, the polytope will be a Minkowski sum of the polytopes from $s$- and $u$- channels. Let us denote the vertices contributed by spin-$\ell$ as
\eq
(x_{\tilde{\ell},k,0},x_{\tilde{\ell},k,2}),
\eqe
where $\tilde{\ell}$ zips together information about spin and channel, for example like $\{(1, s), (2, u), \cdots\}$. Projectively,
\eq
\left(1, x^{(k)}_{\tilde{\ell}}\right)=\left(1, \frac{x_{\tilde{\ell},k,2}}{x_{\tilde{\ell},k,0}}\right),\quad {\rm for}\, k=2,4,6\;,
 \eqe
then we have 
\eq
\text{min}\, x^{(2)}_{\tilde{\ell}}=-\frac{23}{20},\quad \text{min}\, x^{(4)}_{\tilde{\ell}}=-\frac{11}{2},\quad \text{min}\, x^{(6)}_{\tilde{\ell}}=-\frac{165}{16},\quad
\eqe
and hence we choose $\mathcal{W}=(-w,1)$, with $w=-\frac{165}{16}$. Note that again we find that the boundary of the Minkowski sum is given by that of maximal $k$. Now organizing the couplings as  
\eq
\left(\begin{array}{cc} 1 & \frac{a_{2,2}}{a_{2,0}} \\1 & \frac{a_{4,2}}{a_{4,0}} \\1 & \frac{a_{6,2}}{a_{6,0}}\end{array}\right)=\left(\begin{array}{c} \vec{a}_2  \\ \vec{a}_4 \\ \vec{a}_6\end{array}\right)\,,
\eqe 
the constraint in eq.(\ref{AlphaEFT}) then tells us that 
\eqa\label{EFT-hedronstring}
&&(\vec{a}_2\cdot\mathcal{W})(\vec{a}_6\cdot\mathcal{W})- \alpha_{min}(\vec{a}_4\cdot\mathcal{W})^2\nonumber\\
&=&\frac{177}{16}\left(\frac{619}{48}+\frac{2\zeta^3(3)}{3\zeta(9)}\right)-\alpha_{min}\left(\frac{189}{16}\right)^2>0,
\eqae
where $\alpha_{min}$ is defined as the minimum of $\frac{(x^{(6)}_{\tilde{\ell}}-w)(x^{(2)}_{\tilde{\ell}}-w)}{x^{(4)}_{\tilde{\ell}}-w)^2}$.
Direct evaluation shows this is indeed true.
% where $\alpha_{min}[\Delta w]$ is defined as the minimum of $\frac{(u^{(6)}_{\ell}-w)(u^{(2)}_{\ell}-w)}{u^{(4)}_{\ell}-w)^2}$ for a given $\Delta w$. We explicitly plot $\alpha_{min}[\Delta w]$ for our case of spinning polynomials:
% \eq
% \includegraphics[scale=0.4]{aMinString}\,.
% \eqe 
% Using the resulting $\alpha_{min}$, we verify that indeed eq.(\ref{EFT-hedronstring}) is positive for all range of $\Delta w\geq0$,
% \eq
% eq.(\ref{EFT-hedronstring})\quad\vcenter{\hbox{\includegraphics[scale=0.4]{StringPlots}}}\,.
% \eqe
% Note that while the result is positive, one can see that the bound can be tighter when $\Delta w\neq0$, in particular the tightest bound is around $\Delta\sim1.4$.

%%%%%%%%%%%%%%%%%%%%%%%%%%%%%%%%%%%%%%%%
\subsection{Living near the boundary of unitary polytopes}\label{sec:UV}  
%%%%%%%%%%%%%%%%%%%%%%%%%%%%%%%%%%%%%%%% 
Now that we've seen how explicit EFTs satisfy our EFT-hedron bounds, we would like to see where do they actually reside. For example, consider the two dimensional region carved out by $\textbf{X}_{\rm cyc}\cap \textbf{U}_5$ in fig.(\ref{ComboConstraint}), where $\textbf{U}_5$ is the $s$-channel unitary polytope.  Now we consider the following scalar EFTs, each with a distinct known UV completion: 
\begin{itemize}
  \item (a) The tree-level exchange of a massive Higgs in the linear Sigma model 
  \eq
 \left. -\frac{s}{s-m^2}-\frac{t}{t-m^2}\right|_{m\rightarrow\infty}=\cdots+\frac{1}{m^{10}}(s^5+t^5)+\cdots
  \eqe
  \item (b) The one-loop contribution of a massive scalar $X$ coupled to a massless scalar $\phi$ via $X^2\phi$. The one-loop integrand is simply the massive box, whose low energy expansion is: 
  \eq
 \left.   \vcenter{\hbox{\includegraphics[scale=0.4]{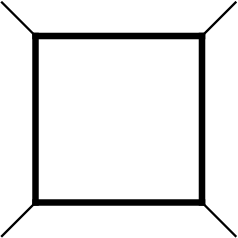}}}\right|_{m\rightarrow\infty}=\cdots+\frac{(s^5+\frac{1}{5}s^4t+\frac{1}{10}s^3t^2+\frac{1}{10}s^2t^3+\frac{1}{5}st^4+t^5)}{1153152 m^{14} \pi^2}+\cdots
  \eqe
  \item (c) The type-I stringy completion of bi-adjoint scalar theory:
  \eqa
  \left. -\frac{\Gamma[{-}\alpha's]\Gamma[{-}\alpha't]}{\Gamma[1{-}\alpha's{-}\alpha't]}\right|_{\alpha'\rightarrow0}=\cdots+\alpha'^5\left[\zeta_7 s^5{+}\left({-}\frac{\pi^4 \zeta_3}{90} {-} \frac{\pi^2\zeta_5}{6}{+} 3 \zeta_7\right)s^4 t\right.\nonumber\\
  \left.{+}\left({-}\frac{\pi^4 \zeta_3}{72} {-} \frac{\pi^2\zeta_5}{3}{+} 5 \zeta_7\right)s^3 t^2 {+}(s\leftrightarrow t)\right]+\cdots
  \eqae 
\end{itemize}
where we've listed the coefficients for $k=5$. Plotting their position with in $\mathbf{X}_{\rm Cyc}\cap\mathbf{U}_5$, we find:
\eq
\includegraphics[scale=0.4]{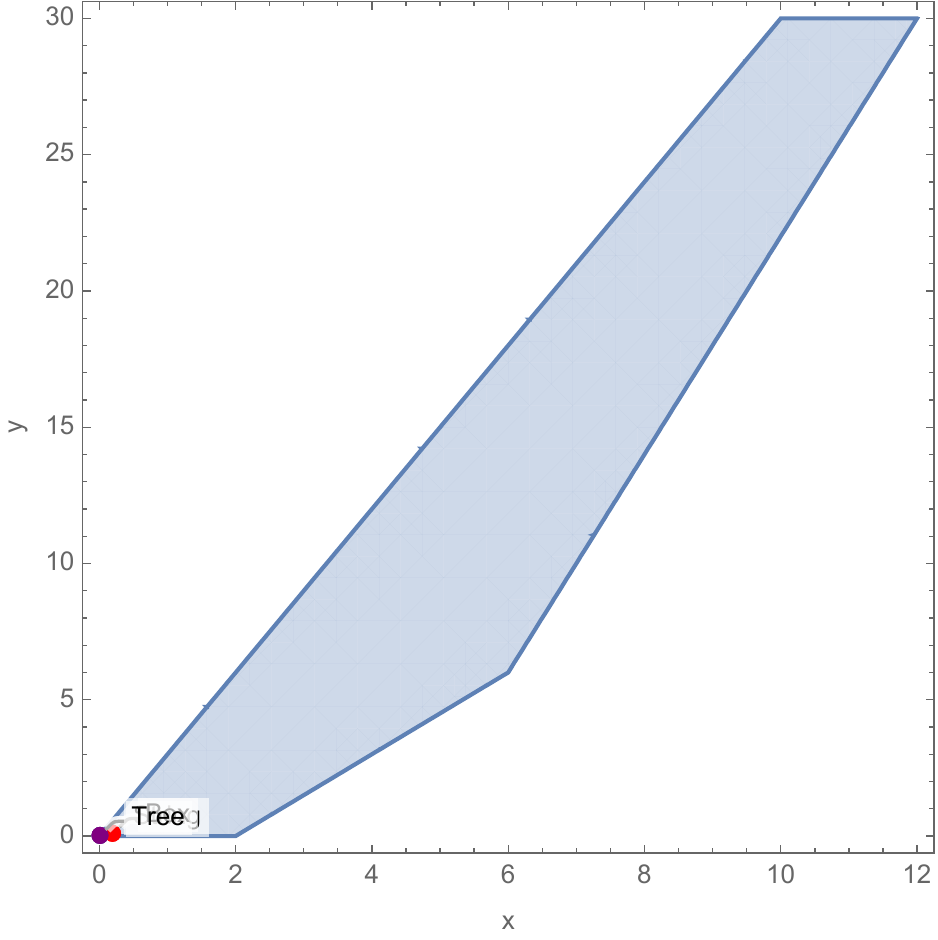}\quad \includegraphics[scale=0.4]{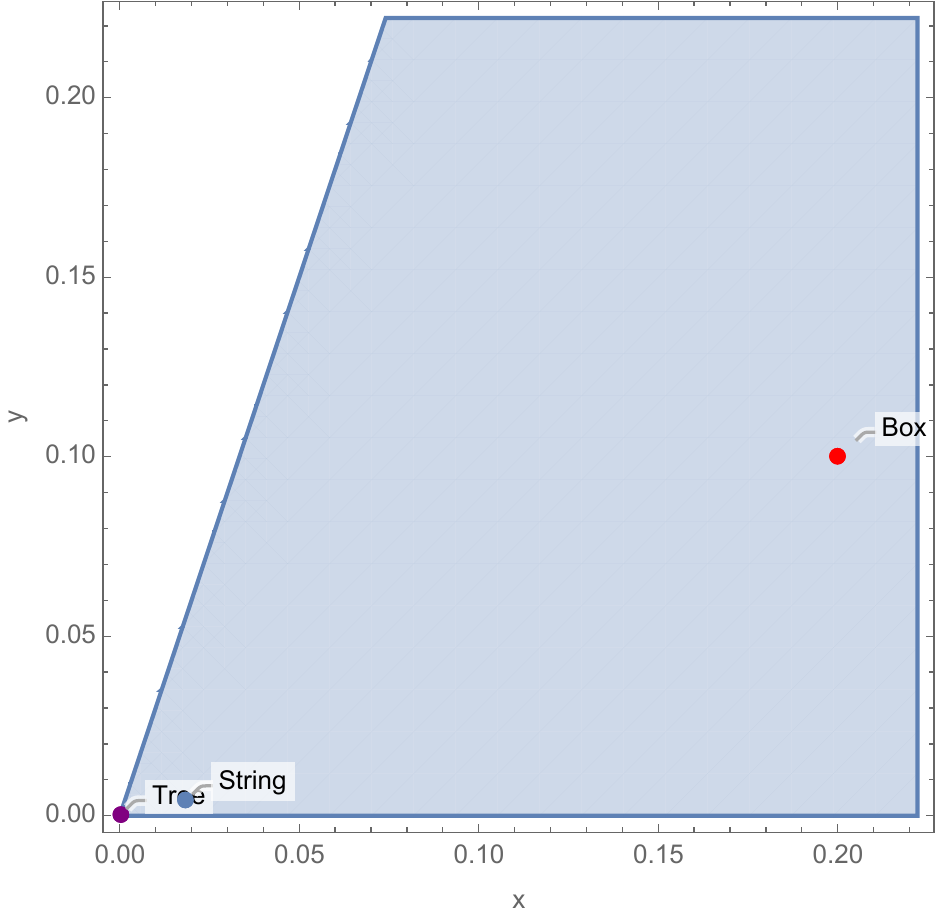}
\eqe  
Note that they are sitting extremely close to the bottom tip of the allowed region!  Let's consider another example for the graviton $s{-}u$ polytope, parameterized for the MHV configuration as:
\eq
\langle24\rangle^4[13]^4\left(\{massless\;poles\}{+}\sum_{k,q}z^{k{-}q}t^q\right)\,.
\eqe
 In the most general case, we can have $R^3$ operator which introduces a $t$-channel obstruction for operators proportional to $z^2$.  Consider the coefficients $(a_{8,0},a_{8,2}, a_{8,4})$ such that the geometry is $\mathbb{P}^2$. In principle the odd power coefficients will also be important for comparing spectral densities contributed from each spin. Here we simply wish to visualize certain coefficients in a convenient way. Two theory points that are nearby on this plot can still have very different spectral densities. 
 
\begin{figure}
\begin{center}
\includegraphics[scale=0.4]{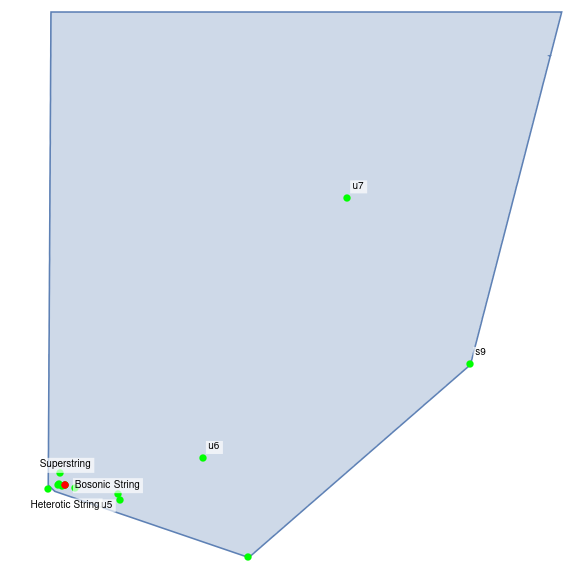}
\caption{The unitary polygon for $(a_{8,0},a_{8,2}, a_{8,4})$ of the graviton EFT. We see that the string theory EFTs are clustered near the low spin boundaries of the polygon.}
\label{StringPlot}
\end{center}
\end{figure}

 We projectively plot the corresponding polygon in the coordinates $(\frac{a_{8,2}}{a_{8,0}},\frac{a_{8,4}}{a_{8,0}})$. The result as well as the positions of the coefficient for Type-II, Heterotic and bosonic strings are presented in fig.\ref{StringPlot}. Labels for lower spin vertices are omitted for clarity. Once again, we see that the three distinct string EFTs are cluttered close to the lowest spins of the entire geometry. 
 
In fact, this behaviour is ubiquitous as we survey other $k$, as well as the $s{-}u$ channel polytopes: all known EFTs sits close to the boundaries characterized by the low-spin vertices. This implies that the residue or discontinuity induced by the UV completion is generically dominated by low spins! For the linear sigma model, we only have a spin zero exchange so this is trivial. Listing the Gegenbauer coefficients for the residue of the open string to level $n$,
\eq
\begin{array}{c|ccccc}\ell \backslash n & 1 & 2 & 3 & 4 &5  \\\hline 0 & 1 & \; & \; & \; &\frac{1}{11880}\\ 1 & \; & \frac{1}{14} & \; & \frac{1}{924}& \;\\ 2 & \; & \; & \frac{1}{84} & \;&\frac{25}{39312} \\ 3  & \; & \; & \; & \frac{2}{693}&\;\\ 4 & \; & \; & \; & \;&\frac{125}{144144} \end{array}\,,
\eqe
we see that the leading scalar coefficient is dominant over the rest. For the box integral, the spinning spectral function for the discontinuity is discussed in detail in appendix \ref{SpinningBox} , see eq.(\ref{spinf}). Plotting the spectral function for spin-$0, 1, 2$ as a function of $s$ we find:
\eq
\textsf{p}_\ell(s)\;\;\vcenter{\hbox{\includegraphics[scale=0.5]{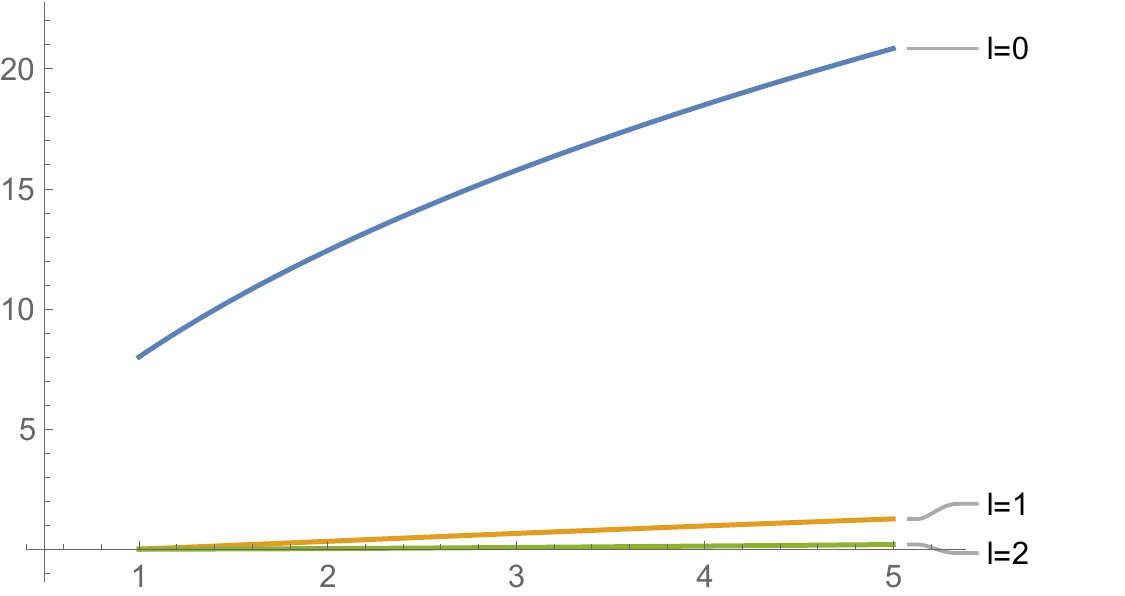}}}_s
\eqe
where $s$ is normalized with respect to $4m^2$, and hence the plot begins only at the branch point $s=1$. Once again the scalar spectral function dominates the contribution from other spins, and the ratio increases as we increase with $s$. Note that the positivity of the six-dimensional $a$-anomaly for a free massive scalar was precisely due to such suppression~\cite{Elvang:2012st}. The suppression of higher spin coefficients can be understood from the polynomial boundedness of the amplitude: as a spin-$\ell$ exchange in the $t$-channel will bring a contribution behaving as $s^\ell$ at large $s$, polynomial boundedness then implies that higher spin contributions must be suppressed. Indeed the suppression at large spins is precisely what led to the Froissart bound as reviewed in appendix~\ref{Froissart}.  Thus in general, we expect physical EFTs to lie near the low spin boundaries of the unitary polytope, although a more quantitative understanding of the implications from such suppression is clearly desired, which we leave to future work.

\begin{figure}
\begin{center}
$\vcenter{\hbox{\includegraphics[scale=0.3]{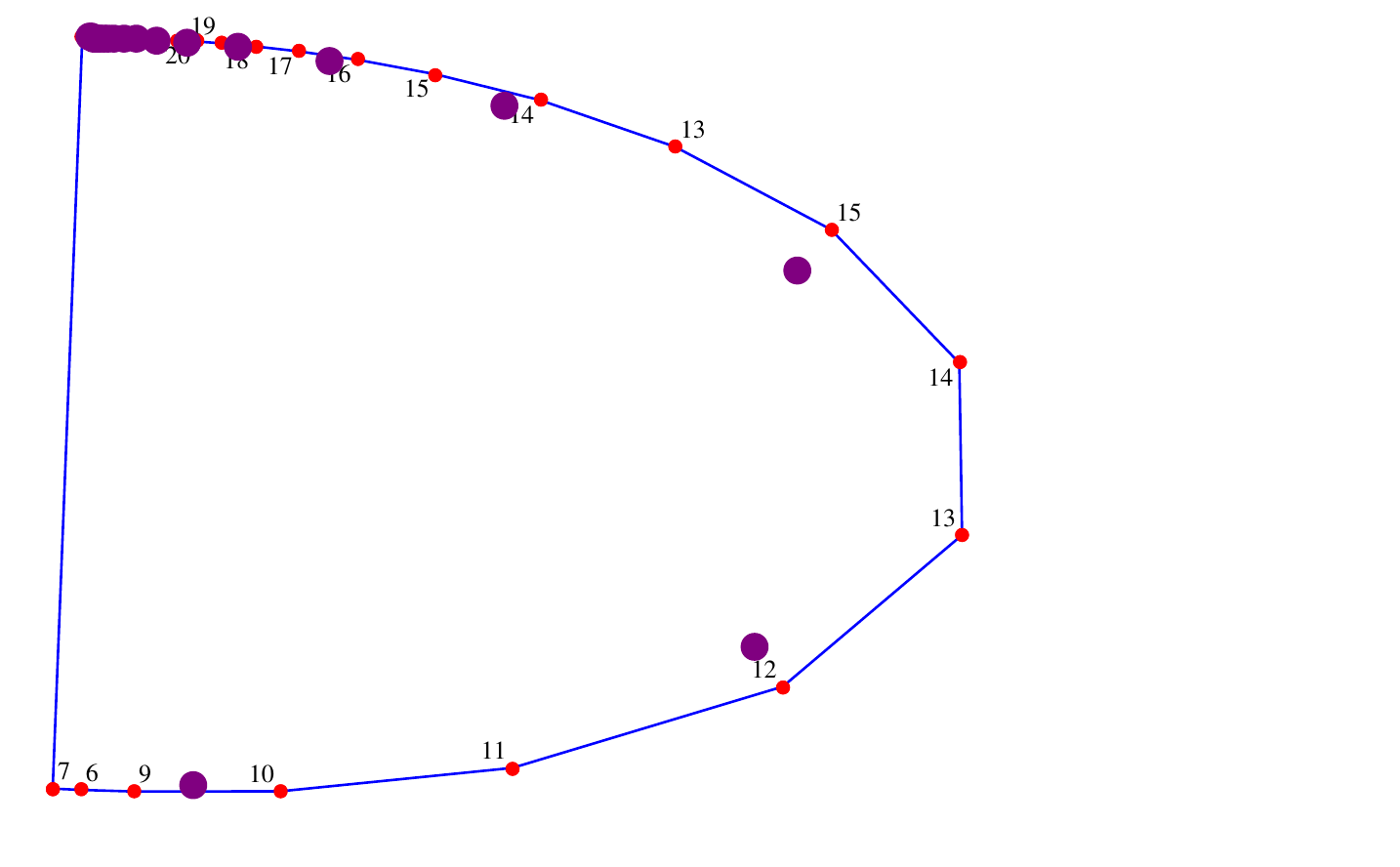}}} \rightarrow\quad\vcenter{\hbox{\includegraphics[scale=0.25]{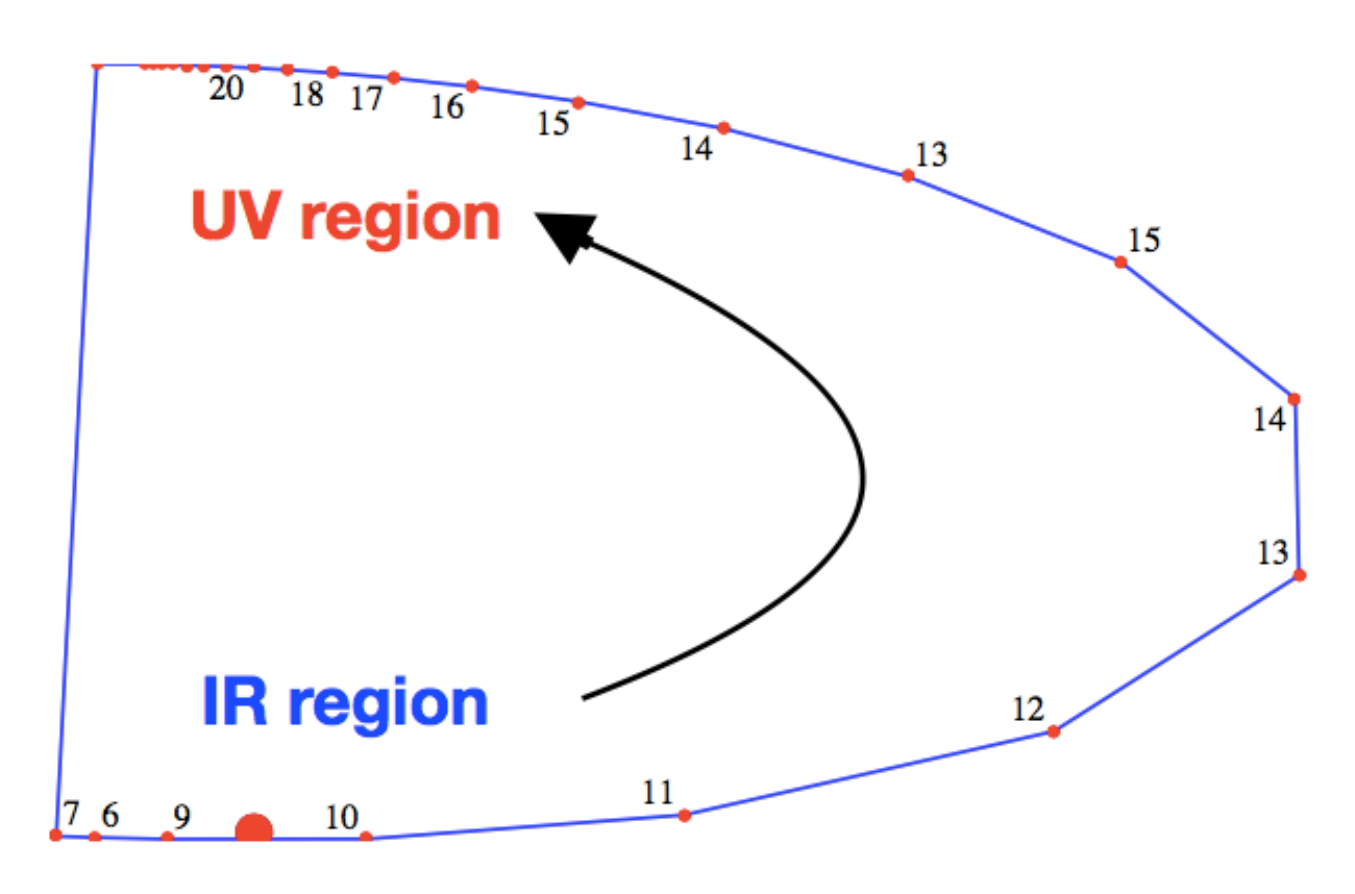}}}$
\caption{Oh the LHS we the purple dots indicate $(x_n,y_n)$ for $n=0,\cdots,20$, representing the position of the type-II string EFT in side the unitary polytope for  $s\sim \frac{n}{\alpha'}$. We see that as we go to large $s$, the EFT tends to the corner with higher spins. This implies that the UV and IR EFTs populate different regions in the polytope, as illustrated on the right. }
\label{fig:UVPlots}
\end{center}
\end{figure}

If EFTs naturally live near the low-spin boundaries of the unitary polytope, what is the purpose of the rest?  Note that for a given UV completion, there exits an entire family of effective theories for which the EFTs discussed above are in the deep IR. Here, the scale dependence under discussion is not from the running generated from the massless loops, which will be the focus in the next section, but rather from the simple fact that different part of the spectrum is visible depending on the energy.    What this means in practice is that at a given energy scale $\Lambda$, the couplings for our higher dimensional operators take the form:
\eq
M(s,t)=\{massless/massive\;poles\}+\sum_{k,q}\;a^{\Lambda}_{k,q}s^{k-q}t^q\,,
\eqe
where the amplitude now contains massless as well as  massive poles for all the massive states below $\Lambda$. When the couplings are defined in such fashion, they naturally becomes $\Lambda$ dependent. Let us consider an explicit example. Imagine that we are studying type
-II string theory at some energy scale and we have discovered the first few massive states up to level $n$. At this scale the amplitude at fixed $t$ should take the form:
\eq
\sum_{a=1}^nR_a(t)\left(\frac{1}{s-a}+\frac{1}{u-a}\right)+\sum_{k,q}a^{(n)}_{k,q}z^{k{-}q}t^q,
\eqe
where $R_a(t)=\frac{1}{(a!)^2}\prod_{i=1}^{a{-}1}(t+i)^2$ is the residue for the resonance $s=a$. The value of the couplings for the higher dimensional operators can be extracted by Taylor expanding both sides of:
\eq
\sum_{k,q}a^{(n)}_{k,q}z^{k{-}q}t^q=\frac{\Gamma[{-}s]\Gamma[{-}t]\Gamma[{-}u]}{\Gamma[1{+}s]\Gamma[1{+}u]\Gamma[{+}t]}-\left[\sum_{a=1}^nR_a(t)\left(\frac{1}{s-a}+\frac{1}{u-a}\right)\right]
\eqe   
Note that by construction, the couplings must reside inside our unitary polytope. Since the massive poles that are ``subtracted" from the full UV completion are precisely the dominating low spin states, we expect that the resulting couplings to \textit{float} towards the upper region of the polytope!  We plotting the coefficients for $(x_n,y_n)=\left(\frac{a^{(n)}_{8,2}}{a^{(n)}_{8,0}+a^{(n)}_{8,4}/10^3},\frac{a^{(n)}_{8,4}}{a^{(n)}_{8,0}+a^{(n)}_{8,4}/10^3}\right)$ in fig.\ref{fig:UVPlots}. We see that indeed as we raise the energy scale the corresponding EFT probes deeper in the unitary polytope.

Thus in summary, the low spin regions of the unitary polytope correspond to the EFTs in the deep IR, while the higher spin region corresponds to the EFTs in the UV. We leave the detailed study of this UV-IR relation to future work.
\newpage
%%%%%%%%%%%%%%%%%%%%%%%%%%%%%%%%%%%%%%%%%%%
\section{Running into the EFT-hedron}\label{RG}
%%%%%%%%%%%%%%%%%%%%%%%%%%%%%%%%%%%%%%%%%%%
Let us now turn to discussing the full amplitude including the massless loops that induce the logarithmic running of the EFT couplings. For example, consider again the linear sigma model, whose tree-amplitude is given in eq.(\ref{LSM4pt}).  At one-loop the coefficients of the $s^4$ starts receiving loop-corrections from the $s^2$ operators: 
\eqa\label{SMLoop}
M^{IR}(s,t)&=&\frac{\bar{a}_2}{m_h^4}(s^2{+}t^2{+}u^2)+\frac{\bar{a}_4}{m_h^8}(s^4{+}t^4{+}u^4){-}\left[\bar{a}_2^2\frac{1}{15(4\pi)^2m_h^8}\left(41s^2{+}u^2{+}t^2\right)s^2\log \frac{s}{s_0}\right.\nonumber\\
&&\left.{+}\left(s\leftrightarrow t\right){+}\left(s\leftrightarrow u\right)\right]{+}\mathcal{O}(p^{10})\,,
\eqae
where $\bar{a}_i$s are to be understood as renormalized couplings at some scale $s_0$. In this paper, we will only consider one-loop effects for EFTs that have a well defined S-matrix. The derivative couplings ensures that expansion near the forward limit is well defined, since the $t$-channel cut appears as $t^n \log t$, as can be seen in the above, and hence there is no singularity at the branch point $t=0$. The presence of the massless logs leads to two pressing issues, 1. there is a massless cut coming all the way to the origin, and thus the low energy couplings, analytically extracted from eq.(\ref{gDef}), are no longer well defined. 2. the fact that coupling runs also brings into question the fate of our previous positivity bounds as the theory flows EFT flows to the IR.

Naively, one can simply introduce a mass regulator,\footnote{This of course can only be consistently done for scalars and vectors, but not  gravity.}  which will allow us to push the massless cut away from the origin of the complex $s$-plane. Since this corresponds to introducing a massive state, all ingredients necessary to the derivation of previous positivity bounds are intact and should hold whenever the EFT is valid. This means that running in the IR will stay within the unitary polytope. However, it is easy to see from explicit examples that this is \textit{not} the case, the massless logs can take us outside of the EFT hedron! This apparent contradiction originated from the fact that the mass deformed theory do not reproduce the correct IR behaviour of the massless loops. It is instructive to see why our intuition was wrong, which in turn, will guide us to defining ``generalized EFT couplings", for which previous positivity constraints apply.

%%%%%%%%%%%%%%%%%%%%%%%%%%%%%%%%%%
\noindent \textbf{Running out of bounds}
%%%%%%%%%%%%%%%%%%%%%%%%%%%%%%
Let's consider the EFT of a single massless scalar with the following higher dimension operators turned on:
\eq\label{ExampleS}
\mathcal{L}_{Int}=\frac{a_2}{\Lambda^4}(\partial \phi )^4+\frac{a_4}{\Lambda^8}(\partial^2 \phi)^4+\frac{a_6}{\Lambda^{12}}(\partial^3\phi)^4\,,
\eqe
The one-loop RG equation is then 
\eq
\mu^2\frac{\partial a_4}{\partial \mu^2}=0,\quad \mu^2\frac{\partial a_4}{\partial \mu^2}=\beta_1 a_2^2, \quad \mu^2\frac{\partial a_6}{\partial \mu^2}=\beta_2 a_2 a_4\,. 
\eqe
With the solution, $a_2=\bar{a}_2$, $a_4=\bar{a}_4+\beta_1 \bar{a}_2^2\log\frac{s_0}{p^2}$ and $a_6=\bar{a}_6+\beta_2 \bar{a}_2 \bar{a}_4\log\frac{s_0}{p^2}$. For simplicity let's consider the forward-limit Hankel matrix constraints, and set  $\bar{a}_i$s be the renormalized couplings at some scale $M^2$ where the constraints hold. For example we have $\bar{a}_i>0$ and  
\eq\label{HankelIR}
\bar{a}_2 \bar{a}_6-\bar{a}^2_4>0\,.
\eqe
Now as we allow the couplings to run in the IR, the determinant of the Hankel matrix becomes:
\eq
Det \left(\begin{array}{cc} \bar{a}_2 & \bar{a}_4+\beta_1 \bar{a}_2^2\delta \\ \bar{a}_4+\beta_1 \bar{a}_2^2\delta & \bar{a}_6+\beta_2 \bar{a}_2\bar{a}_4\delta\end{array}\right)=(\bar{a}_2 \bar{a}_6{-}\bar{a}^2_4)+(\beta_2-2\beta_1)\bar{a}_4 \bar{a}_2^2\delta+\mathcal{O}(\delta^2)\,,
\eqe
where we have used a short-hand notation $\delta=\log\frac{s_0}{p^2}$. If the running couplings were to stay inside the EFT-hedron, we would have a sharp prediction for the one-loop beta functions, namely $(\beta_2-2\beta_1)>0$. Since for our current theory we only have bubble integrals at one-loop, their coefficients can be directly captured from the two-particle cut, which we derive in appendix~\ref{AppBeta}, yielding $\beta_1=\frac{14}{5(4\pi)^2}$ and $\beta_2=\frac{166}{35(4\pi)^2}$. Immediately we see that $\beta_2-2\beta_1<0$ in contradiction to the expectation from the Hankel matrix bounds. In other words, the low energy running drives the couplings outside of the EFT hedron!

Let us see why our intuition from the mass regulated picture failed to yield the correct prediction. Consider the explicit low energy amplitude in the forward limit, which is all that is necessary for eq.(\ref{HankelIR}). We have:
\eq\label{CandidateAmp}
M(s,0)=2\frac{s^2}{\Lambda^4}\bar{a}_2+2 \frac{s^4}{\Lambda^8}\left(\bar{a}_4+\beta_1 \bar{a}_2^2\log\frac{M^2}{s}\right) +2\frac{s^6}{\Lambda^{12}}\left(\beta_2 \bar{a}_2 \bar{a}_4\log\frac{M^2}{s}\right)\,,
\eqe
where we've set $\mu^2=M^2$, representing the scale for which the Hankel constraint holds. Now by deforming the massless loop propagators to be massive, the logs get deformed as:
\eq\label{Modz}
\log\frac{M^2}{s}\rightarrow \log\frac{M^2}{m^2}{-}i\sqrt{\frac{1}{z}{-}1}\log(i\sqrt{z}{+}\sqrt{1{-}z^2}){-}1=\log\frac{M^2}{m^2}-\sum_{n}\frac{(1)_{n{-}1}}{3\left(\frac{5}{2}\right)_{n{-}1}}z^n
\eqe
where $z\equiv \frac{s}{4m^2}$. Thus we see that at low energies, $z\ll 1$, the leading log correction appearing at $s^4$ is $\log\frac{M^2}{m^2}$, reproducing the same running as the massless log if we take $s, m^2\ll M^2$. However the $z$ expansion in eq.(\ref{Modz}) introduces correction to the coefficient of  $s^6,s^8,\cdots$ that dominates over their original logarithms since:
\eq
\frac{1}{m^2}\gg\frac{1}{\Lambda^2}\log\frac{M^2}{m^2}\,
\eqe
as $m^2\rightarrow 0$. Put in another way, \textit{the small mass deformation is no longer ``small" when one considers subleading contributions}. Note that due to these corrections, the Hankel matrix constraint is trivially satisfied for the mass deformed amplitude.  Indeed it is straightforward to check that the Hankel matrix for $a_{n}\equiv \frac{(1)_{n{-}1}}{3\left(\frac{5}{2}\right)_{n{-}1}}$ is total positive, and since the $z$ expansion in eq.(\ref{Modz}) dominates the contributions for $s^6,s^8,\cdots$ couplings, they trivialize the Hankel matrix constraint on the amplitude.

%%%%%%%%%%%%%%%%%%%%%%%%%%%%%%%%%%
\noindent \textbf{Generalized EFT couplings and its dispersive representation}
%%%%%%%%%%%%%%%%%%%%%%%%%%%%%%

\begin{figure}
\begin{center}
\includegraphics[scale=0.6]{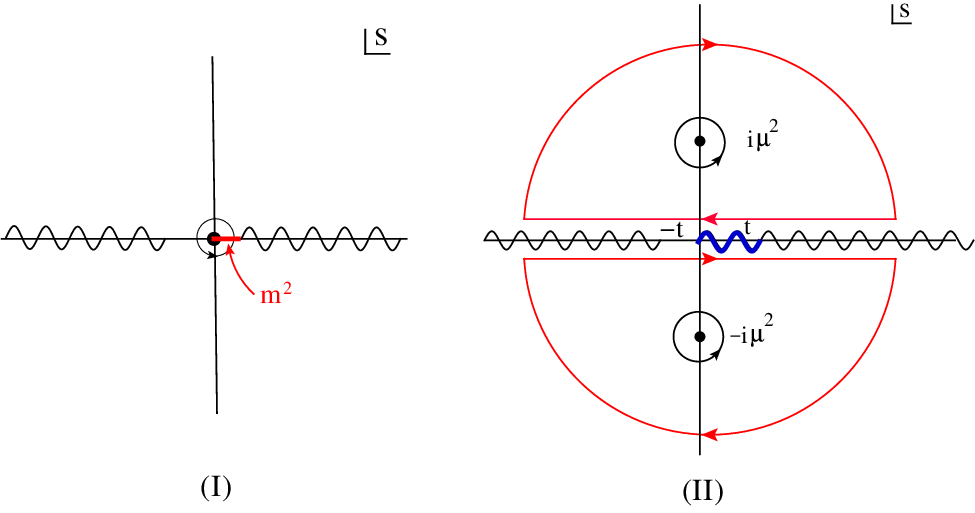}
\caption{In the presence of massless cuts, we can either (I) introduce a small mass regulator and push the cut slightly away from the origin, or (II) we analytically define our generalized couplings by moving the contour at origin onto to the complex plane to $s=\pm i \mu$ in a way that the integration measure is positive definite. After deformation the contour picks up the discontinuity on the real $s$-axes, which for $|t|<|s|$, is controlled by unitarity. We can analytically continue to $|s|<|t|$ for theories with well behaved soft limits. }
\label{mupoles}
\end{center}
\end{figure}

The reasons we've introduced the mass regulated theory is so that the massless cut is pushed off the origin, where the couplings are analytically defined. However, we've just seen that by doing so the EFT no longer captures the correct IR physics beyond leading order. Instead of moving the branch point, lets move the pole itself. For example, consider the following contour integral of the amplitude at fixed $t\ll m^2$:
\eq\label{DefContour}
\frac{1}{2\pi i}\oint \frac{ds\, s}{(s^2+\mu^4)^{n{+}1}}M(s,t)
\eqe 
where the contour encircles the poles at $s=\pm i\mu^2$, and we will take $\mu^2 \ll 1$. Using this contour we can define the following  generalized couplings in the forward limit
\eq\label{DeformLoop}
a^{\mu^2}_{2n,0}\equiv\frac{1}{2\pi i}\oint_{\mathcal{C}_0} \frac{ds\, s}{(s^2+\mu^4)^{n{+}1}}M(s,0)\,.
\eqe 
where the superscript $\mu^2$ on $g^{\mu^2}$ indicates it's the position for which the pole has been moved off the origin.  Note that we've naturally introduced scale dependence into the definition of the coupling. Now in the forward limit, $M(s,0)$ is finite since the $t$-channel cut is suppressed by pre-factors proportional to powers of $t$, guaranteed by the derivative coupling. Again deform the contour $\mathcal{C}_0$ to $\mathcal{C}_{\infty}$, this relates the generalized couplings to the discontinuity of the amplitude on the $s$-axes as illustrated in fig.\ref{mupoles}. In other words, we have 
\eqa\label{gDemo}
a^{\mu^2}_{2n,0}&=&-\frac{1}{2\pi i}\int_{-\infty}^{\infty} \;\frac{ds\, s}{(s^2+\mu^4)^{n{+}1}} \;{\rm Im}\, M(s,0)\,.
\eqae

Once again, let's demonstrate the validity of eq.(\ref{gDemo}) using our linear sigma model amplitude in eq.(\ref{SMLoop}). Since the amplitude behaves as $s^4\log s$ as $s\rightarrow\infty$, we should expect eq.(\ref{DeformLoop}) and  eq.(\ref{gDemo}) to agree for $a^{\mu^2}_{6,0}$. Using eq.(\ref{DeformLoop}) the generalized couplings evaluate to:
\eq\label{Demo}
a^{\mu^2}_{4,0}=\frac{\bar{a}_4}{m_h^8}-\frac{7a_2^2 }{160 \pi^2 m_h^8 }\left(3+2\log \frac{\mu^4}{s_0^2}\right)   , \quad a^{\mu^2}_{6,0}=\frac{7 \bar{a}_2^2}{240\pi^2 m_h^8 \mu^4 }\,.
\eqe
As expected, the $a^{\mu^2}_{4,0}$ is given by the combination of tree coefficient $\bar{a}_4$ and the one-loop log proportional to $\bar{a}^2_2$. Moreover, even though we only consider the amplitude up to $s^4$ terms, all generalized couplings $a^{\mu^2}_{2n,0}$ are nonzero due to the log. Now for eq.(\ref{gDemo}) the imaginary part of the four-point amplitude arising from the $s$- cut is given by:
\eqa\label{bubblecomp}
&&{\rm Im}_{s=[0,\infty]}\, M(s,t)=\;\;\vcenter{\hbox{\includegraphics[scale=0.5]{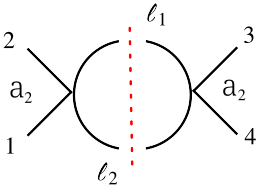}}}\nonumber\\
&&=\frac{\bar{a}^2_2}{(4 \pi)^3 m_h^8}\int d\phi'd\cos\theta' \,(s^2+t^2+u^2)_L(s^2+t^2+u^2)_R={-}\frac{\bar{a}^2_2}{m_h^8}\frac{s^4}{60(4 \pi)^2} (167+\cos 2 \theta)\,,\nonumber\\
\eqae
where $\theta$ is the scattering angle. Taking the forward limit one finds ${\rm Im}_{s=[0,\infty]}\, M(s,0)={-}\frac{7 \bar{a}^2_2 }{m_h^8}\frac{s^4}{40 \pi^2}$, reproducing the coefficient of the $s$-channel logarithm in eq.(\ref{SMLoop}). Using ${\rm Im}_{s=[0,\infty]}\, M(s,0)=-{\rm Im}_{s=[-\infty,0]}\, M(s,0)$, one recovers,  
\eq
\frac{1}{2\pi i}\left(-\int_{-\infty}^{0}{+}\int_{0}^{\infty} \right)\;\frac{ds\, s}{(s^2+\mu^4)^{4}} i\pi\frac{7 \bar{a}^2_2 }{40 \pi^2m_h^8}s^4=\frac{7 \bar{a}^2_2 }{240 \pi^2 \mu^4 m_h^8}\,.
\eqe
In agreement with eq.(\ref{Demo}).

Now deforming the contour one again picks up the discontinuity on the real axes as shown in fig.(\ref{mupoles}) (II). Now the question is whether the discontinuity is given by physical thresholds. For $t<0$, the region $|t|\leq |s|$ corresponds to the physical kinematics and thus it's discontinuity is determined from unitarity. Due to the derivative couplings, there are no new singularities at $t=0$, and  we can analytically continue to positive $t$. Thus the entire $s$-channel discontinuity can be obtained by analytically continuation of that in the physical regime, i.e. it is expressible as a positive sum of the Gegenbauer polynomials in $(D{-}1)$-spatial dimensions:
\eq
Dis_{s>0}[M(s,t)]=\sum_{\ell=0,2,4}\textsf{p}_{\ell}(s)P_\ell(\theta)\,.
\eqe

Let's demonstrate the above in an non-trivial example. The one-loop correction to the scalar theory introduced earlier this section have one-loop logarithm  proportional to $\bar{a}_2^2$, $\bar{a}_2\bar{a}_4$, and  $\bar{a}_4^2$. The first two was computed previously while the latter is given by  
\eqa
\frac{\bar{a}^2_4s^8}{M^{16}20160(4\pi)^2} (39843{+} 988\cos2\theta {+}\cos 4\theta)\,.
\eqae
Summing all three contributions we've obtain the discontinuity on the positive real axes given by the following spinning spectral functions
\eqa
\textsf{p}_{0}(s){=}\frac{ s^4 (25 \bar{a}_2 + 21 \bar{a}_4 s^2)^2}{225(4\pi)^2},\quad\textsf{p}_{2}(s){=}\frac{s^4 (7 \bar{a}_2 + 12 \bar{a}_4 s^2)^2}{2205(4\pi)^2},\quad\textsf{p}_{4}(s){=}\frac{\bar{a}^2_4  s^8}{11025(4\pi)^2}\,,
\eqae
 and indeed they are positive definite.

In conclusion, the generalized coupling constants defined through the contour integral in eq.(\ref{DefContour}), again subject to appropriate boundary behaviour,  will satisfy the same analytic constraint as that before. In the following we will demonstrate with explicit examples that the Hankel matrix constraint is satisfied.

%%%%%%%%%%%%%%%%%%%%%%%%%%%%%%%%%%
\noindent \textbf{ The Hankel matrix constraints}:
%%%%%%%%%%%%%%%%%%%%%%%%%%%%%%%%%%

Let's again take the forward limit four-point amplitude for eq.(\ref{ExampleS})
\eqa
M_4(s,0)&=&2 \,\frac{\bar{a}_2s^2}{\Lambda^4}{+}2\left(\bar{a}_4+\beta_1 \bar{a}_2^2\log\frac{M^2}{s}\right)\frac{s^4}{\Lambda^8}{+}2\left(\bar{a}_6+\beta_1 \bar{a}_2\bar{a}_4\log\frac{M^2}{s}\right)\frac{s^6}{\Lambda^{12}}
\eqae
%\eqa
%M_4(s,0)&=&2 \,\frac{\bar{g}_2s^2}{M^4}+2\left(\bar{g}_4+\beta_1 \bar{g}_2^2\log\frac{M^2}{s}\right)\frac{s^4}{M^8}+2\beta_2 \bar{g}_2 \bar{g}_4\left(\log\frac{M^2}{s}\right)\frac{s^6}{M^{12}}\nonumber\\
%&+&2\left(\beta_3\bar{g}^2_4\log\frac{M^2}{s}\right)\frac{s^8}{M^{16}}\,.
%\eqae
%where $\beta_1,\beta_2$ were given previously, and $\beta_3=\frac{638}{315(4\pi)^2}$ .
The generalized couplings are then given by 
\eqa\label{GenCoup}
&&a^{\mu^2}_{2,0}=\frac{1}{\Lambda^4}\left[\bar{a}_2{+}z^4\left(\beta_1\bar{a}^2_2\left(\frac{1}{2}{-}2\log y\right)-2\bar{a}_4\right){+}\mathcal{O}\left(z^8\right)\right],\nonumber\\
&& a^{\mu^2}_{4,0}=\frac{1}{\Lambda^8}\left[\bar{a}_4{-}\beta_1\bar{a}^2_2\left(\frac{3}{4}{-}\log y\right){+}z^4\left(\beta_2\bar{a}_2\bar{a}_4\left(\frac{5}{4}-3\log y\right)-3 \bar{a}_6\right){+}\mathcal{O}\left(z^8\right)\right],\nonumber\\
 &&a^{\mu^2}_{6,0}=\frac{1}{\Lambda^8\mu^4}\left[\frac{\beta_1\bar{a}^2_2}{6}{+}z^4\left(\bar{a}_6{-}\beta_2\bar{a}_2\bar{a}_4\left(\frac{11}{12}{-}\log y\right)\right){+}\mathcal{O}\left(z^8\right)\right]
\eqae
where $z=\frac{\mu^2}{\Lambda^2}$ and $y=\frac{M^2}{\mu^2}$. First of all, we see that the leading contributions for $a^{\mu^2}_{2,0}$ are given by the tree-level coupling $\bar{a}_2$, where as for $a^{\mu^2}_{4,0}$ the tree-level coupling $\bar{a}_4$ mixes with logarithmic contributions $\beta_1\bar{a}^2_2\log y$ at leading order. However, beyond $a^{\mu^2}_{4,0}$ the original tree-couplings become subdominant to terms that were generated from the logarithms in $a^{\mu^2}_{4,0}$. Indeed for  $a^{\mu^2}_{6,0}$ the tree-level piece $\bar{a}_6$ is \textit{subleading} to a term proportional to $\beta_1\bar{a}^2_2$, which came from the leading logarithm in $a^{\mu^2}_{4,0}$. The dominance of terms induced by the the leading log for all $a^{\mu^2}_{2n,0}$ with $n>2$, is reminiscent of the leading $\frac{1}{m}$ corrections flooding the higher-derivative couplings for the mass regulated case discussed previously.  As we will see, these effects ensures the positivity constraints on the generalized couplings which we now derive.

Now let us consider the dispersive representation: 
\eqa
a^{\mu^2}_{2n,0}&=&-\int_{-\infty}^{\infty} \;\frac{ds\, s}{(s^2+\mu^4)^{n{+}1}} {\rm Im}\;M(s,0)\,,
\eqae
As discussed above, even in the presence of massless cut, the discontinuity is still given by a positive sum of Gegenbauer polynomials. The only modification is that the $s$-channel cut now starts at $s=0$. Incorporating the $u$-channel cut, we then have a branch cut covering the entire real axes leading to 
\eqa
a^{\mu^2}_{2n,0}&=&\left[-\int_{-\infty}^{0}{+}\int_{0}^{\infty}\right] \;\frac{ds\, s}{(s^2+\mu^4)^{n{+}1}} \;\sum_{\ell}\textsf{p}_{\ell}(s) G^{\frac{D-4}{2}}_{\ell}(1)\nonumber\\
&=&\sum_{\ell}\int_{0}^{\infty}\frac{dx}{(x+\mu^4)^{n{+}1}} \;\textsf{p}_{\ell}(x) G^{\frac{D-4}{2}}_{\ell}(1)\,,
\eqae
In other words, it is given by a continuous sum of points on the moment curve:
\eq
\left(\begin{array}{c}a^{\mu^2}_{2,0} \\ a^{\mu^2}_{4,0} \\ a^{\mu^2}_{6,0} \\\vdots \\ a^{\mu^2}_{2n,0}\end{array}\right)=\sum_{i} c_i \left(\begin{array}{c} 1 \\ y_i \\ y_i^2\\ \vdots \\ y_i^{n{-}1}\end{array}\right),\quad c_i>0,\,y_i>\frac{1}{\mu^4}\;\forall i \,.
\eqe
Note that the moment curve is shifted by $\frac{1}{\mu^4}$, and thus the coefficients will obviously satisfy the original Hankel matrix constraint.  

Let us show this in detail for the generalized couplings in eq.(\ref{GenCoup}). First of all in the limit $\mu^2\ll \Lambda^2$, the positivity of $a^{\mu^2}_{2,0}, a^{\mu^2}_{4,0}, a^{\mu^2}_{6,0}$ and $a^{\mu^2}_{2,0}a^{\mu^2}_{6,0}{-}(a^{\mu^2}_{4,0})^2$ is ensured by the positivity of the tree-level coupling and that of the $\beta_i$s. An interesting scenario occur when we deform the position of the pole all the way to the renormalization scale $\mu^2=M^2$, while assuming $M^2\ll \Lambda^2$. The positivity of $a^{\mu^2}_{4,0}$ then requires that 
\eq\label{NewConst}
\bar{a}_4{-}\beta_1\bar{a}^2_2\frac{3}{4}>0
\eqe
where again $\beta_1=\frac{14}{5(4\pi)^2}\sim0.002$. It is easy to see that this imposes further constraint on the couplings beyond that of the tree-level Hankel constraints, i.e. the positivity of $\bar{a}_2,\;\bar{a}_4,\;\bar{a}_6$, and $ \bar{a}_2\bar{a}_6-\bar{a}^2_4$. 

It is interesting to understand why this new constraint arises. First, note that the effective action considered in the beginning of this section, eq.(\ref{ExampleS}), is not the most generic for single scalar theory: it lacks the marginal $\phi^4$ interaction. In general, the lack of $\phi^4$ interaction is associated spontaneous symmetry breaking in the UV, where the resulting EFT respects a shift symmetry. Now due to boundary contributions, for tree-level couplings we are not privy to the information of the constant piece of the amplitude, or $k=0$, which translate to the presence/absence of $\phi^4$ interaction. However, at loop-level, its presence will affect the pattern of IR running for the couplings.  For example, the presence of $\phi^4$ would induce logarithmic running already for the $s^2$ operator, which leads to the modification of  $a^{\mu^2}_{4,0}$ to:
\eq
a^{\mu^2}_{4,0}=\frac{1}{\Lambda_4\mu^4}\left[\frac{\bar{a}_0\bar{a}_2\beta_0}{4}{+}z^4\left(\bar{a}_4{-}\beta_1\bar{a}^2_2\left(\frac{3}{4}{-}\log y\right)\right){+}\mathcal{O}(z^8)\right]\,,
\eqe 
 instead of eq.(\ref{GenCoup}). Here $\bar{a}_0$ is the tree-level coupling for $\phi^4$ and $\beta_0$ is the beta function for $s^2$ operator. We see that the running at $s^2$ now induces corrections for $a^{\mu^2}_{4,0}$ that dominates the original contributions! Now the positivity of $a^{\mu^2}_{4,0}$ simply implies $\bar{a}_0\beta_0>0$, even if we take $\mu$ close to the renormalization scale. 
 
Said in another way, the constraint in eq.(\ref{NewConst}) is a reflection of $\bar{a}_0=0$!  Let's consider an explicit UV completion that realizes such low energy behaviour: the linear sigma model. As discussed previously, the shift symmetry of the EFT ensures that there are no constant piece for the quartic interaction. In IR tree-level couplings can be identified as $\bar{a}_2=\bar{a}_4=\lambda$, where $\lambda$ is the quartic coupling constant of the complex scalar in the UV. Thus we see that in the perturbative regime, where the map between the IR and UV couplings are applicable, eq.(\ref{NewConst}) is trivially satisfied. 

Thus we see that when massless loops are included, the positivity bounds allows us to probe details of the EFT previously hidden behind the ``Froissart horizon" !

%%%%%%%%%%%%%%%%%%%%%%%%%%%%%%%%%%
\noindent \textbf{A peek beyond the forward limit}
%%%%%%%%%%%%%%%%%%%%%%%%%%%%%%%%%%

We now consider the extension away from the forward limit, which correspond to taking a Taylor expansion around $t=0$.  Again due to the $t$-channel log coming in the form $t^n\log t$, the amplitude is finite in the forward limit. Due to the $t$-channel branch cut, once again we deform the $t$ contour  away from the origin to $t=\epsilon$: 
\eqa\label{defanq}
a^{\mu^2}_{k,q}&\equiv& \left(\frac{1}{2\pi i}\right)^2\oint \frac{dt}{(t-\epsilon)^{q{+}1}}\oint \frac{ds\, s^{\frac{1+(-)^{k}}{2}}}{(s^2+\mu^4)^{\lfloor{\frac{k{-}q}{2}}\rfloor {+}1}}M(s,t)\,,\nonumber\\
\eqae 
where $\epsilon>0$. We will be considering the limit where $t$ is much smaller than any massive threshold. Note that since $\epsilon>0$, we are actually analytically continuing $t$ away from the physical regime $t<0$. For theories such as those of interacting goldstones, where the massless amplitudes are soft enough, free of soft/collinear singularities, so that massless amplitudes are well-defined, it is reasonable to expect that discontinuities of the amplitude in the $s-$channel are actually analytic in $t$. Taking this as a working assumption gives us the dispersive representation. We have:
\eq
a^{\mu^2,\epsilon}_{k,q}=\frac{1}{2\pi i}\oint \frac{dt}{(t-\epsilon)^{q{+}1}}\sum_{\ell}\int_{0}^{\infty}\frac{ds\, s^{\frac{1+(-)^{k}}{2}}}{(s^2+\mu^4)^{\lfloor{\frac{k{-}q}{2}}\rfloor {+}1}} \;\textsf{p}_{\ell}(s) G_{\ell}\left(1+2\frac{t}{s}\right)\,,
\eqe 
Evaluating the $t$-integral on the pole then gives the Taylor expansion of the Gegenbauer polynomials $G_{\ell}(x)$ at $x=1+\epsilon$. Now importantly, since we've set $\epsilon>0$, the resulting convex hull is \textit{inside} the Gegenbauer polytope! To see this, recall that under the rescaling $x\rightarrow a x$ with $a>1$, Gegenbauer polynomials rescales to a positive function, i.e. : 
\eq
G_{\ell}((1{+}\epsilon)x)=\sum_{\ell'=0}^{\ell}\quad c_{\ell'}G_{\ell'}(x),\quad c_{\ell'}>0\,.
\eqe
It then follows that the vector $\vec{G}_{\ell}(1+\epsilon)$ is a positive sum of $ \vec{G}_{\ell}(1)$,  and thus  the convex hull of $\vec{G}_{\ell}(1+\epsilon)$ must be inside Gegenbauer polytope! In fact, from eq.(\ref{PositionPos}), we see that the convex hull of $\vec{G}_{\ell}(1+\epsilon)$ is another cyclic polytope. Thus as we increase in $\epsilon$, the couplings must live in a cyclic polytope that is \textit{contained} in the previous ones. In this precise sense, by increasing $\epsilon$ generalized couplings moves deeper inside the original geometry!

%%%%%%%%%%%%%%%%%%%%%%%%%%%%%%%%%%%%%%%%%
\section{Outlook}
%%%%%%%%%%%%%%%%%%%%%%%%%%%%%%%%%%%%%%%%%
We have seen that the constraints on vacuum stability, causality and unitarity place enormously powerful constraints on low-energy effective field theories. There are a large number of obvious open avenues for future work. Most immediately, there is the question of fully understanding the geometry and boundary structure of the EFT-hedron for four-particle scattering; this mathematical problem has been fully solved for the toy example of the $s-$channel only EFT-hedron where it is already rather non-trivial.  We have also bounded the full EFT-hedron for the most general cases of interest, but have still not determined the exact facet structure of the EFT-hedron in complete generality. It would also be interesting to extend the dispersive analysis  beyond $2 \to 2$ scattering. Indeed, if we consider a simple theory with Lagrangian $P(X=(\partial \phi)^2)$, we know that subluminality for small fluctuations around background with $\langle \partial \phi \rangle \neq 0$ demands $P^{\prime \prime}(X)>0$ for all $X$, which enforces positivity conditions on higher-point scattering amplitudes. Another obvious avenue is to systematically explore constraints on scattering for multiple species with general helicities.

It is also important to note that, while the EFT-hedron places powerful constraints on the effective field theory expansion, sensible effective field theories do not appear to populate the entire region allowed by the EFT-hedron, but cluster close to its boundaries. The reason is likely that the physical constraints we have imposed, while clearly necessary,  are still not enough to capture consistency with fully healthy UV theories. In particular, our dispersive representation at fixed $t$, does not make it easy to impose the softness of high-energy, fixed-angle amplitudes where both $s,t$ are large with $t/s$ fixed. It would be fascinating to find a way to incorporate this extra information about UV softness into the constraints, along the lines of the celestial sphere amplitude~\cite{Arkani-Hamed:2020gyp}, which should further reduce the size of the allowed regions for EFT coefficients.

The unexpected power of stability, causality and unitarity in constraining effective field theory raises the specter of a much greater prize, which was in the fact that question that initially motivated this work. Can the same principles be used to strongly constrain, and perhaps with additional conditions actually uniquely determine, consistent UV complete scattering amplitudes? To sharpen this question, we can begin by thinking about UV completions of gravity amplitudes at ``tree-level", assuming the amplitude only has poles. Unlike theories of scalar scattering, which can be UV completed in a myriad of ways such as e.g. glueball scattering in large N-gauge theories, the only consistent tree-gravity scattering amplitudes we know of come from string theory, so it is more likely this question has a unique answer. The four particle tree graviton scattering amplitudes in string theory are essentially unique, independent of any details of compactification and fixed  by the nature of the worldsheet supersymmetry. Indeed the amplitudes differ only by the massless three particles amplitudes in the low-energy theory, with type II theories having only the usual three-graviton vertex, and the heterotic theory also including the $R^2 \phi$ coupling to the dilaton. So it is plausible to conjecture that amplitudes with, say,  only the usual three-graviton amplitude at low-energies, have a unique tree-level UV completion given by the Virasoro-Shapiro amplitude. 

As an easy first step in this direction, it is easy to see that tree-level UV completions of gravity must contain an infinite tower of massive particles of arbitrarily high spin. In fact gravity is not particularly special in this regard. Consider any theory with fundamental cubic interactions, so that four-particle amplitudes already have $\frac{1}{s,t,u}$ poles at tree-level. Suppose we wish to improve the high-energy behavior of the amplitudes relative to what is seen in the low-energy theory, so e.g. for gravity/Yang-Mills/ $\phi^3$ theory, we would like the high-energy limit to drop more quickly that $s^2/s/s^{-1}$ respectively. It is then easy to see that this is impossible unless the UV theory has an infinite tower of particles with arbitrarily large spin. 

Let us briefly sketch the reason for this. It is instructive to contrast the situation with that of simple UV completions for theories whose four-particle interaction begin with contact interactions at low-energies. Consider for instance goldstone scattering in the non-linear sigma model, where the low-energy four-particle amplitude begins as 
${\cal A} = -\frac{1}{f^2}(s + t)$. It is trivial to UV complete this simply by softening $s \to \frac{s}{(1 - s/M^2)},\; t \to \frac{t}{(1 - t/M^2)}$. This is consistent with the causality bounds at large $s$ and fixed $t$, and keeps the fixed-angle amplitude small so long as $M^2 \ll f^2$. And crucially, thanks to the overall negative sign in front of the amplitude, the residues on the massive poles are positive and are interpreted as the production of a scalar particle with positive probability. This is of course nothing but the linear sigma model UV completion of the non-linear sigma model, with the new massive particle identified as the Higgs. Note that had the overall sign of the amplitude been reversed, we would not be able to do this, as the residue on the massive pole would be negative. 

Now, consider instead the amplitude ${\cal} A = g^2(\frac{1}{s} + \frac{1}{t} + \frac{1}{u})$ for $\phi^3$ theory at tree-level, and let us try to add massive poles to make the amplitude decrease faster than $1/s$ at high-energies. It is easy to see that the same strategy used in the goldstone example can't work. For instance if we again attempt to soften $\frac{1}{s} \to \frac{1}{s(1 - s/M^2)}$, the residue on the massive pole will have the opposite sign as that of the (correct, positive) residue on the massless pole at $s=0$! This will happen for any amplitude that is a rational function (finite number of massive poles) in the Mandelstam variables. If the amplitude is softened in the physical region, it is softened everywhere in the $s$-plane; so given that the amplitude vanished faster than $1/s$ at infinity, the sum of all the residues must be zero. But that means that some of the massive residues must be negative, to cancel the positive residue at $s=0$. This can only be avoided if there are infinitely many poles, that allows the function to die in the physical region but blow up elsewhere in the $s$-plane, as familiar in string theory. A small elaboration of this argument also shows the necessity of an infinite tower of spins, and the same arguments apply to gravity and Yang-Mills amplitudes as well. 

It is amusing that theories that only have a life in the UV--such as the weak interactions and the non-linear sigma model, whose low-energy amplitudes are tiny, are ``easy" to UV complete with finitely many massive states. It is theories with IR poles, associated with long-range interactions, that are forced to have much more non-trivial UV completions. This is why the most ancient interaction described by physics--gravity--continues to be the most challenging to UV complete, while the weak interactions were discovered and UV completed within about half a century!  

One can also easily ``discover" the stringy completion of gravity amplitudes, from the bottom-up, as the simplest possible UV completion with an infinite tower of poles satisfying extremely basic consistency conditions, even before imposing the restrictions of causality and unitarity. The tree-level 4-graviton amplitude is ${\cal A}^{+-+-} = G_N \langle 1 3 \rangle^4 [2 4]^4 \times \frac{1}{s t u}$. We know that any tree-level UV completion must have an infinite tower of poles, in the $s,t,u$ channels. Thus, the most general Ansatz for the amplitude would replace $\frac{1}{s t u} \to \frac{N(s,t,u)}{s t u \prod_i (s- m_i^2)(t-m_i^2)(u-m_i^2)}$. Note that this expression has the property that on the $s-$ channel pole at $s=m_j^2$, the residue has poles at $t = m_i^2$ and $u = m_i^2 \to t=  -(m_i^2 + m_j^2)$. These poles must be absent in the physical amplitude, thus the numerator must have zeroes, when $s=m_j^2$, at these values of $t$. It is then natural to make the simple assumption that these are the {\it only} zeroes of the numerator. That tells us that if we write $N(s,t,u) = \prod_j (s+r_i)(t+r_i)(u+r_i)$, that the set of all the roots $\{r_i\}$ must contain all of $\{m_i^2,m_i^2 + m_j^2\}$. And this in turn is most trivially accomplished if $m_j^2 = M_s^2 j$ are just all the integers in the units of a fundamental mass scale $M_s$!

By this simple reasoning, we are led to the infinite product formula for the Virasoro-Shapiro amplitude, putting $\alpha^\prime = M_s^{-2}$:
\begin{eqnarray}
{\cal A}& =& G_N \langle 1 3 \rangle ^4 [2 4]^4 \frac{\prod_{j=1}^\infty (\alpha^\prime s + j)(\alpha^\prime t + j) (\alpha^\prime u + j)}{\prod_{i=0}^\infty (\alpha^\prime s -i)(\alpha^\prime t - i)(\alpha^\prime u - i)} \nonumber \\   
&=& G_N \langle 1 3 \rangle ^4 [2 4]^4 \frac{\Gamma(-\alpha^\prime s) \Gamma(-\alpha^\prime t ) \Gamma(-\alpha^\prime u)}{\Gamma(1 + \alpha^\prime s) \Gamma(1 + \alpha^\prime t) \Gamma(1+ \alpha^\prime u)}\,.
\end{eqnarray}
Of course this is not at all a ``derivation" of the string amplitude, but it nonetheless striking to see how easily the amplitude emerges as the simplest possible way of writing an expression with infinitely many poles that passes even the most basic consistency checks. 

In fact, it is fascinating that directly checking the consistency known string tree amplitudes is high non-trivial. Causality in the form of the correct Regge behavior is readily verified, but unitarity, in the form of the positivity of the Gegenbauer  expansion of the amplitude residues on massive poles, turns into a simple but highly non-trivial statement. For concreteness consider the scattering of colored massless scalars in the type I open superstring theory, where the amplitude is 
\begin{equation}
{\cal A} = s^2 \frac{\Gamma(-s) \Gamma(-t)}{\Gamma(1 - s - t)} 
\end{equation}
The residues on the massive poles at $s=n$ is a polynomial $R_n(x={\rm cos} \theta)$, where $t= - \frac{n}{2}(1 - x)$, given by
\begin{equation}
P_n(x) = \prod_{i=1}^{n-1} \left(x - \frac{(n-2i)}{n} \right)
\end{equation}
Already at $n=3$, we learn something striking: $P_3(x) = (x-\frac{1}{3})(x+\frac{1}{3}) = x^2 - \frac{1}{9}$, which we would like to express as a sum over Gegenbauer polynomial.  The spin 2 Gegenbauer in $d$ spatial dimension is proportional to $x^2 - \frac{1}{d}$, thus by writing $(x^2 - \frac{1}{9}) = (x^2 - \frac{1}{d}) + (\frac{1}{d} - \frac{1}{9})$, we see a massive spin 2 state with positive norm, but also a spin 0 state with norm $(\frac{1}{d}- \frac{1}{9})$, which is $\geq 0$ for $d \leq 9$, but is negative, violating unitarity, for $d>9$. Thus the critical spacetime dimension $D = d+1 = 10$ is hiding in plain sight in the four-particle amplitude, purely from asking for unitarity at on this pole at $s=3$. But of course for unitarity, we must have that 
\begin{equation}
P_n(x) = \sum_s p_{n,s} G^{(d)}_s(x), \; {\rm with} \; p_{n,s} \geq 0 \; {\rm for} \; d \leq 9
\end{equation}
This extremely simple statement turns out to be very difficult to prove directly, indeed are not aware of any direct proof of this fact in the literature! Of course it does follow, more indirectly, from the still rather magical proof of the no-ghost theorem in string theory. 

The miraculous way in with which string amplitudes manage to be consistent make it seem even more plausible that these amplitudes emerge as the unique answer to the question of finding consistent four particle massless graviton amplitudes with only poles. But  some further constraints other than causality, unitarity and good high-energy behavior of just massless graviton scattering, must be imposed to do this, as we have found candidate four-particle amplitudes satisfying all these rules that deform away from the known string amplitudes. Consider again the  Virasoro-Shapiro amplitude for graviton scattering. The residue on the pole at $s=n$ is the square of the open-string residue $P_n(x)^2$, and so the positivity of its Gegenbauer expansion follows directly from the positivity of $P_n(x)$ for the open string. But now consider a deformation by a parameter $\epsilon$ of the form 
\begin{eqnarray}
\frac{\Gamma({-}\alpha^\prime s) \Gamma({-}\alpha^\prime t ) \Gamma({-}\alpha^\prime u)}{\Gamma(1 {+} \alpha^\prime s) \Gamma(1 {+} \alpha^\prime t) \Gamma(1 {+} \alpha^\prime u)} & \to &   \nonumber \\
\frac{\Gamma({-}\alpha^\prime s) \Gamma({-}\alpha^\prime t ) \Gamma({-}\alpha^\prime u)}{\Gamma(1{+} \alpha^\prime s) \Gamma(1 {+}\alpha^\prime t) \Gamma(1{+} \alpha^\prime u)} &  {+} & \epsilon \frac{\Gamma(1{-}\alpha^\prime s) \Gamma(1{-}\alpha^\prime t ) \Gamma(1{-}\alpha^\prime u)}{\Gamma(2 {+} \alpha^\prime s) \Gamma(2 {+} \alpha^\prime t) \Gamma(2 {+} \alpha^\prime u)}
\end{eqnarray}
This deformed amplitude has the same Regge behavior as the usual string amplitude, and the same exponential softness for high-energy fixed-angle scattering. The residue at $s=n$ is given by 
\eq
\frac{1{+}n(1{-}\epsilon)}{n{+}1}\left(\frac{n^{n{-}1}}{2^{n{-}1}n!}\right)^2\left(P_n(x)^2+\frac{4 \epsilon(n{-}1)}{n(1{+}(1{-}\epsilon)n)}P_n(x)P^B_{n{-}4}(x)\right)\,,
\eqe
where $P^B_{n}(x)\equiv \prod_{i=1}^{n+1}\left(x-\frac{n{+}2{-}2i}{n{+}4}\right)$ is the residue of the Veneziano amplitude. It is 
straightforward to see that so long as $0<\epsilon< 1$, the positivity of $P_n(x)$ continues to imply the positivity of the Gegenbauer expansion on the massive poles. Thus this deformed expression satisfies all the constraints we have been imposing on four-particle scattering. It seems very unlikely, however, that this corresponds to amplitudes in some consistent deformation of string theory: the spectrum is exactly the same as the usual (free!) string, and there is no obvious room for an extra parameter $\epsilon$ in the quantization of the string. 

Thus  any claim about consistent UV completion must go beyond merely the consistency of {\it massless} scattering at four particles, and include  consistent expressions for higher-point massless scattering and/or, relatedly, consistent amplitudes for the new massive resonances introduced in the UV completion. This is very reasonable and is after all precisely what happened in the story of the weak interactions, where the four-fermi interaction was UV completed by $W$ particles, which in turn had bad high-energy growth for the scattering of their longitudinal modes that had to be further cured by the Higgs. It is also interesting to note that imposing just a frisson of extra string properties on the four-particle amplitude--such as  the monodromy relations relating different color channels~\cite{Boels:2014dka, Huang:2020nqy}--when combined with the EFT-hedron constraints, {\it do} appear to uniquely fix string amplitudes.  These observations all suggest a number of fascinating open avenues for further exploration at the intersection of unitarity, causality, analyticity, string theory and the UV/IR connection.

%%%%%%%%%%%%%%%%%%%%%%%%%%%%%%%%%%%%%%%%%
\section{Acknowledgements}
%%%%%%%%%%%%%%%%%%%%%%%%%%%%%%%%%%%%%%%%%
We thank Alex Postnikov, Steven Karp and Congkao Wen for stimulating discussions. We thank Zvi Bern, Alexander Zhiboedov, and Dimitrios Kosmopoulos for discussions, and for pointing out an error in the spinning residue polynomials given in the first version of the paper. N.A-H. is supported by DOE grant de-sc0009988. Y-t H is supported by MoST Grant No. 106-2628-M-002-012-MY3. T-c H is supported by the U.S. Department of Energy, Office of Science, Office of High Energy Physics, under Award Number DE-SC0011632.

\appendix

%%%%%%%%%%%%%%%%%%%%%%%%%%%%%%%%%%%%%%%%%%%%%
\section{Causality constraints on amplitudes}\label{Froissart}
%%%%%%%%%%%%%%%%%%%%%%%%%%%%%%%%%%%%%%%%%%%%%

%%%%%%%%%%%%%%%%%%%%%%%%%%%%%%%%%%%%%%%%%%%%%
\subsection{Time delay and positivity bounds}
%%%%%%%%%%%%%%%%%%%%%%%%%%%%%%%%%%%%%%%%%%%%%
It is well-known that causality puts interesting positivity bounds on the amplitude in the low energy effective field theories. Perhaps the simplest example is the case of a single derivatively coupled scalar with lagrangian 
\eq
\mathcal{L}=\frac{1}{2}(\partial \phi)^2+\frac{c}{M^4}(\partial \phi)^4+\cdots\,.
\eqe
The claim is causality demands $c>0$~\cite{Adams:2006sv}. This is slightly surprising at first sight: $c$ reflects unknown physics in the UV, ordinarily we can only probe higher-dimension operators if they violate a symmetry of the low-energy theory, but that is not the case here. And indeed, there is nothing \textit{obviously} wrong with this as an \textbf{Euclidean} EFT. However in the physical Lorentzian world, there \textit{is} something "right-on-the-edge" in the 2-derivative theory: $\phi$ excitations propagate \textit{exactly} on the light-cone. It can happen that in simple backgrounds the coefficient for the higher-dimensional operators push propagation \textit{outside} the light-cone. We can consider for instance the spatially translationally invariant background $\phi=\phi_0+\varphi$ where $\dot{\phi}_0\neq0$. We can make $(\dot{\phi}_0/M^2)$ as tiny as we like such that the background is trustworthy within the EFT. The background breaks Lorentz invariance and small fluctuations propagate with speed $v=(1{-}\frac{c\dot{\phi}_0^2}{M^4})$, so we must have $c>0$ to avoid superluminality.

Note that despite being associated with a higher-dimensional operator, the effect of the superluminality is not ``small. Indeed if we turn on $\dot{\phi}_0\neq0$ inside some bubble of radius $R$, and throw in a $\varphi$ excitation, we get a time advance/delay of $\varphi$ propagation that is $\delta t=\delta_v R=\frac{c\dot{\phi}_0^2}{M^4}R$, which can be made arbitrarily large by increasing $R$. 
$$\includegraphics[scale=0.45]{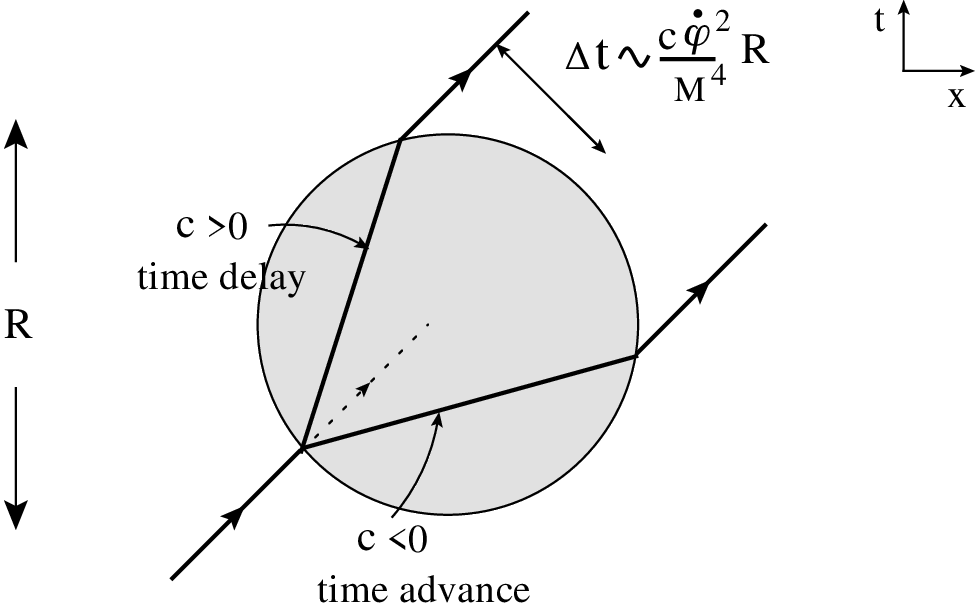}$$
This highlights the fundamental fact that the usual Wilsonian intuition about the decoupling of ``short-distance" from ``long-distance" physics is \textbf{fundamentally Euclidean}. In Euclidean signature, to probe a distance $(x{-}y)^2\sim\frac{1}{\Lambda^2_{\rm UV}}$, one needs probes with wavelength near the UV scale $\Lambda_{\rm UV}$. By contrast in Minkowski space,  ultra-small spacetimes $(x{-}y)^2\sim\frac{1}{\Lambda^2_{\rm UV}}$ can be probed by very long-distance experiments since $(x,y)$ can be separated by huge distances and time but be close to the light-cone, with advances/delays that can be made parametrically large. 

As is also well-known, these positivity constraints can also be derived from unitarity plus dispersion relations, reflecting the historic origin of analytic properties of Green's functions and amplitudes in the investigation of causal propagation!  We will recap this story, but instead of jumping from the classical picture of $ \varphi$ propagation around a background to dispersion relations for the forward $2\rightarrow2$ scattering amplitude, we will connect the two pictures directly, by repeating the above analysis, preformed in the language of classical field theory,  in terms of particle propagation plus scattering. As we will see this will in fact give us more than simply the positivity of the $(\partial \phi)^4$ coefficient; we will see that 
\eq
\frac{\partial}{\partial s}\frac{M(s)}{s}>0\,,
\eqe
where $M(s)$ is the four particle $\varphi$ scattering amplitude in the forward limit as $t\rightarrow 0$.

As is ubiquitous in the quantum particle-classical field theory connection for bosons, we recover the classical field picture of time advance/delay for small fluctuations about the background, by considering the scattering of a single hard $\varphi$ quanta, against a bose condensate of a large number $N$ of soft $\phi_0$ quanta, representing the blob. We begin by recalling familiar undergraduate basics on wave-packets and the connection between amplitude phase shifts and time delays. First free propagation, where we have one particle states with momentum $\vec{p}$. From these we can build good approximation to particles moving with constant momentum trajectories. We can define the state $|\vec{x}_{*},\vec{p}_{*}; t_{*}\rangle$ as 
\eq
|\vec{x}_{*},\vec{p}_{*}; t_{*}\rangle=\int d^dp\;\; e^{i(\vec{p}\cdot \vec{x}_{*}-E(\vec{p})t_{*})}\Psi_{\Delta p}(\vec{p}-\vec{p}_*)
\eqe  
where $\Psi_{\Delta p}(\vec{p}-\vec{p}_*)$ is sharply localized around $\vec{p}=\vec{p}_*$, for example $\Psi_{\Delta p}(\vec{p}-\vec{p}_*)\propto e^{-(\vec{p}-\vec{p}_{*})^2/(\Delta p)^2}$. With this definition, we can compute $|\langle \vec{x}_2,\vec{p},t_2|\vec{x}_1,\vec{p},t_1\rangle|^2$ via stationary phase approximation, giving 
\eq
|\langle \vec{x}_2,\vec{p},t_2|\vec{x}_1,\vec{p},t_1\rangle|^2=e^{-\Delta p^2\left((\vec{x}_1-\vec{x}_2){-}\vec{V}(\vec{p})(t_1{-}t_2)\right)^2}
\eqe
where $\vec{V}(\vec{p})=\frac{\partial E(\vec{p})}{\partial \vec{p}}$; this peaked on the classical constant velocity trajectory $\Delta \vec{x}=\vec{V}\Delta t$ with the unavoidable quantum-mechanical uncertainty of order $\frac{1}{\Delta p}$.

Now let's instead imagine that we are propagating through our blob above. Now in computing the same overlap, we will need the $S$-matrix element for $\varphi$ scattering off the blob, 
\eq
\langle B,\vec{p}|S|B,\vec{p}\rangle=e^{i\delta(E(\vec{p}))}.
\eqe
Note that the momentum uncertainty/transfer associated with the blob is $k\sim\frac{1}{R}$, which we assume to be much smaller than $|\vec{p}|$, so the outgoing momentum is the same as the incoming one. We also assume no other particles were produced, so that this amplitude is just a phase $e^{i\delta(E(\vec{p}))}$. Repeating the stationary phase analysis, we now find that 
\eq
|\langle \vec{x}_2,\vec{p},t_2|\vec{x}_1,\vec{p},t_1\rangle|^2=e^{-\Delta p^2\left(\Delta x{-}\vec{V}(\vec{p})(\Delta t {+}\frac{\partial \delta(E)}{\partial E})\right)^2}\,.
\eqe
Thus, the presence of the blob has given us a time delay/advance given by $\Delta^{ blob}t=\frac{\partial \delta(E)}{\partial E}$. In order for this to be detectable above the quantum uncertainty $\Delta^{quantum}t\sim\frac{1}{\Delta p}\sim \frac{1}{E}$, clearly we must have that the phase $\delta(E)\gg1$ is parametrically large.

Thus to find a situation where the delay/advance is reliably calculable, we must find a setting where $\delta(E)\gg1$ is reliably calculable. Now when we consider few particle scattering in any situation with a weak coupling, where amplitudes are reliably calculable, essentially by definition the phase above will be perturbatively small. However, $\delta(E)\gg1$  is exactly what happens when we scatter $\varphi$ off the condensate ``blob", which we can think of as a large number $N$ of $\varphi$ quanta with $k\sim\frac{1}{R}$. Note that the relation between $N$ and the classical background field $(\partial \phi_0)$ is given e.g. by matching the energy of the blob in the two pictures, as $N\sim (\partial \phi_0)/k^4$. Now let's consider $M=\langle B,E|S|B,E\rangle$ computed in perturbation theory. We can take momentum of order $k$ for the background. At lowest order, we have 
$$M=1+\vcenter{\hbox{\includegraphics[scale=0.35]{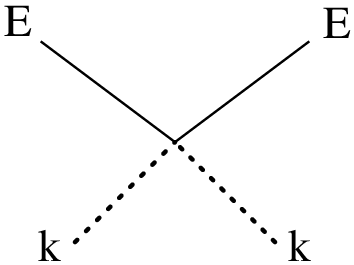}}}+\cdots=1+iA(s{=}kE)+\cdots\,.$$
Again so long as we have weak coupling, $A(s=kE)$ is small. But since $k$ is so small, the corrections from multi-particle scattering are \textit{significantly enhanced} by the $s$-channel propagator $\frac{1}{s}\sim\frac{1}{kE}$. Thus the full amplitude is then the sum over all disconnected graphs, scattering of $0,2,4,\cdots,m$ soft particles
\begin{align}
\label{}
   M=1+&\vcenter{\hbox{\includegraphics[scale=0.35]{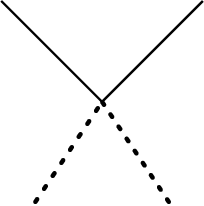}}}&+&\quad\vcenter{\hbox{\includegraphics[scale=0.35]{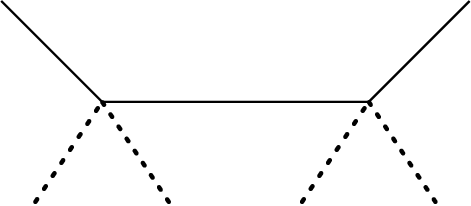}}}&+&\quad\quad\vcenter{\hbox{\includegraphics[scale=0.35]{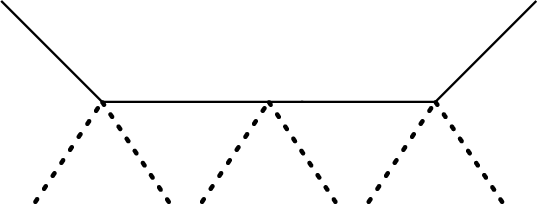}}}+\cdots   \nonumber\\
    &  iA^f(s) & +&\left[iA^f(s)\right] \frac{1}{kE} \left[iA^f(s)\right] & +&\left[iA^f(s)\right]  \frac{1}{kE} \left[iA^f(s)\right]  \frac{1}{kE} \left[iA^f(s)\right] 
\end{align}
where, since we imagine $k\sim\frac{1}{R}$ is tiny, $A^f(s)$ is the forward-limit amplitude. Note these are amplitudes with the conventional relativistic normalization of states: $M=\langle B,E|S|B,E\rangle$ is dimensionless and the units are made up for with powers of $k$. At large $m$ number of scattering, we have 
\eq
M=\sum_m\left(\frac{iA^f(s)k}{E}\right)^m\left(\begin{array}{c}N \\m\end{array}\right)=\left(1+\frac{iA^f(s)k}{E}\right)^N\,.
\eqe
Using $N\sim (\partial \phi_0)^2/k^4$ we have 
\eq
M=exp\left[i\frac{A^f(s)}{Ek}\frac{(\partial \phi_0)^2}{k^2}\right]\,, 
\eqe
and thus we can identify 
\eq
\delta(E)=\left(\frac{A(s)}{s}\right)\frac{(\partial \phi_0)^2}{k^2}\,.
\eqe
So the time delay is 
\eq
\framebox[8cm][c]{$\displaystyle \Delta t=\frac{\partial}{\partial E}\delta(E)=\frac{\partial}{\partial s}\left(\frac{A(s)}{s}\right)(\partial \phi_0)^2R$}\,.
\eqe
From here, we can reproduce the previous result for the $c(\partial \varphi)^4$ theory: there $A(s)=\frac{cs^2}{M^4}$, so $\Delta t=\frac{c}{M^4}(\partial \phi_0)^2R$. 

As another quick check, suppose we had turned on a $\lambda \varphi^4$ interaction. Then $A(s)\sim-\lambda$, and $\Delta t=\frac{\lambda}{s^2}\frac{(\varphi_0)^2}{R^2}R\sim \lambda \frac{(\varphi_0)^2}{E^2}R$. This is again as we'd expect: inside the blob the $\varphi$ particle picks up a mass $m_0^2\sim\lambda\varphi^2_0$. So if the velocity (for $E\gg m_0$) is reduced to $(1-\frac{m_0^2}{E^2})=(1-\lambda\frac{\varphi_0^2}{E^2})$, this leads to a time delay of $\Delta=\lambda\frac{\varphi^2_0}{E^2} R$. Note however that if $\lambda<0$ this does \textit{not} mean we have superluminal propagation; and indeed it is possible to have consistent theories with $\lambda<0$, with vacuum instability on exponentially long time scales $\propto exp(b/|\lambda|)$ as in the Higgs instability in the Standard Model. If $\lambda<0$, turning on $\varphi_0$ destabilizes the vacuum inside the bubble, and so the perturbative assumption of this computation is violated. Strictly speaking then, our arguments says that $\frac{\partial}{\partial s}\left(\frac{A(s)}{s}\right)>0$ so long as $A(s=0)\leq0$ (which allows of course for $A(s=0)$ as for goldstones). 

Thus from consideration of scattering off the blob, we conclude that $\frac{\partial}{\partial s}\left(\frac{A(s)}{s}\right)>0$, a stronger statement than merely the positivity of the coefficient of ($s^2$) in the low energy expansion of $A(s)$. 

We now switch gears to discuss the dispersive representation of the (forward) scattering amplitude, and show how analyticity and unitarity allow us to conclude that  $\frac{\partial}{\partial s}\left(\frac{A(s)}{s}\right)>0$ when $A(s=0)\leq0$. The non-trivial statement that makes this is possible is the Froissart bound, which we will shortly review, following from assumptions of analyticity and a reasonable polynomial boundedness of the forward amplitude. The bound  tells us that $A^f(s)<s\log^2s$ at large $s$, and so Cauchy's theorem allows us to express for a single scalar $\varphi$ (with $s{-}u$ symmetry)
\eq
A(s)=A_0{+}\int dM^2\rho(M^2)\left[\frac{1}{M^2{-}s}+\frac{1}{M^2{+}s}{-}\frac{2}{M^2}\right]
\eqe
where we've separated out the constant piece $A_0=A(s=0)$, since these are not captured by contour integration, and the expression in the brackets vanishes at $s=0$. Of course unitarity tells us that $\rho(M^2)\geq0$. Now we simply note that 
\eq
\frac{\partial}{\partial s} \frac{1}{s}\left[\frac{1}{M^2{-}s}+\frac{1}{M^2{+}s}{-}\frac{2}{M^2}\right]=\frac{2 (M^4 + s^2)}{M^2 (M^4 - s^2)^2}>0\,,
\eqe 
and thus if $A_0<\leq0$ so that $\frac{\partial}{\partial s}\frac{A_0}{s}>0$, we have that  $\frac{\partial}{\partial s}\left(\frac{A(s)}{s}\right)>0$ as desired. This shows quite vividly how unitarity and analyticity in the UV guarantee a rather non-trivial condition needed for IR causality.

We have seen that reliable causality constraints on scattering amplitudes can arise if we can find a background in which small, perturbative amplitude phase-shifts can be calculably exponentiated to large phases, that allow us to look for the presence of a time advance or delay in the scattering process. We have discussed one such background--the ``soft blob" of a scalar condensate, through which we shoot a hard probe. Another limit of this kind arises when we have gravitational long-range forces, and consider the scattering in the Eikonal limit, or equivalently, shooting a probe particle through a gravitational shock wave \cite{Camanho:2014apa}.  In the impact parameter representation, where the impact parameter $\vec{b}$ is fourier-conjugate to the momentum transfer $\vec{q}$ with $t=-\vec{q}^2$, the amplitude again exponentiates to a phase $\delta(s,\vec{b})$ at small $\vec{b}$. If further we assume the UV theory has a weak coupling and so a scale of new physics beneath the Planck scale, as in string theory, at fixed $t$, the leading weak coupling amplitude at large $s$ scales as $a(s)/t$, which maps to an Eikonal phase $\delta(s,\vec{b}) = \frac{a(s)}{s} \log b$. The center of mass energy $s=E_{probe} E_{shock}$; causality and unitarity demand that $|e^{i \delta(E_{probe})}|<1$ everywhere in the upper-half $E_{probe}$ plane, and this tells us that $\delta(E_{probe})$ itself must be bounded by $E_{probe}^1$ at large $E_{probe}$. This in turn tells us that the fixed $t$ amplitude is bounded by $s^2$ at large $s$.  This is easily seen to be satisfied for gravity amplitudes in string theory, which has a Regge behavior at fixed $t$, large $s$ given by $s^{2 + \alpha^\prime t}/t$, giving a power smaller than $s^2$ for physical $t<0$. 

It is amusing that, while the ``small-phase exponentiating backgrounds" are different in these two examples, the final practical constraint on the high-energy behavior of amplitudes is the same. The usual Froissart bound (whose derivation we will review in a moment) tells us that the amplitude at fixed $t$ can grow only logarithmically faster than $s$, while the shockwave arguments applicable for weakly coupled in the UV gravitational theories tells us that the amplitude can't grow as fast as $s^2$. In both cases, we learn that the amplitude is bounded by $s^2$ at fixed $t$. 

%%%%%%%%%%%%%%%%%%%%%%%%%%%%%%%%%%%%%%%%%%%%%%%%%%%%
\subsection{Froissart bound}
%%%%%%%%%%%%%%%%%%%%%%%%%%%%%%%%%%%%%%%%%%%%%%%%%%%%%

Let's recall first the intuition behind the Froissart  bound, going back to an argument by Heisenberg~\cite{Heisenberg:1949kqa}. Consider particles scattering at center of mass energy $E$, involving exchange of a particle with mass $m$. We can imagine the interaction strength grows as $gE^n$, but in position space we also expect the amplitude to behave as $e^{-mR}$:
$$\includegraphics[scale=0.7]{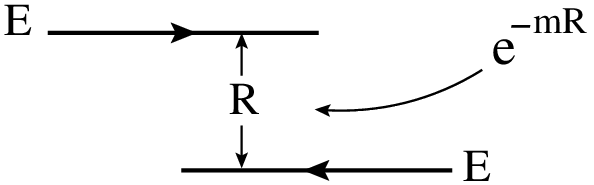}$$
Thus the relevant contributions are given by 
\eq
(g E^n)e^{-mR}\sim 1\quad \rightarrow R\leq\frac{n\log E}{m}\,,
\eqe
so the total cross section should be bounded by 
\eq
\sigma\sim R^2\leq \frac{n^2\log^2 E}{m^2}\,.
\eqe
Now since $\sigma(s)=\frac{Im[M(s,t\rightarrow 0)]}{s}$, this also tells us that 
\eq
Im[M(s,t\rightarrow 0)]\leq \frac{cs\log^2 s}{m^2}\,.
\eqe
for some constant $c$ at large $s$. Note that locality, seen in the finite range of the effective interaction was crucial to this argument. 

We'd like to see how to understand this intuitive result directly from properties of the amplitude. Very naively, one might think that an upper bound on the amplitude would come from unitarity, but this is not enough; as we've seen locality is also crucial, and thus some ``good" analytic properties of the amplitude must also be needed. To begin with, let's write the partial wave expansion of the amplitude
\eq
M(s,\cos\theta)=\sum_{\ell}(2\ell{+}1)\,a_\ell\, P_\ell(\cos\theta)\quad\rightarrow \quad M(s,t)=\sum_{\ell}(2\ell{+}1)\,a_\ell\, P_\ell(1{+}2t/s)\,.
\eqe
Unitarity tells us that $|1{+}ia_\ell|^2\leq1$ so $0\leq |a_\ell|^2\leq 2{\rm Im}\,a_\ell\leq 1$. Note the extremely naive intuition that "unitarity means $A(s)$ can't get too big" is wrong, since unitarity only tells us each $a_\ell$ individually can't get too big. Indeed if we keep all $a_\ell$'s to be $\mathcal{O}(1)$ up to some $\ell\sim\ell_{\rm max}$, we'd have that 
\eq
 M(s,0)=\sum_{\ell\leq \ell_{\rm max}}(2\ell{+}1)\,a_\ell\,\sim \ell^2_{\rm max}\,.
\eqe
Going again to the Heisenberg picture, at the distance $R_{\rm max}\sim\frac{\log E}{m}$, the angular momentum is $\ell_{\rm max}\sim E R_{\rm max} \sim \frac{E\log E}{m}$, so $M\leq \ell^2_{\rm max}\sim s \frac{\log^2 s}{m^2}$ would agree with our Froissart intuition.

So unitarity is not enough, we need an extra argument to tell us that the partial waves above $\ell_{\rm max}(E)\sim E\log E$ are shut off. Let's imagine working at fixed $t$ smaller than any of the thresholds. Importantly we assume that the amplitude at fixed $s$ is analytic in $t$: in other words, we can continue from small negative $t$ (i.e. the physical region) to small positive $t$ smoothly. We will also have at fixed but small $t$, the amplitude is polynomial bounded at large $s$, $M<s^N$. We've already seen heuristic reasons for this from causality, though those are only applicable for physical (negative) $t$. It is our assumption of analyticity in $t$ for small enough $t$ that allows us to continue the bound to positive $t$, which is crucial for the following argument. Now the Legendre polynomials $P_{\ell}(x)$ are wildly oscillating for large $\ell$ when $|x|=\cos\theta<1$, but for $x>1$ they instead are exponentially growing:
\eq
P_{\ell}\left(1{+}\frac{2t}{s}\right)\sim\frac{1}{\sqrt{\ell}}e^{2\ell\sqrt{\frac{2t}{s}}}
\eqe
for $t/s>0$. Now consider Im$M(s,t)=\sum_{\ell}(2\ell{+}1){\rm Im} a_\ell\; P_\ell(1+\frac{2t}{s})$. If we want this to be bounded by $s^N$ at large $s$,  Im$a_\ell$  have to sharply die above some $\ell_{\rm max(s)}$, estimated as
\eq
e^{2\ell_{\rm max}(s)\sqrt{\frac{2t}{s}}}<s^N\quad \rightarrow \quad \ell_{\rm max}(s)\sim N \sqrt{\frac{s}{2t}}\log s\,.
\eqe
Note this is in agreement with what we expect from the Heisenberg picture; taking $t\sim1/R^2$, we have  $ \ell_{\rm max}\sim N R E\log E$ as expected. From here, we recover the Froissart bound. 

Note we can also say slightly more, not just about the imaginary part of the amplitude, but the amplitude itself. We've already seen that Im$a_\ell\rightarrow 0$ for $\ell>\ell_{\rm max}(s)$. But since by unitarity we have $|a_\ell|^2< 2{\rm Im}\,a_\ell$, this means that Re$a_\ell\rightarrow 0$. Thus we learn that for small enough $|t|$
\eq
M(s,t)\leq s \log^{2}s 
\eqe
for large $s$. This is interesting: we began only by assuming $M(s,t)<s^N$ for \textbf{some} power $N$; but analyticity in $t$ for small $t$, and unitarity, then forces upon us the much stronger statement that  $M(s,t)<s\log^2 s$.

%%%%%%%%%%%%%%%%%%%%%%%%%%%%%%%%%%%%%%%%%%%%%
\section{Dispersive representation of loop amplitudes}\label{Lehman}
%%%%%%%%%%%%%%%%%%%%%%%%%%%%%%%%%%%%%%%%%%%%%
In this section, we will show that by integrating out massive states in loops, so long as $t \ll m^2$, the four point amplitude admits the following dispersive representation:
\eq
M(s,t)|_{t\ll m^2}=M^{\rm Sub}+\int^\infty_{M^2_s} \;dM^2\;\frac{\rho_s(M^2)}{s-M^2}+\int^\infty_{M^2_u} \;dM^2\;\frac{\rho_u(M^2)}{u-M^2}
\eqe 
where $M^{\rm Sub}$ is the subtraction terms reproducing with boundary behaviour of  $M(s,t)$ as $s\rightarrow \infty$, and $M^2_s, M^2_u$ are the leading thresholds in the $s$ and $u$- channel. In other words, near the forward limit, the analytic behaviour of the amplitude takes the form
$$\includegraphics[scale=0.5]{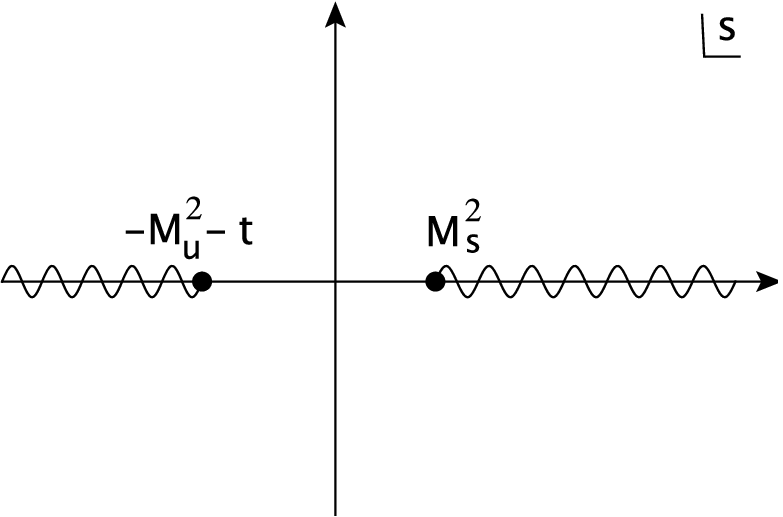}$$
Note that we can say that the loop integral can be represented as a (continuous) sum of tree-exchanges. We will see in generality, that his representation follows directly from the Schwinger parameter representation.

We will illustrate the ideas of the general proof by working through the example of the 1-loop box in $D=4$. But just as an initial warm up, we can consider the bubble in $D=2$
$$\includegraphics[scale=0.5]{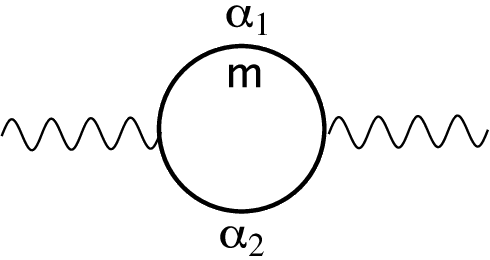}\,,$$
where the parametric representation is
\eq
I(s)=\int \frac{d\alpha_1d\alpha_2}{\rm GL(1)}\frac{1}{(-s)\alpha_1\alpha_2+m^2(\alpha_1+\alpha_2)^2}\,.
\eqe
The important point is that this is manifestly a (continuous) sum over simple poles in $s$ - that is the dispersive representation! More formally, we can write:
\eq
I(s)=\int dM^2\frac{\rho(M^2)}{-s+M^2}
\eqe
 where 
 \eqa
 \rho(M^2)&=&\int \frac{d\alpha_1d\alpha_2}{\rm GL(1)}\frac{1}{\alpha_1\alpha_2}\delta\left(M^2-\frac{m^2(\alpha_1+\alpha_2)^2}{\alpha_1\alpha_2}\right)\nonumber\\
 &=&\int \frac{d\alpha_1}{\alpha_1}\delta\left(M^2-\frac{m^2(\alpha_1+1)^2}{\alpha_1}\right)\,.
 \eqae
 Since the $\alpha_i$s are integrated over $R^+$, $min\;\frac{(1+\alpha_1)^2}{\alpha_1}=4$, and thus $\rho(M^2)=0$ when $M^2<4m^2$. For $M^2>4m^2$ the integral is localized by the delta function and one has:
 \eq
  \rho(M^2)=\frac{2}{\sqrt{M^2(M^2-4m^2)}}\Theta(M^2-4m^2)\,.
 \eqe
 This manifests the position of the branch point at $s=4m^2$.

We now turn to the $D=4$
$$\includegraphics[scale=0.5]{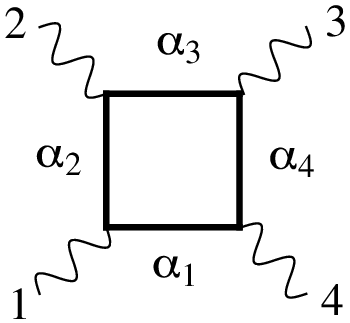}\,.$$
We will see that 
\eq\label{BoxDisperse}
I(s,t)=\int_{4m^2}^\infty dM^2\;\frac{\rho(M^2,t)}{-s+M^2}
\eqe
where $\rho(M^2,t)$ is analytic in $t$ around $t=0$, with a cut at large positive $t\sim m^2$, but finite for $t<0$. Note that importantly the starting point of the integral is at $4m^2$ which is independent of $t$. If this had then say $4m^2-t$, then we would not have an analytic expression in $t$. Now let's look at the the box integral in Schwinger parameter space:
\eq\label{BoxScwinger}
I(s,t)=\int \frac{d\alpha_1\cdots d\alpha_4}{\rm GL(1)}\frac{1}{(\underbrace{ (-s)\alpha_1\alpha_3+(-t)\alpha_2\alpha_4+m^2(\alpha_1+\alpha_2+\alpha_3+\alpha_4)^2}_{\Delta})^2}
\eqe
We begin in the Euclidean regime where $-s, -t >0$, the denominator $\Delta$ is positive and the integral is perfectly analytic.  In fact, even if $(-s)$ and $(-t)$ are negative, as long as they are small with respect to $m^2$ we are fine, since $\Delta$ can be rewritten as
\eq
\Delta=(4m^2-s)\alpha_1\alpha_3+(4m^2-t)\alpha_2\alpha_4+m^2\left((\alpha_1-\alpha_3)^2+(\alpha_2-\alpha_4)^2+2(\alpha_1+\alpha_3)(\alpha_2+\alpha_4)\right)\,.
\eqe
Now let's keep $t$ fixed and small but increase $s$. Clearly $\Delta>0$ for any $s<4m^2$. But note that for \textbf{any} positive $\epsilon$, we can make $\Delta<0$ at $s=4m^2+\epsilon$. Naively one might worry about $(-t)$ being positive, but simply by considering the limit $(\alpha_1,\alpha_3)\rightarrow \infty$ while $(\alpha_2,\alpha_4)$ held fixed, we can make $\Delta<0$ for any value of positive  $\epsilon$. So, we see that we hit a branch point singularity at $s=4m^2$ independent to the value of $t$. 

Now let's first get the dispersive representation starting in the forward limit $t\rightarrow0$. Fixing the GL(1) symmetry by setting $\alpha_1=1$, we have
\eqa
I(s,t{=}0)&=&\int d\alpha_2d\alpha_3 d\alpha_4\frac{1}{(\underbrace{ (-s)\alpha_3+m^2(1+\alpha_2+\alpha_3+\alpha_4)^2}_{\Delta})^2}\nonumber\\
&=&\int dM^2\frac{\tilde{\rho}(M^2)}{(M^2-s)^2}
\eqae 
where 
\eq
\tilde{\rho}(M^2)=\int d\alpha_2d\alpha_3 d\alpha_4\frac{1}{\alpha^2_3}\delta\left(M^2-\frac{m^2(1+\alpha_2+\alpha_3+\alpha_4)^2}{\alpha_3}\right).
\eqe
Note that since the minimum of $(1+\alpha_2+\alpha_3+\alpha_4)^2/\alpha_3$ is at $4$, $\tilde{\rho}(M^2)$ will vanish when $M<4m^2$ so we have 
$I(s,t{=}0)=\int_{4m^2}^\infty dM^2\frac{\tilde{\rho}(M^2)}{(M^2-s)^2}$. Integrating by parts, we have 
\eq\label{BoxForward}
I(s,t{=}0)=-\int_{4m^2}^\infty dM^2\frac{\partial}{\partial M^2}\frac{\tilde{\rho}(M^2)}{(M^2-s)}+\int_{4m^2}^\infty dM^2\frac{1}{(M^2-s)}\frac{\partial}{\partial M^2}\tilde{\rho}(M^2)
\eqe
The boundary term at $M^2= \infty$ vanishes. Importantly, for $M^2=4m^2$, $\tilde{\rho}(4m^2)$ itself also vanishes. This can be explicitly confirmed, but it must be: if $\tilde{\rho}(M^2\rightarrow 4m^2)=const.$, then the integral near 
\eq
\int_{4m^2} dM^2\frac{1}{(M^2-s)^2}\sim \frac{1}{4m^2-s}
\eqe 
gives a pole in $s=4m^2$, while we can see easily that one can at most get a branch cut there. Let us explicitly compute $\tilde{\rho}(M^2)$:
\eq
\tilde{\rho}(M^2)=\int d\alpha_2d\alpha_3 d\alpha_4\;\frac{1}{m^2\alpha_3}\delta\left((\alpha_3-\alpha^+)(\alpha_3-\alpha^-)\right)
\eqe
where $\alpha^{\pm}={-}(1{+}\alpha_2{+}\alpha_4){+}\frac{x}{2}(1\pm\sqrt{1{-}\frac{4}{x}(1{+}\alpha_2{+}\alpha_4)})$ and $x=\frac{M^2}{m^2}$. We use the delta functions to localize $\alpha_3$, while the integration over $\alpha_2$ and $\alpha_4$ is bounded by $1+\alpha_2+\alpha_4\leq \frac{x}{4}$ to ensure that $\alpha^{\pm}$ stays real. In the end we find: 
\eq
\tilde{\rho}(M^2)=\frac{\log(1+\sqrt{1-\frac{4}{x}} )-\log(1-\sqrt{1-\frac{4}{x}})-2\sqrt{1-\frac{4}{x}}}{m^2}\,,
\eqe
which indeed vanishes when $M^2=4m^2$. Substituting the result back into eq.(\ref{BoxForward}), we find
\eq
I(s,t{=}0)=\int_{4m^2}^\infty dM^2\frac{\rho(M^2)}{(M^2-s)}\quad \rho(M^2)=\frac{\sqrt{1-\frac{4}{x}}}{M^2m^2}\,.
\eqe

We can proceed in the same way to compute the $t$-expansion. We simply Taylor expand eq.(\ref{BoxScwinger}) where we have:
\eq
\sum_{q}\;(-t)^q \int dM^2\; \frac{\tilde{\rho}^{(q)}(M^2)}{(M^2-s)^{q+2}},\quad \tilde{\rho}^{(q)}(M^2){=}\int d\alpha_2d\alpha_3 d\alpha_4\frac{(\alpha_2\alpha_4)^q}{\alpha^{2+q}_3}\delta\left(M^2{-}\frac{m^2(1{+}\alpha_2{+}\alpha_3{+}\alpha_4)^2}{\alpha_3}\right)\,.
\eqe
Again the $\alpha_3$ integral localizes and we are restricted to $\alpha_2+\alpha_4<\frac{M^2}{4m^2}-1$. Note that this shows that due to the $(\alpha_2\alpha_4)^q$ factor, $ \tilde{\rho}^{(q)}(M^2)$ and all $q$ of its derivatives with respect to $M^2$ vanishes at $M^2\rightarrow 4m^2$. Thus we can write the coefficient of $(-t)^q$ as 
\eq
\int_{4m^2}^{\infty} dM^2\; \frac{\rho^{(q)}(M^2)}{(M^2-s)},\quad \rho^{(q)}(M^2)=\frac{\partial^{q+1}}{\partial (M^2)^{q+1}} \tilde{\rho}^{(q)}(M^2)
\eqe
This leads to the dispersive representation for the box integral around $t=0$:
\eq\label{BoxDisperse}
I(s,t)=\int_{4m^2}^\infty dM^2\frac{\sum_q(-t)^q\rho^{(q)}(M^2)}{(M^2-s)}\,.
\eqe
As an example we can explicitly compute $\rho^{(1)}(M^2)$. Starting with:
\eq
\tilde{\rho}^{(1)}(M^2)=\frac{3(M^2{+}6m^2)\left(\log(1{+}\sqrt{1{-}\frac{4}{x}} ){-}\log(1{-}\sqrt{1{-}\frac{4}{x}})\right){-}(11M^2{+}16m^2)\sqrt{1{-}\frac{4}{x}}}{18m^4}\,,
\eqe
we find:
\eq
\rho^{(1)}(M^2)=\frac{\partial^2 }{\partial (M^2)^2} \tilde{\rho}^{(1)}(M^2)=\frac{(1{-}\frac{4}{x})\sqrt{1{-}\frac{4}{x}}}{6 M^2 m^4}\,.
\eqe
Finally, we note that due to the increase in $M^2$ derivatives, $\rho^{(q)}(M^2)$ are increasingly suppressed for larger $q$ as $M^2\rightarrow \infty$. We will come back to this point when we study the partial wave expansion of the numerator in eq.(\ref{BoxDisperse}).

Having seen all the relevant ideas in the 1-loop examples, let's now consider the general story. Consider any integral associated with a graph $G$, as far as the analytic structure is concerned we can just take scalar graphs with numerator $=1$. The integral in general takes the form 
\eq
I=\Gamma\left(E{-}LD/2\right) \int \frac{d^E\alpha}{GL(1)}\frac{1}{\mathcal{U}^{D/2}}\left(\frac{\mathcal{U}}{\mathcal{F}}\right)^{E{-}LD/2}\,,
\eqe
where $\mathcal{U}$, $\mathcal{F}$ are the Symanzik polynomials  given as 
\eqa
\mathcal{U}=\sum_{\substack{T\,\in \rm spanning \\ {\rm tree} }}\;\left(\prod_{i \notin T}\; \alpha_i\right),\quad \mathcal{F}=\mathcal{F}^0+(\sum_i m_i^2\alpha_i)\mathcal{U}\nonumber\\
\mathcal{F}^0=\sum_{\substack{T_2\,\in\rm spanning \\ {\rm 2-tree} }} \;\left(\prod_{i \notin T_2}\; \alpha_i\right)(\sum_{j\in L} p_j)^2
\eqae
In particular all the dependence on the external Mandelstams is in the $\mathcal{F}$-polynomial. Specializing to four-points, we have that 
\eq
\mathcal{F}=(-s)\mathcal{F}_s^0{+}(-t)\mathcal{F}_t^0{+}(-u)\mathcal{F}_u^0+(\sum_i m_i^2\alpha_i)\mathcal{U}
\eqe
Note that every $2$-tree that contributes to $\mathcal{F}^0$ must appear in $\mathcal{U}$, so every monomial in $\mathcal{F}^0$ also occurs in $(\sum_i m_i^2\alpha_i)\mathcal{U}$; this makes it manifest that $\mathcal{F}>0$, so long as $(-s)$, $(-t)$, $(-u)$ are small enough.

Now we'd like to show that, at fixed $t$, we have some branch point singularity at $s\rightarrow M_s^2$ (independent of $t$), and $u\rightarrow M_u^2$ (again independent of $t$). Of course at general loops, there can be many ``thresholds", but one of them will occur at smallest $s$; for example 
$$\includegraphics[scale=0.5]{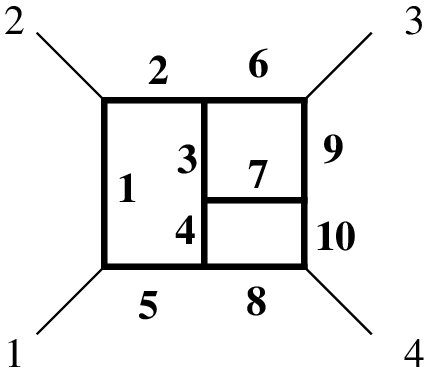}$$
we can have thresholds at $s=(m_2+m_5)^2$ or $(m_6+m_7+m_8)^2$. We can systematically identify these as follows. Pick any monomial \textit{m}$^{(s)}$ in $\mathcal{F}_s^0$, since these monomials do not appear in $\mathcal{F}_t^0$ or $\mathcal{F}_u^0$, they will dominate if we scale those $\alpha$s $\rightarrow \infty$. So for each monomial we will have some threshold $M^2_{\textit{m}^{(s)}}$. The minimum of those over all monomials \textit{m}$^{(s)}$ is some \textit{m}$^{*(s)}$, and the branch point is at $M^2_s\equiv M^2_{\textit{m}^{*(s)}}$. Similarly for $M^2_u$. Furthermore, for any $\epsilon>0$, by scaling all of the $\alpha$s in \textit{m}$^{*(s)}$ to infinity, we see that we can always make $\mathcal{F}<0$ for $s=M_s^2+\epsilon$, so the branch point sits at $s=M_s^2$ independent of $t$, and similarly for $M^2_u$.

Now in general the four-point loop integral takes the form 
\eqa
I(s,t)&=& \int \frac{d^E\alpha}{GL(1)}\frac{1}{\mathcal{U}^{a}}\frac{1}{\left((-s)\mathcal{F}_s^0{+}(-t)\mathcal{F}_t^0{+}(-u)\mathcal{F}_u^0+(\sum_i m_i^2\alpha_i)\mathcal{U}\right)^{b}}\nonumber\\
&=&\int \frac{d^E\alpha}{GL(1)}\frac{1}{\mathcal{U}^{a}}\frac{1}{\left[(-s)(\mathcal{F}_s^0{-}\mathcal{F}_u^0){+}(-t)(\mathcal{F}_t^0{-}\mathcal{F}_u^0){+}(\sum_i m_i^2\alpha_i)\mathcal{U}\right)^{b}}\nonumber\\
&=& \int dM^2 \frac{\sum_q (-t)^q \tilde{\rho}^{(q)}(M^2)}{(M^2-s)^b}
\eqae 
where 
\eq
\tilde{\rho}^{(q)}(M^2)=\int \frac{d^E\alpha}{GL(1)}\frac{1}{\mathcal{U}^{a}}\frac{(\mathcal{F}_t^0{-}\mathcal{F}_u^0)^q}{(\mathcal{F}_s^0{-}\mathcal{F}_u^0)^{b+q}}\delta\left(M^2-\frac{(\sum_i m_i^2\alpha_i)\mathcal{U}}{(\mathcal{F}_s^0{-}\mathcal{F}_u^0)}\right)\,.
\eqe
Now, the point is again that the $\delta$ function constraint forces either that for $M^2>0$, $M^2>M^2_s$, or for $M^2<0$, that $M^2<-M_u^2-t$, so that we can write 
\eqa
I(s,t)&=& \int^\infty_{M_s^2} dM^2 \frac{\sum_q (-t)^q \tilde{\rho}_s^{(q)}(M^2)}{(M^2-s)^b}+ \int^\infty_{M_u^2} dM^2 \frac{\sum_q (-t)^q \tilde{\rho}_u^{(q)}(M^2)}{(M^2-u)^b}\,.
\eqae
By the same integration by parts idea, we arrive at our final form:
\eq
I(s,t)= \int^\infty_{M_s^2} dM^2 \frac{\rho_s(M^2,t)}{(M^2-s)}+ \int^\infty_{M_u^2} dM^2 \frac{\rho_u(M^2,t)}{(M^2-u)}\,.
\eqe
%%%%%%%%%%%%%%%%%%%%%%%%%%%%%%%
\section{Partial wave expansion of unitarity cuts}\label{ApennA}
%%%%%%%%%%%%%%%%%%%%%%%%%%%%%%%

As stressed in the main text, near the forward limit the singularities of the four-point amplitude are associated with threshold productions. We would like to demonstrate that contributions from these singularities, which are the imaginary part of the amplitude on the real $s$-axes, is given by a positive expansion on the Gegenbauer polynomial.  We begin by considering scalar scattering in the C.O.M frame, with the spatial momenta of the incoming and out going particles given by $\hat{p}_{in}=p_1{-}p_2$ and $\hat{p}_{out}=p_3-p_4$ respectively, which span a $D{-}1$-dimensional space. As the singularites are associated with threshold production, in the C.O.M frame these are all single or multi-particle states forming  irreducible representations under SO($D{-}1$). To this end, let us first build up general irreps of SO($n{+}1$), latter identifying $n=D{-}2$.

For a system with rotational SO($n{+}1$) symmetry, it is useful to consider operators as matrix elements on the Hilbert space of states that form irreducible representations of SO($n{+}1$). To this end, we introduce $n{+}1$-dimensional unit vectors $x$, i.e. points on an $n$-sphere. The states in the Hilbert space will be functions of these vectors, in particular we have states $|x\rangle$ equipped with the inner product $\langle x|y\rangle=\delta(x,y)$. To integrate these functions, we introduce the SO($n{+}1$) invariant measure $\langle x d^n x\rangle\equiv\frac{1}{\Omega_{n}} \varepsilon(x dx\cdots dx)$, where it is normalized with the solid angle $\Omega_{n}$.

Now we will like to construct states that transforms as irreps under SO($n{+}1$), i.e. they transform linearly. To draw an analogy, consider the state labeled by coordinate $X$, $|X\rangle$. Under translations $T_a$, it transforms non-linearly, $T_a|X\rangle=|X+a\rangle$. For \textit{linear} representations, we know we can define the Fourier transformed state $|k\rangle$ which transforms under translation as:
\eq
|k\rangle =\int dX e^{ikX}|X\rangle \rightarrow  T_a|k\rangle=e^{-ika}|k\rangle\,.
\eqe
We would like a similar representation for SO(n{+}1). Now clearly the state 
\eq
|\;\rangle = \int \langle xd^nx\rangle\; |x\rangle
\eqe 
is invariant as $|x\rangle\rightarrow|Rx\rangle$, where $R$ is a SO($n{+}1$) rotation, while 
\eq
|i\rangle = \int \langle xd^nx\rangle\; x^i|x\rangle
\eqe  
transforms as a vector. For $| i j\rangle$ we cannot simply use $\int \langle xd^nx\rangle\; x^ix^j|x\rangle$ since it is not reducible and contains a trace piece. This tells us that we should use $| i j\rangle=\int \langle xd^nx\rangle\;\left( x^ix^j-\frac{\delta^{ij}}{n+1}\right)|x\rangle$. Going onward it is clear that that this is the same task we've encountered previously in deriving the Gegenbauer polynomial from tree-exchanges. Borrowing from that experience, we see that the irreducible states can be simply generated by expanding:
\eq
\int \frac{\langle x d^nx\rangle}{|x-y|^{n-1}}|x\rangle=\sum_{\ell} y^{i_1}\cdots y^{i_\ell}|i_1\cdots i_\ell\rangle\,.
\eqe
The states $|i_1\cdots i_\ell\rangle$ are now irreps: symmetric traceless tensors of SO(n{+}1). Note that the Gegenbauer polynomials in this language is simply 
\eq
G^{\frac{n-1}{2}}_\ell (\cos\theta)=\mathcal{A}_{n,\ell}y_{i_1}\cdots y_{i_\ell}\langle x|i_1\cdots i_\ell\rangle,\quad \cos\theta=y\cdot x\,
\eqe
where $\mathcal{A}_{n,\ell}:=2^\ell\frac{\Gamma(\ell+\frac{n-1}{2})}{\Gamma(\frac{n-1}{2})\ell!}$. The orthogonality property of Gegenbauer polynomials is then simply:
\eqa\label{InnerProduct}
\nonumber
\int \langle zd^nz\rangle \frac{G^{\frac{n-1}{2}}_\ell(y\cdot z)}{\mathcal{A}_{n,\ell}}\frac{G^{\frac{n-1}{2}}_{\ell'} (w \cdot z)}{\mathcal{A}_{n,\ell'}}
& = &\int \langle zd^nz\rangle\;y_{i_1}\cdots y_{i_\ell}\langle i_1\cdots i_\ell |z\rangle\langle z| j_1\cdots j_{\ell'}\rangle w^{j_1}\cdots w^{j_{\ell'}}\\
%\nonumber
%& = & y_{i_1}\cdots y_{i_\ell}\left(\int \langle zd^nz\rangle w^{j_1}\cdots w^{j_{\ell'}} \langle j_1\cdots j_{\ell'} |z\rangle\langle z|\right)  |i_1\cdots i_\ell\rangle\\
%\nonumber
%& = & \delta_{\ell,\ell'}y_{i_1}\cdots y_{i_\ell}\left(\mathcal{B}_{n,\ell}\langle w |\right)|i_1\cdots i_\ell\rangle\\
&=&\mathcal{B}_{n,\ell}\delta_{\ell,\ell'}\frac{G^{\frac{n-1}{2}}_\ell(y\cdot w)}{\mathcal{A}_{n,\ell}}\,,
\eqae
where we've used that the states $| i_1\cdots i_{\ell}\rangle$ and $| j_1\cdots j_{\ell'}\rangle$ are orthogonal to each other if $\ell\neq \ell'$ since they have different quantum numbers, and here $\mathcal{B}_{n,\ell}=\frac{2^{-\ell}\Gamma(n+\ell-1)\Gamma\left(\frac{n+1}{2}\right)}{\Gamma(n-1)\Gamma\left(\ell+\frac{n+1}{2}\right)}$. If we let $y=w$ and replace $\langle zd^nz\rangle$ by 
\eq
\frac{\Omega_{n-1}}{\Omega_n} \sin^{n-2}\theta d\cos\theta \,,
\eqe
we get the usual normalization factor for Gegenbauer polynomials:
\eq
\int G_\ell(\cos\theta)G_{\ell'}(\cos\theta)\sin^{n-2}\theta d\cos\theta=\mathcal{N}_{n,\ell}\delta_{\ell,\ell'}
\eqe
with
\eq
\mathcal{N}_{n,\ell}=\frac{\Omega_n}{\Omega_{n-1}}\mathcal{A}_{n,\ell}\mathcal{B}_{n,\ell}=\frac{\pi 2^{2-n}\Gamma[\ell{+}n{-}1]}{\ell!\left(\ell{+}\frac{n{-}1}{2}\right)\Gamma^2\left[\frac{n{-}1}{2}\right]}.
\eqe
The orthogonality relation also implies that:
\eqa
\langle x|i_1\cdots i_\ell\rangle\langle i_1\cdots i_\ell|y\rangle&=&\mathcal{B}^{-1}_{n,\ell}\int \langle zd^nz\rangle \left( \langle x|i_1\cdots i_\ell\rangle z^{i_1}\cdots z^{i_\ell}\right)\left(z^{j_1}\cdots z^{j_\ell}\langle j_1\cdots j_\ell|y\rangle\right)\nonumber\\
&=&\mathcal{B}^{-1}_{n,\ell}\int \langle zd^nz\rangle x^{i_1}\cdots x^{i_\ell} \langle i_1\cdots i_\ell|z\rangle \langle z|j_1\cdots j_\ell\rangle y^{j_1}\cdots y^{j_\ell}=\mathcal{A}^{-1}_{n,\ell}G^{\frac{n-1}{2}}_\ell(x\cdot y)\,,\nonumber\\
\eqae
where the first equality holds since the SO($n{+}1$) invariant integration of $ z^{i_1}\cdots z^{j_\ell}$ yields a polynomial of products of Kronecker deltas, and when acting on the irreps, only $i,j$ contractions yield contributions as any trace pieces vanish.

Finally, these irreducible states also provides a basis for operators. A general operator can be expanded as:
\eq
\mathcal{O}=\sum \mathcal{O}^{i_1\cdots i_\ell;j_1\cdots j_{\ell'}}|i_1\cdots i_\ell\rangle\langle j_1\cdots j_{\ell'}|\,.
\eqe
However, for SO(n{+}1) invariant ones, the operator $\mathcal{O}^{i_1\cdots i_\ell;j_1\cdots j_{\ell'}}$ can only be comprised of Kronecker deltas and since $\delta_{i_a i_b}$ contracted with the states $| i_1\cdots i_{\ell}\rangle$ vanishes, it can only be polynomials of $\delta_{i_aj_b}$. This tells us that $\ell=\ell'$, i.e. it is diagonal in spin space. In the last equality we've used eq.(\ref{InnerProduct}).  
Thus we conclude that SO(n{+}1) invariant operators can be written as 
\eq
\langle y | \mathcal{O}^{Inv}|x\rangle=\sum_\ell \mathcal{N}_{n,\ell}\textsf{p}_\ell G^{\frac{n-1}{2}}_\ell(x\cdot y)\,,
\eqe
i.e. it is expandable on the Gegenbauer polynomials.

\begin{figure}
\begin{center}
\includegraphics[scale=0.5]{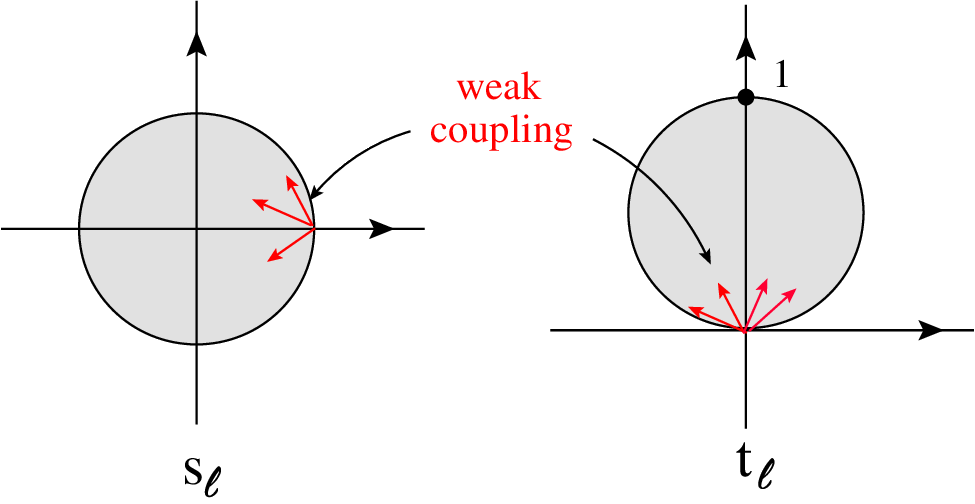}
\caption{The region allowed for $\mathbf{s}_\ell$ and $\mathbf{t}_\ell$ by unitarity. At weak coupling this constraint is only reflected in Re$[\mathbf{s}_\ell]\leq1$ and Im$[\mathbf{t}_\ell]\geq0$.}
\label{figUnitCircle}
\end{center}
\end{figure}

Now let's consider $\mathbb{S}$, the $s$-matrix of the full theory. Restricting ourselves to the $2\rightarrow 2$ elastic scattering, we can define the ``little" matrix $\mathbf{s}$
\eq
\langle \hat{p}_{out}|\mathbf{s}|\hat{p}_{in}\rangle=_{\footnotesize{out}}\langle p_3,p_4|\mathbb{S}|p_1,p_2\rangle_{in}\,.
\eqe
In other words $\mathbf{s}$ is only defined only on the $2\rightarrow 2$ states. The full $s$-matrix satisfy $\mathbb{S}^\dagger \mathbb{S}=\mathbb{I}$, while the small $s$-matrix satisfy 
\eq
\mathbf{s}^\dagger \mathbf{s}\leq\mathbb{I}\,,
\eqe
as an operator statement, i.e. for any state $|\psi\rangle$, we have $\langle \psi |\mathbf{s}^\dagger \mathbf{s}|\psi\rangle\leq\langle \psi |\psi\rangle$. Now since $\mathbf{s}$ is rotationally invariant, we can write 
\eq
\mathbf{s}=\sum_{\ell}\;\; \mathbf{s}_{\ell}\,|i_1i_2\cdots i_\ell\rangle\langle  i_1i_2\cdots i_\ell|\,,
\eqe
then $\mathbf{s}^\dagger \mathbf{s}\leq 1$ implies $|\mathbf{s}_{\ell}|\leq1$. If we write $\mathbf{s}=1+i \mathbf{t}$, then this implies $|1+i\mathbf{t}_{\ell}|\leq1$. Note that  
\eq
\mathbf{t}=\sum_{\ell}\;\; \mathbf{t}_{\ell}\,|i_1i_2\cdots i_\ell\rangle\langle  i_1i_2\cdots i_\ell|\;\rightarrow \langle \hat{p}_{out}|\mathbf{t}|\hat{p}_{in}\rangle=\mathcal{N}_{n,\ell}\sum_{\ell} \; \mathbf{t}_{\ell} G^{\frac{n-1}{2}}_\ell(\hat{p}_{out}\cdot \hat{p}_{in})\
\eqe
where $\langle \hat{p}_{out}|\mathbf{t}|\hat{p}_{in}\rangle$ is the four-point amplitude of interest. Since $|1+i\mathbf{t}_{\ell}|\leq1$, 
\eq\label{Imag}
1+i(\mathbf{t}_{\ell}-\mathbf{t}^*_{\ell})+|\mathbf{t}_{\ell}|^2\leq 1\;\rightarrow\;i(\mathbf{t}^*_{\ell}-\mathbf{t}_{\ell})\geq |\mathbf{t}_{\ell}|^2\,.
\eqe
More explicitly we have $1+i\mathbf{t}_{\ell}=\eta_\ell e^{i\delta_\ell}$ with $\eta_\ell\leq1$. Note that in a weakly coupled theory, eq.(\ref{Imag}) just tells us that $i(\mathbf{t}_{\ell}-\mathbf{t}^*_{\ell})\geq0$, i.e. the imaginary part is positive.  The full non-linear constraint is only present at strong coupling see fig.\ref{figUnitCircle}. Since the imaginary part is positive, we have
\eq
\mathrm{Im}[\langle \hat{p}_{out}|\mathbf{t}|\hat{p}_{out}\rangle]=\mathcal{N}_{n,\ell}\sum_{\ell} \; \mathrm{Im}[\mathbf{t}_{\ell}] G^{\frac{n-1}{2}}_\ell(\hat{p}_{out}\cdot \hat{p}_{in})
\eqe 
i.e. the imaginary part of the amplitude is positively expandable on the Gegenbauer polynomials.

%%%%%%%%%%%%%%%%%%%%%%%%%%%%%%%
\section{The spinning-spectral function for massive box}\label{SpinningBox}
%%%%%%%%%%%%%%%%%%%%%%%%%%%%%%%
From appendix \ref{Lehman} we've seen that near the forward limit, the four-point amplitude admits a  K\"all\'en-Lehman representation representation, where the ``spectral function" depends on $t$, i.e. $\rho(M^2,t)$. Since the spectral function is a polynomial in $t$ near the forward limit, it has a partial wave expansion. Now from appendix \ref{ApennA}, we've seen that the discontinuity for $A,B\rightarrow A,B$ type scattering should be positively expandable on the Gegenbauer polynomials. Since the discontinuity in the dispersive representation is the spectral functions, we conclude that the ``spinning spectral function"  should be a positive function. Here we will use the massive box to demonstrate this fact.

Let us consider an explicit example, the discontinuity for the box-integral with massive internal propagators in four-dimensions. The integrand in the phase space integral is simply given by the product of two tree-propagators:
\eq
\vcenter{\hbox{\includegraphics[scale=0.5]{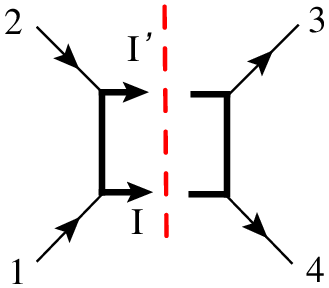}}}=\frac{1}{2(p_1\cdot p_I)}\frac{1}{2(p_4\cdot p_I)}=\frac{4}{s^2}\frac{1}{1-\sqrt{1-\frac{4m^2}{s}}\hat{p}_1\cdot \hat{p_I}}\frac{1}{1-\sqrt{1-\frac{4m^2}{s}}\hat{p}_4\cdot \hat{p_I}}\,,
\eqe
where we are again considering the kinematics in center of mass frame. The discontinuity is now given as:
\eq\label{PhaseSpace}
 \langle \hat{p}_{in} |T^\dagger T| \hat{p}_{out} \rangle =\int^\infty_{4m^2} ds \frac{4J_s}{s^2}\;\int \langle \hat{p}_{I}d^2\hat{p}_{I}\rangle F^*(\hat{p}_1\cdot \hat{p}_I)F(\hat{p}_4\cdot \hat{p}_I)
\eqe
where $F(x)=\left(1-\sqrt{1-\frac{4m^2}{s}}x\right)^{-1}$, and 
$J_s$ is the dimensionless Jacobian factor stemming from the phase space integral: 
\eq
\int d^D\ell\delta\left(\ell^2-m^2\right)\delta\left((\ell-p_{12})^2-m^2\right)=\frac{(s-4m^2)^{\frac{D-3}{2}}}{\sqrt{s}}\int d\Omega_{D{-}2}\,,
\eqe
which for $D=4$ is simply $J_s=\sqrt{1{-}\frac{4m^2}{s}}$\,.

Let us write $F(x)$ as an expansion on the Gegenbauer polynomial with coefficient $f_{\ell}$, $F(x)=\sum_\ell f_\ell(s) G^{\frac{1}{2}}_\ell(x)$. Then the two-dimensional angular integral simply reduces the corresponding product of $G^{\frac{1}{2}}_\ell(x)$s in eq.(\ref{PhaseSpace}) to $\sum_\ell |f_\ell(s)|^2 \frac{2}{2\ell+1}G^{\frac{1}{2}}_\ell(\hat{p}_1\cdot\hat{p}_4 )$, where $\theta$ is precisely the scattering angle. Thus we conclude that the discontinuity is simply 
\eq\label{Check1}
\langle \hat{p}_{in} |T^\dagger T| \hat{p}_{out} \rangle =\int_{4m^2}^\infty ds \frac{4J_s }{s^2}\;\sum_\ell \textsf{p}_\ell(s)\frac{2}{2\ell+1} G^{\frac{1}{2}}_\ell(\cos\theta ),\quad
\eqe
where $\textsf{p}_\ell(s)\equiv|f_\ell(s)|^2$ is the positive definite ``spinning" spectral function. Let us compute the $f_\ell(s)$s explicitly.

Using the generating function and the orthogonality of the Gegenbauer polynomials, we can write down the following generating function for $f_\ell(s)$,
\eq\label{Check2}
\int_{-1}^1 dx\;\frac{1}{(1-ax)} \frac{1}{(1-2rx+r^2)^{\frac{1}{2}}}=\sum_{\ell} r^\ell  \frac{2}{2\ell+1}f_{\ell}(s)\,,
\eqe
where $a=\sqrt{1-\frac{4m^2}{s}}$. A straight forward integration yields for the LHS:
\eq\label{Check3}
\frac{1}{ab}\log\left[\frac{(1-r+b)}{(1-r-b)}\frac{(1+r-b)}{(1+r+b)}\right],\quad\quad b=\sqrt{1+r^2-2\frac{r}{a}}
\eqe
As the generating function is non-polynomial in $r$, we have an infinite tower of spin in the expansion. The coefficient for the first few spins are:
\eq\label{spinf}
f_0=\frac{1}{2}\frac{\log \delta}{a},\quad f_1=\frac{3}{2}\frac{(-2 a + \log \delta)}{a^2},\quad f_2=\frac{5}{2}\frac{(-6 a + 3 \log \delta- a^2\log \delta)}{2a^3}\,,
\eqe
where $\delta=\frac{1+a}{1-a}$. Since $a$ takes value between 0 and 1, one can straightforwardly see that the coefficient decreases for increasing spin.

\begin{figure}
\begin{center}
$f_{\ell}(s)\quad\vcenter{\hbox{\includegraphics[scale=0.4]{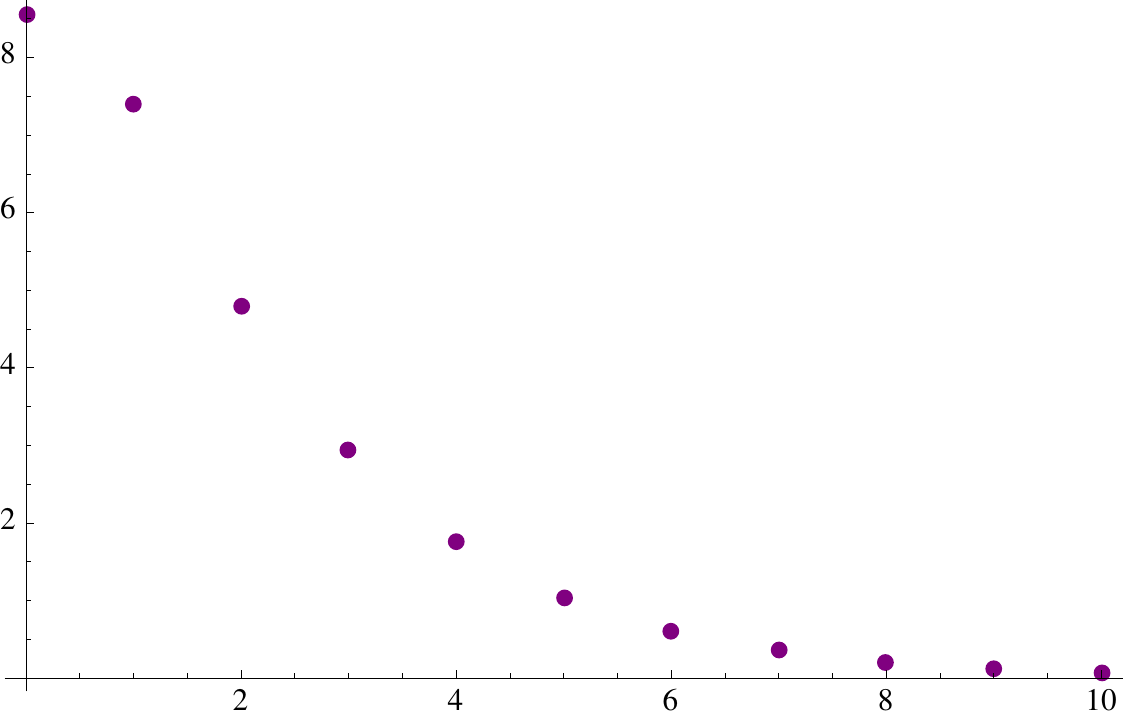}}}_\ell$
\caption{We plot the coefficients $f_{\ell}(s)$ for $s=14$. We see that the coefficients are suppressed for higher spins }
\label{Coeff}
\end{center}
\end{figure}

\begin{figure}
\begin{center}
\includegraphics[scale=0.4]{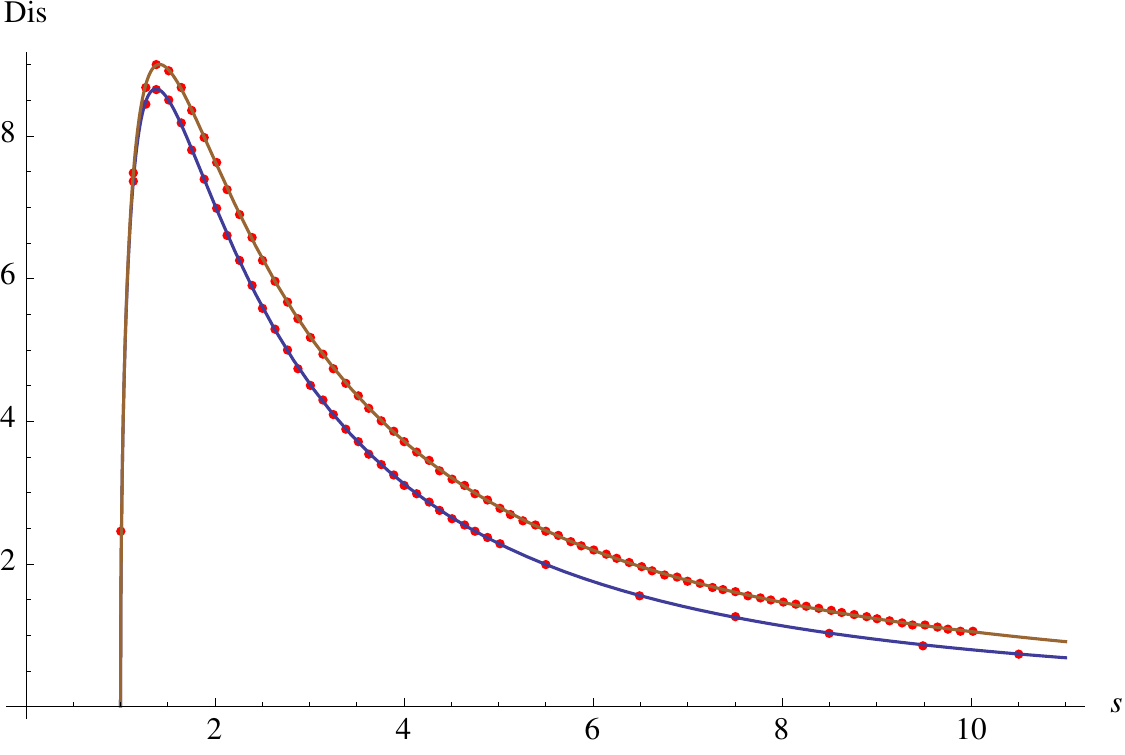}
\caption{We compare our Gegenbauer sum expression in eq.(\ref{Check1}), truncating at $\ell=10$, with the explicit discontinuity in eq.(\ref{DisMatch}). We've normalized $\frac{s}{4m^2}\rightarrow s$, so that the discontinuity begins at $s=1$ to $\infty$. We've compared the result of eq.(\ref{Check1}), in red dots, to eq.(\ref{DisconT}) which is the colored curve. The brown curve is for $\cos\theta=\frac{1}{2}$, and the blue curve for $\cos\theta=1/6$. Both exhibit perfect matching. }
\label{DisMatch}
\end{center}
\end{figure}

Let us verify that eq.(\ref{Check1}), combined with (\ref{Check2}) and (\ref{Check3}), indeed  reproduces the correct discontinuity of eq.(\ref{BoxResult})
\eq\label{DisconT}
I_4[s,t]-\left.I_4[s,t]\right|_{\beta_{u}\rightarrow -\beta_{u}}\,.
\eqe
To compare, we first note that the coefficients $f_{\ell}(s)$ is suppressed for higher spin, see. fig \ref{Coeff}. Thus we should find a good approximation by truncating at $\ell=10$. Indeed summing eq.(\ref{Check1}) up to spin-$10$ the result matches with that of eq.(\ref{DisconT}) as shown in fig.(\ref{DisMatch}), thus confirming eq.(\ref{Check1}).

%%%%%%%%%%%%%%%%%%%%%%%%%%%%%%%
\section{Positivities of the Gegenbauer matrix}\label{PositiveProof}
%%%%%%%%%%%%%%%%%%%%%%%%%%%%%%%
The results on the total positivity of Gegenbauer polynomials follow from general theorems connecting total positivity to orthogonal polynomials with positive measure discovered in the 1960s \cite{GeneralProof}. Here, we will give elementary and explicit computations that show the positivity properties explicitly for the Gegnebauer polynomial case of immediate interest to us. For the simplest case of $d=2$, where we just have Fourier expansion in cos$(\theta)$, we will give an especially simple argument for positivity going back essentially to Chebyshev. We will then give a simple explicit computation of the determinants associated with the Taylor expansion of Gegenbauer polynomials, where they can explicitly be seen to be positive

%%%%%%%%%%%%%%%%%%%%%%%%%%%%%%%%%%%%%%%
\subsection{Total positivity of Chebyshev matrix}\label{PositionProof}
%%%%%%%%%%%%%%%%%%%%%%%%%%%%%%%%%%%%%%%
Let us consider a general strategy in proving the positivity of the determinant of matrices constructed from specific functions $V_{\ell}(y)$. In particular, the columns of the matrix is given by evaluating the function at $n$ distinct  ordered points $y_1<y_2<\cdots<y_n$, i.e. $\mathbf{V}_{\ell}=(V_{\ell}(y_1),\,V_{\ell}(y_2),\cdots,V_\ell(y_n))$. Our task is to prove that for a collection of $n$ such vectors, 
\eq\label{Target}
{\rm Det}(\mathbf{V}_{\ell_1},\mathbf{V}_{\ell_2},\cdots,\mathbf{V}_{\ell_n})= {\rm Det} \left(\begin{array}{cccc} V_{\ell_1}(y_1) & V_{\ell_2}(y_1) & \cdots & V_{\ell_n}(y_1) \\ V_{\ell_1}(y_2) & V_{\ell_2}(y_2) & \cdots & V_{\ell_n}(y_2) \\ \vdots & \vdots & \cdots & \vdots \\V_{\ell_1}(y_n) & V_{\ell_2}(y_n) & \cdots & V_{\ell_n}(y_n)\end{array}\right)>0\,.
\eqe
The general strategy, as also discussed in~\cite{CFTHedra}, is to show that the above can never be zero for any choice of distinct $y_i$s. In other words, the sign of the determinant is fixed. Then the vanishing of the determinant implies that the column vectors are now linearly dependent, or 
\eq\label{Target2}
\sum_{i=1}^n c_i V_{\ell_i}(y_j)=0\;\,.
\eqe
for $j=1,2\cdots,n$. Said in another way, the function $\sum_{i=1}^n c_i V_{\ell_i}(y)$ have $n$ roots on the real axes. Thus proving the definite sign of eq.(\ref{Target}) amounts to proving  that eq.(\ref{Target2}) cannot have $n$ real solutions.

Before considering Chebyshev polynomials, let's first begin with $V_{\ell}(y)=e^{\ell y}$.  Choose a sets of $n$ $\ell_i$s conveniently labelled with $\ell_1<\ell_2<\cdots< \ell_n$, the goal is to show that 
\eq
f_n(y)=\sum_{i=1}^n c_i e^{\ell_i y}
\eqe
cannot have $n$ real roots for any $c_i$. We will prove this by induction. First for n=1, indeed $f_1(y)=e^{\ell_1 y}$ does not have a root. Next, lets assume that there are at most $n{-}2$ roots for $f_{n{-}1}(y)$, but $f_{n}(y)$ has $n$ roots. We will show that this leads to a contradiction. If $f_{n}(y)$ has $n$ roots, then multiplied by $e^{-\ell_1 y}$ will not change that. That is,  
\eq
e^{-\ell_1 y}f_{n}(y)= c_1+c_2 e^{(\ell_2-\ell_1) y}+\cdots, +c_n e^{(\ell_n-\ell_1) y}
\eqe 
will also have $n$ roots. Now the derivative of a function with $n$ roots on the real axes must have at least $n{-}1$ real roots. Taking the derivative we find, 
\eq
\left(e^{-\ell_1 y}f_n(y)\right)'=c_2 (\ell_2-\ell_1) e^{(\ell_2-\ell_1) y}+\cdots, +c_{n}(\ell_{n}-\ell_1)e^{(\ell_{n}-\ell_1) y}\,.
\eqe
But this is nothing but $f_{n{-}1}$ with another set of ordered $\ell_i$, which now has $n{-}1$ real roots, a contradiction to our initial assumption! Thus we conclude that $f_n(y)$ cannot have $n$-roots and the determinant in eq.(\ref{Target}) can never be zero. Note that if one replaces $y=\log x$, then the functions we are considering are simply moments $x^\ell$. As we assume that $y$ is real,  we have $x>0$ and thus the positivity of eq.(\ref{Target}) also leads to the total positivity of the Vandermonde matrix for half moment curves. 

We are interested in the Chebyshev polynomials $\cos \ell y$. Since we will be interested in cases where $\cos y>1$,  $y$ is purely imaginary and the Chebyshev polynomial becomes $\cosh \ell y$ with $y$ being real. Now we want to show that 
\eq
\sum_{i=1}^nc_i \cosh \ell_i y =0
\eqe 
cannot have $2n$ real roots (or $n$ positive roots since its a even function). But we've already shown that any linear combination of $2n$ distinct $e^{ \ell y}$ cannot have $2n$ roots, thus a contradiction! Thus this proves that 
\eq
{\rm Det} \left(\begin{array}{cccc} \cosh \ell_1 y_1 & \cosh \ell_2 y_1 & \cdots & \cosh \ell_n y_1 \\ \cosh \ell_1 y_2 & \cosh \ell_2 y_2 & \cdots & \cosh \ell_n y_2 \\ \vdots & \vdots & \cdots & \vdots \\ \cosh \ell_1 y_n & \cosh \ell_2 y_n & \cdots & \cosh \ell_n y_n\end{array}\right)\neq 0.
\eqe
i.e. it has a definite sign. Finally since all that we assumed for our Chebyshev matrix is that the spin is ordered, the minors of a given matrix obviously satisfies the same criteria, and hence we conclude that the  Chebyshev matrix is a totally positive matrix.

%%%%%%%%%%%%%%%%%%%%%%%%%%%%%%%
\subsection{Positivity of the Taylor scheme Gegenbauer matrix }
%%%%%%%%%%%%%%%%%%%%%%%%%%%%%%%
Here we analytically prove that the determinant of the Gegenbauer matrix in the derivative scheme. Starting with the Taylor coefficients defined in eq.(\ref{TaylorDef}), first we reorganize the analytic expression as:
\begin{equation}
v^{\textrm{\tiny D}}_{\ell,q}=\frac{1}{q!(\ell-q)!}\frac{(\Delta)_{\ell+q}}{\prod_{a=1}^q(\Delta+2a-1)}=\frac{(\Delta)_{ \ell}}{(q!)(\ell !)}\frac{1}{\prod_{a=1}^q(\Delta+2a-1)}\left[(\ell)_{- q}(\ell+\Delta)_{ q}\right]\,,
\end{equation}
where $\Delta=D{-}3$, $(a)_{-q}=a(a-1)\cdots(a-q+1)$ and $(a)_0=1$. Now consider the determinant of  $n{+}1$ Taylor vectors. Due to our rearrangement, the determinant can be written in a factorized form:
\eqa
{\rm Det}\left[\begin{array}{ccc} v^{\textrm{\tiny D}}_{\ell_1,0} & v^{\textrm{\tiny D}}_{\ell_2,0} & \cdots  \\   v^{\textrm{\tiny D}}_{\ell_1,1} & v^{\textrm{\tiny D}}_{\ell_2,1} & \cdots \\ \vdots & \vdots & \cdots \end{array}\right]=\left(\prod^{n{+}1}_{i=1}\frac{(\Delta)_{ \ell_i}}{\ell_i!}\frac{1}{\prod_{a=1}^{i-1}(\Delta+2a-1)a!}\right)\nonumber\\ 
\times {\rm Det}
\begin{pmatrix}
(\ell_1)_{0}(\ell_1+\Delta)_{0}&(\ell_1)_{-1}(\ell_1+\Delta)_{1} &...\\
(\ell_2)_{0}(\ell_2+\Delta)_{ 0}&(\ell_2)_{-1}(\ell_2+\Delta)_{1} &...\\
... & ... & ...
\end{pmatrix}.
\eqae
Now we know that the remaining determinant must have the factor $\prod_{i<j}(\ell_j-\ell_i)$ since the result vanishes if $\ell_i=\ell_j$. Furthermore, using 
\begin{equation}
(-a)_{ b}=(-a)(-a+1)...(-a+b-1)=(-1)^b(a)_{-b}\,,
\end{equation}
we can see that the remaining determinant is invariant under $\ell\rightarrow -\ell-\Delta$. This together with power counting leads to 
\eq
{\rm Det}
\begin{pmatrix}
(\ell_1)_{0}(\ell_1+\Delta)_{0}&(\ell_1)_{-1}(\ell_1+\Delta)_{1} &...\\
(\ell_2)_{0}(\ell_2+\Delta)_{0}&(\ell_2)_{-1}(\ell_2+\Delta)_{1} &...\\
... & ... & ...
\end{pmatrix}=\prod_{i<j}(\ell_j-\ell_i)(\Delta+\ell_j+\ell_i).
\eqe
Thus we find that 
\eq
(\prod_{i}v^{\textrm{\tiny D}}_{\ell_i,\sigma_i})\epsilon^{\sigma_1\sigma_2\cdots}=\left(\prod^{n{+}1}_{i=1}\frac{(\Delta)_{ \ell_i}}{\ell_i!}\frac{1}{\prod_{a=1}^{i-1}(\Delta+2a-1)a!}\right)\prod_{i<j}(\ell_j-\ell_i)(\Delta+\ell_j+\ell_i).
\eqe
As one can see, the result is positive so long as $\ell_1<\ell_2<\cdots<\ell_{n{+}1}$!

%%%%%%%%%%%%%%%%%%%%%%%%%%%%%%%
\section{The true boundary of the $\mathbb{P}^1$ EFT-hedron}\label{EFTP1}
%%%%%%%%%%%%%%%%%%%%%%%%%%%%%%%
The EFT-hedron constraint relies on two aspects, the wall $\vec{\mathcal{W}}_I$ and the resulting deformation parameters $\{\alpha_i\}$. Let us consider dotting $\vec{a}$ in to some wall $\mathcal{W}=\left(-w, 1\right)$, then the RHS of eq.(\ref{1Dsu}) then tells us that:
\eq
\left(\begin{array}{c}\vec{a}_2\cdot\mathcal{W}\\\vec{a}_4\cdot\mathcal{W}\\\vec{a}_6\cdot\mathcal{W}\end{array}\right)=\left(\begin{array}{c}a_2(\beta_2-w) \\ a_4(\beta_4-w) \\ a_6(\beta_6-w)\end{array}\right)=\sum_{a}\textsf{p}_{a} \left(\begin{array}{c} (u^{(2)}_{\ell_a}-w) \\ (u^{(4)}_{\ell_a}-w)y_a \\ (u^{(6)}_{\ell_a}-w)y^2_a\end{array}\right)
\eqe 
where we absorbed factors of $x_a$ into $\textsf{p}_{a}$, and $y_a=x_a^2$. We see that the inner product lives in the hull of multiple deformed curves. To ensure that the hull is non-trivial, we would like to ensure all entires of the deformed moment curves to be non-negative. In other words we want $\mathcal{W}$ to satisfy
\eq\label{MinkSum}
 (u^{(k)}_{\ell}-w) >0,\quad \forall \ell.
\eqe
As the minimum of $u^{(k)}_{\ell}$ listed in eq.(\ref{Umin}) is $-\frac{21}{4}$, we write $w=-\frac{21}{4}-\Delta w$ with $\Delta w\geq0$. Since we have a collection of deformed curves, the constraint for $\vec{a}_k\cdot\mathcal{W}$ should be derived from a curve that encapsulate all the other curves. i.e. the master moment curve. In other words, we want to find $(1,  x, \alpha x^2)$ such that its convex hull contains all the individual moment curves, or,   
\eq
\frac{(u^{(2)}_{\ell}-w)(u^{(6)}_{\ell}-w)}{\alpha}-(u^{(4)}_{\ell}-w)^2\geq0,\quad\forall \ell 
\eqe
This tells us that there is an upper bound for $\alpha$, corresponding to the the minimum of $\frac{(u^{(6)}_{\ell}-w)(u^{(2)}_{\ell}-w)}{(u^{(4)}_{\ell}-w)^2}$, which we denote as $\alpha_{min}[\Delta w]$, reflecting the fact that it is a function of $\Delta w$. Explicitly plotting $\alpha_{min}[\Delta w]$ we find:
$$\includegraphics[scale=0.35]{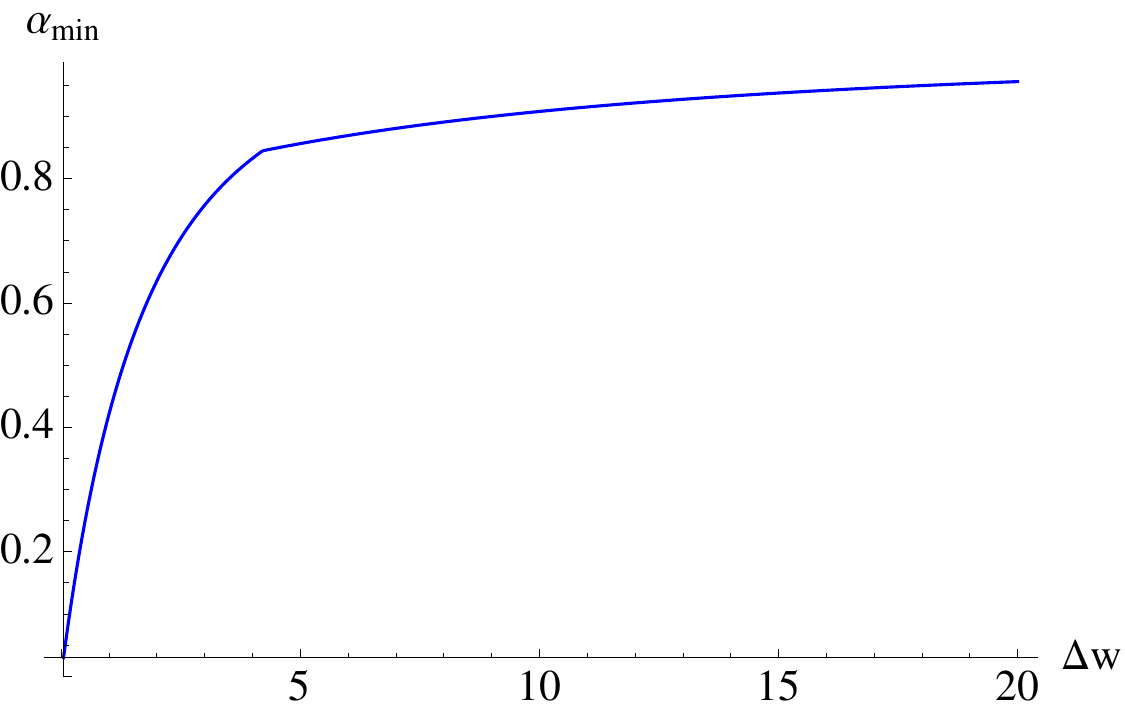}$$
We see that $\alpha_{min}$ rises approximately linear with $\Delta w$ up to around $\Delta w\sim 5$, after which $\alpha_{min}\sim1$ for all $\ell$.

Equipped with $\alpha_{min}[\Delta w]$ we can now write down the non-linear constraint for $\vec{a}_k\cdot\mathcal{W}$:
\eq\label{Wbound}
(\vec{a}_2\cdot\mathcal{W})(\vec{a}_6\cdot\mathcal{W})- \alpha_{min}[\Delta w](\vec{a}_4\cdot\mathcal{W})^2>0
\eqe
It is important to see if above gives constraints that go beyond those in eq.(\ref{Bounds}). To this end we write $\beta_2=-\frac{3}{4}+\hat{\beta}_2$, $\beta_4=-\frac{3}{2}+\hat{\beta}_4$ and $\beta_6=-\frac{21}{4}+\hat{\beta}_6$, so that the original polytope bound is simply that  $\hat{\beta}_i\geq0$. In terms of these new parameters, eq.(\ref{Wbound}) becomes, 
\eq\label{Nonlinear}
\left(\hat{\beta}_6+\Delta w\right)\left(\hat{\beta}_2+\frac{9}{2}+\Delta w\right)-\alpha_{min}[\Delta w] \frac{a^2_4}{a_2 a_6}\left(\hat{\beta}_4+\frac{15}{4}+\Delta w\right)^2\geq0\;.
\eqe
If the above leads to any constraint for $\hat{\beta}_i$ beyond that it is non-negative, or $\epsilon\equiv\frac{a^2_4}{a_2 a_6}<1$ then we have found new constraints beyond eq.(\ref{Bounds}). For example, for $\Delta w=0$, $\alpha_{min}[0]=0$ and eq.(\ref{Nonlinear}) does not implement anything new.

However, for non-zero $\alpha_{min}[\Delta w]$ we will always obtain new constraints! For example, since $\hat{\beta}_4\geq0$, eq.(\ref{Nonlinear}) implies 
\eq\label{Nonlinear3}
\framebox[10cm][c]{$\displaystyle\frac{\left(\hat{\beta}_6+\Delta w\right)\left(\hat{\beta}_2+\frac{9}{2}+\Delta w\right)}{\alpha_{min}[\Delta w]\epsilon}\geq (\frac{15}{4}+\Delta w)^2$}~\,,
\eqe
and we see that $(\hat{\beta}_2, \hat{\beta}_6)$ is bounded from below. Let's set $\hat{\beta}_2, \hat{\beta}_6=0$, and consider
\eq\label{Cep}
j(\Delta w)=\frac{\Delta w\left(\frac{9}{2}+\Delta w\right)}{\alpha_{min}[\Delta w]\epsilon}- \left(\frac{15}{4}+\Delta w\right)^2\;.
\eqe
We have non-trivial lower bounds for $(\hat{\beta}_2, \hat{\beta}_6)$ if $j(\Delta w)<0$. Plotting $j(\Delta w)$ for fixed $\epsilon$ with respect to $\Delta w$ we find 
$$\epsilon=0.4:\;\vcenter{\hbox{\includegraphics[scale=0.3]{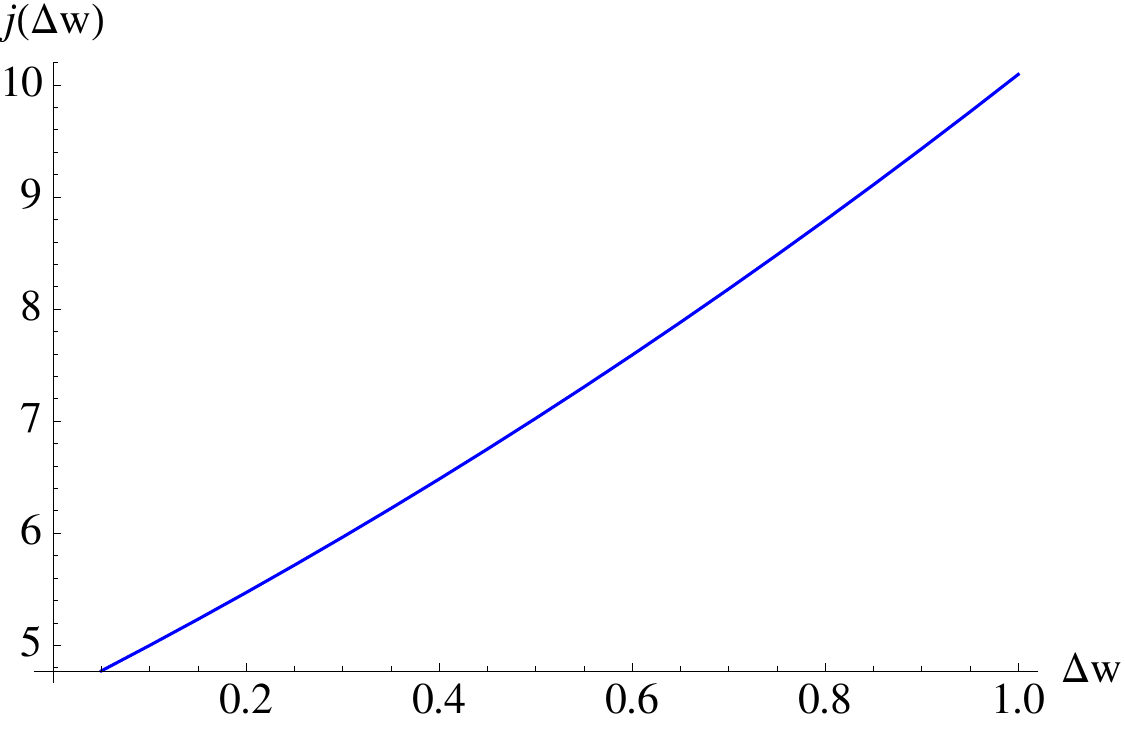}}}\quad\quad\epsilon=0.55:\;\vcenter{\hbox{\includegraphics[scale=0.3]{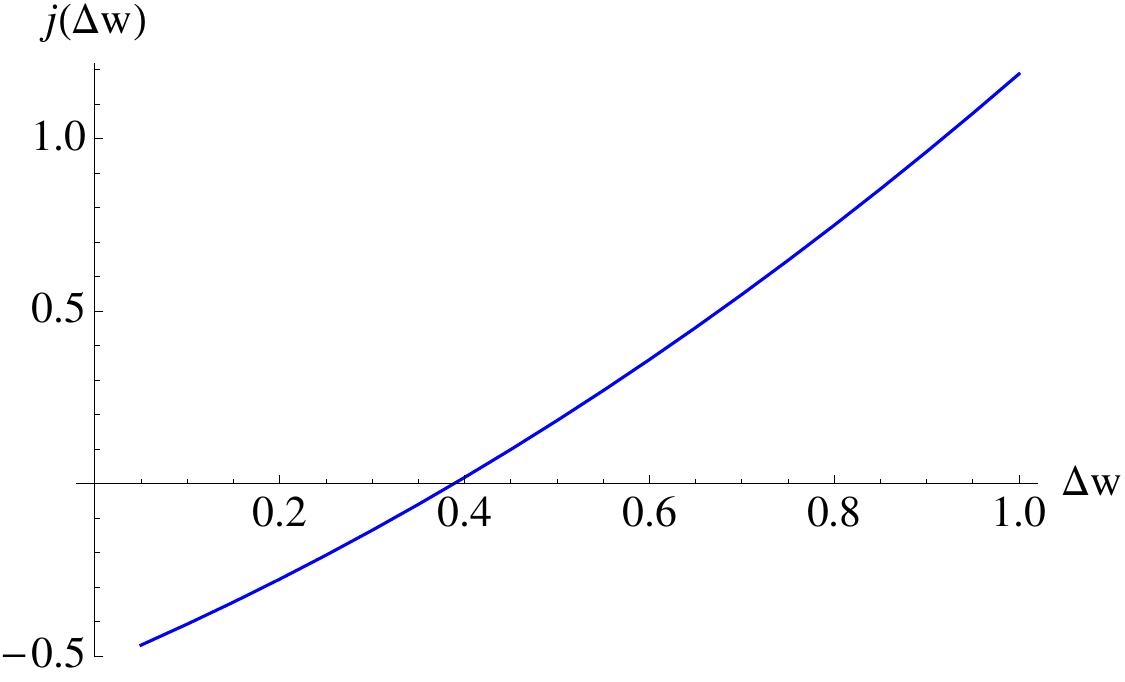}}}$$
$$\epsilon=0.85:\;\vcenter{\hbox{\includegraphics[scale=0.3]{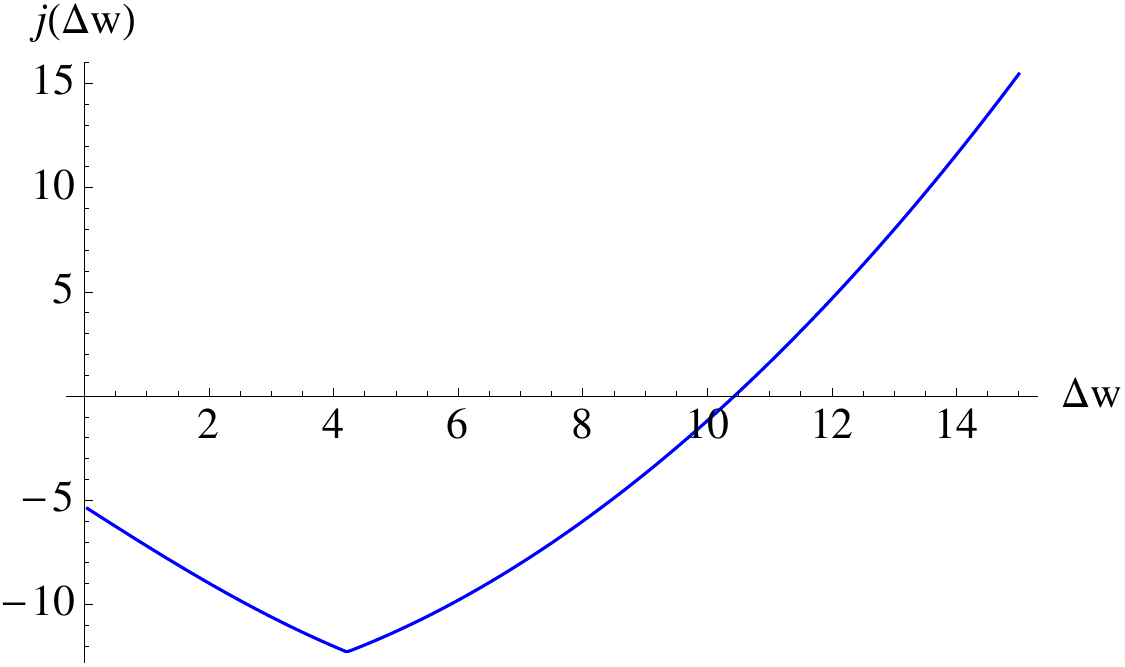}}}$$
We see that if $\epsilon$ is above a critical value $\epsilon_c=0.54$, there are ranges of $\Delta w$ where the constraint is non-trivial. Thus we either have a non-trivial lower bound for $(\hat{\beta}_2,\hat{\beta}_6)$, or that we have an upper bound for $\epsilon<\epsilon_c$. Note that these non-trivial bounds are derived from walls that are not the walls of the original polytopes.

From eq.(\ref{Nonlinear}) one can also derive an upper bound for $\hat{\beta}_4$:
\eq\label{Nonlinear2}
\framebox[10cm][c]{$\displaystyle\sqrt{\frac{\left(\hat{\beta}_6+\Delta w\right)\left(\hat{\beta}_2+\frac{9}{2}+\Delta w\right)}{\alpha_{min}[\Delta w]\epsilon}}-\frac{15}{4}-\Delta w\geq  \hat{\beta}_4$}~\,
.\eqe
Obviously, the bound is most stringent when $\hat{\beta}_6=\hat{\beta}_2=0$. Thus we consider 
\eq
j_{\beta_4}(\Delta w)=\sqrt{\frac{\Delta w\left(\frac{9}{2}+\Delta w\right)}{\alpha_{min}[\Delta w]\epsilon}}-\frac{15}{4}-\Delta w\,.
\eqe 
We plot the above function with respect to $\Delta w$ and look for the upper bound for $\hat{\beta}_4$ as the minimum of $j_{\beta_4}(\Delta w)$. The result depends on $\epsilon$:
$$\epsilon=0.4:\;\vcenter{\hbox{\includegraphics[scale=0.3]{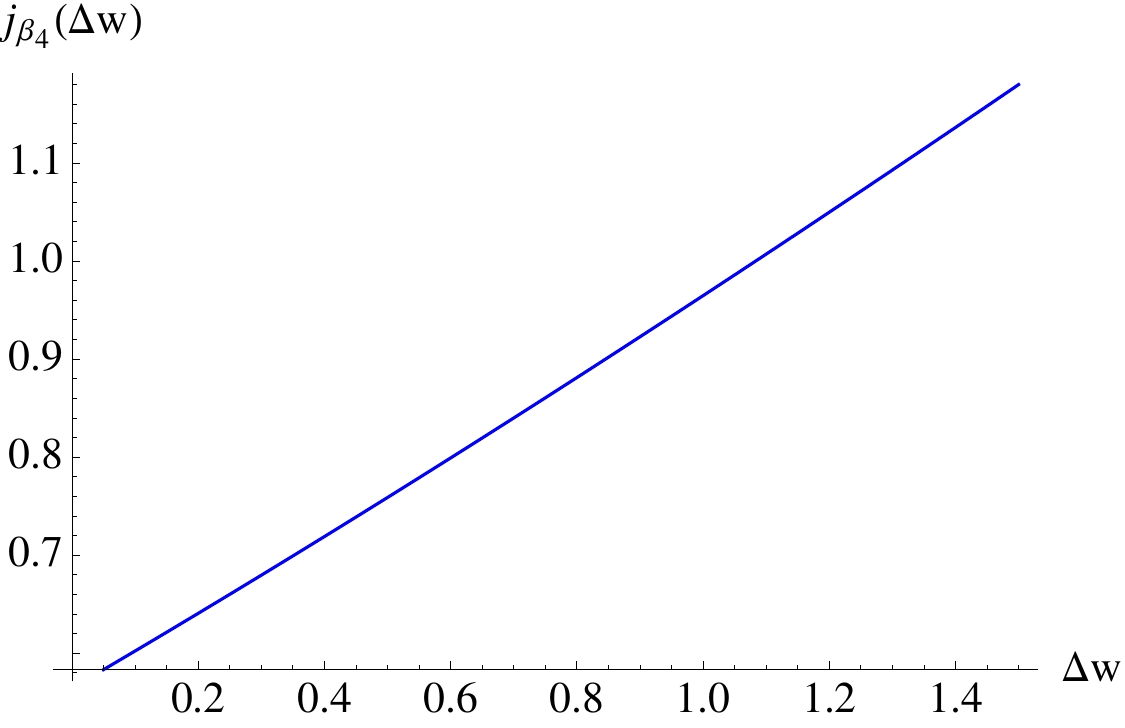}}},\;\epsilon=0.5:\;\vcenter{\hbox{\includegraphics[scale=0.3]{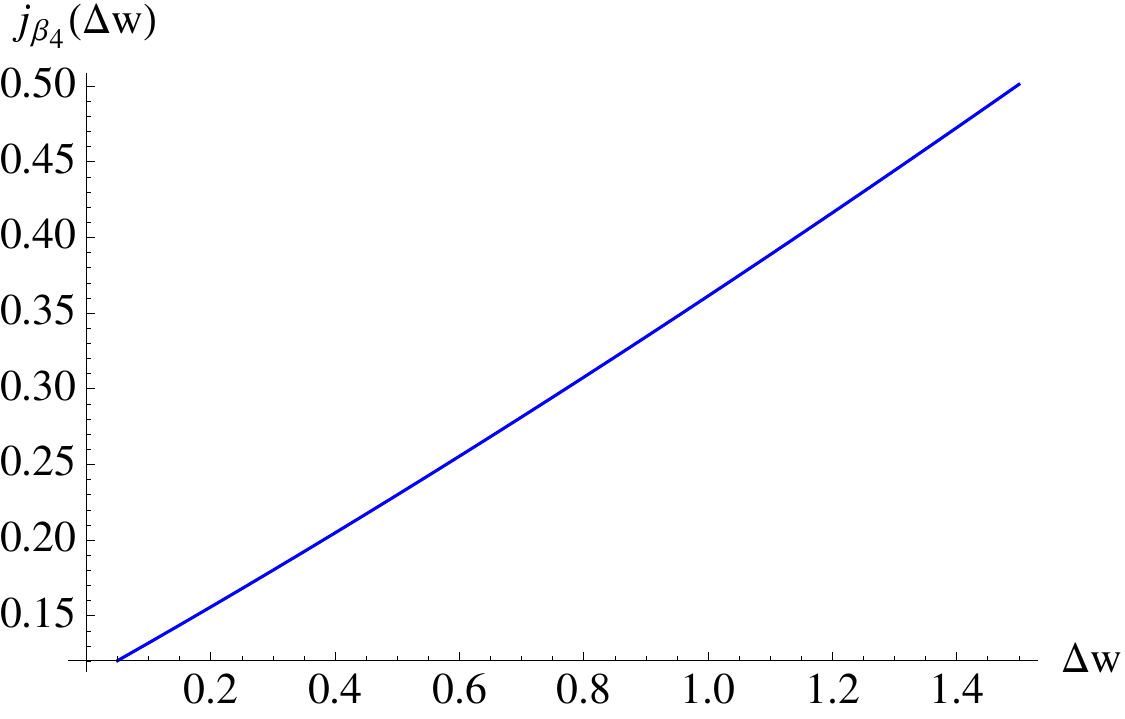}}}$$
$$\epsilon=0.6:\;\vcenter{\hbox{\includegraphics[scale=0.3]{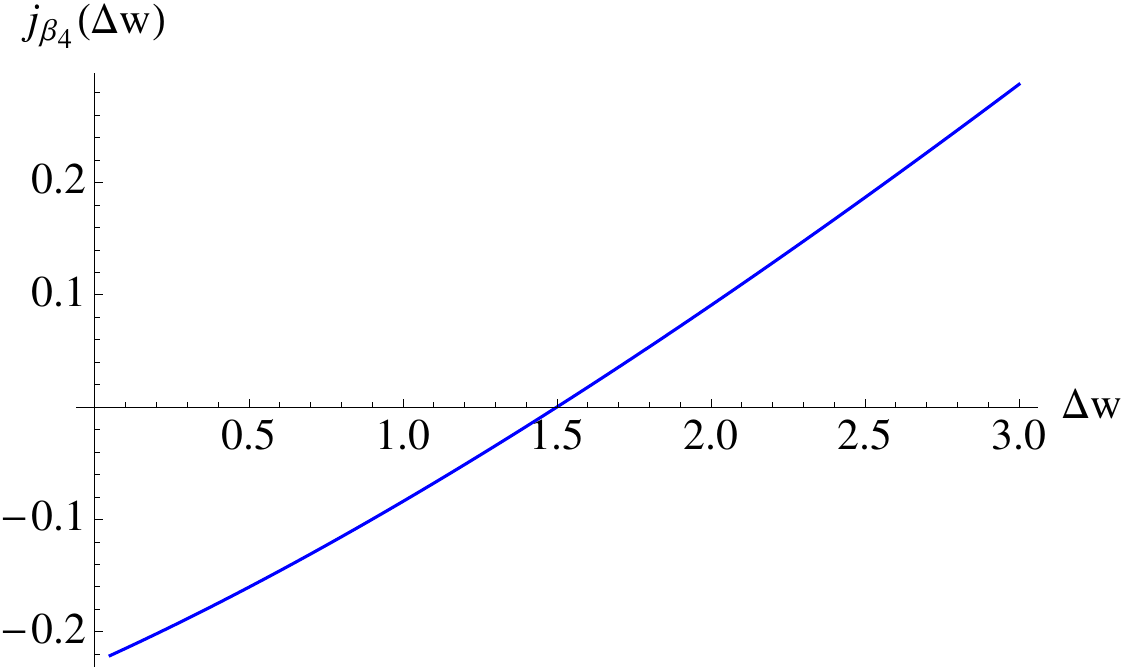}}}$$
For the first two graphs we consider $\epsilon<\epsilon_c$,  where no lower bounds on $(\hat{\beta}_6,\hat{\beta}_2)$ were imposed from eq.(\ref{Nonlinear3}), we see that there is always an upper bound for $\hat{\beta}_4$. For $\epsilon>\epsilon_c$, we have a region of walls, $\Delta w<15$, where there's no new bounds on $\hat{\beta}_4$, however for these cases, there are lower bounds on $\hat{\beta}_6,\hat{\beta}_2$. 

In summary, we find that using walls that are ``outside" the walls of $Conv[\vec{u}_{\ell,k}]$, imposes further constraint through eq.(\ref{Nonlinear}) either as a upper bound on $\hat{\beta}_4$, or lower bound on $(\hat{\beta}_2, \hat{\beta}_6)$, depending on whether $\epsilon$ is above or below $\epsilon_c$. Thus eq.(\ref{Nonlinear3}) and eq.(\ref{Nonlinear2}) characterizes the $\mathbb{P}^1$ EFT-hedron.
%%%%%%%%%%%%%%%%%%%%%%%%%%%%%%%%%%%%%%%%%%%%%%%%%%%%%%%%%%
\section{Beta function for eq.(\ref{ExampleS})}\label{AppBeta}
%%%%%%%%%%%%%%%%%%%%%%%%%%%%%%%%%%%%%%%%%%%%%%%%%%%%%%%%%%
Here we present the details for the computation of the beta functions from two-particle cuts in eq.(\ref{ExampleS}).

\begin{figure}
\begin{center}
\includegraphics[scale=0.55]{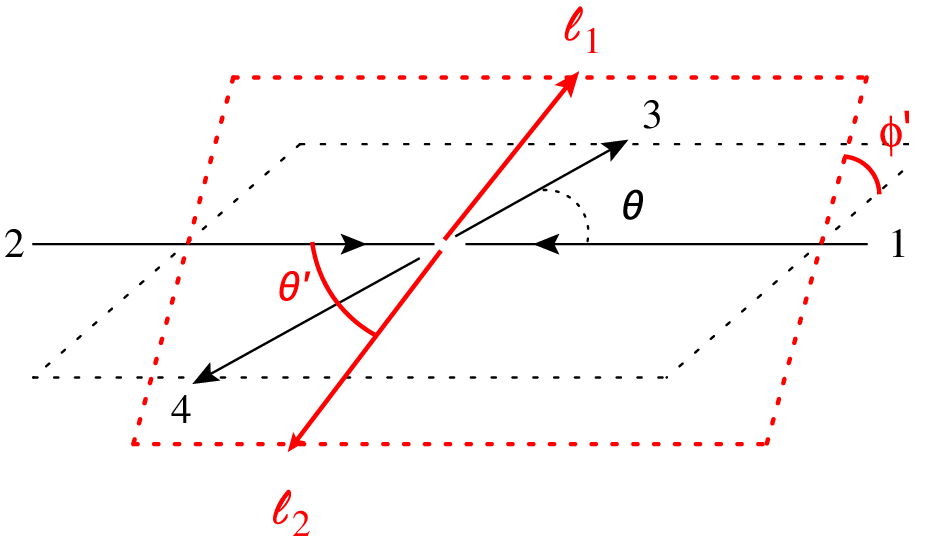}
\caption{We represent the internal loop momentum in the center of mass frame. The angle between the loop momentum and $\vec{p}_1=-\vec{p}_2$ is $\theta'$, while the angle between the plane spanned by ($\vec{p}_1,\vec{\ell}_1$) and the plane ($\vec{p}_1,\vec{p}_2$) is $\phi'$. $\theta$ is then the usual scattering angle.}
\label{phase1}
\end{center}
\end{figure}

We compute the two-particle cut by taking the product of the two tree-amplitudes parameterized in the center of mass frame as illustrated in fig.(\ref{phase1}) : $(\theta',\phi')$ is the angular dependence of the phase space for the cut propagators, and $\theta$ is the scattering angle for the external momenta. For example, for the $\bar{a}^2_2$ coupling the bubble coefficient is given by: 
\eqa
\vcenter{\hbox{\includegraphics[scale=0.55]{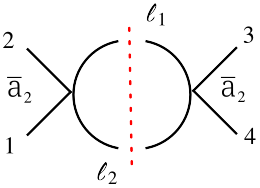}}}\;=\;{-}\frac{\bar{a}^2_2}{\Lambda^8(4\pi)}\int  d\phi'd\cos\theta' \;\;(s^2+u^2+t^2)_L (s^2+u^2+t^2)_R\nonumber\\
={-}\frac{\bar{a}^2_2}{\Lambda^8}s^4\int  d\phi'd\cos\theta' \;\;F_{2,L}F_{2,R}
=\frac{\bar{a}^2_2}{\Lambda^8}\frac{s^4}{60} (167+\cos 2 \theta)
\eqae
where we've defined the short hand notation:
\eqa
F_{n,L}&=&\left(1{+}\left(\frac{1{+}\cos\theta'}{2}\right)^n{+}\left(\frac{1{-}\cos\theta'}{2}\right)^n\right)\nonumber\\
F_{n,R}&=&\left(1{+}\left(\frac{1{+}\cos\theta'\cos\theta{+}\sin\theta'\cos\phi'\sin\theta}{2}\right)^n{+}\left(\frac{1{-}\cos\theta'\cos\theta{-}\sin\theta'\cos\phi'\sin\theta}{2}\right)^n\right)\,.\nonumber\\
\eqae
Changing back to Mandelstam variables we find the coefficient for the $\frac{\bar{a}^2_2}{15}(41s^2 + t^2 + u^2)s^2$ for the $s$-channel coefficient. Summing over the three channels we obtain ${-}\frac{14\bar{a}^2_2}{5\Lambda^8(4\pi)^2}(s^4{+}t^4{+}u^4)\log\frac{p^2}{\mu^2}$ and hence $\beta_1=\frac{14}{5(4\pi)^2}$. Similarly for $s^6$ we have:
\eqa
\vcenter{\hbox{\includegraphics[scale=0.5]{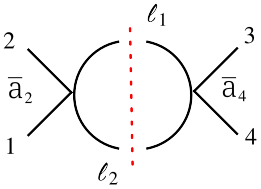}}}\;=\;{-}\frac{\bar{a}_2\bar{a}_4s^6}{\Lambda^{12}(4\pi)}\int  d\phi'd\cos\theta' \;\;F_{2,L}F_{4,R}+F_{4,L}F_{2,R}\nonumber\\
=\frac{2 \bar{a}_2\bar{a}_4}{\Lambda^{12}35}s^6 (82+ \cos2\theta)= \frac{2\bar{a}_2\bar{a}_4}{35\Lambda^{12}}  s^4 (83 s^2 + 8 s t + 8 t^2)\,.
\eqae
Again summing over all three channels we obtain ${-}\frac{2\bar{a}_2\bar{a}_4}{35\Lambda^{12}(4\pi)^2}\left(83(s^6{+}t^6{+}u^6){-}24(s t u)^2\right)\log\frac{p^2}{\mu^2}$, and $\beta_2=\frac{166}{35(4\pi)^2}$.

\end{document}